\documentclass[
11pt, % The default document font size, options: 10pt, 11pt, 12pt
english,%ngerman,
singlespacing, % Single line spacing, alternatives: onehalfspacing or doublespacing
headsepline, % Uncomment to get a line under the header
]{book} % The class file specifying the document structure

\usepackage[numbers,sort&compress]{natbib}
\input{packages/packages.tex}

\definecolor{VertexColor}{RGB}{163,53,101}
\definecolor{vertexcolor}{RGB}{163,53,101}

\definecolor{SEcolor}{RGB}{255,110,182}
\definecolor{SEColor}{RGB}{255,110,182}

\definecolor{GluonColor}{RGB}{0,163,180}
\definecolor{gluoncolor}{RGB}{0,163,180}

\definecolor{phisix}{RGB}{255,161,122}

\tikzset{
corner/.style={line width=1pt,dashed,draw=black,dash pattern=on 6pt off 4pt},
scalar/.style={line width=1pt,draw=black},
gluon/.style={line width=1pt,decorate, draw=GluonColor,
    decoration={complete sines,aspect=0,amplitude=1.25mm,segment length=1.5mm,start up,end up}},
gluontwo/.style={line width=1pt,decorate, draw=GluonColor,
    decoration={complete sines,aspect=0,amplitude=.7mm,segment length=1mm,start up,end up}},
ghost/.style={line width=1pt,loosely dotted,draw=black},
wilson/.style={line width=2pt,draw=black},
 }
\NewDocumentCommand\semiloop{O{black}mmmO{}O{above}}
{%
\draw[#1] let \p1 = ($(#3)-(#2)$) in (#3) arc (#4:({#4+180}):({0.5*veclen(\x1,\y1)})node[midway, #6] {#5};)
}

\newcommand{\SO}[1]{\ensuremath{\mathrm{SO}(#1)}}
\newcommand{\SU}[1]{\ensuremath{\mathrm{SU}(#1)}}
\newcommand{\SL}[1]{\ensuremath{\mathrm{SL}(#1)}}

\newcommand{\U}[1]{\ensuremath{\mathrm{U}(#1)}}

\newcommand{\e}{\mathrm{e}}
\newcommand{\I}{\mathrm{i}}

\newcommand{\com}[2]{\lbrack #1, #2\rbrack}

\newcommand{\dd}{\mathrm{d}}

\newcommand{\sumint}[1]{\mathrlap{\displaystyle\int_{#1}}\mathrlap{\textstyle\sum}}

\newcommand{\RNum}[1]{\uppercase\expandafter{\romannumeral #1\relax}}

\newcommand{\Log}[1]{\text{log}\left( #1 \right)}

\newcommand{\tr}[1]{\mathrm{tr}\left[ #1 \right]}
\newcommand{\corr}[1]{\left< #1 \right>}
\newcommand{\x}{\ensuremath{\times}}

\RequirePackage{geometry}
\input{./figures/images.tex}

\begin{document}

\pagestyle{plain} % Default to the plain heading style until the thesis style is called for the body content

\vspace{1cm}

\vspace*{.005\textheight}

\begin{center}
\noindent\textcolor{Black}{\rule{14.5cm}{.5mm}}
%\hfill
\vspace{0.4cm}
\begin{spacing}{1.2}
	\LARGE\bfseries  \textcolor{Black}{Integrable systems: From the ice rule to supersymmetric fishnet Feynman diagrams} 
\end{spacing}
\noindent\textcolor{Black}{\rule{14.5cm}{.5mm}}

\vspace{1cm}
 
\large \textit{Dissertation  zur Erlangung des akademischen Grades} \par
\vspace{0.5cm}
\textsc{doctor rerum naturalium (Dr. rer. nat.)} \par
\vspace{0.7cm}
\textit{im Fach:} \textsc{Physik}  \par
\textit{Spezialisierung:} \textsc{Theoretische Physik}  \par
\vspace{0.2cm}
\textit{eingereicht am} \par
10.\ M\"{a}rz 2025 \par
\vspace{0.2cm}
\textit{an der} \par
\vspace{0.7cm}
\large Mathematisch-Naturwissenschaftlichen Fakultät  \par
\large der Humboldt-Universität zu Berlin  \par
\vspace{1cm}
\textit{von} \par
\vspace{0.5cm}
\textcolor{PineGreen}\Large\textsc{\textcolor{Black}{Moritz Kade}}\par
\vspace{0.5cm}

\end{center}
\vspace{2cm}

Präsidentin der Humboldt-Universität zu Berlin: \par
\vspace{0.1cm}
Prof. Dr.  Julia von Blumenthal

\vspace{0.5cm}

Dekan der Mathematisch-Naturwissenschaftlichen Fakultät: \par
\vspace{0.1cm}
Prof. Dr. Emil List-Kratochvil

\vspace{0.5cm}

Gutachter: \par
\vspace{0.1cm}
1. Prof. Dr.  Matthias Staudacher \par
2. Prof. Dr.  Zoltan Bajnok \par
3. Prof. Dr.  Agostino Patella \par

\vspace{1cm}
Tag der mündlichen Prüfung: \, 9.\ Mai 2025

\thispagestyle{empty}

\vfill

% TITLE GRAPHICS OPTION 2
\begin{comment}
\raisebox{-.5\height}{\includegraphics[width=15cm]{Line}}
\hfill
\begin{spacing}{1.2}
	\huge\bfseries Conformal Correlators in the AdS/CFT correspondence???? \par
\end{spacing}
\raisebox{-.5\height}{\includegraphics[width=15cm]{Line}}

\vspace{1cm}
 
\textsc{Dissertation}  zur Erlangung des akademischen Grades \par
\vspace{0.5cm}
\textsc{\textcolor{NavyBlue}{doctor rerum naturalium (Dr. rer. nat.)} }\par
\vspace{0.7cm}
\textsc{im Fach: Physik}  \par
\textsc{Spezialisierung: Theoretische Physik}  \par
\vspace{0.2cm}
\text{eingereicht an der} \par
\vspace{0.7cm}
\textsc{\textcolor{NavyBlue}{Mathematisch-Naturwissenschaftlichen Fakultät}}  \par
\textsc{\textcolor{NavyBlue}{der Humboldt-Universität zu Berlin}}  \par
\vspace{0.5cm}
\text{von} \par
\vspace{0.5cm}
\textsc{\textcolor{NavyBlue}{Giulia Peveri}}
\end{comment}
%%%%%%%%%%%%%%%%%%%%%%%%%%%%

%----------------------------------------------------------------------------------------
%L

\newpage
\thispagestyle{empty}
\mbox{}

%----------------------------------------------------------------------------------------
%	DEDICATION - R
%----------------------------------------------------------------------------------------

\frontmatter % Use roman page numbering style (i, ii, iii, iv...) for the pre-content pages

\newpage
\vspace*{.1\textheight}

\begin{flushright}
\textit{In Erinnerung an meinen Vater Thomas.\\ 
Du hast mir gezeigt wie spannend es ist die Natur zu verstehen,\\ diese Arbeit ist Dir gewidmet.}
\end{flushright}

%\thispagestyle{empty}

%----------------------------------------------------------------------------------------
%L
\newpage
\mbox{}

%----------------------------------------------------------------------------------------
%	DECLARATION PAGE - R
%----------------------------------------------------------------------------------------

\newpage

\begin{center}
\huge \textbf{Eidesstattliche Erkl\"arung} \par
\end{center}
\normalsize \noindent Ich erkläre, dass ich die Dissertation selbständig und nur unter Verwendung der von mir gemäß $\S 7$ Abs. 3 der Promotionsordnung der Mathematisch-Naturwissenschaftlichen Fakultät, veröffentlicht im Amtlichen Mitteilungsblatt der Humboldt-Universität zu Berlin Nr. 42/2018 am 11.07.2018 angegebenen Hilfsmittel angefertigt habe.
\vspace{1cm}

\noindent Ort,\,Datum: Berlin, 7. M\"{a}rz 2025\\
%\rule[0.5em]{25em}{0.5pt}
 
\noindent Unterschrift: \textsc{Moritz Kade}\\
%\rule[0.5em]{25em}{0.5pt} 
%\end{declaration}

%\thispagestyle{empty}

%----------------------------------------------------------------------------------------
%L
\newpage
\mbox{}

%----------------------------------------------------------------------------------------
%	ABSTRACT PAGE - R
%----------------------------------------------------------------------------------------

\newpage

\begin{center}
\huge \textbf{Abstract} \par
\end{center}
\normalsize \noindent
This thesis examines the correspondence between models of statistical physics and Feynman graphs of quantum field theories by a common property: integrability.
We review integrable structures for periodic boundary conditions on both sides, while focusing on the eight- and six-vertex model and the bi-scalar fishnet theory.
The latter is a double-scaled $\gamma$-deformations of $\mathcal{N} = 4$ super Yang-Mills theory.
Interesting applications of integrability existing in the literature that we reconsider are the computation of the free energy in the thermodynamic limit and its quantum field theory (QFT) counterpart, the critical coupling.
In addition, we provide a detailed overview of the calculation of exact anomalous dimensions and operator product expansion (OPE) coefficients in the conformal bi-scalar fishnet theory.

The original contributions of this work comprise the results of the critical coupling for models with fermions, the brick wall theory, and the fermionic fishnet theory.
Additionally, we extend the study of integrable Feynman graphs to supersymmetric diagrams in superspace.
By establishing an efficient graphical formalism, we obtain the critical coupling of double-scaled $\beta$-deformations of $\mathcal{N} = 4$ super Yang-Mills theory and Aharony-Bergman-Jafferis-Maldacena theory, the super brick wall and superfishnet theory, respectively.
Moreover, we apply superspace methods to the superfishnet theory and find results for anomalous dimensions and an OPE coefficient, which are all-loop exact in the coupling.
In addition, we study boundary integrability in the six-vertex model and for Feynman diagrams.
We present new box-shaped boundary conditions for the six-vertex model and conjecture a closed form for its partition function at any lattice size.
On the QFT side, we find integrable boundary scattering matrices in the form of generalized Feynman diagrams by graphical methods.

%\thispagestyle{empty}

%----------------------------------------------------------------------------------------
%L
\newpage
\mbox{}
%---------------------------------------------------------------------

\newpage

\begin{center}
\huge \textbf{Zusammenfassung} \par
\end{center}
\normalsize \noindent
Diese Dissertation untersucht den Zusammenhang zwischen Modellen der statistischen Physik und Feynman-Diagrammen in Quantenfeldtheorien, anhand einer gemeinsamen Eigenschaft, der Integrabilit\"{a}t.
In beiden F\"{a}llen betrachten wir integrable Strukturen f\"{u}r periodische Randbedingungen und setzen dabei unseren Fokus auf das Acht- und Sechs-Vertex-Modell, sowie die bi-skalare Fischnetz-Theorie. 
Letztere ist eine gewisse deformierte Version der $\mathcal{N} = 4$ supersymmetrischen Yang-Mills-Theorie. 
Wir geben einen \"{U}berblick \"{u}ber eine bekannte Anwendung von Integrabilit\"{a}t in diesen Theorien, n\"{a}mlich die Berechnung der freien Energie im thermodynamischen Limes und ihr Gegenst\"{u}ck in Quantenfeldtheorien (QFT), die kritische Kopplungsst\"{a}rke.
Auch wiederholen wir die Berechnung anomaler Dimensionen und Koeffizienten der Operator-Produkt Entwicklung (OPE) in der bi-skalaren, konformen Fischnetz-Feldtheorie, dessen Ergebnis zu beliebiger Schleifenordnung bestimmt werden kann.

Die neuen Erkenntnisse dieser Arbeit umfassen die Ergebnisse zur kritischen Kopplungsst\"{a}rke für gewisse Modelle mit Fermionen.
Genauer handelt es sich dabei um die Brick-Wall-Theorie (Ziegel-Theorie) und die fermionische Fischnetz-Theorie. 
Dar\"{u}ber hinaus verallgemeinern wir die Studien integrabler Feynman-Diagramme f\"{u}r supersymmetrische Diagramme.
Dank der Entwicklung eines effizienten graphischen Formalismus sind wir in der Lage die kritische Kopplungsst\"{a}rke von starken Deformationen der $\mathcal{N} = 4$ Super-Yang-Mills-Theorie sowie der Aharony-Bergman-Jafferis-Maldacena-Theorie zu bestimmen – die sogenannte Super-Brick-Wall- und Superfishnet-Theorie (Superziegel- und Superfischnetz-Theorie).
Desweitern leiten wir n\"{u}tzliche Superraumtechniken her, die uns erlauben die anomale Skalendimension und einen OPE-Koeffizienten zu beliebiger Schleifenordung zu bestimmen.
Zudem untersuchen wir die Rand-Integrabilit\"{a}t im Sechs-Vertex-Modell und in Feynman-Diagrammen. 
Wir pr\"{a}sentieren neue, kastenf\"{o}rmige Randbedingungen f\"{u}r das Sechs-Vertex-Modell und postulieren eine geschlossene Form der Zustandssumme bei beliebiger Gittergr\"{o}\ss e. 
Auf der Seite der QFT finden wir integrable Streumatrizen f\"{u}r den Rand, die, dank graphischer Methoden, in Form verallgemeinerter Feynman-Diagramme ausgedr\"{u}ckt werden k\"{o}nnen.

%\thispagestyle{empty}

%----------------------------------------------------------------------------------------
%L
\newpage
\mbox{}

%----------------------------------------------------------------------------------------
%	STATEMENT OF ORIGINALITY - R
%----------------------------------------------------------------------------------------

\newpage

\begin{center}
\huge \textbf{Statement of Originality} \par
\vspace{0.3cm}
\end{center}
\normalsize \noindent This thesis is based on original research in collaboration with various researchers whose work we gratefully acknowledge. In the following, we list the papers by the author on which this work is based, some still under completion. 
\bibliographystyle{bib/JHEP}
\nobibliography*{}
\begin{itemize}
%\vspace{0.3cm}
\item[\cite{Kade:2023xet}] \bibentry{Kade:2023xet} 
\vspace{0.3cm}\\
I present the content of this paper in section \ref{sec:IntegrableVacuumDiagramsAndTheCriticalCoupling}.
It comprises the detailed derivation of inversion relations for Feynman diagrams, as well as their application on the brick wall theory and the fermionic fishnet theory.
The latter theory and the results for the critical couplings are a completely new result and were first proposed in the paper.
\vspace{0.3cm}
\item[\cite{Kade:2024ucz}] \bibentry{Kade:2024ucz}

An $\mathcal{N}=1$ superspace formulation of the double-scaled $\beta$-deformation of $\mathcal{N}=4$ SYM is presented in this article.
Furthermore, it was observed that the supersymmetric Feynman graphs obey a regular brick wall structure and that the inversion relations were generalized to supersymmetric vacuum graphs.
This yielded the novel result of the theory's critical coupling. 
In this dissertation, I present the supergraph analysis underlying the inversion relations in section \ref{sec:TheGeneralizedSuperspacePropagatorAsLatticeWeight}.
The super brick wall theory is derived in section \ref{sec:TheSuperBrickWallTheory_sec} and the application on supergraph inversion relations is contained in section \ref{sec:VacuumGraphsThermoLimit_susy}.
\vspace{0.3cm}
\item[\cite{Kade:2024lkc}] \bibentry{Kade:2024lkc} 

In this article, the superspace analysis was extended to the three-dimensional $\mathcal{N}=2$ superspace.
The supergraphs of double-scaled $\beta$-deformation of ABJM theory were examined.
They are of fishnet shape and allow for the application of the method of inversion relations.
The new results are the critical coupling, and the all-loop exact scaling dimensions of spinless zero- and two-magnon operators.
In this dissertation, I present these findings in sections \ref{sec:TheGeneralizedSuperspacePropagatorAsLatticeWeight}, \ref{sec:VacuumGraphsThermoLimit_susy} and \ref{sec:DoubleScaledBetaDeformationOfABJM}, respectively.
\vspace{0.3cm}
\item[\cite{ToAppAlessandro}] \bibentry{ToAppAlessandro} 

We introduce novel integrable boundary conditions for the six-vertex model.
These restrict the lattice to a box, with integrable boundaries all around.
We conjecture an expression for the exact partition function of any size of the lattice.
Our findings are contained in section \ref{sec:TheBoxBoundaryCondition} of this thesis.
\vspace{0.3cm}
\item[\cite{ToAppChangrim}] \bibentry{ToAppChangrim} 

In this article, we examine generalized scalar Feynman diagrams from the perspective of boundary integrability.
It yields new integrable K-matrices built from scalar propagators, which are used to build commuting double-row transfer matrices.
In this thesis, I present these results in chapter \ref{chpt:BoundaryIntegrabilityInFeynmanGraphs}.
\end{itemize}

\normalsize \noindent 
I have made crucial and indispensable contributions to all of the above publications.\\
This dissertation was typeset with the programs TexMaker, Overleaf and Writefull.

%\thispagestyle{empty}

%----------------------------------------------------------------------------------------
%L
\newpage
\mbox{}

%----------------------------------------------------------------------------------------
%	ACKNOWLEDGEMENTS - R
%----------------------------------------------------------------------------------------

\newpage

\begin{center}
\huge \textbf{Acknowledgments} \par
\end{center}
\small \noindent 
Firstly, I would like to thank my supervisor Matthias Staudacher for his guidance, vast support, and numerous discussions where he shared his limitless expertise with me.
I am very grateful that he sparked my affection for integrable models and that he introduced me to many outstanding researchers at conferences around the world.
In addition, his comments on the manuscript of this thesis were highly appreciated.
Thank you very much, Matthias, for our collaboration and the amazing time in your research group!

I would like to thank Changrim Ahn, who taught me countless insights on integrability.
His support and guidance contributed greatly to my research.
Thank you for our many discussions, while drawing blackboards full of boundary diagrams or having a delicious Korean dinner.
I am looking forward to visit you in Seoul!

I would like to thank my second supervisor, Emanuel Malek, for our interesting discussions about string theory and his support and advice on becoming a researcher.

I am thankful to Zoltan Bajnok and Agostino Patella for agreeing to referee my thesis, as well as for helpful discussions on conferences, retreats and in the hallway.
Furthermore, I would like to thank Valentina Forini and J\"{u}rgen Rabe for being part of the committee.

I would like to thank Patrick Vaudrevange, Hans Peter Nilles, Saul Ramos-Sanchez, and Alexander Baur for our collaboration in Munich and for helping me with the first steps into academia.

I am thankful to Alessandro Sfondrini, Dmitri Sorokin, and Roberto Volpato for offering me a position in their research group.
I am looking forward to work with you in Padua!

I have enjoyed a lot my time at the institute in Adlershof and especially being part of the RTG helped tremendously to learn from many outstanding people about various fields of physics and provided a great workspace.
Special thanks to Jan Plefka, Agostino Patella, and Valentina Forini for this opportunity!
I would like to thank Alessandro, Anne, Burkhard, Camilla, Daniele, Davide, Felipe, Florian, Gabriel, Giulia, Gustav, Ilaria, Jenny, Julien, Luke, Maria, Matthias, Michele, Mika, Olaf, Peppe, Rob, Roberto, Stijn, Sylvia, Tim, and Tomas for this great time.

Finally, I could not have made it through these years without my friends and family and the infinite support of Trixi, Lorenz, and Ciup.
Thank you for always being there for me!

This research is funded by the Deutsche Forschungsgemeinschaft (DFG, German Research Foundation) - Projektnummer 417533893/GRK2575 "Rethinking Quantum Field Theory".

%\thispagestyle{empty}

%----------------------------------------------------------------------------------------
%L
\newpage
\mbox{}

%----------------------------------------------------------------------------------------
%	LIST OF CONTENTS/FIGURES/TABLES PAGES - R
%----------------------------------------------------------------------------------------

\newpage

\pagenumbering{gobble}

\hypersetup{linkcolor=black}
\tableofcontents % Prints the main table of contents

%----------------------------------------------------------------------------------------
%L
%\newpage
%\thispagestyle{empty}
%\mbox{}

%-----------------------------------------------------------------
%	GRAPHICAL SETTINGS
%----------------------------------------------------------------------------------------
%\newpage

\mainmatter

\pagestyle{fancy} 
\fancyhf{}
\fancyhead[LE]{\thepage}
\fancyhead[RE]{\fontsize{9}{12}\selectfont\itshape\nouppercase{\leftmark}}
\fancyhead[RO]{\thepage}
\fancyhead[LO]{\fontsize{9}{12}\selectfont\itshape\nouppercase{\rightmark}}
% \fancyfoot[CE,CO]{}
%\fancyfoot[LE,RO]{}
\renewcommand{\chaptermark}[1]{\markboth{\chaptername \ \thechapter.\ \ #1}{}}  %on the left it displys the chapter
\renewcommand{\sectionmark}[1]{\markright{\thesection.\ \ #1}{}} %on the right the section

%%%----------------------------------------------------------------------------------------
%%%	THESIS CONTENT - PART 1 - Introduction 
%%%----------------------------------------------------------------------------------------
\chapter{Introduction}
\label{chpt:Introduction}

\begin{figure}[t]
$\adjincludegraphics[valign=c,scale=1]{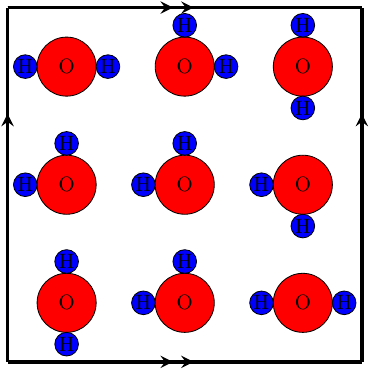} 
\rightarrow
\adjincludegraphics[valign=c,scale=1]{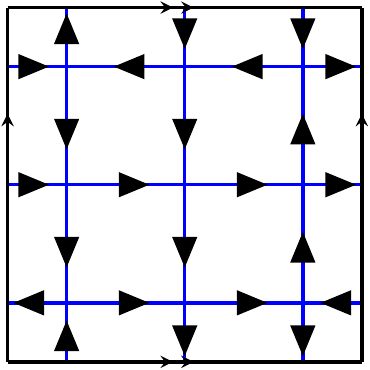} $
\centering
\caption{The left picture shows a configuration of two-dimensional square-ice with doubly-periodic boundary condition on a $3\times 3$ lattice. Hence, the left- and right boundary, and the top- and bottom boundary are identified. This is signaled by the small arrows. The right picture shows the same configuration but illustrated in the conventions of the six-vertex model: an ingoing arrow indicates that the hydrogen atom of the bond is associated with the water molecule, which is sitting at the vertex where the arrow points to.}
\label{fig:SquareIce6V}
\end{figure}

Ice is a cool subject of study, not only because water is a fundamental element of life on Earth.
At very low temperatures, its molecules $\mathrm{H}_2 \mathrm{O}$ form a crystalline structure, where each oxygen atom is surrounded by four other oxygen atoms, which are arranged in a tetrahedron \cite{Pauling}.
The entropy of ice is closely related to important thermodynamic properties, such as the heat capacity.
It was measured almost 100 years ago and a small discrepancy to the expected value due to its crystalline structure was found \cite{Giauque1,Giauque2}.
The residual entropy contributed to the many different configurations of the hydrogen atoms in the crystal and Pauling \cite{Pauling} proposed the idea of a two-dimensional ice model to study their effect, see e.\,g.\ \cite{nagle1966lattice}.
It assumes that a hydrogen atom sits between each pair of oxygen atoms and has two states, where it binds to either of the two oxygen atoms more closely.
This became known as the \textit{ice rule}.
The dimensionally reduced, two-dimensional ice model, where oxygen atoms are arranged in a square lattice, sparked the formulation and study of the six-vertex model \cite{Lieb:1967zz}.
Around each oxygen atom at the vertices of the lattice, the two hydrogen atoms of water molecules can be placed in six different ways, while only one hydrogen atom occupies an edge of the lattice, representing an oxygen-oxygen bond.
Typically, the configuration of the hydrogen atoms is illustrated by arrows drawn on the edges, which point to the associated oxygen atom; see fig.\ \ref{fig:SquareIce6V}.
In this picture, the ice rule is equivalent to conservation of the number of ingoing and outgoing arrows at each vertex. 
Remarkably, Lieb \cite{Lieb:1967zz} solved the model for periodic boundary conditions and showed that the residual entropy of ice can be explained by the thermodynamic limit of the partition function of the six-vertex model.
Not only was he able to determine the partition function of the model exactly in the thermodynamic limit, but his use of the Bethe ansatz \cite{bethe1931theorie} uncovered the integrable structure of the model.

Subsequently, the six-vertex model became a subject of study in its own right as a celebrated two-dimensional integrable model.
The integrability of the model becomes even more apparent when placed in an external electric field \cite{Sutherland1}.
In the partition function, the electric field weights the six vertices differently and results in the necessary generalization of the ice model by two parameters, termed are called \textit{spectral parameter} and \textit{crossing parameter}.
The integrable structure is traced down to a Yang-Baxter equation (YBE) that the generalized vertex weights fulfill.
After its early appearance in the solution of the Ising model \cite{Ising1925BeitragZT,Onsager} and in many-body models \cite{mcguire1964study,Yang}, Baxter \cite{BAXTER1972193} showed that the YBE is also behind the integrability of the eight-vertex model \cite{sutherland1970two,fan1970general}.
This model generalizes the six-vertex model even further by including also vertices with all arrows outgoing or ingoing.
The YBE implies the existence of commuting transfer matrices, which can be expanded in the spectral parameter to obtain a set of commuting charges, resulting in quantum integrability \cite{Faddeev:1996iy}.
This is in analogy to classical integrability, which requires the existence of a maximal set of Poisson-commuting invariants \cite{arnol2013mathematical}.

There are many relations of the six- and eight-vertex model to other integrable models of statistical physics.
In particular, the eight-vertex model contains the Hamiltonian of the XYZ Heisenberg spin chain \cite{heisenberg1928theorie} as one of its conserved charges and, by this correspondence, the six-vertex model is related to the XXZ spin chain \cite{Baxter:1989ct}.
Furthermore, for a special choice of parameters the six-vertex model may be reformulated as a Potts model \cite{Potts_1952,baxter1982critical,Jacobsen:2005xz}, which is generalizing the Ising model by allowing multiple states at each lattice site.
Driven by the spirit of finding further generalizations, the chiral Potts model was formulated in \cite{AUYANG1987219,BAXTER1988138}, where the interactions between the generalized spins of the Potts model no longer need to be symmetric.
It turned out \cite{Au-Yang:1999kid} that in a particular limit, where the number of spin states goes to infinity in a controlled way, the chiral Potts model connects to another field of theoretical physics, quantum field theory (QFT).
In the limit, the spin states at each site become continuous variables, and to calculate the partition function of the models one has to integrate over these \cite{Bazhanov:2016ajm}.
Accordingly, the partition function of the chiral Potts model in this limit describes a vacuum Feynman graph of square-lattice shape.

Twenty years earlier, Zamolodchikov had described such an integrable model reminiscent of fishing-net vacuum Feynman graphs \cite{Zamolodchikov:1980mb}.
He realized that the uniqueness relation that appears in conformal field theories (CFT) \cite{DEramo:1971hnd,Symanzik:1972wj}, which allows one to turn a Feynman diagram of the shape of a three-spiked star into a triangle-shaped one, allows the construction of commuting transfer matrices.
Therefore, the star-triangle relation hides a YBE ensuring the integrability of the vacuum diagrams.
The spectral parameter appears in the form of a general exponent of the propagators.
Hence, for an unspecified spectral parameter, the diagrams are called generalized Feynman diagrams.
Using the method of inversion relations \cite{Baxter:1982xp,StroganovInvRelLatticeModels}, the partition function in the thermodynamic limit was calculated for doubly periodic boundary conditions.
This worked even for triangular and hexagonal lattices \cite{Zamolodchikov:1980mb}.
In contrast to the residual entropy of a statistical model, the thermodynamic limit of the vacuum diagrams corresponds to the critical coupling of the QFT producing the diagrams.
This is the radius of convergence of the QFT's free energy.
Typically, perturbative expansions in QFTs are asymptotic series, which may not even converge.
%However, they serve as an arbitrarily accurate approximation of the observables of the theory.
Hence, the graphs considered by Zamolodchikov must be related to a QFT with very controlled perturbative expansions to be able to study their convergence properties.
At the time, the action of a QFT leading to the regular, lattice-like, toroidal vacuum graphs was unknown.

Meanwhile, the discovery of another incarnation of integrability in QFT led to the formulation of an action for these fishing-net like theories.
Some years after the discovery of the AdS/CFT-correspondence between $\mathcal{N} = 4$ super Yang-Mills theory (SYM) and type $\mathrm{IIB}$ superstrings on $\mathrm{AdS}_5 \times S^5$  \cite{Maldacena:1997re}, it was first shown that $\mathcal{N} = 4$ SYM possesses another remarkable property in the planar limit, quantum integrability \cite{Minahan:2002ve,Beisert:2003tq,Beisert:2003yb,Beisert:2004ry,Beisert:2010jr}.
The integrable structure was proposed in analogy with integrable spin chains:
the fields of $\mathcal{N} = 4$ SYM are in the adjoint representation of the gauge group $\SU{\mathrm{N}}$ and thus $\mathrm{N} \times \mathrm{N}$ matrix-valued.
The object corresponding to the spin chain are gauge invariant single-trace operators, such that the periodicity of the chain is implemented by the trace over multiple fields.
Next to gauge symmetry and supersymmetry, $\mathcal{N} = 4$ SYM enjoys also a global $\SU{4} \cong \SO{6}$ R-symmetry, $\SO{1,3}$ Lorentz symmetry, and conformal symmetry.
Since all of these are also present after quantization, the fields organize themselves into representations of the $\mathcal{N} = 4$ superconformal algebra in four dimensions $\mathfrak{psu} (2 ,2 \vert 4)$, see e.\,g.\ \cite{Minahan:2010js}.
Instead of $\mathfrak{su} (2)$ spins in the case of the Heisenberg spin chain, one has $\mathfrak{psu} (2 ,2 \vert 4)$ spins sitting in the single trace operators of $\mathcal{N} = 4$ SYM.
In this picture, the Hamiltonian is the dilatation operator of the superconformal algebra, and its eigenvalues correspond to the scaling dimensions of the single-trace operators.
The spectral problem can be formally solved by a Bethe ansatz, which was developed for the Heisenberg spin chain.
Technically, solving the Bethe ansatz is hard and does not account for wrapping corrections \cite{Sieg:2005kd,Bajnok:2008bm,Bajnok:2008qj}.
Thus, alternative integrability-based methods were developed, see \cite{Bombardelli:2016rwb,Beisert:2010jr} for references.
Notably, in further developments, starting from the asymptotic Bethe ansatz \cite{Beisert:2005fw}, a TBA/Y-system was proposed \cite{Gromov:2009tv,Bombardelli:2009ns,Gromov:2009bc,Bajnok:2010ke} and yielded the quantum spectral curve (QSC) \cite{Gromov:2013pga,Gromov:2017blm,Levkovich-Maslyuk:2019awk}.
It provides a powerful numerical tool to access the spectrum at very high orders.

However, at the lowest order, the dilatation operator is derived from one-loop calculations and generally relies on perturbative input.
A window into the strong-coupling regime is offered by the AdS/CFT-correspondence.
On the supergravity side of the correspondence, integrability manifests itself by the fact that the string theory can be formulated as a supersymmetric sigma model on the Lie supergroup of the isometries of $\mathrm{AdS}_5 \times S^5$, which is a coset space.
The Maurer-Cartan form of the supergroup, together with a $\mathbb{Z}_4$ automorphism, implies the existence of a flat Lax-connection, which is a sufficient criterion for the existence of integrals of motion in involution and thus for classical integrability \cite{Bena:2003wd,Frolov:2003qc,Arutyunov:2003uj}.
To enter the realm of quantum integrability and to quantize the sigma model, one has to expand the embedding fields around a classical solution \cite{Frolov:2002av,Ahn:2010ka}.
It is again a perturbative approach, this time accessing the higher orders of the strong-coupling expansion of the correspondence.
It is still an open question how integrability manifests itself at a non-perturbative level and how it bridges the gap between the weak- and strong-coupling regime in the AdS/CFT-correspondence.

To tackle this issue, the study of integrable deformations became an important topic.
This is analogous to deformations of ordinary spin chains, which are achieved by switching on an external electric field.
This leads to the concept of $q$-deformed Lie algebras, the so-called quantum groups , see e.\,g.\ \cite{Faddeev:1996iy,klimyk2012quantum}. 
In the context of the $\mathcal{N}=4$ SYM spin chain, analogously, one can think of deforming the global symmetries \cite{Berenstein:2004ys,Beisert:2005if}.
For the sake of avoiding non-commutative QFTs, one focuses on deformations of the $\SU{4}$ R-symmetry, which potentially break some or all supersymmetries.
On the string theory side of the AdS/CFT-correspondence, this amounts to deforming the $S^5$-sphere into a higher-dimensional ellipsoid.
The Lie algebra $\mathfrak{su} (4)$ has three Cartan generators that may be subject to the deformation.
Deforming all of them simultaneously yields a three-parameter deformation of $\mathcal{N}=4$ SYM, the so-called $\gamma$-deformation \cite{Lunin:2005jy,Frolov:2005dj,Frolov:2005iq}.
When all three deformation parameters are equal, $\mathcal{N} = 1$ supersymmetry is left, and the one-parameter deformation is called $\beta$-deformation \cite{Leigh:1995ep,Lunin:2005jy,Aharony:2002hx}.
On the level of the action, the deformation is implemented by replacing the point-wise products of fields by a non-commutative star-product, which depends on the R-charge of its two factors.
This way, the deformation introduces different phases in front of the various interaction terms in the Lagrangian.
When computing correlation functions, on the one hand, this yields cancellations between different contributions; on the other hand, contributions which cancel in the undeformed theory no longer cancel.
Due to the latter, one is forced to introduce double-trace counterterms and their couplings run with the renormalization scale. 
To maintain conformal symmetry on the quantum level, one has to tune the double-trace couplings to a fixed-point \cite{Fokken:2013aea,Fokken:2014soa}.
Thanks to the integrability, many techniques for concrete calculations carry over to the $\gamma$-deformation \cite{Gromov:2010dy,deLeeuw:2012hp,Fokken:2013mza,Kazakov:2015efa,Zoubos:2010kh}.

Simpler theories may be obtained when one performs the so-called double-scaling limit on the $\gamma$-deformation of $\mathcal{N}=4$ SYM \cite{Gurdogan:2015csr}.
This is the limit where the deformation parameters are considered to tend to (imaginary) infinity, while the 't Hooft coupling is scaled to zero.
The two limits are taken in a controlled way, such that the products of the divergent deformation parameters with the vanishing 't Hooft coupling, the effective couplings, stay at a finite value.
This procedure is called the double-scaling limit.
Its consequences are, that some interaction terms disappear from the action, which implies that some fields decouple from the theory.
Moreover, the resulting theories have a non-unitary Lagrangian.
From a phenomenological standpoint, this is highly concerning, because norms of the theory's states might be negative, which spoils unitarity.
Yet, the simplicity of the double-scaled $\gamma$-deformations and its potential to uncover the integrable structure hidden in $\mathcal{N}=4$ SYM justify the study of these toy models. 
The outcome of the double-scaling limit results in multiple distinct theories, depending on the number of effective couplings that are set to zero.
In any case, the gauge fields decouple and one is left with a theory of solely Yukawa- and quartic scalar interactions.
If none of the effective couplings is set to zero, the so-called $\chi$-CFT is obtained, while setting one effective coupling to zero results in the $\chi_0$-CFT  \cite{Kazakov:2018gcy}.

If two effective couplings are set to zero, the double-scaling limit yields a particularly simple theory, the bi-scalar fishnet theory \cite{Gurdogan:2015csr}.
Its two scalars are the only remaining degrees of freedom and they are coupled in a single non-unitary quartic interaction term.
However, since it is derived from a $\gamma$-deformation, additional double-trace terms are produced by radiative corrections and their coupling has to be tuned to a fixed-point to obtain a four-dimensional CFT \cite{Sieg:2016vap,Grabner:2017pgm,Korchemsky:2018hnb}.
The theory's Feynman diagrams have a remarkably regular fishing-net topology.
In the planar limit, where $\mathrm{N}$ approaches infinity, toroidal diagrams dominate and the vacuum diagrams are exactly of the form considered by Zamolodchikov \cite{Zamolodchikov:1980mb} and mentioned above.
Hence, we reconnect here with the interpretation of QFT vacuum diagrams as a partition function of a particular generalization of the Potts model and eventually the six-vertex model.
It allows us to study the emergence of integrability \cite{Chicherin:2017cns,Chicherin:2017frs,Gromov:2017cja} at an arbitrary order in perturbation theory, aside from asymptotic series.
Notably, it is possible to extract exact all-loop anomalous dimensions of various operators and to determine the coefficients of the operator product expansion (OPE) to all orders in the coupling \cite{Grabner:2017pgm,Gromov:2018hut}.
Furthermore, a special class of four-point correlation functions has only one Feynman diagram contributing.
These are the Basso-Dixon diagrams and they can be computed to arbitrary loop order, which was proven by using the theory's integrable structure \cite{Basso:2017jwq,Derkachov:2018rot,Derkachov:2019tzo,Derkachov:2020zvv,Basso:2021omx}.
Moreover, the toolbox that was established for $\mathcal{N}=4$ SYM was adapted to the fishnet theory, e.\,g.\ the QSC \cite{Kazakov:2018ugh,Cavaglia:2021mft}, the thermodynamic Bethe ansatz \cite{Basso:2018agi,Basso:2019xay} and the Yangian bootstrap \cite{Loebbert:2019vcj,Corcoran:2021gda}.
The integrable structure relies on the existence of an R-matrix in a non-compact representation of the conformal group \cite{Derkachov:2021rrf,Chicherin:2012yn}, which takes the form of a certain generalized subgraph of the fishnet Feynman diagrams.
The spin-chain analogy was made explicit by constructing the one-loop dilatation operator, which contains Jordan blocks and signals properties of a logarithmic CFT \cite{Ipsen:2018fmu,Ahn:2020zly,Ahn:2021emp}.
Massive and $D$-dimensional generalizations of the fishnet theory were considered as well \cite{Kazakov:2018qbr,Loebbert:2020hxk,Loebbert:2020tje,Loebbert:2020glj,Loebbert:2021qef}.
Following the AdS/CFT-correspondence to the strong-coupling regime, the holographic dual of the fishnet theory, the fishchain, was found \cite{Gromov:2019aku,Gromov:2019bsj,Gromov:2019jfh}.
Due to the fruitful study of the bi-scalar fishnet theory, theories with a similar regularity in their Feynman graphs, but without a connection to $\mathcal{N}=4$ SYM, were studied \cite{Mamroud:2017uyz,Pittelli:2019ceq,Kazakov:2022dbd,Alfimov:2023vev}.

Next to $\mathcal{N} = 4$ SYM, the interplay between holography and integrability of QFTs was also examined on the example of the Aharony-Bergman-Jafferis-Maldacena (ABJM) theory \cite{Aharony:2008ug}.
It is a three-dimensional $\mathcal{N}=6$ superconformal Chern-Simons theory with gauge group $\U{\mathrm{N}} \otimes \U{\mathrm{N}}$ and bi-fundamental matter, which couples to the gauge fields and a superpotential.
Its global superconformal algebra is $\mathfrak{osp}(6 \vert 4)$ and in the planar limit, the single-trace operators can be associated to integrable spin chains \cite{Minahan:2008hf,Gaiotto:2008cg,Gromov:2008qe,Klose:2010ki}, similar to $\mathcal{N} = 4$ SYM.
There exists also a holographic dual, which is type $\mathrm{IIA}$ superstring theory on $\mathrm{AdS}_4 \times \mathbb{CP}^3$ \cite{Aharony:2008ug}.
The spin chain integrability allows to obtain the spectrum of single-trace operators and the QSC was established to facilitate its calculation \cite{Cavaglia:2014exa,Bombardelli:2017vhk,Bombardelli:2018bqz,Lee:2018jvn}.
R-symmetry deformations \cite{Imeroni:2008cr} and its integrable structure were studied in \cite{He:2013hxd,Chen:2016geo}.
Finally, the double-scaling limit of $\gamma$-deformed ABJM was performed in \cite{Caetano:2016ydc} and its spin chain integrability was showcased. 
Similar to the $\chi$-CFT, $\chi_0$-CFT and bi-scalar fishnet theory, the double-scaled $\gamma$-deformation yields different theories for different values of the effective couplings.
The fishnet-equivalent, where two of the three effective couplings vanish, is a three-dimensional scalar theory with a single sextic interaction term.
Its vacuum diagrams have the form of partition functions on a triangular lattice, also studied in Zamolodchikov's original work \cite{Zamolodchikov:1980mb}.
This motivates the study of its Feynman diagrams as a partition function in analogy to models in statistical physics.

\subsubsection{Results and outline}
In this thesis, we advocate the existence of an integrable structure common to statistical lattice models and the (super) Feynman diagrams of integrable QFTs.
Accordingly, throughout this work, we review the following important concepts of this correspondence:
We present integrability in lattice models on the example of the six- and eight-vertex model on a square lattice in section \ref{sec:8Vmodel}.
In the case of doubly-periodic boundary conditions, we show in section \ref{sec:ToroidalBoundaryConditions} how the YBE implies the existence of commuting transfer matrices and derive the free energy in the thermodynamic limit by the method of inversion relations \cite{StroganovInvRelLatticeModels,Baxter:1989ct}.
In sections \ref{sec:TheGeneralizedScalarPropagatorAsLatticeWeight}, we showcase the relation between generalized bosonic and fermionic Feynman diagrams and integrable statistical lattice models.
Zamolodchikov \cite{Zamolodchikov:1980mb} first discovered that one can describe integrable Feynman graphs and statistical edge interaction models using a common framework of spectral lines.
Later, this idea was extended to more models, statistical ones \cite{Bazhanov:2016ajm} and QFTs \cite{Kazakov:2022dbd}.
On the QFT side, we review the most relevant example of a QFT that produces integrable Feynman diagrams, the bi-scalar fishnet theory.
In section \ref{sec:TheBiScalarFishnetTheory} we present its derivation from $\mathcal{N} = 4$ SYM \cite{Gurdogan:2015csr}, discuss its renormalization and conformal properties \cite{Sieg:2016vap,Grabner:2017pgm} and repeat the remarkable exact calculation of a particular anomalous dimension of a length-two operator and the associated OPE coefficients \cite{Grabner:2017pgm,Gromov:2018hut}.
Appendix \ref{app:PerturbativeCalculationsInFishnetTheory} accompanies this part with detailed perturbative calculations of the beta function and an exact anomalous dimension \cite{Korchemsky:2018hnb}.
Furthermore, in section \ref{sec:IntegrableVacuumDiagramsAndTheCriticalCoupling}, we repeat the original computation of the critical couplings of the bi-scalar fishnet theory, the triangular fishnet theory of ABJM and a six-dimensional QFT with cubic interactions by the method of inversion relations \cite{Zamolodchikov:1980mb}.

Next to the presentation of known instances of the correspondence between lattice models and integrable Feynman graphs from the literature, this thesis aims at a generalization to different boundary conditions and to fermionic- and supersymmetric Feynman diagrams.
The original contributions of this thesis, which have been only partially published so far, are the following:
\begin{itemize}
\item
We consider novel boundary conditions for the six-vertex model in section \ref{sec:TheBoxBoundaryCondition}.
Instead of toroidal boundary conditions, we assume that the lattice on which the six-vertex model is placed has the form of a rectangular box with integrable boundaries in the form of K-matrices \cite{Sklyanin:1988yz} on all four walls.
Thanks to the discovery of recursion relations of the partition function in this setting, we can conjecture a closed formula for the square-shaped box partition function,
\begin{equation*}
    \begin{split}
    Z_M 
    \left( 
    \lbrace x_i \rbrace %_{i=1, ..., M}
    \vert
    \lbrace y_i \rbrace %_{i=1, ..., M}
    \right)
\propto
    \underset{1 \leq i,j \leq M}{\mathrm{det}}
    \left[
        \frac{
			c^2 
			a(x_i y_j) a(\frac{1}{x_i y_j}) 
			F^{\mathrm{LU}} (x_i)
			F^{\mathrm{DR}} (y_j)         
        }{
        \left(
            \frac{x_j}{y_i} - \frac{y_i}{x_j}
        \right)
        W (x_i , y_j)
        }
    \right] 
     ~.
    \end{split}
\end{equation*}
The full result is presented in \eqref{eq:6V_BoxPartitionFunctionSquareSolution}.
It was obtained in collaboration with A.\ Cotellucci \cite{ToAppAlessandro}.
\item
The extension of the correspondence between lattice models and Feynman graphs to boundary integrability is presented in chapter \ref{chpt:BoundaryIntegrabilityInFeynmanGraphs}.
We translate the boundary Yang-Baxter equation (bYBE) of Sklyanin \cite{Sklyanin:1988yz} into the language of generalized Feynman diagrams in \eqref{eq:bYBE_fused_type1}.
We find that the bYBE holds if a boundary weight exists that solves one of the two so-called boundary-star-triangle relations (bSTR).
These are presented in \eqref{eq:bSTR_type1And2}.
Based on generalized propagators, we find two solutions that yield two K-matrices.
We prove the existence of commuting double-row matrices graphically.
This work is done in collaboration with C.\ Ahn and M.\ Staudacher \cite{ToAppChangrim}.
\item
The critical coupling of brick wall theory \cite{Pittelli:2019ceq} and fermionic fishnet theory \cite{Kade:2023xet} is calculated.
To do so, we establish the map between integrable lattice models and fermionic generalized Feynman graphs and derive a useful integral relation, the x-unit relation, in section \ref{sec:TheGeneralizedFermionicPropagatorAsLatticeWeight}.
It allows us to derive the inversion relations for fermionic theories and solving them gives the values for the critical coupling, 
\begin{equation*}
\rho_\mathrm{cr}^\mathrm{BW} 
=
\frac{1}{2 \pi^{12} \, \eta (\I)^4 }
 ~~~~ \mathrm{and} ~~~~
\rho_\mathrm{cr}^\mathrm{FFN}
=
\frac{27}{ 2^{8/3} \pi ^{12} | \eta ( \e^{\frac{i \pi }{3}} ) | ^8} ~,
\end{equation*}
which is derived in section \ref{sec:IntegrableVacuumDiagramsAndTheCriticalCoupling} and presented in \eqref{eq:ResultsCriticalCoupling_NonSusy}.
This work was published in \cite{Kade:2023xet} in collaboration with M.\ Staudacher.
\item
We propose a superspace formulation of the double-scaled $\beta$-deformations of $\mathcal{N} = 4$ SYM and ABJM theory.
These were first derived in the works by the author, \cite{Kade:2024ucz} and \cite{Kade:2024lkc}, respectively.
Amazingly, their supersymmetric Feynman graphs obey a regular brick wall and fishnet pattern, and they are presented in sections \ref{sec:TheSuperBrickWallTheory_sec} and \ref{sec:DoubleScaledBetaDeformationOfABJM}, respectively.
In section \ref{sec:TheGeneralizedSuperspacePropagatorAsLatticeWeight}, we show how a type of integral relation, the chain relations, can be generalized to superspace.
However, we fail to find a superspace STR that would allow us to make integrability manifest.
The chain relations are a specialization of the STR and, therefore, we still consider generalized supergraphs to be integrable.
We derive a super-x-unity relation and calculate the critical coupling for the super brick wall theory and the superfishnet theory,
\begin{equation*}
\xi_\mathrm{cr}^\mathrm{SBW}
=
\frac{3^{9/8}}{4\pi^3\, | \eta ( \e^{\frac{i \pi }{3}} )|}
 ~~~~ \mathrm{and} ~~~~
\xi_\mathrm{cr}^\mathrm{SFN}
=
\frac{1}{ \pi^3 \sqrt{2}\, | \eta ( \I )| ^{2} } ~,
\end{equation*}
as presented in \eqref{eq:Result_CriticalCoupling_SUSY} and published in \cite{Kade:2024ucz} and \cite{Kade:2024lkc}, respectively.
The work on the super brick wall theory was done in collaboration with M.\ Staudacher.
Furthermore, we can use the super chain relations to derive the exact all-loop spectrum of various operators and use them to calculate an OPE coefficient to all loop orders in the superfishnet theory.
Notably, we present the explicit and exact scaling dimensions
\begin{equation*}
\Delta
=
1 
+
\frac{1}{2}
\left(
-1
\pm
2
\sqrt{
(S+\tfrac{1}{2})^2 - 4 \pi^4 \xi^2
}
\right) 
\end{equation*}
of the zero-magnon operator $\mathrm{tr}[ \Phi_1 \partial^S \Phi_3^\dagger ]$ in \eqref{eq:0Magnon_result_scalingDim_spinning}.
The latter is an unpublished result for $S > 0$.
In \cite{Kade:2024lkc}, the case $S=0$ was first obtained, together with the implicit result of the all-loop scaling dimensions of the two-magnon operators $\square^n \, \mathrm{tr}[ \Phi_1^\dagger \Phi_2^\dagger \Phi_1^\dagger \Phi_2^\dagger ]$ in \eqref{eq:2Magnon_ImplicitScalingDims}.
Furthermore, we were able to calculate the OPE coefficient in the zero-magnon case to all orders in the coupling.
The result is 
\begin{equation*}
\mathsf{C}_{\Delta ,S}
=
- 
2^{S-1-2 \Delta}
\pi
\frac{
\Gamma (S+\frac{3}{2})
\Gamma (\Delta )  
\Gamma (\frac{S-\Delta +2}{2} ) 
\Gamma (\frac{S+\Delta }{2} )
}{
\Gamma (S+1) 
\Gamma (\Delta +\frac{1}{2} ) 
\Gamma (\frac{S-\Delta +3}{2} ) 
\Gamma (\frac{S+\Delta +1}{2} )
} 
\end{equation*}
and the derivation is exclusively presented in \eqref{eq:Superfishnet_exactOPEcoefficient} of this thesis.
\end{itemize}
This thesis is structured as follows.
In chapter \ref{chpt:SquareIceAndThe8V} we present the eight-vertex model as our reference integrable model and introduce the box-boundary conditions.
The transition to quantum field theory relies on the contents of chapter \ref{chpt:AuxiliaryRelations}, where we present the necessary chain-, star-triangle- and x-unity relation for generalized bosonic and fermionic propagators, and the super chain relations and the super x-unity relation for generalized superspace propagators.
Furthermore, we introduce a graphical notation which facilitates and illustrates the following calculations.
Chapter \ref{chpt:BoundaryIntegrabilityInFeynmanGraphs} is solely about generalized bosonic propagators and introduces the concept of boundary integrability to Feynman diagrams.
In chapter \ref{chpt:NonSusyFishnetTheoriesAndItsRelatives}, we derive the $\chi$-CFT, $\chi_0$-CFT and bi-scalar fishnet theory from the superspace formulation of $\mathcal{N}=4$ SYM.
We review its diagrammatics and its renormalization.
Additionally, we showcase the computation of exact anomalous dimensions and OPE coefficients, as well as the calculation of critical couplings for various integrable theories involving bosons and fermions.
The chapter \ref{chpt:SupersymmetricDoubleScaledDeformations} is about the superfishnet and super brick wall theory.
We put the super chain- and super x-unity relation into work and determine the spectrum of different classes of single-trace operators as well as an OPE coefficient in the superfishnet theory, and the critical coupling of both supersymmetric models.
The concluding chapter \ref{chpt:Conclusions} gives an outlook on future research directions, based on the results of this thesis.

The work is completed with four appendices.
Appendix \ref{app:ProofOfUniquenessRelations} gives a detailed proof of the basic integral relations that are the scalar STR, the fermionic STR, the superspace chain rule and the super x-unity relation.
The second appendix \ref{app:PerturbativeCalculationsInFishnetTheory} reviews in detail the perturbative calculation of the double-trace beta function and the spinless zero-magnon scaling dimension of bi-scalar fishnet theory from the literature.
In appendix \ref{sec:SolvingInversionRelations}, we present the iterative procedure leading to the solution of generic inversion relations.
Additionally, we define the theta functions and the elliptic gamma functions and comment on their appearance in the free energy of the eight-vertex model.
The final appendix \ref{app:SuperspaceNotations} accompanies the superspace calculations by presenting our conventions and other useful relations.

%%%----------------------------------------------------------------------------------------
%%%	THESIS CONTENT - PART 2 - Square ice and the six-vertex model
%%%----------------------------------------------------------------------------------------
\chapter{Square ice and the eight-vertex model}
\label{chpt:SquareIceAndThe8V}
The six-vertex model is a statistical model that consists of two-state spins (usually denoted as arrows) on the edges of a lattice.
The lattice has to have four-valent vertices and requiring arrow-conservation at each vertex constrains the arrow distribution by the following six allowed configurations:
\begin{equation}
\adjincludegraphics[valign=c,scale=1.4]{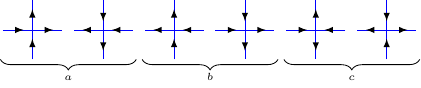} ~.
\label{eq:6V_VertexWeights}
\end{equation}
These have an equal number of incoming and outgoing arrows, the so-called \textit{ice rule}.
Depending on the configuration, we assign to each of the six vertices a weight $a$, $b$ or $c$, which is a complex number that specifies the contribution of a single vertex to the partition function.
In the introduction, chapter \ref{chpt:Introduction}, we mentioned the motivation for the six-vertex model, which is the study of the residual entropy of ice.
Here, however, we spotlight the integrability of the model and its generalization, the eight-vertex model.
We return to the six-vertex model in the last section to introduce the new box boundary conditions and conjecture an exact expression for its partition function.
The six-vertex model and its many applications in mathematics and physics are reviewed in e.\,g.\ \cite{zinnjustin2009sixvertexlooptilingmodels,Lamers:2015dfa,jacobsen2019integrability}. 

\section{The integrable eight-vertex model}
\label{sec:8Vmodel}
The six-vertex model can be generalized by allowing two more vertices, which explicitly break the ice rule.
One, where all arrows point into the vertex, a \textit{sink} for arrows, and one where all arrows point outward, the \textit{source}.
Altogether, the generalized model consists of the eight configurations 
\begin{equation}
\adjincludegraphics[valign=c,scale=1.4]{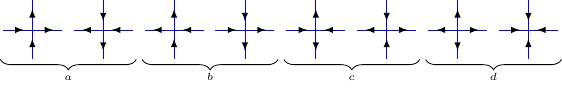} 
\label{eq:8V_VertexWeights}
\end{equation}
and is consequently called the \textit{eight-vertex model}.
The weight $d$ corresponds to the newly introduced vertices and we note that choosing $d=0$ leads us back to the six-vertex model \eqref{eq:6V_VertexWeights}.

The eight-vertex model is defined on a lattice $\Lambda$, which must have four-valent vertices.
A canonical choice is the doubly periodic lattice with horizontal $M$ and vertical $N$ lines.
In the case of the six-vertex model, a non-periodic, bounded lattice is studied in section \ref{sec:TheBoxBoundaryCondition}.
The partition function is defined as the sum of all possible configurations, and a valid configuration $\Omega$ must respect the lattice $\Lambda$ and must consist of vertices \eqref{eq:8V_VertexWeights}.
If this is the case, we write $\Omega \in \Gamma (\Lambda )$.
The weights $a$, $b$, $c$ and $d$ determine how the eight different vertices contribute to the partition function of the eight-vertex model, which reads \cite{sutherland1970two,fan1970general}
\begin{equation}
Z (a,b,c,d)
=
\sum_{\Omega \in \Gamma (\Lambda )}
a^{n_a} b^{n_b} c^{n_c} d^{n_d} ~.
\label{eq:8VPatitionFunction}
\end{equation}
Thereby, the integers $n_a$, $n_b$, $n_c$ and $n_d$ count how often the vertices of type $a$, $b$, $c$ and $d$ appear in the configuration $\Omega$.
In the partition function $Z ( \rho\cdot a, \rho\cdot b, \rho\cdot c, \rho\cdot d)$, we can factor $\rho^V$, where $V = n_a + n_b + n_c + n_d$ is the total number of vertices of the lattice.
Hence, we have the relation $Z ( \rho\cdot a, \rho\cdot b, \rho\cdot c, \rho\cdot d) = \rho^V Z (a,b,c,d)$.
Therefore, the problem of finding \eqref{eq:8VPatitionFunction} is equally hard when scaling all the weights simultaneously.
This implies that the space of different eight-vertex models is projective $\mathbb{CP}^3$.
In \cite{Baxter:1989ct}, additional symmetry properties were examined, which yield the result that $Z (a,b,c,d)$ is symmetric with respect to multiplication by $-1$ in each of its arguments.

The calculation of the partition function \eqref{eq:8VPatitionFunction}, correlation functions and other observables becomes hard very fast as the size of the lattice gets larger.
However, in the space of the eight-vertex models $\mathbb{CP}^3$ exists a subspace, where the powerful property of integrability allows computations for any lattice sizes and even for the thermodynamic limit in which the size tends to infinity \cite{baxter1971eight,Baxter:1989ct}.
The integrability of the eight-vertex model arises when one considers the weights as local, inhomogeneous quantities on the lattices.
Therefore, one considers the lattice with its four-valent vertices as a set of so-called \textit{spectral lines}, which intersect each other, but only two lines at a time.
Additionally, the spectral lines carry a \textit{spectral parameter}.
Whenever two lines intersect, the weights depend on the quotient of the two spectral parameters upon distribution of the arrows on the lattice.
The R-matrix encodes the relation between spectral parameters of the spectral lines and the vertex weights \eqref{eq:6V_VertexWeights} at their intersection,
\begin{equation}
R (\tfrac{x}{y} ) 
=
\adjincludegraphics[valign=c,scale=1]{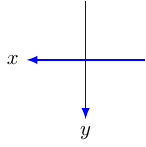} 
=
\adjincludegraphics[valign=c,scale=1]{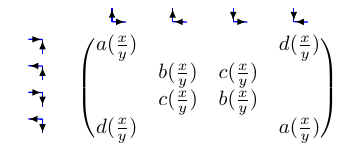} ~.
\label{eq:8V_Rmatrix}
\end{equation}
Here, the spectral parameters of the horizontal and vertical spectral lines are $x$ and $y$, respectively.
We denote spectral lines in blue with a blue arrow indicating their direction, while black arrows represent the spin configuration on an edge of the lattice.
Whenever we are not explicitly showing the black spin arrows, we mean a general, matrix-valued object.
An R-matrix is generally an endomorphism of the Kronecker product of two vector spaces, $R(\sfrac{x}{y}) \in \mathrm{End} (V_\mathrm{aux} \otimes V_\mathrm{phys})$.
The vector space $V_\mathrm{aux}$, associated with the spectral parameter $x$ and the horizontal spectral line, is called the \textit{auxiliary space}. 
The vector space $V_\mathrm{phys}$ is called the \textit{physical space} and is associated with the vertical spectral line, which carries the spectral parameter $y$.
This jargon was established in the study of integrable spin chains \cite{Faddeev:1996iy} but, also in the context of integrable lattice models, it is helpful to view the spectral lines as individual vector spaces, which are ``scattered'' in the R-matrix.
The weights $a$, $b$, $c$ and $d$ are generalized to functions of the quotient of the spectral parameters.
These functions are constrained by requiring the central equation of integrability, the Yang-Baxter equation (YBE).
Diagrammatically, it can be represented by
\begin{equation}
\adjincludegraphics[valign=c,scale=1]{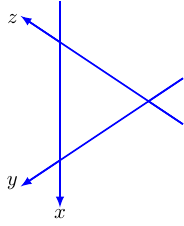}
~~~=~~~ 
\adjincludegraphics[valign=c,scale=1]{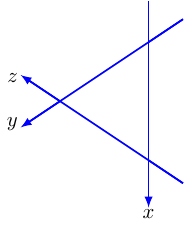} ~.
\label{eq:8V_YBE}
\end{equation}  
Algebraically, this is a matrix equation between three R-matrices on the l.\ h.\ s. and the r.\ h.\ s.\ each.
We can solve the Yang-Baxter equation for the weight-functions $a(x)$, $b(x)$, $c(x)$ and $d(x)$.
In practice, one considers \eqref{eq:8V_YBE} component by component by consecutively specifying the external arrow configurations.
For example, the external configuration 
\begin{equation}
\adjincludegraphics[valign=c,scale=0.8]{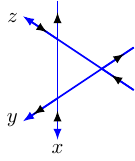}
~~~=~~~ 
\adjincludegraphics[valign=c,scale=0.8]{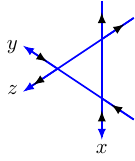} 
\label{eq:8V_YBE_config}
\end{equation}  
leads to the equation $b(\sfrac{y}{x}) a(\sfrac{z}{x}) c(\sfrac{y}{z}) + d(\sfrac{y}{x}) d(\sfrac{z}{x}) b(\sfrac{y}{z}) = a(\sfrac{y}{x}) b(\sfrac{z}{x}) c(\sfrac{y}{z}) + c(\sfrac{y}{x}) c(\sfrac{z}{x}) b(\sfrac{y}{z})$, when we sum over all possible internal arrow configurations and weight them according to \eqref{eq:8V_VertexWeights}.
Varying the external arrow configuration, we deduce five more equations of this type and together they enforce the weights $\left[a: b: c: d\right]$ to parameterize an elliptic curve in the projective model-space \cite{Baxter:1989ct}.
Jacobi theta functions $\vartheta_i (z \vert q)$, for $i=1,2,3,4$ and with the \textit{elliptic nome} $q$, provide an immersion of the elliptic curve into $\mathbb{CP}^3$ \cite{mumford2007tata}.
Therefore, it is natural to parameterize the weights with the help of theta functions, which are presented in appendix \ref{sec:ThetaFctEllipticGammaFctAndFreeEnergy8V}.
Additionally, we define theta values by $\vartheta_i := \vartheta_i (0 \vert q)$.
Furthermore, we employ the functions $H(u)$ and $\Theta (u)$, which are part of Jacobi's earlier convention for elliptic functions.
They are related to the theta functions by
\begin{equation}
\begin{array}{ccc}
H (u)
=
\vartheta_1
\left(
\frac{u}{\sqrt{\vartheta_3}} \vert q
\right) &
\mathrm{and} &
\Theta (u)
=
\vartheta_4
\left(
\frac{u}{\sqrt{\vartheta_3}} \vert q
\right) ~,
\end{array}
\label{eq:Def_HTheta}
\end{equation}
where the first argument is rescaled by the third theta value.
The functions are antisymmetric and symmetric, $H (-u) = - H(u)$ and $\Theta (-u) = \Theta (u)$, respectively.
They satisfy useful addition formulas, which read \cite{Baxter:1989ct}
\begin{subequations}
\begin{align}
H (u - v) H (u + v) \Theta ( 0 )^2
=
H ( u )^2 \Theta ( v )^2
-
\Theta ( u )^2 H ( v )^2 ~, \\
\Theta (u - v) \Theta (u + v) \Theta ( 0 )^2
=
\Theta ( u )^2 \Theta ( v )^2
-
H ( u )^2 H ( v )^2 ~.
\end{align}\label{eq:HTheta_AdditionFormulas}%
\end{subequations}
Moreover, they are related to the Jacobian elliptic function by
\begin{equation}
\mathrm{sn} (u) 
=
\frac{1}{\sqrt{k}}
\cdot
\frac{H (u)}{\Theta (u)}
~~~~\mathrm{with} ~~~~
k
:=
\frac{\vartheta_2^2}{\vartheta_3^2}
\label{eq:Def_sn}
\end{equation}
and the dependence on the elliptic nome is completely implicit here.
The combination $k$ is known as the \textit{elliptic modulus}.

Employing these functions, the weights as a function of $x = \e^u$, are expressed as \cite{Baxter:1989ct}
\begin{equation}
\begin{array}{cc}
a
=
- \I \rho 
\cdot
\Theta (\I \eta) \,
H (\I (\eta - u)) \,
\Theta (\I u) ~, &
c
=
- \I \rho 
\cdot
H (\I \eta) \,
\Theta (\I (\eta - u)) \,
\Theta (\I u) ~, \\
b
=
- \I \rho 
\cdot
\Theta (\I \eta) \,
\Theta (\I (\eta - u)) \,
H (\I u) ~, &
d
=
\phantom{-}\I \rho 
\cdot
H (\I \eta) \,
H (\I (\eta - u)) \,
H (\I u) ~.
\end{array}
\label{eq:8V_weights_HTheta}
\end{equation}
Using the parametrization in terms of the additive spectral parameter $u$, the additive crossing parameter $\eta$ and the elliptic nome $q$, ensures that the point $\left[ a:b:c:d \right] \in \mathbb{CP}^3$ lies on the integrable elliptic curve. 
The parameter $\rho$ is an overall scale factor and reminds us that weights are considered projective.
Equivalently to \eqref{eq:8V_weights_HTheta}, we can represent the integrable elliptic curve in model-space as
\begin{equation}
\left[ a:b:c:d \right]
=
\left[
\I\, \mathrm{sn} ( \I (\eta - u) ) :
\I\, \mathrm{sn} ( \I u ) :
\I\, \mathrm{sn} ( \I \eta ) :
- \I\, k\cdot \mathrm{sn} ( \I \eta ) \mathrm{sn} ( \I u ) \mathrm{sn} ( \I (\eta - u) ) 
\right] ~.
\label{eq:8V_weights_sn}
\end{equation}
If one is interested in a particular model, one has to specify the numerical values for $a$, $b$, $c$ and $d$.
Finding the right parameters $u$, $\eta$ and $q$ means inverting the elliptic functions in \eqref{eq:8V_weights_sn}, which might be cumbersome and potentially even impossible, if it is not integrable.

However, we want to connect to the six-vertex model, which corresponds to $d=0$.
Therefore, the parameterization \eqref{eq:8V_weights_sn} is useful, since we can study the limit $k \rightarrow 0$, which is equivalent to $q\rightarrow 0$, to obtain the six-vertex model when $\Delta > 1$ or $-1 > \Delta$ for $\Delta = \frac{a^2 + b^2 - c^2}{2 a b}$.
This is the ferroelectric and antiferroelectric phase of the six-vertex model.
The remaining disordered phase $1 > \Delta > -1$ is obtained in the limit $k \rightarrow 1$, which is equivalent to $q \rightarrow 1 $.
As it is common for q-functions like the theta functions, evaluation at special values of $q$ should be performed with care.
In the limit $q \rightarrow 0$, it holds that the elliptic function turns into a hyperbolic one \cite{Baxter:1989ct}: $\mathrm{sn} (u) \rightarrow \mathrm{sinh} (u)$.
Applying the limit to the eight-vertex model weights \eqref{eq:8V_weights_sn} gives the six-vertex model weights 
\begin{equation}
\left[ a:b:c \right]
=
\left[
p\, x - p^{-1}x^{-1} :
x - x^{-1} :
p - p^{-1} 
\right] ~,
\label{eq:6V_WeightsParameterization}
\end{equation}
which we expressed in multiplicative spectral- and crossing parameter, $x=\e^u$ and $p = \e^{-\eta}$, respectively.
Thus, the six-vertex model is called integrable if the weights describe a point $\left[a: b: c\right] \in \mathbb{CP}^2$ on the curve parameterized by $x$, $p$ according to \eqref{eq:6V_WeightsParameterization}.

An important consequence of the YBE is unitarity.
It amounts to the relation 
\begin{equation}
\adjincludegraphics[valign=c,scale=1.2]{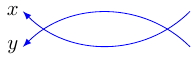} 
~~~= 
\left[
a(\tfrac{x}{y}) a(\tfrac{y}{x}) 
+
d(\tfrac{x}{y}) d(\tfrac{y}{x}) 
\right]
\cdot
\adjincludegraphics[valign=c,scale=1.2]{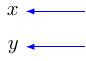}
~~~= 
w(\tfrac{x}{y}) w(\tfrac{y}{x}) 
\cdot
\mathbb{1} ~,
\label{eq:6V_Unitarity}
\end{equation}
which the R-matrix \eqref{eq:8V_Rmatrix} satisfies.
We use the abbreviation 
\begin{equation}
w(x)
:=
- \I \rho
\cdot
\Theta (0)
\Theta (\I (\eta + u) )
H (\I (\eta + u) )
\stackrel{d=0}{\rightarrow}
a(x)\vert_{d=0}
=
p\, x - p^{-1}x^{-1}
\label{eq:8V_unitarity_seed}
\end{equation}
with, again, $x=\e^u$ and $p = \e^{-\eta}$ and $\mathbb{1}$ is the four-times-four unit matrix.
The expression for $w(x)$ and \eqref{eq:6V_Unitarity} is obtained using the addition formulas \eqref{eq:HTheta_AdditionFormulas}.
Furthermore, the R-matrix \eqref{eq:8V_Rmatrix} has another useful property.
When setting the spectral parameter of the R-matrix to $1$, which is equivalent to $u=0$ in \eqref{eq:8V_weights_HTheta} and \eqref{eq:8V_weights_sn}, we observe that $a(1) = c$ and $b(1) = 0$.
This follows from $H(0) = 0$ and $\mathrm{sn} (0) = 0$, which is a consequence of \eqref{eq:Def_thetafunctions}, \eqref{eq:Def_HTheta} and \eqref{eq:Def_sn}.
Graphically, we can display this evaluation as
\begin{equation}
R(1)
=
c\cdot 
\begin{pmatrix}
1 & & & \\
& & 1 & \\
& 1 & & \\
& & & 1 
\end{pmatrix}
=
c\cdot \mathbb{P}
=
c \cdot \adjincludegraphics[valign=c,scale=1]{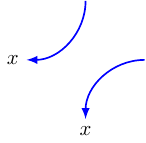} ~,
\end{equation}
and the R-matrix permutes the vertical with the horizontal spectral line.
Its action is the one of a permutation matrix $\mathbb{P}$.

\section{Toroidal boundary conditions}
\label{sec:ToroidalBoundaryConditions}
The integrability of the eight-vertex model is studied in section \ref{sec:8Vmodel} on a local level, where we considered a restricted patch of the full lattice.
To complete the definition of the partition function \eqref{eq:8VPatitionFunction}, we have to fix the boundary conditions and the shape of the lattice.
Here, we restrict the lattice to lie on a torus with periodic boundary conditions in both directions of the torus.
The example of a $3\times 3$ toroidal lattice is displayed in fig.\ \ref{fig:SquareIce6V} at the beginning of the introduction.
When we consider a doubly-periodic lattice with toroidal boundary conditions and $M$ horizontal- and $N$ vertical lines, we can express the partition function by a homogeneous periodic transfer matrix as 
\begin{equation}
Z_{MN}(a,b,c,d)
=
\mathrm{tr}
\left[
T_N( z )^M
\right] 
=
\sum_i
\Lambda_{N,i} (z)^M
~,
\label{eq:8V_partitionfunction_ev}
\end{equation}
where ``$\mathrm{tr}$'' means the trace over the total physical space $V_\mathrm{phys}^{\otimes N}$.
Graphically, it identifies the vertical spectral lines and $z$ is the ratio of the spectral parameters carried by the horizontal and vertical lines, $z = \sfrac{x}{y}$.
The homogeneous periodic transfer matrix is a special case of the transfer matrix \eqref{eq:8V_TransfeMatrix_periodic} below, where all vertical spectral parameters are set to the same value, $\forall j =1,...,N : ~ y_j = y$.
The partition function is related to the weights $a$, $b$, $c$ and $d$ by \eqref{eq:8V_weights_HTheta} or \eqref{eq:8V_weights_sn} by $z=\e^u$ and the dependence of the transfer matrix on the global crossing parameter $\eta$ and the elliptic nome $q$ is implicit.
The second equality in \eqref{eq:8V_partitionfunction_ev} uses the fact that the trace is a sum over the eigenvalues $\Lambda_{N,i}$ of the transfer matrix.

\subsection{Commuting transfer matrices}
After studying the implications of the YBE \eqref{eq:8V_YBE} on the weights, we turn to another important consequence: the existence of commuting transfer matrices.
To be more specific, we consider periodic transfer matrices, in contrast to open ones, which appear in the case of boundary integrability in \eqref{eq:6V_CommutingRows}.
Periodic transfer matrices are a sensible object to study when at least one of the directions of the lattice $\Lambda$ is periodic.
The inhomogeneous transfer matrix is defined as
\begin{equation}
\begin{split}
T_N( \lbrace \tfrac{x}{y_j} \rbrace)
&=
\mathrm{Tr}
\left[
R (\tfrac{x}{y_1})
R (\tfrac{x}{y_2})
R (\tfrac{x}{y_3})
\cdots
R (\tfrac{x}{y_N})
\right] \\
&=
\adjincludegraphics[valign=c,scale=1]{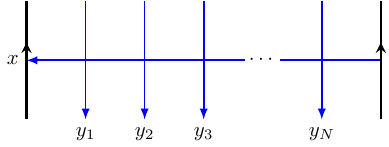}  ~.
\end{split}
\label{eq:8V_TransfeMatrix_periodic}
\end{equation}
We denote that the product between the R-matrices $R(\sfrac{x}{y_j}) \in \mathrm{End} ( V_\mathrm{aux} \otimes V_{\mathrm{phys},j} )$ from \eqref{eq:8V_Rmatrix} is not the ordinary matrix product between matrices of the form \eqref{eq:8V_Rmatrix}, but rather the product in the common auxiliary space $V_\mathrm{aux}$.
The trace ``Tr'' refers to the trace over the auxiliary space $V_\mathrm{aux}$, which is indicated by periodical identification of the horizontal spectral line.
As usual, we draw thick black lines with an arrow to indicate that we should identify the sides with this marking.
Furthermore, we use the shorthand notation $\lbrace y_j \rbrace$ for the set of all $y_j$, $\lbrace y_j \rbrace_{j=1,...,N}$.

Next, we show the commutativity of two transfer matrices $T_N( \lbrace \sfrac{x_1}{y_j} \rbrace)$ and $T_N( \lbrace \sfrac{x_2}{y_j} \rbrace)$ with respect to the product on $\otimes_{j=1}^N V_{\mathrm{phys},j}$, which identifies the spectral lines corresponding to the $y_j$'s.
We denote it by $\circ$.
We employ a standard technique of integrability, known as the \textit{train track argument} \cite{Faddeev:1996iy}.
Generally, it is an immediate implication of the existence of a YBE and an unitarity relation.
For the eight-vertex model, the two equations are the YBE \eqref{eq:8V_YBE} and the unitarity relation \eqref{eq:6V_Unitarity}.
We present the train track argument in a graphical form,
\begin{subequations}
{\allowdisplaybreaks
\begin{align}
T_N( \lbrace \tfrac{x_1}{y_j} \rbrace)
\circ
T_N( \lbrace \tfrac{x_2}{y_j} \rbrace)
&=
\adjincludegraphics[valign=c,scale=1]{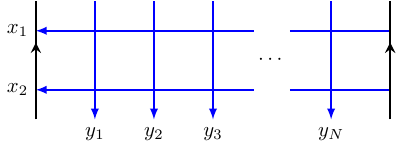} \\
& \stackrel{\eqref{eq:6V_Unitarity}}{=}
\adjincludegraphics[valign=c,scale=1]{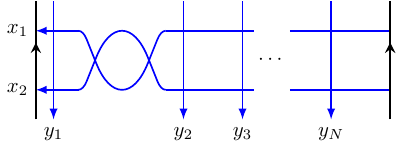} 
\cdot \frac{1}{w(\tfrac{x_1}{x_2}) w(\tfrac{x_2}{x_1})} \label{eq:8V_periodicTransferMatrix_comm1}\\
& \stackrel{\eqref{eq:8V_YBE}}{=}
\adjincludegraphics[valign=c,scale=1]{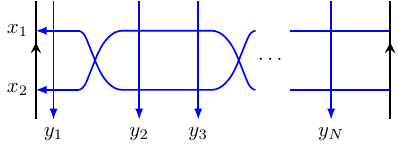}
\cdot \frac{1}{w(\tfrac{x_1}{x_2}) w(\tfrac{x_2}{x_1})} \\
& \stackrel{\eqref{eq:8V_YBE}}{=}
\adjincludegraphics[valign=c,scale=1]{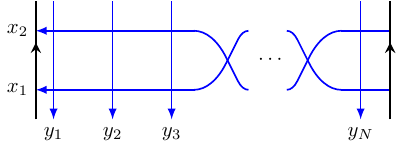} 
\cdot \frac{1}{w(\tfrac{x_1}{x_2}) w(\tfrac{x_2}{x_1})} \\
& \stackrel{\eqref{eq:8V_YBE}}{=}
\adjincludegraphics[valign=c,scale=1]{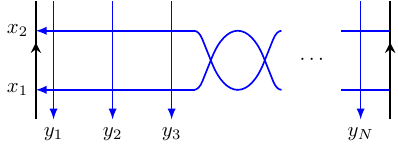}
\cdot \frac{1}{w(\tfrac{x_1}{x_2}) w(\tfrac{x_2}{x_1})} \\
& \stackrel{\eqref{eq:6V_Unitarity}}{=}
\adjincludegraphics[valign=c,scale=1]{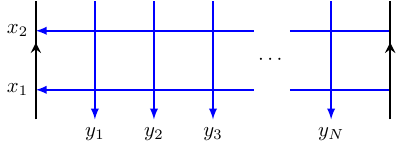} \\
& =
T_N( \lbrace \tfrac{x_2}{y_j} \rbrace)
\circ
T_N( \lbrace \tfrac{x_1}{y_j} \rbrace) ~,
\end{align}\label{eq:8V_periodicTransferMatrix_comm}%
}%
\end{subequations}
which implies $\com{T_N( \lbrace \sfrac{x_1}{y_j} \rbrace)}{T_N( \lbrace \sfrac{x_2}{y_j} \rbrace)} = 0$.
In \eqref{eq:8V_periodicTransferMatrix_comm1} we have inserted two intertwining R-matrices by using the unitarity relation \eqref{eq:6V_Unitarity} from right to left.
The next steps show how each of the R-matrices is transported in opposite directions along the product of the two transfer matrices using the YBE \eqref{eq:8V_YBE}.
Since it is periodic, the intertwiners will meet after they have collectively passed all the vertical spectral lines and annihilate again by unitarity \eqref{eq:6V_Unitarity}.
In the process, the intertwiners have exchanged the vertical positions of the two horizontal spectral lines, and the product of two transfer matrices is obtained with the orders of the factors interchanged however.

Let us turn to the homogeneous case where we identify all spectral parameters of the vertical lines, $\forall j=1,...,N: y_j = y$.
Then, we denote the homogeneous transfer matrix as $T_N( z )$ with $z = \sfrac{x}{y}$ and we can expand it in terms of the argument $z$, 
\begin{equation}
T_N( z )
=
\sum_{n=0}^\infty
Q_n z^n ~.
\end{equation} 
The expansion coefficients $Q_n$ are operator-valued, and the commutation \eqref{eq:8V_periodicTransferMatrix_comm} implies that they form a commuting set, $\com{Q_n}{Q_m} = 0$.
It gives rise to an infinite set of conservation laws and the construction of functionally-independent conserved charges, proving the integrability of the eight-vertex model.
% https://home.itp.ac.ru/~lashkevi/lectures/ariel21/lec02-talk.pdf claims the conserved charges are not just the coefficients of T but T(0)^{-1} T ( z )
In principle, the conservation laws can be used to determine the observables of the model.
Thanks to the integrability of the eight-vertex model, it is possible to use powerful techniques like the Bethe ansatz to determine the eigenvalues of the transfer matrix and access the full partition function, even though it might still be computationally demanding \cite{Faddeev:1996iy}.

\subsection{Free energy in the thermodynamic limit}
\label{subsec:FreeEnergyInThermoDynLimit_8V}
Applications like the entropy of square ice, c.\,f.\ fig.\ \ref{fig:SquareIce6V}, motivate the limit where the dimensions of the lattice become infinite.
In this limit, the partition function is called free energy and it is related to the residual entropy of the model.
We write 
\begin{equation}
\kappa (a,b,c,d)
=
\lim_{M,N \rightarrow \infty}
Z_{MN}(a,b,c,d)^{\frac{1}{MN}}
\label{eq:8V_partitionfunction_thermodynamicLimit}
\end{equation}
and in the case of the integrable eight vertex model we also use the compact notation $\kappa (u)$, which is implicitly dependent on the crossing parameter $\eta$ and the nome $q$.
Baxter \cite{baxter1971eight,Baxter:1989ct} was the first who computed the free energy exactly using integrability.
We aim at the same goal by using the method of inversion relations \cite{StroganovInvRelLatticeModels}.
Therein, one starts by deriving two functional relations for $\kappa (u)$.
The first one we can find one immediately by considering the function $\kappa (\eta - u)$.
On the level of the weights \eqref{eq:8V_weights_sn} in the R-matrix elements \eqref{eq:8V_Rmatrix}, the operation exchanges $a$ with $b$.
We realize that rotating the whole lattice by $90$ degrees has the same effect, by looking at the vertices \eqref{eq:8V_VertexWeights}.
The other weights and vertices remain invariant.
The partition function \eqref{eq:8V_partitionfunction_ev}, and consequently its thermodynamic limit \eqref{eq:8V_partitionfunction_thermodynamicLimit}, should remain invariant under the crossing transformation $u \rightarrow \eta - u$ and we find the functional relation $\kappa (\eta - u) = \kappa (u)$.
We present the second relation ad-hoc by requiring that $\kappa (u)$ should inherit the unitarity properties of the R-matrix \eqref{eq:6V_Unitarity} and satisfy $\kappa (u) \kappa (-u) = \mathrm{w} (u) \mathrm{w} (-u)$.
Thereby, we denote $\mathrm{w}(u) = w (\e^u)$ and $w(x)$ is the function in \eqref{eq:8V_unitarity_seed}.
Therefore, the inversion relations to solve are
\begin{equation}
\kappa (\eta - u)
=
\kappa (u)
~~~~ \mathrm{and} ~~~~
\kappa (u) \,
\kappa (- u)
=
\mathrm{w} (u) \mathrm{w} (-u) ~.
\label{eq:InvRel_8V}
\end{equation}
In appendix \ref{sec:ThetaFctEllipticGammaFctAndFreeEnergy8V}, we solve them with the additional assumption that $\kappa (u)$ must not have a pole in the strip $\left[ 0, \eta \right)$.
We find Baxter's result \cite{Baxter:1989ct}, which can be expressed in terms of elliptic gamma functions \cite{Felder_2000} and reads\footnote{
To make contact with Baxter's original result or the representation in terms of elliptic gamma functions in \cite{Felder_2000}, one has to make the redefinitions $\mathrm{p} = x_\mathrm{B}^4$ and $\mathrm{x} = x_\mathrm{B} z_\mathrm{B}$.
}
\begin{equation}
\kappa (u) 
=
c
\cdot
\frac{
\Gamma^{(1)} (\mathrm{p}^\frac{1}{2} \vert q^2) \,
\Gamma^{(1)} (q \vert q^2)
}{
\Gamma^{(1)} (\mathrm{p}^\frac{1}{2}\, \mathrm{x}^{-1} \vert q^2) \,
\Gamma^{(1)} (\mathrm{x} \vert q^2)
}
\frac{
\Gamma^{(2)} (\mathrm{p} \, \mathrm{x}^{-1} \vert q , \mathrm{p}) \,
\Gamma^{(2)} (\mathrm{p}^\frac{1}{2} \mathrm{x} \vert q , \mathrm{p})
}{
\Gamma^{(2)} (\mathrm{p}^\frac{1}{2} \mathrm{x}^{-1} \vert q , \mathrm{p}) \,
\Gamma^{(2)} (\mathrm{x} \vert q , \mathrm{p})
} ~.
\label{eq:8V_freeEnergy_kappa}
\end{equation}
The elliptic gamma functions are defined in \eqref{eq:Def_ellipticGamma} according to the conventions of \cite{rains2007limits}.
The result \eqref{eq:8V_freeEnergy_kappa} corresponds to the weights \eqref{eq:8V_weights_HTheta}.
The multiplicative spectral parameter $\mathrm{x}$ and the crossing parameter $\mathrm{p}$ are therein redefined to include the theta value, which appears in the argument of $H (u)$ and $\Theta (u)$ in \eqref{eq:Def_HTheta}.
We have
\begin{equation}
\mathrm{x} 
=
\e^{ - \frac{2 u}{\sqrt{\vartheta_3}} }
~~~~ \mathrm{and} ~~~~
\mathrm{p} 
=
\e^{ - \frac{4 \eta}{\sqrt{\vartheta_3}} } ~.
\end{equation}
The free energy \eqref{eq:8V_freeEnergy_kappa} holds for the large parameter space of integrable eight-vertex models, however, as we are interested in the six-vertex model at $d=0$, we have to take the limit $q \rightarrow 0$ or $q \rightarrow 1$ carefully.
The behavior of the elliptic gamma functions in the limit $q \rightarrow 0$ is straightforward.
For $q \rightarrow 1 $, however, also the parameters $\mathrm{x}$ and $\mathrm{p}$ tend to one due to their dependence on $\vartheta_3$.
A careful evaluation of this limit is presented in \cite{rains2007limits} and in \cite{Baxter:1989ct} the values for $u$ and $\eta$ are given to land on the different phases of the six-vertex model.
In the disordered phase, which we obtain by $q \rightarrow 1$, one finds two important results of the free energy of the six-vertex model \cite{duminilcopin2022},
\begin{equation}
\kappa (1,1,1,0) 
=
\left(\frac{4}{3} \right)^{\frac{3}{2}} 
~~~~ \mathrm{and} ~~~~
\kappa (1,1,2,0) 
=
\left(\frac{1}{4} \frac{\Gamma (\frac{1}{4})}{ \Gamma (\frac{3}{4})} \right)^{2} ~. 
\end{equation}
The first result is the famous Lieb's constant \cite{Lieb:1967zz}, who first computed the residual entropy of water square ice.
The second result is interesting because it involves a ratio of gamma function values at $\tfrac{1}{4}$ and $\tfrac{3}{4}$.
In chapters \ref{chpt:NonSusyFishnetTheoriesAndItsRelatives} and \ref{chpt:SupersymmetricDoubleScaledDeformations}, we will encounter similar numbers as the critical coupling, after we adapted the calculation of the free energy to integrable quantum field theories.

\section{The box boundary condition}
\label{sec:TheBoxBoundaryCondition}
After focusing on periodic boundary conditions in section \ref{sec:ToroidalBoundaryConditions}, we consider another class of boundary conditions, the box.
Therefore, we have to introduce a wall where the spectral lines get redirected.
We study a rectangular square lattice bounded by walls to all sides, where pairs of spectral lines spawn and end at opposite sides.
We call this boundary condition the box boundary condition, and in this section we restrict ourselves to the six-vertex model $d=0$ with the bulk weights \eqref{eq:6V_WeightsParameterization}.
Graphically, the inhomogeneous partition function of the six vertex model on such a lattice with $2M$ horizontal- and $2N$ vertical spectral lines can be represented as
\begin{equation}
Z_{MN} (\lbrace x_i \rbrace_{i=1, ... , M} \vert \lbrace y_j \rbrace_{j=1, ... , N})
=
\adjincludegraphics[valign=c,scale=1]{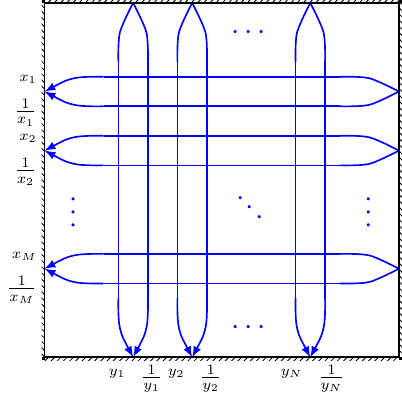} ~.
\label{eq:6V_box_MNPartitionFunction}
\end{equation}
For the sake of computing it, we need to specify another object appearing in the partition function, the K-matrix.
It encodes the boundary weights on the four walls surrounding the bulk, which consists of R-matrices of the type \eqref{eq:8V_Rmatrix}.
The components of the K-matrix are the four possible configurations of arrows when two spectral lines meet at the boundary walls.
As an example, the K-matrix with its four components for the left wall in \eqref{eq:6V_box_MNPartitionFunction} is
\begin{equation}
K_\mathrm{L} (x)
=
\adjincludegraphics[valign=c,scale=1.3]{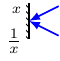} 
= \left( \begin{array}{cc}
\adjincludegraphics[valign=c,scale=1.3]{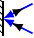} &
\adjincludegraphics[valign=c,scale=1.3]{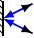} \\
\adjincludegraphics[valign=c,scale=1.3]{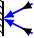} &
\adjincludegraphics[valign=c,scale=1.3]{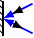}
\end{array}\right)
=
\begin{pmatrix}
k_\mathrm{L}^+ (x) & k_\mathrm{L}^r (x) \\
k_\mathrm{L}^l (x) & k_\mathrm{L}^- (x)
\end{pmatrix}
\end{equation}
and analogously for the other three boundaries.
Depending on the arrow configuration of the bounded lattice, the sixteen boundary weights $k_{\mathrm{L},\mathrm{R},\mathrm{U},\mathrm{D}}^{+,-,l,r}$ enter the partition function.
We would like to keep the model integrable, which means that we have to impose the boundary Yang-Baxter equation (bYBE) for all four K-matrices.
Again, taking the left wall as an example, the bYBE diagrammatically takes the form
\begin{equation}
\adjincludegraphics[valign=c,scale=1]{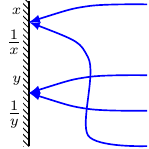}
\hspace{1cm} = \hspace{1cm}
\adjincludegraphics[valign=c,scale=1]{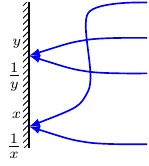} ~.
\label{eq:6V_bYBE}
\end{equation} 
Similar to the bulk case for the components of the R-matrix, the bYBE constrains the boundary weights.
We do not attempt to solve for the boundary weights in full generality, but in section \ref{subsec:ArrowReflectingBox} below we find solutions for some simplified, yet nontrivial case.

The bYBE is a necessary condition for integrability because in combination with the bulk YBE \eqref{eq:8V_YBE} it allows to prove the commutativity of transfer matrices, the so-called \textit{double-row transfer matrices} $T_N (x)$.
The derivation can be carried out graphically,
\begin{subequations}
{\allowdisplaybreaks
\begin{align}
T_N (x_1) \circ T_N (x_2)
\stackrel{ \phantom{\eqref{eq:6V_Unitarity}} }{=}
\adjincludegraphics[valign=c,scale=0.9]{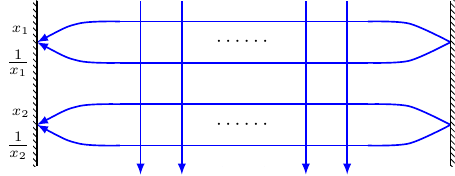} & \\
\stackrel{\eqref{eq:6V_Unitarity}}{=}
\adjincludegraphics[valign=c,scale=0.9]{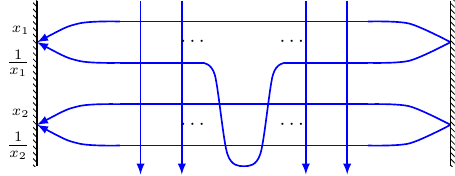} &
\cdot
\frac{1}{ W(x_1, x_2)}\\
\stackrel{\eqref{eq:8V_YBE}}{=}
\adjincludegraphics[valign=c,scale=0.9]{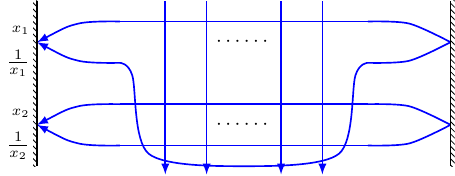} & 
\cdot
\frac{1}{ W(x_1, x_2)}\\
\stackrel{\eqref{eq:6V_bYBE}}{=}
\adjincludegraphics[valign=c,scale=0.9]{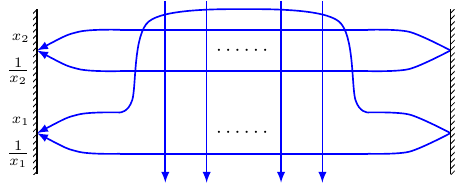} &
\cdot
\frac{1}{ W(x_1, x_2)}\\
\stackrel{\eqref{eq:8V_YBE}}{=}
\adjincludegraphics[valign=c,scale=0.9]{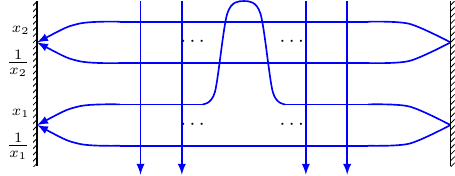} &
\cdot
\frac{1}{ W(x_1, x_2)}\\
\stackrel{\eqref{eq:6V_Unitarity}}{=}
\adjincludegraphics[valign=c,scale=0.9]{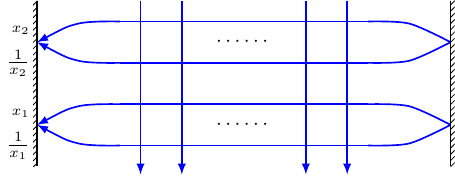} &
=
T_N (x_2) \circ T_N (x_1) ~,
\end{align}
\label{eq:6V_CommutingRows}%
}%
\end{subequations}
and the commutativity $ \left[ T_N (x_1) , T_N (x_2) \right] = 0$ follows.
We define the introduced factor related to unitarity as 
\begin{equation}
W(x,y) := a(xy) a(\tfrac{1}{xy}) a(\tfrac{x}{y}) a(\tfrac{y}{x})
\label{eq:6V_WFactor}
\end{equation}
and it is symmetric, i.\,e.\ $W(x,y) = W(y,x)$, and invariant under inversion of its arguments, i.\,e.\ $W(x,\tfrac{1}{y}) = W(x,y)$.
Similarly to \eqref{eq:6V_CommutingRows}, the same steps can be used to show the commutativity of double-column transfer matrices.
Furthermore, together, the commutativity of rows and columns implies that the box partition function \eqref{eq:6V_box_MNPartitionFunction} is a symmetric function under the exchange of horizontal and vertical spectral parameters, $\lbrace x_i \rbrace_{i=1, ... , M}$ and $\lbrace y_j \rbrace_{j=1, ... , N}$, respectively.

We will now turn to the box partition function and derive a recursion relation.
We will present a solution which is symmetric under the exchange of the horizontal and vertical spectral parameters.
The recursion relation holds for the case when one horizontal spectral parameter coincides with a vertical one, i.\,e.\ $x_m = y_n$.
Since the partition function is symmetric in both $\lbrace x_i \rbrace$ and $\lbrace y_j \rbrace$, the specific value of $m \in \lbrace 1 , ... , M \rbrace$ and $n \in \lbrace 1 , ... , N \rbrace$ is not important.
At the intersection of the two relevant pairs of spectral lines, we have
\begin{equation}
\adjincludegraphics[valign=c,scale=1]{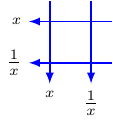} 
~~ = ~~
\adjincludegraphics[valign=c,scale=1]{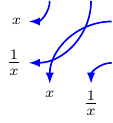} 
\cdot 
c^2 
~~ = ~~
\adjincludegraphics[valign=c,scale=1]{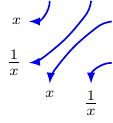} 
\cdot
c^2 \; a(x^2) a(\tfrac{1}{x^2}) ~.
\label{eq:6V_PP}
\end{equation}
Within the partition function, the spectral lines therefore disentangle at the position of the $m$-th row and the $n$-th column.
As an example, we can consider the $5\times 5$ partition function $Z_{5,5}$.
If we set $x_3 = y_3$, we obtain according to \eqref{eq:6V_PP}
\begin{equation}
Z_{5,5} (\lbrace x_i \rbrace \vert \lbrace y_j \rbrace)\vert_{x_3 = y_3}
=
\adjincludegraphics[valign=c,scale=0.8]{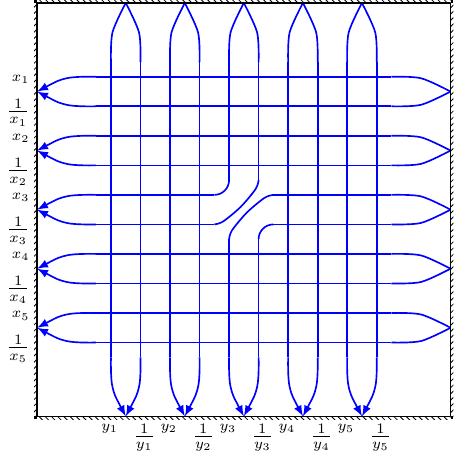}
\cdot c^2 \; a(x_3^2) a(\tfrac{1}{x_3^2})\vert_{x_3 = y_3}
\label{eq:6V_RecursionStep1}
\end{equation}
The next step towards the recursion relation is to use the YBE to move the $x_3 = y_3$ lines to the boundary.
It yields 
\begin{equation}
Z_{5,5} (\lbrace x_i \rbrace \vert \lbrace y_j \rbrace)\vert_{x_3 = y_3}
=
\adjincludegraphics[valign=c,scale=0.6]{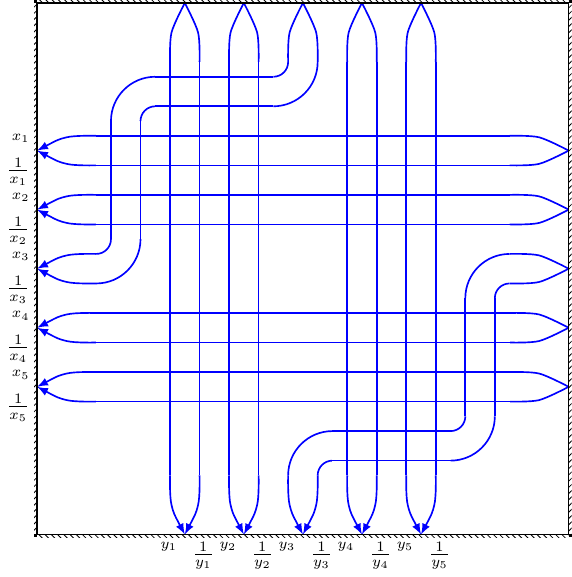} 
\cdot c^2 \; a(x_3^2) a(\tfrac{1}{x_3^2})\vert_{x_3 = y_3} ~.
\label{eq:6V_RecursionStep2}
\end{equation}
We wish to isolate the $x_3 = y_3$ dependent K-matrices in the top-left and bottom-right corner of the box.
Therefore, we establish an auxiliary relation.
Let us illustrate them at the example of the left wall:
\begin{equation}
\adjincludegraphics[valign=c,scale=1]{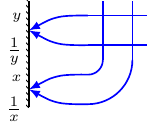} 
=
\adjincludegraphics[valign=c,scale=1]{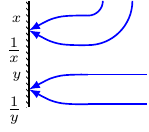} 
\cdot
W (x,y)
\label{eq:6V_bMove}
\end{equation}
The relation holds thanks to the bYBE \eqref{eq:6V_bYBE} and unitarity \eqref{eq:6V_Unitarity}.
Similar relations also hold for the top, right, and bottom walls, all of them have the same proportionality factor $W(x,y)$ with $x$ and $y$ being the spectral parameters of the two involved K-matrices.

Next, we apply these auxiliary relations to \eqref{eq:6V_RecursionStep2} to move the $x_3$ and $y_3$ lines to the corners and decouple them from the remaining lattice.
We obtain 
\begin{equation}
\begin{split}
Z_{5,5} (\lbrace x_i \rbrace & \vert \lbrace y_j \rbrace)\vert_{x_3 = y_3}
= 
\adjincludegraphics[valign=c,scale=0.6]{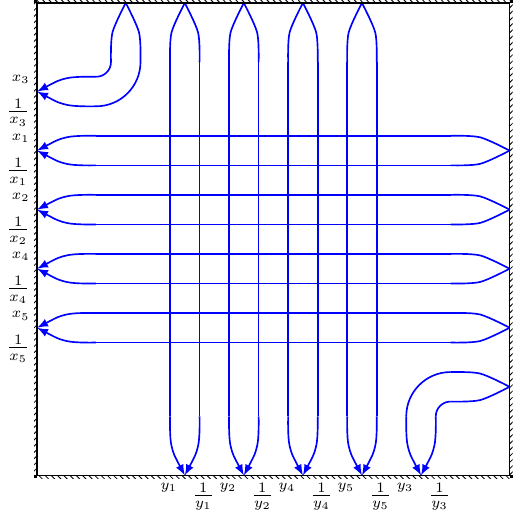}
\cdot c^2 \; a(x_3^2) a(\tfrac{1}{x_3^2}) \\
&\cdot
\left[
\prod_{i = 1}^{2} W (x_3, x_i)
\right]
\left[
\prod_{j = 1}^{2} W (x_3, y_j)
\right]
\cdot
\left[
\prod_{i = 4}^{5} W (y_3, x_i)
\right]
\left[
\prod_{j = 4}^{5} W (y_3, y_j)
\right]
\vert_{x_3 = y_3} ~.
\label{eq:6V_RecursionStep2}
\end{split}
\end{equation}
We observe that the lattice splits into three unrelated parts, two traces of two K-matrices in the top-left and bottom-right corners of the box, and an ordinary inhomogeneous $4\times 4$ partition function.
The trace depends on the integrable choice of boundary weights and we give an explicit expression below.
We will use the abbreviations
\begin{equation}
\begin{array}{cc}
F^{\mathrm{LU}} (x)
:=
\adjincludegraphics[valign=c,scale=1]{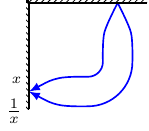} ~,
&
F^{\mathrm{DR}} (y)
:=
\adjincludegraphics[valign=c,scale=1]{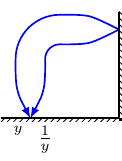} 
\end{array}
\label{eq:6V_bTraceAuxiliaryFunctions}
\end{equation}
and obtain the recursion relation for our $5\times 5$ example,
\begin{equation}
\begin{split}
Z_{5,5} (\lbrace x_i \rbrace \vert \lbrace y_j \rbrace)\vert_{x_3 = y_3}
~=~ &
c^2 \; a(x_3^2) a(\tfrac{1}{x_3^2})
\cdot
\left[
\prod_{\substack{i = 1\\ i\neq 3}}^{5} W (x_3, x_i)
\right]
\left[
\prod_{\substack{j = 1\\ j\neq 3}}^{5} W (x_3, y_j)
\right]
F^{\mathrm{LU}} (x_3)
F^{\mathrm{DR}} (x_3) \\
&\cdot 
Z_{4,4} (\lbrace x_i \rbrace_{\substack{i=1, ... , 5 \\ i\neq 3} } \vert \lbrace y_j \rbrace_{\substack{j=1, ... , 5 \\ j\neq 3}})
~.
\end{split}
\end{equation}
The generalization to the general $M\times N$ case, where $x_m$ and $y_n$ are identified, works analogously.
The recursion relation for the box boundary condition partition function is 
\begin{equation}
\begin{split}
Z_{M,N} (\lbrace x_i \rbrace \vert \lbrace y_j \rbrace)\vert_{x_m = y_n}
~=~ &
c^2 \; a(x_m^2) a(\tfrac{1}{x_m^2})
\cdot
\left[
\prod_{\substack{i = 1\\ i\neq m}}^{M} W (x_m, x_i)
\right]
\left[
\prod_{\substack{j = 1\\ j\neq n}}^{N} W (x_m, y_j)
\right]
F^{\mathrm{LU}} (x_m)
F^{\mathrm{DR}} (x_m) \\
&\cdot 
Z_{M-1,N-1} (\lbrace x_i \rbrace_{\substack{i=1, ... , M \\ i\neq m} } \vert \lbrace y_j \rbrace_{\substack{j=1, ... , N \\ j\neq n}})
~.
\end{split}
\end{equation}
Next to the symmetry in the horizontal and vertical spectral parameters, the recursion relation is the second requirement for a solution of the partition function.
Inspired by the partition function of the domain-wall boundary condition of the six-vertex model \cite{izergin1987partition,AGIzergin_1992}, which satisfies Korepin's recursion relation \cite{korepin1982calculation} and similar symmetry properties, we propose the solution for the box partition function in the square case $M=N$,
\begin{equation}
    \begin{split}
    Z_M 
    \left( 
    \lbrace x_i \rbrace %_{i=1, ..., M}
    \vert
    \lbrace y_i \rbrace %_{i=1, ..., M}
    \right)
    =
    \frac{
        \prod_{i,j=1}^M
        \left(
            \frac{x_i}{y_j} - \frac{y_j}{x_i}
        \right)
        W(x_i, y_j)
    }{
        \prod_{1\leq i < j \leq M}
        \left(
        \frac{x_j}{x_i} - \frac{x_i}{x_j}
        \right)
        \left(
        \frac{y_i}{y_j} - \frac{y_j}{y_i}
        \right)
    }
    \cdot
    \underset{1 \leq i,j \leq M}{\mathrm{det}}
    \left[
        \frac{
			c^2 
			a(x_i y_j) a(\frac{1}{x_i y_j}) 
			F^{\mathrm{LU}} (x_i)
			F^{\mathrm{DR}} (y_j)         
        }{
        \left(
            \frac{x_j}{y_i} - \frac{y_i}{x_j}
        \right)
        W (x_i , y_j)
        }
    \right] ~.
    \end{split}
\label{eq:6V_BoxPartitionFunctionSquareSolution}
\end{equation}
The function $W(x,y)$ is given in \eqref{eq:6V_WFactor} and the functions $F^{\mathrm{LU}}$ and $F^{\mathrm{DR}}$ will be determined in the following, based on a particular solution of the bYBE.

\subsection{Arrow-reflecting box}
\label{subsec:ArrowReflectingBox}
If we require the wall to maintain the ice rule, i.\,e.\ the number of arrows pointing to the walls is the same as the number of arrows pointing out, we assume that the K-matrices at all four walls are diagonal.
We have to solve the bYBE \eqref{eq:6V_bYBE} for the left wall, as well as the bYBE for the three remaining walls to determine the nonzero boundary weights $k_{\mathrm{L},\mathrm{R},\mathrm{U},\mathrm{D}}^{+,-,l,r}$.
Illustrated along the $1\times 1$ partition function, the K-matrices with the solutions of the bYBE are 
\begin{equation}
\adjincludegraphics[valign=c,scale=1]{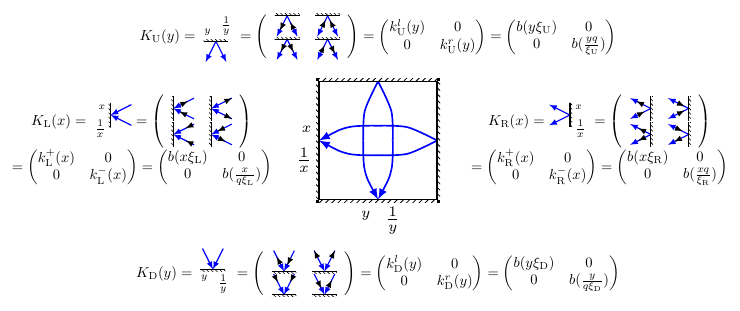}
\label{eq:6V_ArrowReflectingKMatrices} 
\end{equation}
where the boundary weights contain four free boundary parameters, $\xi_\mathrm{L}$, $\xi_\mathrm{U}$, $\xi_\mathrm{R}$, and $\xi_\mathrm{D}$, one for each wall.
The K-matrices shown in \eqref{eq:6V_ArrowReflectingKMatrices} are related to Sklyanin's solution of the bYBE.

To determine the partition function \eqref{eq:6V_BoxPartitionFunctionSquareSolution}, we need to compute the trace-functions \eqref{eq:6V_bTraceAuxiliaryFunctions}.
We find them by summing over all possible configurations,
\begin{subequations}
\begin{align}
F^\mathrm{LU} (x)
&=
\adjincludegraphics[valign=c,scale=1]{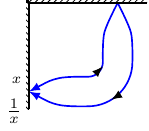} 
+
\adjincludegraphics[valign=c,scale=1]{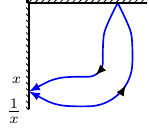} 
=
b(x \xi_\mathrm{L}) b(\tfrac{xq}{\xi_\mathrm{U}})
+
b(\tfrac{x}{q \xi_\mathrm{L}}) b(x \xi_\mathrm{U}) ~, \\
F^\mathrm{DR} (y)
&=
b(y \xi_\mathrm{D}) b(\tfrac{yq}{\xi_\mathrm{R}})
+
b(\tfrac{y}{q \xi_\mathrm{D}}) b(y \xi_\mathrm{R}) ~.
\end{align}
\end{subequations}
It completes the conjectured form of the partition function \eqref{eq:6V_BoxPartitionFunctionSquareSolution} for the arrow-reflecting walls.
It will also be interesting to study other possible boundary configurations such as walls, which source or swallow the arrows.
This is left for future explorations, and in the next chapter, we turn to the study of integrable Feynman diagrams.
However, in chapter \ref{chpt:BoundaryIntegrabilityInFeynmanGraphs} we return to the important concept of boundary integrability and construct boundary-integrable Feynman graphs.

%%%----------------------------------------------------------------------------------------
%%%	THESIS CONTENT - PART 3 - Auxiliary relations
%%%----------------------------------------------------------------------------------------
\chapter{Feynman graphs as an integrable model}
\label{chpt:AuxiliaryRelations}

Zamolodchikov's fundamental insight \cite{Zamolodchikov:1980mb} was to interpret propagators in quantum field theory (QFT) as weights, similar to statistical lattice models.
A priori, this raises the question of whether a star-triangle relation (STR) exists and how the Yang-Baxter equation (YBE) and integrability manifest themselves in Feynman graphs.
In \cite{Zamolodchikov:1980mb} it was shown that the propagators can be generalized to satisfy a STR and the research on statistical edge-interaction models, see e.\,g.\ \cite{Bazhanov:2011mz}, established a procedure to fuse four propagators into an R-matrix, which satisfies a YBE.
In this chapter, we will see that one can associate generalizations of Feynman graphs to specific configurations of spectral lines which satisfy a YBE and, therefore, describe an integrable model \cite{Zamolodchikov:1980mb,Bazhanov:2016ajm,Kazakov:2022dbd}.
The generalization of the Feynman graphs is parameterized by spectral parameters, which have to be fixed to particular values such that the map between the Feynman graphs and the spectral lines corresponds to a specific QFT.

Moreover, we will find that the integrability condition, the YBE, is a consequence of the star-triangle relation, which only holds for Feynman graphs invariant under conformal transformations.
Hence, the QFT producing the integrable Feynman graphs in its perturbative expansion has to be a conformal field theory (CFT).
An important special case of the star-triangle relation for integrable QFTs are the so-called chain relations.
They are related to unitarity of the generalized propagators and are a crucial tool for gaining the benefits of a QFT's integrable structure.

Furthermore, we will derive another auxiliary relation for generalized Feynman graphs, the so-called x-unity relation \cite{Kade:2023xet,Kade:2024ucz,Kade:2024lkc}.
It is the pivotal relation for the calculation, which yields the critical coupling of the integrable QFT under examination.
The x-unity relation has the form of a so-called star-square relation, which played an important role in the usage of integrability in various statistical models \cite{Jungling_1974,Jungling_1975} and the gauge/YBE correspondence \cite{Mullahasanoglu:2023nes,Catak:2024ygo}.

This chapter aims at introducing the generalized scalar QFT propagator as a lattice model weight by showing its relation to spectral lines. Then, we will establish the chain relations, the star-triangle relation, and the x-unity relation, as well as the R-matrix solving a YBE.
In two subsequent sections, we will repeat the derivation for fermionic generalized propagators and the superspace propagators, respectively.
However, in the latter case, a star-triangle relation is still speculative.

In this thesis, we work in Euclidean space $\mathbb{R}^D$ instead of Minkowski spacetime $\mathbb{R}^{1,D-1}$.
The reason is that the two spaces are related by a Wick rotation, which complexifies the time coordinate, and one uses the $\I \varepsilon$-prescription for propagators to ensure time-ordered products and causality.
It is a common lore that the Wick rotation allows for the translation of Euclidean results into Minkowski spacetime, however, the precise procedure is subtle \cite{Corcoran:2020akn}.
Especially in the context of supersymmetry, switching between Euclidean and Minkowski space is non-trivial, since supersymmetry in Euclidean space is very different from Minkowski space due to the existence or non-existence of Majorana-Weyl spinors \cite{Nicolai:1978vc}.
Here, we use the Euclidean metric for non-supersymmetric theories, while the Clifford algebra of Minkowski space is used for supersymmetric models, see appendix \ref{app:SuperspaceNotations}.
However, we still use the term ``spacetime'' for Euclidean space.

\section{The generalized scalar propagator as a lattice weight}
\label{sec:TheGeneralizedScalarPropagatorAsLatticeWeight}
The propagator of a complex scalar field is derived from the kinetic part of the action of a QFT, which is in $D$ dimensions $\int \dd^D x ~ \phi \square \phi^*$, with $\square := \partial^\mu \partial_\mu$.
It is the inverse of the kinetic operator $\delta^{(D)} (x_{12}) \square_2$, which can be read off from the kinetic term as
\begin{equation}
\int \dd^D x ~ \phi \square \phi^*
=
\int \dd^D x_1 \dd^D x_2 ~ \phi (x_1) \delta^{(D)} (x_{12}) \square_2 \phi^*(x_2) ~,
\label{eq:ScalarUndefKinTerm}
\end{equation}
where the abbreviations $x_{ij} := x_i - x_j$ and $\square_i := \partial_i^\mu \partial_{i,\mu}$ with $\partial_{i,\mu} = \frac{\partial}{\partial x_i^\mu}$ are used.

The emergence of a spectral parameter $v$ can be motivated by a non-local deformation of the kinetic term, which is $\int \dd^D x ~ \phi \square^v \phi^*$.
Note that in the limit $v \rightarrow 1$, we recover the undeformed kinetic term \eqref{eq:ScalarUndefKinTerm}.
The corresponding kinetic operator is $\delta^{(D)} (x_{12}) \square^v_2$ and its functional inverse is the propagator $G_v (x_{12}) = \left\langle \phi(x_1) \phi(x_2)^* \right\rangle$ that we regard as a weight of an integrable lattice model,
\begin{equation}
\int \dd^D x_2 ~
\delta^{(D)} (x_{12}) \square^v_2
G_v (x_{23})
=
\delta^{(D)} (x_{13}) ~.
\label{eq:GenPropagatorInverseOfKinOperator}
\end{equation}
Equivalently, performing the integration via the delta function yields $\square^v_1
G_v (x_{13}) = \delta^{(D)} (x_{13})$, which can be solved via Fourier transform.
The generalized propagator is related to its Fourier transform $\tilde{G}_v (p)$ as
\begin{equation}
G_v (x) 
=
\frac{1}{(2\pi)^{D/2}}
\int \dd^D p ~
\e^{- \I p x}\;
\tilde{G}_v (p)
\label{eq:GenPropagatorFourierTrafo}
\end{equation}
and the delta function is represented by $\delta^{(D)} (x) = (2\pi)^{- D/2} \int \dd^D p ~ \e^{- \I p x}$.
In Fourier space, the derivatives are algebraic, and thus \eqref{eq:GenPropagatorInverseOfKinOperator} takes the form $(-p^2)^v \tilde{G}_v (p) = 1$, which is equivalent to $\tilde{G}_v (p) = (- p^2)^{-v}$.
We find, following \eqref{eq:GenPropagatorFourierTrafo} and performing the Gaussian integral after usage of the Schwinger trick \eqref{eq:SchwingerTrick}, the generalized propagator in coordinate space to be 
\begin{equation}
G_v (x) 
=
\frac{1}{(2\pi)^{D/2}}
\int \dd^D p ~
\e^{- \I p x}
\left[
\frac{-1}{p^2}
\right]^v
=
- (-1)^v
2^{D/2 - 2 v}
\frac{\Gamma (\frac{D}{2} - v)}{\Gamma (v)}
\frac{1}{\left[ x^2 \right]^{D/2 - v}} ~.
\label{eq:GenPropagatorGWeight}
\end{equation}
The expression consists of a factor depending on the spectral parameter, which contains a unique combination of $\Gamma$-functions.
It will be called $a_0 (v)$, that is a special case of the abbreviation
\begin{equation}
a_\ell (v) 
~ := ~
\frac{\Gamma (\frac{D}{2} - v + \ell)}{\Gamma (v + \ell)} ~,
\label{eq:Factor_aEll}
\end{equation}
which is a factor we will encounter ample times in the following and which satisfies $a_\ell (v) a_\ell (\frac{D}{2} - v) = 1$.
Eventually, one considers Feynman graphs and for each of them the propagators' factors, which are not spacetime dependent, can be collectively multiplied in the end. 
We call them $c_0 (v)$ and define the abbreviation 
\begin{equation}
c_\ell (v)
~ := ~
- (-1)^v 2^{D/2 - 2 v} a_\ell (v)
\label{eq:Factor_cEll}
\end{equation}
for later convenience.
Note that the abbreviations \eqref{eq:Factor_aEll} and \eqref{eq:Factor_cEll} depend implicitly on the spacetime dimension.
The other part of the expression \eqref{eq:GenPropagatorGWeight} is the spacetime dependent part $( x^2 )^{D/2 - v}$, which will be affected by the spacetime integration at vertices of Feynman graphs in an interacting theory.
Performing those integrals is the major challenge in evaluating a Feynman diagram. 
Thus, we introduce the following new weight function
\begin{equation}
W_u (x_{10})
~ := ~
\frac{1}{\left[ x_{10}^2 \right]^u}
~ = ~
\adjincludegraphics[valign=c,scale=1]{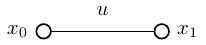}
\label{eq:GenPropagatorWWeight}
\end{equation}
which solely describes the spacetime-dependent part.
Furthermore, we introduced the graphical Feynman diagram notation for the weight, where the spectral parameter written next to the propagator line specifies the weight's exponent.

In the following subsections, we first introduce the chain relations and then map the scalar, generalized propagator weights to a lattice made of spectral lines, which satisfy a YBE and show the integrability of generalized Feynman graphs.

\subsection{Scalar chain relation}
\label{subsec:ScalarChainRelations}
The chain relations are beneficial auxiliary relations which, in terms of momentum-space Feynman diagrams, correspond to bubble integrals with arbitrary exponents of the propagators.
The chain relations ensure that the convolution of two weights of the form \eqref{eq:GenPropagatorWWeight} with two arbitrary spectral parameters $u_1$ and $u_2$ yield a single weight with spectral parameter $u_1 + u_2 - \frac{D}{2}$ times a factor depending on $u_1$ and $u_2$.
It reads
\begin{subequations}
\begin{align}
\int \dd^{D} x_0  ~~ 
\frac{1}{\left[ x_{10}^2 \right]^{u_1}}
\frac{1}{\left[ x_{20}^2 \right]^{u_2}}
~=~&
r_0(D - u_1 - u_2 , u_1 , u_2)
\cdot
\frac{1}{\left[ x_{12}^2 \right]^{u_1 + u_2 - \frac{D}{2}}} ~, \\
\adjincludegraphics[valign=c,scale=1]{figures/chainrelations/scalar/scalarscalarLHS.pdf}
~=~&
r_0(D - u_1 - u_2 , u_1 , u_2)
\cdot
\adjincludegraphics[valign=c,scale=1]{figures/chainrelations/scalar/scalarscalarRHS.pdf} ~.
\end{align}\label{eq:ChainRuleScalar}%
\end{subequations}
In section \ref{subsec:BosonicSTR} below, we derive the chain relation as a limiting case of the star-triangle relation, which in turn is proved in section \ref{appsec:ProofOfTheScalarUniquenessRelation}.
The constant $r_0$ is the $\ell = 0$ case of
\begin{equation}
r_\ell(u_1,u_2,u_3)
~ := ~
\pi^{\frac{D}{2}} \cdot
a_0( u_1 )\, a_\ell( u_2 )\, a_\ell( u_3 ) ~,
\label{eq:Factor_rEll}
\end{equation}
where the $a_\ell ( v )$ were defined in \eqref{eq:Factor_aEll}.
This notation will be used frequently in the following.

\subsection{Bosonic star-triangle relation}
\label{subsec:BosonicSTR}
The relation that allows the identification of generalized scalar propagators \eqref{eq:GenPropagatorWWeight} as solutions to a star-triangle relation is the CFT uniqueness relation.
It describes the equality \cite{DEramo:1971hnd}
\begin{equation}
\int \dd^D x_0
\frac{1}{\left[ x_{01}^2 \right]^{u_1}}
\frac{1}{\left[ x_{02}^2 \right]^{u_2}}
\frac{1}{\left[ x_{03}^2 \right]^{u_3}}
\stackrel{u_1 + u_2 + u_3 = D}{=}
r_0 \left( u_1, u_2 , u_3 \right)
\frac{1}{\left[ x_{12}^2 \right]^{\frac{D}{2} - u_3}}
\frac{1}{\left[ x_{13}^2 \right]^{\frac{D}{2} - u_2}}
\frac{1}{\left[ x_{23}^2 \right]^{\frac{D}{2} - u_1}} ~,
\label{eq:CFTUniqueness_Scalar}
\end{equation}
which holds if the exponents of the l.\,h.\,s.\ add up to the dimension of spacetime, $u_1 + u_2 + u_3 = D$.
We provide a detailed proof of the uniqueness relation \eqref{eq:CFTUniqueness_Scalar} in appendix \ref{appsec:ProofOfTheScalarUniquenessRelation}.
In terms of Feynman diagrams, \eqref{eq:CFTUniqueness_Scalar} is represented as
\begin{equation}
\adjincludegraphics[valign=c,scale=0.9]{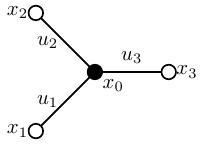} 
\stackrel{u_1 + u_2 + u_3 = D}{=}
r_0(u_1,u_2,u_3)\cdot
\adjincludegraphics[valign=c,scale=0.9]{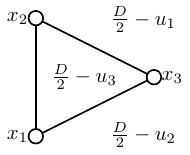} ~.
\label{eq:CFTUniqueness_Scalar_Diagram}
\end{equation}
It is a special case of a star-triangle relation (STR) and, hence, a sufficient condition for the ability to construct R-matrices satisfying a YBE.
Let us explain the conventions for Feynman diagrams, using \eqref{eq:CFTUniqueness_Scalar} and \eqref{eq:CFTUniqueness_Scalar_Diagram} as examples.
Internal spacetime points, which are integrated over, are denoted by a filled circle, while external spacetime points are indicated by an unfilled circle.
Furthermore, the exponents of the r.\,h.\,s.\ of \eqref{eq:CFTUniqueness_Scalar_Diagram} add up to $\frac{D}{2}$ to ensure conformality.

There is an important special case of \eqref{eq:CFTUniqueness_Scalar}, namely the limit when one of the star's exponents approaches $\frac{D}{2}$, i.\,e.\ $u_1 = \frac{D}{2} - \varepsilon$ in the limit $\varepsilon \rightarrow 0$.
Then, the vertical side of the triangle on the r.\,h.\,s.\ of \eqref{eq:CFTUniqueness_Scalar_Diagram} effectively disappears and on the l.\,h.\,s.\ of the remaining equation \eqref{eq:CFTUniqueness_Scalar_Diagram} the weight $W_{u_1}(x_{10})$ divided by $r_0(u_1,u_2,u_3)$ acts like annihilating the integral in the limit $\varepsilon \rightarrow 0$. 
Accordingly, one interprets this combination as a representation of the $D$-dimensional delta function
\begin{equation}
\delta^{(D)}\left( x_{10} \right) 
~=~
\lim_{\varepsilon \rightarrow 0} ~ \pi^{ - \frac{D}{2}} a_0 (\varepsilon)
\cdot W_{\frac{D}{2} -\varepsilon}(x_{10})
%\frac{1}{\left( x_{10}^2 \right)^u}
~=~
\adjincludegraphics[valign=c,scale=1]{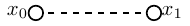},
\label{eq:Delta_bosonic}
\end{equation}
which is denoted by a dotted line in a Feynman graph.

Furthermore, observe that setting $u_1 = u$ and $u_2 = D - u -\varepsilon$ in \eqref{eq:ChainRuleScalar} and taking the limit $\varepsilon \rightarrow 0$ gives the prescription of the delta-distribution \eqref{eq:Delta_bosonic} on the r.\,h.\,s.\,. 
We may then write
\begin{equation}
\lim_{\varepsilon\rightarrow 0} 
\int \dd^{D} x_0  ~~ 
W_{u}(x_{10})W_{D - u - \varepsilon}(x_{20}) 
~=~
\pi^D
a_0 (u)
a_0 (D - u)
\cdot
\delta^{(D)} (x_{12}) ~,
\label{eq:Inversion_WWeights}
\end{equation}
which could also be obtained from the uniqueness relation \eqref{eq:CFTUniqueness_Scalar} by setting one of the spectral parameters to $\varepsilon$ and then taking the limit $\varepsilon \rightarrow 0$.

We still did not prove the chain relation \eqref{eq:ChainRuleScalar}.
As mentioned above, we can obtain it from the uniqueness relation \eqref{eq:CFTUniqueness_Scalar} by sending one external point to infinity, $x_i \rightarrow \infty$. 
Collecting the divergent factors of \eqref{eq:CFTUniqueness_Scalar} yields $\left[x_i^2\right]^{- u_i}$ on each side and canceling them leaves us with the chain relation \eqref{eq:ChainRuleScalar}.

\subsection{Bosonic x-unity relation}
\label{subsec:BosonicXUnity}
The x-unity relation is the central tool in calculating the critical coupling of the integrable QFT under investigation.
Here we derive the bosonic version in detail, based on the special case of the chain relation \eqref{eq:Inversion_WWeights} and the STR \eqref{eq:CFTUniqueness_Scalar}.
Since the chain relation is a consequence of a star-triangle relation, the x-unity relation holds whenever the weights satisfy a STR.
The converse may not be true, as we will see later.

Let us make two observations before deriving the x-unity relation. 
\begin{itemize}
\item 
Consider a three-spiked star integral, where the weights add up to $D$, similar to the l.\,h.\,s.\ of \eqref{eq:CFTUniqueness_Scalar_Diagram}, and where we take one propagator exponent to be $\varepsilon$ and then take the limit $\varepsilon \rightarrow 0$.
The propagator then disappears in the limit, and we obtain the integral relation \eqref{eq:Inversion_WWeights}.
This yields the relation
\begin{equation}
\lim_{\varepsilon\rightarrow 0}
\adjincludegraphics[valign=c,scale=0.9]{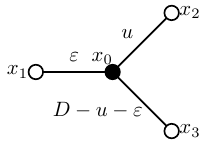}
=
\pi^D
a_0(u)\, a_0(D - u) ~\cdot
\adjincludegraphics[valign=c,scale=0.9]{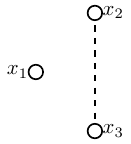}  ~.
\label{eq:3ptFctnWithOneVanishingParameter_Scalar}
\end{equation}

\item
We obtain another helpful relation when integrating one external point of a three-spiked star integral, in the case where the propagator weights add up to $D$.
We can then evaluate the central star integral by the STR \eqref{eq:CFTUniqueness_Scalar} and afterwards the integral over the external point can be performed by the chain relation \eqref{eq:ChainRuleScalar}.
The remaining diagram contains two weights, $W_{\frac{D}{2} - u_1}(x_{23}) W_{u_1 - \frac{D}{2}}(x_{23})$, which is one.
Thus, we are only left with a factor which can be further simplified by $a_0(u)a_0(\frac{D}{2} - u) = 1$.
We obtain the equality
\begin{equation}
\adjincludegraphics[valign=c,scale=1]{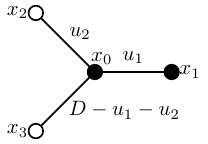} 
=
\pi^D
a_0(u_1)\, a_0(D - u_1)
\adjincludegraphics[valign=c,scale=1]{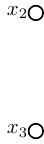} ~.
\label{eq:3ptFctnOnePointIntegrated_Scalar}
\end{equation}
We observe that the whole expression reduces to a factor independent of $u_2$.
\end{itemize}
The derivation of the x-unity relation based on \eqref{eq:3ptFctnWithOneVanishingParameter_Scalar} and \eqref{eq:3ptFctnOnePointIntegrated_Scalar} is 
\begin{subequations}
{\allowdisplaybreaks
\begin{align}
\adjincludegraphics[valign=c,scale=0.8]{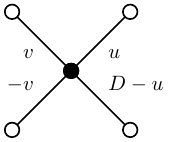} 
&\stackrel{\eqref{eq:Inversion_WWeights}}{=}
\lim_{\substack{\varepsilon\rightarrow 0\\\delta\rightarrow 0}}
\frac{1}{\pi^D a_0(\varepsilon) a_0(D - \varepsilon)}
\adjincludegraphics[valign=c,scale=0.8]{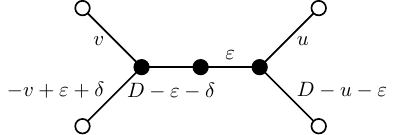} 
\\
&\stackrel{\eqref{eq:3ptFctnWithOneVanishingParameter_Scalar}}{=}
\left[
\lim_{\varepsilon\rightarrow 0}
\frac{a_0(u) a_0(D - u)}{a_0(\varepsilon) a_0(D - \varepsilon)}
\right]
\cdot
\lim_{\delta\rightarrow 0}
\adjincludegraphics[valign=c,scale=0.8]{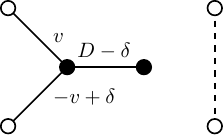} \\
&\stackrel{\eqref{eq:3ptFctnOnePointIntegrated_Scalar}}{=}
\left[
\lim_{\substack{\varepsilon\rightarrow 0\\\delta\rightarrow 0}}
\pi^D a_0(u) a_0(D - u)
\frac{a_0(\delta) a_0(D - \delta)}{a_0(\varepsilon) a_0(D - \varepsilon)}
\right]
\cdot
\adjincludegraphics[valign=c,scale=0.8]{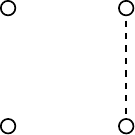} \\
&\stackrel{\phantom{\eqref{eq:3ptFctnOnePointIntegrated_Scalar}}}{=}
\pi^D
a_0(u) a_0(D - u) \hspace{0.5cm}
\cdot
\adjincludegraphics[valign=c,scale=0.8]{figures/xunity/scalar/derivation/RHS/xunity_derivation3.pdf}  ~.
\end{align}
\label{eq:XUnity_Scalar_derivation}%
}%
\end{subequations}
To summarize, the x-unity relation allows to replace a Feynman integral with four external points and a particular choice of propagator exponents by a single delta function.
It is the unity integration kernel, therefore the name x-unity.
For later reference, the scalar x-unity is finally
\begin{equation}
\adjincludegraphics[valign=c,scale=1]{figures/xunity/scalar/derivation/LHS/xmove1.pdf} 
~ = ~
\pi^D
a_0(u) a_0(D - u) \hspace{0.5cm}
\cdot
\adjincludegraphics[valign=c,scale=1]{figures/xunity/scalar/derivation/RHS/xunity_derivation3.pdf}
\label{eq:XUnity_Scalar}
\end{equation}
and we will also use the version of \eqref{eq:XUnity_Scalar} where the left and right external points are interchanged.

\subsection{Integrable scalar Feynman graphs}
\label{subsec:IntegrableScalarFeynmanGraphs}
So far, this section presented integral identities for Feynman graphs with generalized propagators, which are inspired from a non-local deformation of the kinetic term in some action.
Now, we present how they encode YBEs and what R-matrices solve them.
Next, we will explain how these propagators are related to the YBE, as well as to the R-matrices solving the latter.

\subsubsection{The medial lattice}
\label{subsubsec:TheMedialLattice}
Let us get back to the spectral lines introduced in section \ref{sec:8Vmodel}.
The generalized propagator \eqref{eq:GenPropagatorWWeight} may be associated to an intersection of two spectral lines in two ways,
\begin{equation}
W_u (x_{12})
=
\adjincludegraphics[valign=c,scale=1]{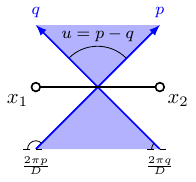} 
\quad \mathrm{and} \quad
\bar{W}_u (x_{12})
=
W_{\frac{D}{2} - u} (x_{12})
=
\adjincludegraphics[valign=c,scale=1]{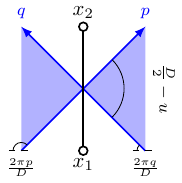}  ~.
\label{eq:GenPropagatorWWeight_medial}
\end{equation}
The exponent of the propagator is related to the difference of the spectral parameters, carried by each of the two spectral lines, $u = p - q$.
There are two options on how the propagator is related to the intersecting spectral lines.
If both spectral lines point to the same side of the propagator, we assign the weight $W_u(x_{12})$.
If the spectral lines point to different sides of the propagator, this indicates the assignment of the so-called crossed weight $\bar{W}_u(x_{12}) = W_{\frac{D}{2} - u}(x_{12})$.
The term \textit{crossing} refers to the flipping of one spectral line involved in the intersection and we see that $W_u(x_{12})$ and $\bar{W}_u(x_{12})$ differ by such a transformation.
It effectively exchanges the spectral parameter difference $u$ by $\frac{D}{2} - u$, which is an example of the general crossing transformation of a spectral parameter $u \rightarrow \eta - u$ with the crossing parameter $\eta$.
Thus, we can identify the crossing parameter in integrable, generalized Feynman graphs as 
\begin{equation}
\eta = \frac{D}{2}
\end{equation}
We denote the crossed spectral parameter by $\bar{u} = \frac{D}{2} - u$, which implies $\bar{\bar{u}} = u$.

In \eqref{eq:GenPropagatorWWeight_medial}, the angle to the horizontal line is indicated as $\frac{2\pi}{D} p$ and $\frac{2\pi}{D} q$ for the spectral line carrying the spectral parameters $p$ and $q$, respectively.
The spectral parameter can be thought of as proportional to the slope of the corresponding spectral line.
The proportionality factor is the ratio of the two crossing factors, $\pi$ for the geometric angle picture and $\frac{D}{2}$ for the generalized Feynman graphs.
Thus, the exponent of the propagator weights \eqref{eq:GenPropagatorWWeight_medial} is proportional to the angle under which the two spectral lines intersect.
The geometric interpretation of the spectral parameters as angles is very natural in 2-dimensional integrable scattering, where the spectral lines correspond to trajectories and their intersections to the scattering described by the S-matrix satisfying the YBE \cite{Zamolodchikov:1978xm}.
However, for illustrative reasons, we will not draw the medial lines with the slope corresponding to their spectral parameter.
 
Another important graphical feature of \eqref{eq:GenPropagatorWWeight_medial} is the shading of those faces of the graph that do not contain an external point the propagator is attached to \cite{Zamolodchikov:1980mb,Bazhanov:2016ajm}.
It hints towards the general relation between a planar generalized Feynman graph $\mathscr{F}$ and the graph spanned by multiple spectral lines $\mathscr{G}$. 
The latter is assumed to be a planar graph, which implies that the spectral lines can be imagined to live on a two-dimensional surface.
This is instrumental, since integrability in QFTs appears in the context of $\mathcal{N} = 4$ SYM in the planar limit, where its leading-order Feynman diagrams often have a toroidal topology.
The map \eqref{eq:GenPropagatorWWeight_medial} tells us to associate with every edge in $\mathscr{F}$ a vertex in $\mathscr{G}$, to every vertex in $\mathscr{F}$ a face in $\mathscr{G}$, and with every face in $\mathscr{F}$ a face in $\mathscr{G}$, as well.
Thus, some of the faces of $\mathscr{G}$ correspond to a vertex of $\mathscr{F}$ and others correspond to a face in $\mathscr{F}$.
The latter faces are exactly the ones we will shade, when drawing the graph $\mathscr{G}$ corresponding to $\mathscr{F}$.
Note that the two sets of faces in $\mathscr{G}$ are alternating, this means that two faces from the same set touch each other at most at one single vertex, while two faces, which are not in the same set, touch each other along an edge in $\mathscr{G}$.
Hence, two shadings are possible for a graph $\mathscr{G}$.

A graph $\mathscr{G}$ that is constructed from $\mathscr{F}$ in such a way is known as the \textit{medial} graph of $\mathscr{F}$ and we denote it $\mathscr{G} = \mathscr{M} (\mathscr{F})$. 
In the following, we will also use the term \textit{medial lines} for the spectral lines drawn in blue, c.\,f.\ \eqref{eq:GenPropagatorWWeight_medial}.
The medial map $\mathscr{M}$ is two-to-one, since the dual\footnote{Remember that the dual graph is obtained by replacing each vertex of the original planar graph by a face and each face by a vertex, while the dual graph's edges correspond to a pair of faces separated by an edge in the original graph.} graph of $\mathscr{F}$, $\mathscr{F}^*$, admits the same medial graph, $\mathscr{G} = \mathscr{M} (\mathscr{F}) = \mathscr{M} (\mathscr{F}^*)$.
If we equip the medial map with a shading, we obtain an one-to-one map.
Our convention is to shade the faces of $\mathscr{G}$, which correspond to the faces in $\mathscr{F}$, so that we can draw external and internal vertices of $\mathscr{F}$ into unshaded faces of $\mathscr{G}$.
We denote the shaded medial graph of $\mathscr{F}$ as $\mathscr{G}_\mathscr{F}$.
If we invert the shading, we obtain $\mathscr{G}_{\mathscr{F}^*}$.

In summary, the shaded medial map is a one-to-one relation of a planar Feynman graph $\mathscr{F}$ to a shaded medial graph $\mathscr{G}_\mathscr{F}$. 
If we take the exponents of the propagators $\left\lbrace u_i \right\rbrace$ and the spectral parameters of the medial lines $\left\lbrace p_i \right\rbrace$ into account we have the invertible, one-to-one map \cite{Zamolodchikov:1980mb,Bazhanov:2016ajm}
\begin{equation}
\mathscr{F} (\left\lbrace u_i \right\rbrace)
\longleftrightarrow
\mathscr{G}_\mathscr{F} (\left\lbrace p_i \right\rbrace) ~,
\label{eq:Map_FeynmanMedial}
\end{equation}
where the exponents are differences of various spectral parameters, $u_i = p_j - p_k$.
In the context of integrable QFTs, $\mathscr{G}_\mathscr{F} (\left\lbrace p_i \right\rbrace)$ is sometimes called \textit{Baxter lattice} or \textit{loom} \cite{Kazakov:2022dbd,Kazakov:2023nyu}.
A minimal illustration of these concepts is \eqref{eq:GenPropagatorWWeight_medial}: the equation on the left shows the Feynman graph $\mathscr{F} ( u )$ in black consisting of a single propagator with two external vertices connected by one edge and exponent $u = p - q$.
The corresponding shaded medial graph is $\mathscr{G}_\mathscr{F} (p, q)$ in blue on the r.\,h.\,s.\ of the left equation in \eqref{eq:GenPropagatorWWeight_medial}.
The equation on the right therein shows the dual Feynman graph in black and the corresponding inversely shaded medial graph in blue.

The power of the map \eqref{eq:Map_FeynmanMedial} manifests itself when applying it on the STR \eqref{eq:CFTUniqueness_Scalar_Diagram} by using \eqref{eq:GenPropagatorWWeight_medial},
\begin{equation}
\adjincludegraphics[valign=c,scale=1]{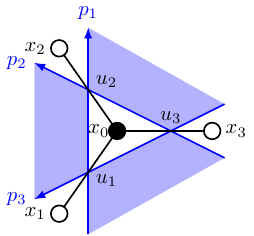} 
\stackrel{\substack{u_1 = \frac{D}{2} - (p_2 - p_3) \\ u_2 = p_1 - p_3 \\ u_3 = \frac{D}{2} - (p_1 - p_2)}}{=}
r_0(u_1,u_2,u_3)\cdot
\adjincludegraphics[valign=c,scale=1]{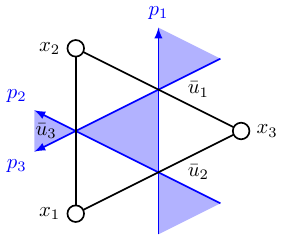} ~,
\label{eq:STR_Scalar_medial}
\end{equation}
from which we gain multiple insights.
First, we observe that the constraint $u_1 + u_2 + u_3 = D$ is automatically satisfied when we understand the exponents coming from the spectral parameters of the medial lines.
This implies that a generalized Feynman graph, which is constructed from its shaded medial graph, has conformal internal vertex integrations by construction, stemming from a CFT.
Furthermore, \eqref{eq:STR_Scalar_medial} illustrates that closed faces in $\mathscr{G}_\mathscr{F}$ correspond to internal, integrated vertices in the generalized Feynman graph.
Additionally, only planar Feynman graphs can be generated from a shaded medial graph.
In addition, inverting the shading just switches the r.\,h.\,s.\ and l.\,h.\,s.\ of the equation.

Secondly, equation \eqref{eq:STR_Scalar_medial} crucially shows that the uniqueness relation \eqref{eq:CFTUniqueness_Scalar} on the Feynman graph $\mathscr{F}(u_1, u_2, u_3)$ hides a YBE under the map \eqref{eq:Map_FeynmanMedial} on the level of the shaded medial graph $\mathscr{G}_\mathscr{F} (p_1, p_2, p_3)$!
Therefore, the shaded medial graph uncovers the integrable structure of generalized Feynman graphs.
In contrast to the usual form of the YBE, \eqref{eq:STR_Scalar_medial} involves the factor $r_0(u_1,u_2,u_3)$, which should not be forgotten in concrete calculations\footnote{In fact, the factor $r_0(u_1,u_2,u_3)$ is absent if we use redefined weights in \eqref{eq:GenPropagatorWWeight_medial}. 
Therein, the left diagram should correspond to $\Gamma (u) W_u (x_{12})$ and the diagram on the right to $\Gamma (\frac{D}{2} - u) \bar{W}_u (x_{12})$. The $\pi^{\frac{D}{2}}$ can be associated to a rescaled integration measure.}. 
Yet, \eqref{eq:STR_Scalar_medial} allows us to transfer a plethora of methods and techniques developed for integrable models, using the central equation of integrability, the YBE, to the medial graph.
Thus, by the map \eqref{eq:Map_FeynmanMedial}, and practically by \eqref{eq:GenPropagatorWWeight_medial}, we can study the implications of integrability on generalized Feynman graphs.
In a next step, one can fix the spectral parameters to specific values where the propagators \eqref{eq:GenPropagatorWWeight} describe physical propagators of the theory under investigation.

As a first application of this strategy, we can ask for the form of unitarity as a generalized Feynman graph.
In short, unitarity refers to the fact that two spectral lines, which intersect each other twice, can be untangled.
After translating the medial picture back to the Feynman graph, depending on the shading, we find two relations,
\begin{subequations}
\begin{align}
\adjincludegraphics[valign=c,scale=1]{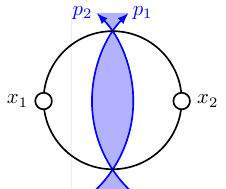} 
& =
\adjincludegraphics[valign=c,scale=1]{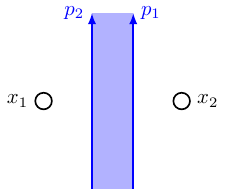} 
 ~, \label{eq:Unitarity_Scalar_medial_One} \\
\adjincludegraphics[valign=c,scale=1]{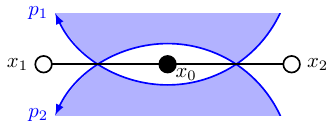} 
& \stackrel{u = \frac{D}{2} - (p_1 - p_2)}{=}
\pi^D
a_0 (u)
a_0 (D - u)
\cdot
\adjincludegraphics[valign=c,scale=1]{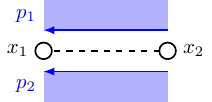} 
 ~. \label{eq:Unitarity_Scalar_medial_Delta}%
\end{align}\label{eq:Unitarity_Scalar_medial}%
\end{subequations}
Regarding \eqref{eq:Unitarity_Scalar_medial_One}, we observe that unitarity translates to the trivial fact $W_u (x_{12}) W_{-u} (x_{12}) = 1$ on the generalized Feynman graph, with $u = p_1 - p_2$.
When we invert the shading, we find the Feynman graphs corresponding to \eqref{eq:Inversion_WWeights} and therefore we have to include the factor $\pi^D a_0 (u) a_0 (D - u)$ in \eqref{eq:Unitarity_Scalar_medial_Delta}.
In this case we have $u = \frac{D}{2} - (p_1 - p_2)$.

\subsubsection{The R-matrices}
\label{subsubsec:TheRmatrices}
In \eqref{eq:STR_Scalar_medial} we saw that a YBE appears graphically when we translate the STR \eqref{eq:CFTUniqueness_Scalar_Diagram} to the shaded medial lattice.
By going a step back and forgetting the shading for the moment, we present the corresponding algebraic YBE and the R-matrix solving it.
However, the shading procedure allows us to study single Feynman graphs, and hence reinstalling it leads us to a fused R-matrix, which satisfies a staggered YBE.
The fusion leads to objects, which depend on staggered medial lines.

Disregarding the shading in $\mathscr{G}_\mathscr{F}$ yields the medial graph $\mathscr{G}$.
This has the drawback that one deals with the Feynman graph $\mathscr{F}$ and its dual $\mathscr{F}^*$ at the same time.
Therefore, on the level of the weights \eqref{eq:GenPropagatorWWeight_medial}, we have to combine the weight and its crossed counterpart into a single object.
This is the R-matrix
\begin{equation}
R^{x_2 y_2}_{x_1 y_1}(u)
=
W_u (y_{12})
\bar{W}_u (x_{12})
\stackrel{u = p_1 - p_2}{=}
\adjincludegraphics[valign=c,scale=1]{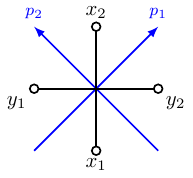} ~,
\label{eq:Rmatrix_facetype_WWbar}
\end{equation}
where the weight $W_u (y_{12})$ corresponds to the horizontal propagator representing $\mathscr{F}$, and $\bar{W}_u (x_{12})$ to the vertical one representing $\mathscr{F}^*$.
Within the R-matrix $R^{x_2 y_2}_{x_1 y_1}(u)$ the two weights do not interfere with each other, giving the factorized form in the spacetime coordinates $x_i$, $y_i$.
The YBE is obtained when we remove the shading of \eqref{eq:STR_Scalar_medial} and draw the dual graph as well,
\begin{equation}
\adjincludegraphics[valign=c,scale=1.1]{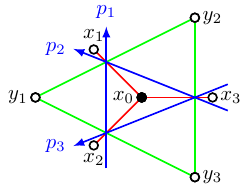} 
~ = ~
\adjincludegraphics[valign=c,scale=1.1]{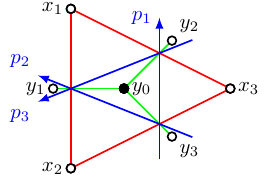} .
\label{eq:YBE_facetype_Diagram2}
\end{equation} 
On the level of the individual Feynman graphs $\mathscr{F}$ (red) and $\mathscr{F}^*$ (green), the STR \eqref{eq:CFTUniqueness_Scalar_Diagram} yields \eqref{eq:YBE_facetype_Diagram2}.
While the red graph represents \eqref{eq:CFTUniqueness_Scalar_Diagram} in the displayed order, the green graph represents \eqref{eq:CFTUniqueness_Scalar_Diagram} with left- and right hand side interchanged.
In contrast to \eqref{eq:CFTUniqueness_Scalar_Diagram}, we notice the absence of the factor $r_0$.
It appears in the relations of both graphs $\mathscr{F}$ and $\mathscr{F}^*$ and cancels.
Informally, one could write $\mathrm{YBE} = (\mathrm{STR})^2$ to describe the relationship between YBE and STR.
In terms of the R-matrix \eqref{eq:Rmatrix_facetype_WWbar}, the YBE \eqref{eq:YBE_facetype_Diagram} reads
\begin{equation}
\int \dd^D x_0 ~
R^{x_2 y_3}_{x_0 y_1} (p_{12})
R^{y_1 x_0}_{y_2 x_3} (p_{13})
R^{x_0 y_3}_{x_1 y_2} (p_{23})
=
\int \dd^D y_0 ~
R^{y_0 x_1}_{y_2 x_3} (p_{12})
R^{x_2 y_3}_{x_1 y_0} (p_{13})
R^{y_1 x_2}_{y_0 x_3} (p_{23}),
\label{eq:YBE_facetype}
\end{equation}
where we write for short $p_{ij} = p_i - p_j$.
The equation \eqref{eq:YBE_facetype} is known as a facetype-YBE, which can be brought to the usual vertex-type by intertwining vectors \cite{Isaev:2022mrc}.
Moreover, unitarity of the R-matrix \eqref{eq:Rmatrix_facetype_WWbar} is a combination of the relation \eqref{eq:Unitarity_Scalar_medial_One} and \eqref{eq:Unitarity_Scalar_medial_Delta}. 

The R-matrix \eqref{eq:Rmatrix_facetype_WWbar} is less useful for QFT applications due to the construction from the medial graph $\mathscr{G}$, since practically one is interested in a single Feynman graph instead of two, $\mathscr{F}$ and its dual $\mathscr{F}^*$.
Therefore, the R-matrix should be constructed from a shaded medial graph $\mathscr{G}_\mathscr{F}$.
The way to go is through a procedure called \textit{fusion}.
Graphically, it amounts to considering the intersection of a pair of medial lines with another pair (we may consider such a pair as a kind of bound-state of two elementary, partonic excitations).
We call the fused R-matrix $\mathbb{R}$.
Of course, a two-times-two intersection can be decomposed into four one-times-one intersections of medial lines of type \eqref{eq:GenPropagatorWWeight_medial}, but the main advantage of $\mathbb{R}$ is that it can be composed, while respecting the shading.
Hence, if we consider the shaded intersection of two pairs of medial lines and if we use the weights \eqref{eq:GenPropagatorWWeight_medial} for the one-times-one intersections, we obtain
\begin{equation}
\mathbb{R}^{x_2 y_2}_{x_1 y_1}( p_1, p_2 \vert p_3 ,p_4 ) 
= 
\adjincludegraphics[valign=c,scale=1]{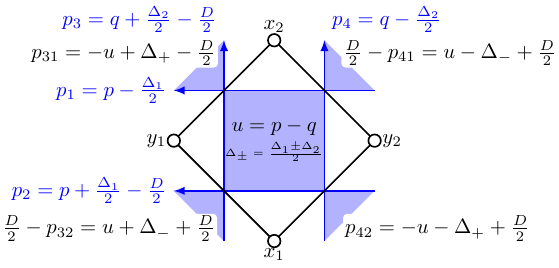} ~.
\label{eq:Rmatrix_Scalar_Fused}
\end{equation}
Note that the R-matrix depends on three free parameters, since the four exponents satisfy the constraint $p_{31} + \bar{p}_{41} + \bar{p}_{32} + p_{42} = D$.
The parameterization in terms of the three free parameters $\Delta_1$, $\Delta_2$ and $u$ is taken from the original derivation \cite{Chicherin:2012yn} and we use the notation $\mathbb{R}^{x_2 y_2}_{x_1 y_1}(u; \Delta_1 , \Delta_2 ) = \mathbb{R}^{x_2 y_2}_{x_1 y_1}( p_1, p_2 \vert p_3 ,p_4 )$.
The scaling dimensions $\Delta_1$ and $\Delta_2$ are the labels of the two representations of the scalar principal series\footnote{
The scalar principle series representation of the conformal group in $D$ dimensions is acting on the vector space of square-integrable functions
\begin{equation*}
V_\Delta
=
\mathrm{L}^2
\left(
\mathbb{R}^D ,
(1 + x^2)^{2 \mathrm{Re}(\Delta) - D}
\dd^D x
\right)
\end{equation*}
and the second argument is the integral measure.
The representation can be generalized to account for spinning fermionic propagators as well.
Accordingly, the representation acts on the tensor product of $V_\Delta$ with the complex vector space of symmetric traceless tensors of rank $\ell$.
We denote it by $V_{\Delta , \ell , \bar{\ell}}$ and the R-matrix \eqref{eq:Rmatrix_Scalar_Fused} is an element of $\mathrm{End} (V_{\Delta , \ell , \bar{\ell}} \otimes V_{\Delta , \ell , \bar{\ell}})$ \cite{Dobrev:1977qv,Ferrando:2021yek}.
}, which are associated with the pairs of medial lines.
The R-matrix $\mathbb{R}$ describes the intersection of these two representations.
For the QFT applications in section \ref{chpt:NonSusyFishnetTheoriesAndItsRelatives}, the R-matrix \eqref{eq:Rmatrix_Scalar_Fused} is of high importance, since tuning the spectral parameter and conformal dimensions to suited values yields parts of physical Feynman diagrams, which build up the entire Feynman graph.
Therefore, these evaluated R-matrices are called graph builders, see e.\,g.\ \cite{Gromov:2017cja}.

There exists an interesting limit of the R-matrix \eqref{eq:Rmatrix_Scalar_Fused}, namely $u= \varepsilon$ and $\Delta_1 = \Delta_2$.
When we multiply the R-matrix with a compensation factor and eventually take the limit $\varepsilon \rightarrow 0$, we get
\begin{equation}
\lim_{\varepsilon \rightarrow 0} 
\pi^{-D}
a_0 (\varepsilon)^2
\cdot
\mathbb{R}^{x_2 y_2}_{x_1 y_1}(\varepsilon ; \Delta_1, \Delta_1)
~ = ~
\adjincludegraphics[valign=c,scale=1]{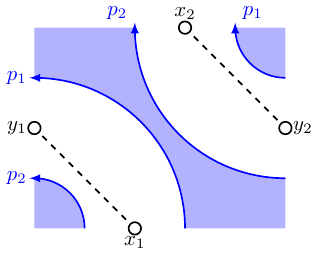} 
~ = ~
\mathbb{P}_{xy} ~.
\label{eq:Rmatrix_spectralParameterZero}
\end{equation}
In this limit, the spectral parameters of the horizontal medial lines are the same as the spectral parameter of the vertical ones. 
Hence, we see that for a vanishing spectral parameter, the R-matrix becomes the permutation operator $\mathbb{P}_{xy}$, which interchanges the two vector spaces it is acting on.
Here, they are labeled by spacetime points $x$ and $y$.

Naturally, the R-matrix \eqref{eq:Rmatrix_Scalar_Fused} solves a YBE, which we call $\mathbb{YBE}$ to distinguish it from \eqref{eq:YBE_facetype_Diagram}.
The $\mathbb{YBE}$, graphically displayed as the generalized Feynman graph $\mathscr{F}$ together with the shaded medial graph $\mathscr{G}_\mathscr{F}$, is (the exponents of the weights are suppressed for clarity)
\begin{equation}
\adjincludegraphics[valign=c,scale=1]{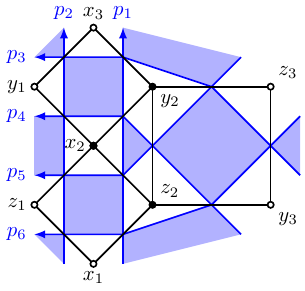} 
~ = ~
\adjincludegraphics[valign=c,scale=1]{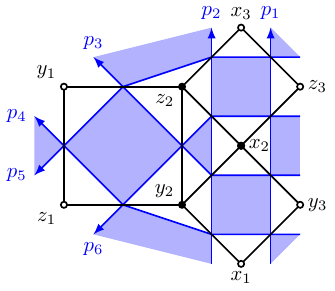} ~.
\label{eq:YBE_Scalar_Fused}
\end{equation}
Contrary to the Feynman graph, proving it up to a factor is straightforward on the corresponding graph $\mathscr{G}_\mathscr{F}$: starting from the l.\,h.\,s.\,, the STR \eqref{eq:STR_Scalar_medial} allows to shift the medial lines $p_1$ and $p_2$ to the very right side and yields the r.\,h.\,s.\,.
However, it is left to show that all the factors $r_0$, \eqref{eq:Factor_rEll}, which are produced or annihilated by the STR, cancel.
We demonstrate the complete proof of \eqref{eq:YBE_Scalar_Fused} at the level of the generalized Feynman diagram $\mathscr{F}$.
However, a priori it is not clear where and in which direction the STR \eqref{eq:CFTUniqueness_Scalar_Diagram} should be used.
That is why the medial picture and the map \eqref{eq:Map_FeynmanMedial} are so helpful.
Having figured out, which medial line gets pushed through which medial vertex in $\mathscr{G}_\mathscr{F}$, shows us which star (if exponents add up to $D$) should be turned into a triangle and which triangle (if exponents add up to $\frac{D}{2}$) should be turned into a star within $\mathscr{F}$.
The diagram part of the proof is performed in multiple steps by usage of \eqref{eq:CFTUniqueness_Scalar_Diagram} in both ways,
\begin{equation}
\begin{gathered}
\adjincludegraphics[valign=c,scale=0.8]{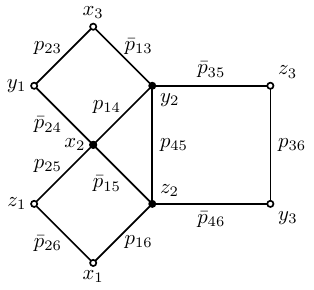} 
\stackrel{\eqref{eq:CFTUniqueness_Scalar_Diagram}}{\longrightarrow}
\adjincludegraphics[valign=c,scale=0.8]{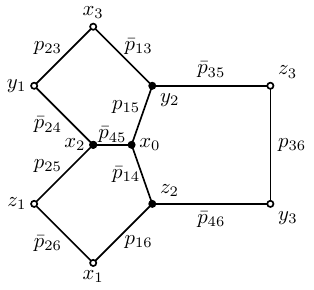} \\
\stackrel{3 \times \eqref{eq:CFTUniqueness_Scalar_Diagram}}{\longrightarrow}
\adjincludegraphics[valign=c,scale=0.8]{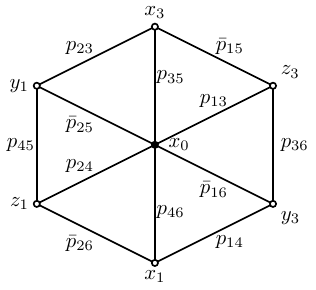}
\stackrel{3 \times \eqref{eq:CFTUniqueness_Scalar_Diagram}}{\longrightarrow} \\
\adjincludegraphics[valign=c,scale=0.8]{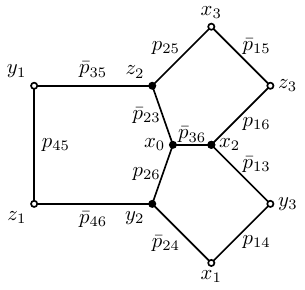}
\stackrel{\eqref{eq:CFTUniqueness_Scalar_Diagram}}{\longrightarrow}
\adjincludegraphics[valign=c,scale=0.8]{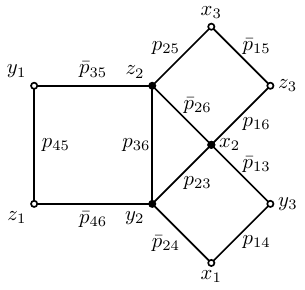} ~.
\end{gathered}
\label{eq:YBE_Scalar_Fused_proof}
\end{equation}
Along the steps, we pick up multiple $r_0$ factors, according to \eqref{eq:CFTUniqueness_Scalar_Diagram}.
After the final step, we have to multiply the last diagram in \eqref{eq:YBE_Scalar_Fused_proof} by
\begin{equation}
\frac{
r_0 ( \bar{p}_{35} , \bar{p}_{13} , p_{15})\;
r_0 ( \bar{p}_{45} , \bar{p}_{24} , p_{25})\;
r_0 ( \bar{p}_{46} , \bar{p}_{14} , p_{16})
\cdot
r_0 ( \bar{p}_{36} , \bar{p}_{23} , p_{26})
}{
r_0 ( \bar{p}_{45} , \bar{p}_{14} , p_{15})
\cdot
r_0 ( \bar{p}_{35} , \bar{p}_{23} , p_{25})\;
r_0 ( \bar{p}_{36} , \bar{p}_{13} , p_{16})\;
r_0 ( \bar{p}_{46} , \bar{p}_{24} , p_{26})
}
=
1
\label{eq:YBE_Scalar_Fused_proof_factor}
\end{equation}
Remember that the factor $r_0$ (defined in \eqref{eq:Factor_rEll}) factorizes in $a_0$ factors (defined in \eqref{eq:Factor_aEll}) of their arguments.
And since the factors in \eqref{eq:YBE_Scalar_Fused_proof_factor} in the numerator have the same arguments as the ones in the denominator, we obtain one.
This proves the $\mathbb{YBE}$ \eqref{eq:YBE_Scalar_Fused}.

Naturally, the R-matrix \eqref{eq:Rmatrix_Scalar_Fused} satisfies a unitarity relation,
\begin{equation}
\adjincludegraphics[valign=c,scale=0.75]{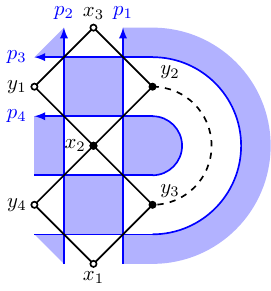} 
~ = ~
\pi^{2D}
a_0 (\bar{p}_{13})\, a_0 (D - \bar{p}_{13})
a_0 (\bar{p}_{24})\, a_0 (D - \bar{p}_{24})
\cdot
\adjincludegraphics[valign=c,scale=0.75]{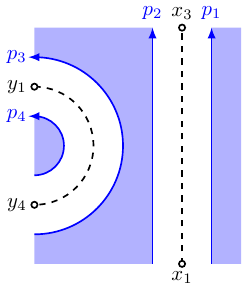}
\label{eq:Unitarity_Rmatrix_scalar_fused} 
\end{equation}
which is a consequence of the unitarity of the weights \eqref{eq:Unitarity_Scalar_medial}.

\section{The generalized fermionic propagator as a lattice weight}
\label{sec:TheGeneralizedFermionicPropagatorAsLatticeWeight}
The propagator of spin-$\frac{1}{2}$ Weyl fermions $\psi_\alpha$ has, analogously to the scalar case in section \ref{sec:TheGeneralizedScalarPropagatorAsLatticeWeight}, a generalization based on a non-local deformation of the kinetic term.
Accordingly, its derivation is similar to the scalar case.
In Euclidean space, the deformed kinetic term reads
\begin{equation}
- \I \int \dd^D x ~ 
\bar{\psi} 
\square^v 
\bar{\partial}
\psi
=
\int \dd^D x_1 \dd^D x_2 ~ 
\bar{\psi}_{\dot{\alpha}} (x_1)
\left[
- \I
\delta^{(D)} (x_{12})\,
\bar{\sigma}^{\mu , \dot{\alpha}\alpha}
\square_2^v 
\partial_{2,\mu}
\right]
\psi_\alpha (x_2) ~,
\label{eq:FermionicUndefKinTerm}
\end{equation}
where we use the notations $x_{\alpha \dot{\alpha}} = x_\mu \sigma^\mu_{\alpha \dot{\alpha}}$ and $\bar{x}^{\dot{\alpha}\alpha} = x_\mu \bar{\sigma}^{\mu ,\dot{\alpha}\alpha}$.
Here, we use the standard sigma matrices $\sigma^\mu$ and $\bar{\sigma}^\mu$, appropriate for even D-dimensional spacetime.
The odd $D$-dimensional case should be described by using gamma matrices $\gamma^\mu$ instead.
Importantly, the matrices satisfy a Clifford algebra and their properties for $D=3$ and $D=4$ are shown in appendix \ref{app:SuperspaceNotations}.
In this section, we use the symbol of sigma matrices for a representation of the $D$ dimensional Clifford algebra, knowing that in spacetime dimensions other than four, the actual matrix representation has to be adjusted.
However, all our findings carry over to other spacetime dimensions as well; see e.\,g.\ \cite{Preti:2018vog}.
The propagator $G_{v,\alpha \dot{\beta}} (x_{12}) = \langle \psi_\alpha (x_1) \bar{\psi}_{\dot{\beta}} (x_2) \rangle$ as the inverse of the kinetic operator satisfies
\begin{equation}
\left[
- \I \,
\bar{\sigma}^{\mu , \dot{\alpha}\alpha}
\square^v 
\partial_{\mu}
\right]
G_{v,\alpha \dot{\beta}} (x)
=
\delta^{(D)} (x)
\cdot
\delta^{\dot{\alpha}}_{\dot{\beta}} ~.
\label{eq:GenFermionicPropagatorInverseOfKinOperator}
\end{equation}
As in the scalar case, we can solve this equation via Fourier transform and integration by parts, which gives the relation $\bar{\sigma}^{\mu , \dot{\alpha}\alpha} p_\mu \tilde{G}_{v,\alpha \dot{\beta}} (p) = - (p^2)^{-v} \delta^{\dot{\alpha}}_{\dot{\beta}}$, where $\tilde{G}_{v,\alpha \dot{\beta}} (p)$ is the Fourier transform of $G_{v,\alpha \dot{\beta}} (x)$.
We can solve the relation for $\tilde{G}_{v,\alpha \dot{\beta}} (p)$ by use of the Clifford algebra of the sigma matrices, see \label{eq:SigmaMatricesRelations}.
Then, we find $\tilde{G}_{v,\alpha \dot{\beta}} (p) = - \sigma^\nu_{\alpha \dot{\beta}}\, p_\nu (p^2)^{- v - 1}$ and we can reverse the Fourier transform.
Eventually, we obtain
\begin{equation}
G_{v,\alpha \dot{\beta}} (x)
=
- \I\,
\sigma^\mu_{\alpha \dot{\beta}}
\partial_\mu
G_{v + 1} (x)
=
c_\ell (v + \tfrac{1}{2})\cdot
\frac{x_{\alpha \dot{\beta}}}{\left[ x^2 \right]^{\frac{D}{2} - v}}
\label{eq:GenFermionicPropagatorGWeight}
\end{equation}
where $G_{v + 1} (x)$ is the scalar generalized propagator from \eqref{eq:GenPropagatorGWeight} and $c_\ell (v)$ is the factor defined in \eqref{eq:Factor_cEll}.
Note that we obtain the conventional local propagator in \eqref{eq:GenFermionicPropagatorGWeight}, if we set $v \rightarrow 0$.
Its scaling dimension is then the conventional $1-D$, which is twice the scaling dimension for fermions in $D$ dimensions, $\frac{1-D}{2}$.
We define the corresponding fermionic lattice weight by neglecting the factor in
\eqref{eq:GenFermionicPropagatorGWeight}, which is not spacetime dependent,
\begin{equation}
W_{u, \alpha \dot{\beta}}^{\frac{1}{2}}(x_{10}) 
= 
\frac{1}{\left( x_{10}^2 \right)^u}
\frac{x_{10, \alpha \dot{\beta}}}{\left|x_{10}\right|}
=
\adjincludegraphics[valign=c,scale=1]{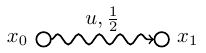} ~.
\label{eq:GenFermionicPropagatorWWeight}
\end{equation}
In $W_u^{\frac{1}{2}}(x_{10})$, we suppress the spinor indices.
Furthermore, the notation stems from the fact that we can combine the scalar weight \eqref{eq:GenPropagatorWWeight} and the fermionic one \eqref{eq:GenFermionicPropagatorWWeight} into a single object
\begin{equation}
W_{u, \alpha \dot{\beta}}^\ell(x_{10}) 
=
\frac{1}{\left( x_{10}^2 \right)^u}
\left[\frac{x_{10, \alpha \dot{\beta}}}{\left|x_{10}\right|}\right]^{2\ell}
=
\adjincludegraphics[valign=c,scale=1]{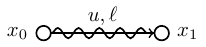} 
=
\left\lbrace
\begin{array}{l}
\adjincludegraphics[valign=c,scale=0.8]{figures/propagators/scalar/propagator_weight_scalar.pdf} ,~ \ell = 0\\
\adjincludegraphics[valign=c,scale=0.8]{figures/propagators/fermion/propagator_weight_fermion.pdf} ,~ \ell = \frac{1}{2}
\end{array}
\right. .
\end{equation}
Note that higher spinning generalized propagators were studied in \cite{Derkachov:2018rot,Derkachov:2019tzo,Derkachov:2020zvv}.
Switching to Lorentz indices, we can represent them as 
\begin{equation}
W_{u}^{\ell, \mu_1 \cdots \mu_{2\ell}}(x_{10}) 
=
\frac{1}{\left( x_{10}^2 \right)^u}
\left[
\prod_{i=1}^{2\ell}
\frac{x_{10}^{\mu_i}}{\left|x_{10}\right|}
\right]
=
\adjincludegraphics[valign=c,scale=1]{figures/propagators/scalarfermion/propagator_weight_scalarfermion.pdf}  .
\label{eq:GenFermionicPropagatorWWeight_higherSpin}
\end{equation}
Here, we focus on $\ell = 0$ and $\ell = \frac{1}{2}$.
Furthermore, we introduce a graphical way to represent a conjugated weight, which is contracted with the Pauli matrix $\bar{\sigma}^\mu$,
\begin{equation}
W_{u}^{\ell, \dot{\alpha} \beta }(x_{10}) 
=
\frac{1}{\left( x_{10}^2 \right)^u}
\left[\frac{\bar{x}_{10}^{\dot{\alpha} \beta }}{\left|x_{10}\right|}\right]^{2\ell}
=
\adjincludegraphics[valign=c,scale=1]{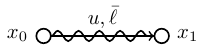} 
=
\left\lbrace
\begin{array}{l}
\adjincludegraphics[valign=c,scale=0.8]{figures/propagators/scalar/propagator_weight_scalar.pdf} ,~ \ell = 0\\
\adjincludegraphics[valign=c,scale=0.8]{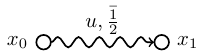} ,~ \ell = \frac{1}{2}
\end{array}
\right. .
\end{equation}
Note that the conjugated weight is identical to the ordinary one, when we use gamma matrices in odd dimensional spacetime.
Furthermore, it is antisymmetric in $x_0 \leftrightarrow x_1$ for $\ell = \frac{1}{2}$ and we have to give the graphical representation of the generalized propagator a direction by adding a little arrow at the end of the propagator.

With these new lattice weights, we repeat now the derivation of the STR, unitarity, chain rule, and x-unity and show how to construct a fermionic R-matrix that satisfies a YBE.
The situation is almost identical to the scalar case; however, we have to take into account the spin.
Concerning the map \eqref{eq:Map_FeynmanMedial}, there is a new feature.
In the fermionic case, the medial lines carry a spin label $\ell$, as well as a spectral parameter $p$.
We collectively denote the parameter of a medial line by $\mathbf{p} = (p, \ell)$.
The map between generalized Feynman graph and shaded medial graph is given by the graphical representation of the weight \eqref{eq:GenFermionicPropagatorWWeight} and its crossed counterpart,
\begin{equation}
\begin{array}{cc}
W_{u, \alpha \dot{\beta}}^\ell (x_{12})
=
\adjincludegraphics[valign=c,scale=1]{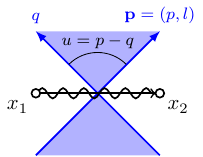} , &
\bar{W}_u^{\ell, \dot{\alpha}\beta } (x_{12})
=
\adjincludegraphics[valign=c,scale=1]{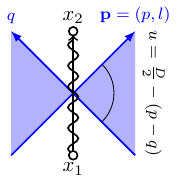}  \\
W_u^{\ell, \dot{\alpha}\beta } (x_{12})
=
\adjincludegraphics[valign=c,scale=1]{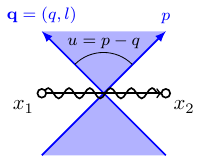} ,  &
\bar{W}_{u, \alpha \dot{\beta}}^\ell (x_{12})
=
\adjincludegraphics[valign=c,scale=1]{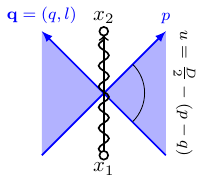} ,
\end{array}
\label{eq:GenFermionicPropagatorWWeight_medial}
\end{equation}
depending if the fermionic line is to the right or the left, respectively.
Again, we denote the crossed weights by $\bar{W}_{u, \alpha \dot{\beta}}^\ell (x_{12}) = W_{\frac{D}{2} - u, \alpha \dot{\beta}}^\ell (x_{12})$ and $\bar{W}_u^{\ell, \dot{\alpha}\beta } (x_{12}) = W_{\frac{D}{2} - u}^{\ell, \dot{\alpha}\beta } (x_{12})$.

In appendix \ref{appsec:ProofOfTheFermionicUniquenessRelation}, we show that the fermionic weights \eqref{eq:GenFermionicPropagatorWWeight_medial} satisfy a fermionic version of the STR \cite{Symanzik:1972wj},
\begin{equation}
\adjincludegraphics[valign=c,scale=1]{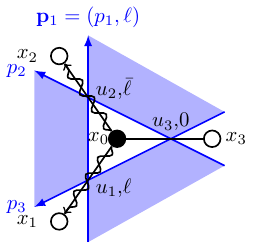} 
~ = ~
r_\ell (u_3 , u_1 , u_2)
\adjincludegraphics[valign=c,scale=1]{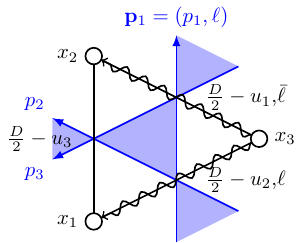} ~,
\label{eq:STR_Fermion_medial}
\end{equation}
which has the bosonic STR \eqref{eq:STR_Scalar_medial} as a special case for $\ell = 0$.
For $u_3 \rightarrow \frac{D}{2}$, we obtain the corresponding generalizations of the unitarity relations 
\begin{subequations}
{\allowdisplaybreaks
\begin{align}
\adjincludegraphics[valign=c,scale=0.9]{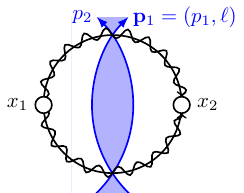} 
& =
\adjincludegraphics[valign=c,scale=0.9]{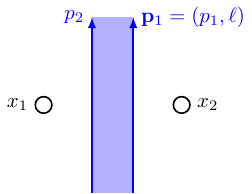} 
\cdot
\mathbb{I}^{(\ell)}
 ~, \label{eq:Unitarity_Fermion_medial_One} \\
\adjincludegraphics[valign=c,scale=0.9]{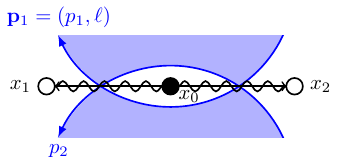} 
& =
\pi^D
a_\ell (u)
a_\ell (D - u)
\cdot
\adjincludegraphics[valign=c,scale=0.9]{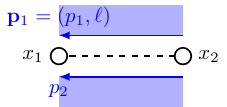} 
\cdot
\mathbb{I}^{(\ell)}
 ~. \label{eq:Unitarity_Fermion_medial_Delta}%
\end{align}\label{eq:Unitarity_Fermion_medial}%
}%
\end{subequations}
Note that we used the property of Clifford algebras, which is here $\sigma^\mu \bar{\sigma}^\nu + \sigma^\nu \bar{\sigma}^\mu = - 2 g^{\mu \nu} \mathbb{I}$ with $\mathbb{I}$ either $\delta_\alpha^{\beta}$ or $\delta_{\dot{\alpha}}^{\dot{\beta}}$.
Hence, the unit matrix $\mathbb{I}^{(\ell)}$ is the one appropriate to the spin structure on the right-hand side of the $D$-dimensional Clifford algebra: One has $\mathbb{I}_{2^{D/2-1}}$ for even $D$ and $\mathbb{I}_{2^{(D-1)/2}}$ for odd $D$ if $\ell = \frac{1}{2}$. 
In the scalar case $\ell = 0$, the unit matrix becomes a scalar factor, $\mathbb{I}^{(0)} = 1$.
The relation \eqref{eq:Unitarity_Fermion_medial_One} is sometimes called \textit{merging relation} \cite{Preti:2019rcq}, since it allows to merge propagators, which start and end at the same spacetime point.
While the merging leads to an annihilation in \eqref{eq:Unitarity_Fermion_medial_One}, there is also the constructive case $W_{u_1}^{\ell, \mu_1 \cdots \mu_{2\ell}}(x_{12}) \cdot W_{u_2}^{0}(x_{12}) = W_{u_1 + u_2}^{\ell, \mu_1 \cdots \mu_{2\ell}}(x_{12})$, which follows directly from the definition of the weights \eqref{eq:GenFermionicPropagatorWWeight_higherSpin}.

Furthermore, we can consider the limits $x_3 \rightarrow \infty$ and $x_2 \rightarrow \infty$ (or equivalently $x_1 \rightarrow \infty$).
This way we find the chain relations
\begin{subequations}
\begin{align}
\adjincludegraphics[valign=c,scale=1]{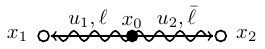} 
& =
r_\ell (D - u_1 - u_2 , u_1 , u_2) 
\cdot
\adjincludegraphics[valign=c,scale=1]{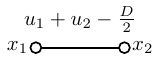}
\cdot
\mathbb{I}^{(\ell)}
 ~, \label{eq:ChainRuleFermionFermion} \\
\adjincludegraphics[valign=c,scale=1]{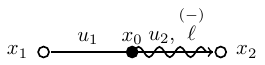} 
& =
r_\ell (u_1 , u_2 , D - u_1 - u_2) 
\cdot
\adjincludegraphics[valign=c,scale=1]{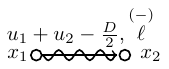} ~, \label{eq:ChainRuleScalarFermion}
\end{align}\label{eq:ChainRuleWithFermions}%
\end{subequations}
respectively.
The fermionic chain relation \eqref{eq:ChainRuleScalarFermion} also holds for $\ell > \frac{1}{2}$ and the higher spin weights \eqref{eq:GenFermionicPropagatorWWeight_higherSpin}.
However, the Lorentz indices have to be contracted with a symmetric traceless tensor\footnote{The higher-spin chain relation \eqref{eq:ChainRuleScalarFermion} can be shown by using the scalar chain relation \eqref{eq:ChainRuleScalar} and the useful representation of the higher-spin weight \eqref{eq:GenFermionicPropagatorWWeight_higherSpin} by derivatives acting on a scalar one \eqref{eq:GenPropagatorWWeight},
\begin{equation}
C_{\mu_1 \cdots \mu_{2\ell}}
W_{u}^{\ell, \mu_1 \cdots \mu_{2\ell}}(x_{12}) 
=
\frac{1}{(-2)^{2\ell}}
\frac{\Gamma (u - \ell)}{\Gamma (u + \ell)}
C_{\mu_1 \cdots \mu_{2\ell}}
\partial_1^{\mu_1} \cdots \partial_1^{\mu_{2\ell}}
W_{u - \ell} (x_{12}) ~.
\end{equation} 
Here, $C_{\mu_1 \cdots \mu_{2\ell}}$ has to be a traceless symmetric tensor \cite{Ferrando:2021yek}.}.

Next, we will generalize the scalar x-unity relation \eqref{eq:XUnity_Scalar} to include fermionic generalized propagators.
Again, we have two auxiliary relations, which are the generalizations of \eqref{eq:3ptFctnWithOneVanishingParameter_Scalar} and \eqref{eq:3ptFctnOnePointIntegrated_Scalar}.
They read
\begin{subequations}
\begin{align}
\lim_{\varepsilon\rightarrow 0}
\adjincludegraphics[valign=c,scale=0.8]{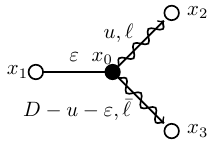}
& \stackrel{\eqref{eq:ChainRuleFermionFermion}}{=}
\pi^D
a_\ell(u)\, a_\ell(D - u) ~\cdot
\adjincludegraphics[valign=c,scale=0.8]{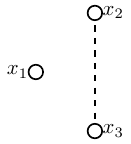} 
\cdot
\mathbb{I}^{(\ell)} ~,
\label{eq:3ptFctnWithOneVanishingParameter_Fermion} \\
\adjincludegraphics[valign=c,scale=0.8]{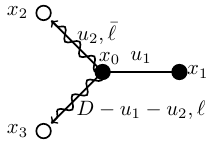} 
& \stackrel{\eqref{eq:GenFermionicPropagatorWWeight_medial}}{=}
\pi^D
a_\ell(u_1)\, a_\ell(D - u_1)
\adjincludegraphics[valign=c,scale=0.8]{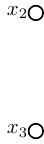} 
\cdot 
\mathbb{I}^{(\ell)} ~.
\label{eq:3ptFctnOnePointIntegrated_Fermion}
\end{align}
\end{subequations}
Then, we derive the x-unity relation from \eqref{eq:3ptFctnWithOneVanishingParameter_Fermion} and \eqref{eq:3ptFctnOnePointIntegrated_Fermion} as 
\begin{subequations}
{\allowdisplaybreaks
\begin{align}
\adjincludegraphics[valign=c,scale=0.8]{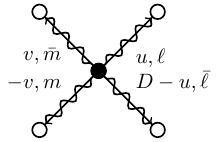} 
&\stackrel{\eqref{eq:Inversion_WWeights}}{=}
\lim_{\substack{\varepsilon\rightarrow 0\\\delta\rightarrow 0}}
\frac{1}{\pi^D a_0(\varepsilon) a_0(D - \varepsilon)}
\adjincludegraphics[valign=c,scale=0.8]{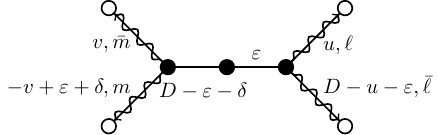} 
\\
&\stackrel{\eqref{eq:3ptFctnWithOneVanishingParameter_Fermion}}{=}
\left[
\lim_{\varepsilon\rightarrow 0}
\frac{a_\ell (u) a_\ell (D - u)}{a_0(\varepsilon) a_0(D - \varepsilon)}
\right]
\cdot
\lim_{\delta\rightarrow 0}
\adjincludegraphics[valign=c,scale=0.8]{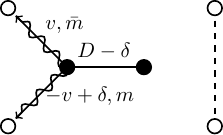}
\cdot 
\mathbb{I}^{(\ell)}
\\
&\stackrel{\eqref{eq:3ptFctnOnePointIntegrated_Fermion}}{=}
\left[
\lim_{\substack{\varepsilon\rightarrow 0\\\delta\rightarrow 0}}
\pi^D a_\ell (u) a_\ell (D - u)
\frac{a_0(\delta) a_0(D - \delta)}{a_0(\varepsilon) a_0(D - \varepsilon)}
\right]
\cdot
\adjincludegraphics[valign=c,scale=0.8]{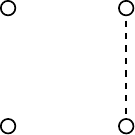} 
\cdot
\mathbb{I}^{(m)}
\mathbb{I}^{(\ell)}
\\
&\stackrel{\phantom{\eqref{eq:3ptFctnOnePointIntegrated_Fermion}}}{=}
\pi^D
a_\ell(u) a_\ell(D - u) \hspace{0.5cm}
\cdot
\adjincludegraphics[valign=c,scale=0.8]{figures/xunity/fermion/derivation/RHS/xunity_derivation3.pdf}
\cdot
\mathbb{I}^{(m)}
\mathbb{I}^{(\ell)}
 ~.
\end{align}
\label{eq:XUnity_Fermion_derivation}%
}%
\end{subequations}
In summary, the fermionic x-unity relation allows, as in the scalar case, to annihilate a four-point integral into a single delta function.
Compactly, we write
\begin{equation}
\adjincludegraphics[valign=c,scale=1]{figures/xunity/fermion/derivation/LHS/xmove1.pdf} 
~ = ~
\pi^D
a_\ell (u) a_\ell (D - u) \hspace{0.5cm}
\cdot
\adjincludegraphics[valign=c,scale=1]{figures/xunity/fermion/derivation/RHS/xunity_derivation3.pdf} 
\cdot
\mathbb{I}^{(m)}
\mathbb{I}^{(\ell)}
~.
\label{eq:XUnity_Fermion}
\end{equation}
Lastly, we construct the generalization of the fused R-matrix \eqref{eq:Rmatrix_Scalar_Fused}.
Therefore, we consider the intersection of a bosonic pair of medial lines with another pair subject to the fermionic generalization.
By \eqref{eq:GenFermionicPropagatorWWeight_medial}, we find
\begin{equation}
\mathbb{R}^{x_2 y_2}_{x_1 y_1}(p_1, p_2 \vert \mathbf{p}_3 , \mathbf{p}_4) = \adjincludegraphics[valign=c,scale=1]{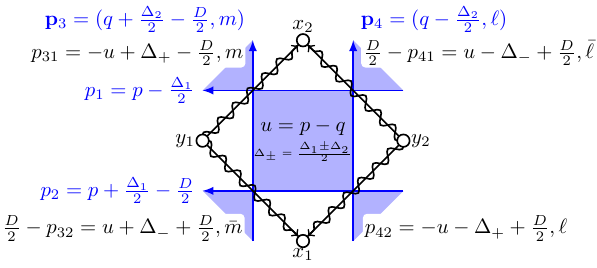} ~.
\label{eq:Rmatrix_Fermion_Fused}
\end{equation} 
The fermionic R-matrix was derived in \cite{Chicherin:2012yn}, we can adapt their parametrization of the exponents by writing $\mathbb{R}^{x_2 y_2}_{x_1 y_1\, m\ell}(u; \Delta_1, \Delta_2) =\mathbb{R}^{x_2 y_2}_{x_1 y_1}(p_1, p_2 \vert \mathbf{p}_3 , \mathbf{p}_4)$, and it first used in the QFT context of the dynamical fishnet theory in \cite{Kazakov:2018gcy}.
Note that in the case $\ell = m = 0$, we obtain the all-scalar R-matrix \eqref{eq:Rmatrix_Scalar_Fused}.
The generalized fermionic R-matrix solves a YBE, similar to \eqref{eq:YBE_Scalar_Fused},
\begin{equation}
\adjincludegraphics[valign=c,scale=1]{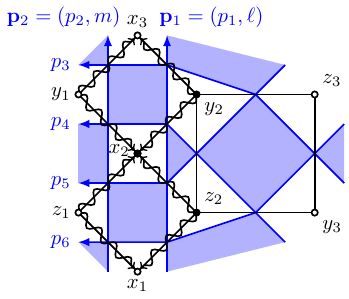} 
~ = ~
\adjincludegraphics[valign=c,scale=1]{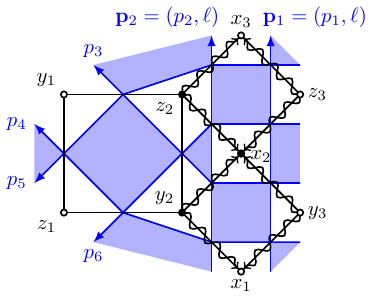} ~.
\label{eq:YBE_Fermion_Fused}
\end{equation}
Its proof works analogously to the scalar case \eqref{eq:YBE_Scalar_Fused_proof}, instead of the scalar STR we use the generalized fermionic one \eqref{eq:STR_Fermion_medial}.
Also, the composite factor made up from the usage of all the individual STRs cancels to one.
It is the same as \eqref{eq:YBE_Scalar_Fused_proof_factor} after replacing $r_0$ by $r_\ell$ or $r_m$.
Finally, as in the scalar case \eqref{eq:Unitarity_Rmatrix_scalar_fused}, the R-matrix \label{eq:Rmatrix_Fermion_Fused} satisfies unitarity relations, where the generalized propagators are either fermionic or scalar.

\section{The generalized superspace propagator as a lattice weight}
\label{sec:TheGeneralizedSuperspacePropagatorAsLatticeWeight}

\subsection{The generalized superpropagator}
\label{subsec:TheGeneralizedSuperpropagator}
Similar to the starting points of sections \ref{sec:TheGeneralizedScalarPropagatorAsLatticeWeight} and \ref{sec:TheGeneralizedFermionicPropagatorAsLatticeWeight}, the generalized superpropagator is motivated from the non-local deformation of the kinetic term for chiral superfields, which is called the K\"{a}hler potential.
We restrict ourselves to the canonical one, see e.\,g.\ \cite{Wess:1992cp,Buchbinder:1998qv,Gates:1983nr}, which is 
\begin{equation}
\int \dd^4 x \; \dd^2 \theta \, \dd^2 \bar{\theta} ~
\Phi^\dagger \square^{v} \Phi 
=
\int \dd^4 x\;
\left[ 
\phi^\dagger \square^{1+v} \phi -
\I \bar{\psi} \bar{\sigma}^\mu \square^{v} \partial_\mu \psi +
F^\dagger \square^{v} F
\right] ~,
\label{eq:Superfield_KinAction}
\end{equation}
with $\square = \eta^{\mu \nu} \partial_\mu \partial_\nu$.
The fermionic coordinates $\theta^\alpha$ and $\bar{\theta}^{\dot{\alpha}}$ are complex conjugated, spinor-valued Gra\ss mann numbers that anti-commute, $\lbrace \theta^\alpha, \theta^\beta \rbrace = 0$.
In our applications, the chiral superfields and their components eventually are matrix-valued fields in the adjoint representation of a gauge group, at this point of the derivation, however, we treat them like scalars.
With the notation of the $^\dagger$ we prepare for the use of derivation to superfields in the adjoint, yet a $^*$ would be more precise here.
Furthermore, we showcase the derivation of the generalized superpropagator in four-dimensional $\mathcal{N} = 1$ superspace, but the derivation for three-dimensional $\mathcal{N} = 2$ superspace is very similar.
It however differs in some signs due to the spinor conventions, e.\,g.\ see appendix \ref{appsec:TheThreeDimensionalN2Superspace}.
The chiral superfields are defined by the constraints $\bar{D}_{\dot{\alpha}} \Phi = 0$ and $D_\alpha \Phi^\dagger = 0$.
Consequently, in four-dimensional $\mathcal{N} = 1$ superspace, they admit an superspace expansion
\begin{subequations}
\begin{align}
\Phi
&=
\phi (x_+) + \sqrt{2}\; \theta \psi (x_+) + \theta^2 F(x_+)
=
\e^{\I \theta \sigma^\mu \bar{\theta} \partial_\mu}
\left[
\phi (x) + \sqrt{2}\; \theta \psi (x) + \theta^2 F (x)
\right]\\
&=
\phi (x)  +  \I \theta \sigma^\mu \bar{\theta} \partial_\mu \phi (x)  +   \frac{1}{4} \theta^2 \bar{\theta}^2 \square \phi (x)  +  \sqrt{2}\; \theta \psi (x)  -  \frac{\I}{2} \theta^2 \partial_\mu \psi(x) \sigma^\mu \bar{\theta}  +  \theta^2 F (x)\label{eq:Superfield_Components_chiral} \\
\Phi^\dagger 
&=
\phi^\dagger (x_-) + \sqrt{2}\; \bar{\theta} \bar{\psi} (x_-) + \bar{\theta}^2 F^\dagger(x_-)
=
\e^{-\I \theta \sigma^\mu \bar{\theta} \partial_\mu}
\left[
\phi^\dagger (x) + \sqrt{2}\; \bar{\theta} \bar{\psi} (x) + \bar{\theta}^2 F^\dagger(x)
\right]\\
&=
\phi^\dagger (x)  -  \I \theta \sigma^\mu \bar{\theta} \partial_\mu \phi^\dagger (x)  +   \frac{1}{4} \theta^2 \bar{\theta}^2 \square \phi^\dagger (x)  -  \sqrt{2}\; \bar{\theta} \bar{\psi} (x)  +  \frac{\I}{2} \bar{\theta}^2 \theta \sigma^\mu \partial_\mu \bar{\psi}(x)  +  \bar{\theta}^2 F^\dagger (x) \label{eq:Superfield_Components_antichiral}
\end{align}\label{eq:Superfield_Components}%
\end{subequations}
with $x^\mu_\pm = x^\mu \pm \I \theta \sigma^\mu \bar{\theta}$. 
The expansions \eqref{eq:Superfield_Components} lead to the r.\,h.\,s.\ of the kinetic component action \eqref{eq:Superfield_KinAction}.
Here, the $\phi$ denotes a complex scalar field, the $\psi$ a Weyl fermion and $F$ is a non-propagating auxiliary field.
In appendix \ref{appsec:TheFourDimensionalN1Superspace} and \ref{appsec:TheThreeDimensionalN2Superspace}, the suppression of spinor indices is explained in detail.
The generalized superfield propagator can be deduced from the individual generalized propagators of the component fields.
In fact, using \eqref{eq:Superfield_Components}, we find 
\begin{equation}
\begin{split}
\left\langle \Phi(z_1) \Phi^\dagger(z_2) \right\rangle 
=&~
\e^{
\I\theta_1 \sigma^\mu \bar{\theta}_1 \partial_{1,\mu} -
\I\theta_2 \sigma^\mu \bar{\theta}_2 \partial_{2,\mu}} \cdot \\
& \left[
\left\langle \phi(x_1) \phi^\dagger(x_2) \right\rangle 
+ 2 \theta_1^\alpha \bar{\theta}_2^{\dot{\alpha}}
\left\langle \psi_{\alpha}(x_1) \bar{\psi}_{\dot{\alpha}}(x_2) \right\rangle 
+ \theta_1^2 \bar{\theta}_2^2
\left\langle F(x_1) F^\dagger(x_2) \right\rangle
\right] ~.
\end{split}\label{eq:superPropagator_in_components}
\end{equation}
Here, $z_n = (x_n, \theta_n, \bar{\theta}_n)$ are again supercoordinates.
The generalized propagators of the components are derived from the inverse of the kinetic operators in the generalized action \eqref{eq:Superfield_KinAction}.
For the fermion, it was done in \eqref{eq:GenFermionicPropagatorGWeight} and for the boson and the auxiliary field we use the result of \eqref{eq:GenPropagatorGWeight} to find
\begin{subequations}
\begin{align}
\left\langle \phi(x_1) \phi^\dagger(x_2) \right\rangle 
& =
\phantom{-\I} \; G_{1 + v}(x_{12}) ~, \\
\left\langle \psi_{\alpha}(x_1) \bar{\psi}_{\dot{\alpha}}(x_2) \right\rangle
& =
- \I \, \sigma^\mu_{\alpha \dot{\alpha}} \partial_{1,\mu}\; G_{1 + v}(x_{12}) ~, \\
\left\langle F(x_1) F^\dagger(x_2) \right\rangle
& =
\phantom{-\I} \square_1\; G_{1 + v}(x_{12}) ~.\label{eq:gen_Propagators_aux}
\end{align}\label{eq:gen_Propagators_components}%
\end{subequations}
Here we used the notation $x_{12}^\mu = x_1^\mu - x_2^\mu$ and $G_u (x)$ is the generalized scalar propagator defined in \eqref{eq:GenPropagatorGWeight} for $D=4$.
For $u \rightarrow 0$, $G_{u}(x)$ becomes proportional to a delta function, see the representation \eqref{eq:Delta_bosonic} in combination with \eqref{eq:GenPropagatorGWeight}. 
This is in agreement with the expected delta function propagator for the non-dynamical auxiliary field \eqref{eq:gen_Propagators_aux} in the undeformed theory.
Finally, by plugging \eqref{eq:gen_Propagators_components} into \eqref{eq:superPropagator_in_components}, we obtain the generalized propagator of a chiral superfield
\begin{equation}
\begin{split}
\left\langle \Phi(z_1) \Phi^\dagger(z_2) \right\rangle
=
(-4)^{- v} \frac{\Gamma (1 - v)}{\Gamma (1 + v)}
\e^{\I \left[ 
\theta_1 \sigma^\mu \bar{\theta}_1 +
\theta_2 \sigma^\mu \bar{\theta}_2 -
2 \theta_1 \sigma^\mu \bar{\theta}_2 \right]\partial_{1,\mu}}
\frac{1}{\left[ x_{12}^2 \right]^{1 - v}} =
\frac{c_0 (1 + v) }{\left[x_{1\bar{2}}^2 \right]^{1 - v}} ~.
\label{eq:gen_superPropagator}
\end{split}
\end{equation}
The exponential in \eqref{eq:gen_superPropagator} is a shift operator and produces the superconformal covariant interval 
$x_{1\bar{2}}^\mu := x_{12}^\mu+\I \left[ 
\theta_1 \sigma^\mu \bar{\theta}_1 +
\theta_2 \sigma^\mu \bar{\theta}_2 -
2 \theta_1 \sigma^\mu \bar{\theta}_2 \right]$.
Note that even though we are in Minkowski spacetime we did not explicitly include the $\I\varepsilon$ in the denominators of \eqref{eq:gen_superPropagator} for conciseness of notation. 

Graphically the generalized superpropagator is represented by 
\begin{equation}
\left\langle \Phi(z_1) \Phi^\dagger(z_2) \right\rangle
=
c_0 (1 + v) \cdot
\adjincludegraphics[valign=c,scale=1]{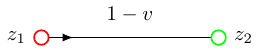} ,
\label{eq:gen_superPropagator_graphical}
\end{equation}
where the little arrow indicates the chiral end and $a_0(u)$ is a normalization defined in \eqref{eq:Factor_aEll}.
Observe that by tuning all spectral parameters to $v = 0$, the generalized superpropagator \eqref{eq:gen_superPropagator} reduces to the ordinary superpropagator $\sfrac{1}{x_{1\bar{2}}^2}$ derived in \cite{Wess:1992cp} because for $D=4$ we have $c_0 (1) = 1$.
Similar to the bosonic and fermionic cases, the pre-factor in \eqref{eq:gen_superPropagator_graphical} is just the proper normalization of the super Feynman diagrams under investigation.
Accordingly, we will focus in the following on the weight function
\begin{equation}
\frac{1}{\left[x_{1\bar{2}}^2 \right]^{u}}
=
\adjincludegraphics[valign=c,scale=1]{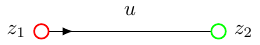} ~,
\label{eq:gen_superWeight_graphical}
\end{equation}
which happens to be the superspace generalization of \eqref{eq:GenPropagatorWWeight}.
Here $u\in \mathbb{C}$ is the spectral parameter needed to exploit the integrability of superspace Feynman diagrams.

The derivation of the $D=3$ $\mathcal{N} = 2$ superpropagator works analogously and we state the result
\begin{equation}
\begin{split}
\left\langle \Phi(z_1) \Phi^\dagger(z_2) \right\rangle
= 
(-1)^{v} 
2^{- \frac{1}{2} -2v} 
\frac{\Gamma (\frac{1}{2} - v)}{\Gamma (1 + v)}
\frac{1}{\left[x_{1\bar{2}}^2 \right]^{\frac{1}{2} - v}}
= 
\frac{c_0 ( 1 + v )}{\left[x_{1\bar{2}}^2 \right]^{\frac{1}{2} - v}} ~.
\label{eq:gen_superPropagator_3DN2}
\end{split}
\end{equation}
In three-dimensional $\mathcal{N} = 2$ superspace, the superconformal covariant interval is
$x_{1\bar{2}}^\mu := x_{12}^\mu+\I \left[ 
\theta_1 \gamma^\mu \bar{\theta}_1 +
\theta_2 \gamma^\mu \bar{\theta}_2 -
2 \theta_1 \gamma^\mu \bar{\theta}_2 \right]$, where the sigma matrices are replaced by three-dimensional gamma matrices, see appendix \ref{appsec:TheThreeDimensionalN2Superspace}.
Note that in the undeformed superfishnet theory, we will encounter factors $c_0(1) = \sqrt{\frac{\pi}{2}}$.
Also for the three-dimensional superspace, we employ the graphical notation \eqref{eq:gen_superWeight_graphical} for the superspace-dependent part of the propagator and write the spectral parameter next to it, while the dependence on the dimension $D$ is implicit.

In both superspaces, whenever the spectral parameter next to a superpropagator \eqref{eq:gen_superWeight_graphical} is not specified in the superspace diagrams below, we understand it to be set to the default value, which is $u=1$ and $u=\frac{1}{2}$ in four- and three-dimensional bosonic spacetime, respectively.
Partially, we will also affix a flavor index $i$ to the superpropagator; we hope that the difference between a spectral parameter and a flavor index will be clear from the context.

Moreover, we stress that the superpropagator in both superspaces is invariant under the global $\U{1}$ R-symmetry.
It acts on fermionic coordinates as $\theta \rightarrow \e^{\I \alpha}\theta$ and $\bar{\theta} \rightarrow \e^{-\I \alpha} \bar{\theta}$, and on the chiral superfields as
\begin{equation}
\begin{array}{cc}
\Phi \rightarrow \e^{-\frac{\I}{2} \alpha}\, \Phi (x,\e^{\I \alpha}\theta ,\e^{-\I \alpha} \bar{\theta}) ~,
& \Phi^\dagger \rightarrow \e^{\frac{\I}{2} \alpha}\, \Phi^\dagger (x,\e^{\I \alpha}\theta ,\e^{-\I \alpha} \bar{\theta}) ~.
\end{array}
\label{eq:U1_Rsymmetry_transformation}
\end{equation}
Consequently, the chiral superfield and its conjugate are defined to have R-charge $+\tfrac{1}{2}$ and $-\tfrac{1}{2}$, respectively.

Besides the propagator, we will encounter two other formal superspace two-point functions, 
\begin{equation}
\frac{\theta_{12}^2}{\left[x_{12}^2 \right]^{u}} =
\adjincludegraphics[valign=c,scale=0.7]{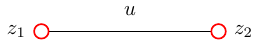} ~,
\hspace{1cm}
\frac{\bar{\theta}_{12}^2}{\left[x_{12}^2 \right]^{u}} =
\adjincludegraphics[valign=c,scale=0.7]{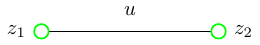} ~,
\label{eq:FeynmanRules_AuxTwoPointFctns}
\end{equation}
which have R-charge $-2$ and $2$, and represent chiral and anti-chiral delta functions on the fermionic subspace, respectively.
Here we introduced the abbreviation $\theta_{12} = \theta_1 - \theta_2$.
The dependence on the bosonic coordinates is that of a bosonic generalized propagator \eqref{eq:GenPropagatorWWeight}.
Therefore, there is the possibility that the spectral parameter approaches $\frac{D}{2}$.
Assuming the proper normalization, by the representation of the $D$-dimensional bosonic delta function \eqref{eq:Delta_bosonic}, the two-point functions \eqref{eq:FeynmanRules_AuxTwoPointFctns} can turn into chiral and anti-chiral superspace delta functions, respectively.
Graphically, we denote them with a dashed line, 
\begin{equation}
\delta^{(2)}\left(\theta_{12}\right) \delta^{(D)}\left( x_{12} \right) =
\adjincludegraphics[valign=c,scale=0.9]{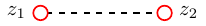} ~,
\hspace{0.2cm}
\delta^{(2)}\left(\bar{\theta}_{12}\right) \delta^{(D)}\left( x_{12} \right) =
\adjincludegraphics[valign=c,scale=0.9]{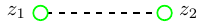} ~.
\label{eq:SuperspaceDeltaDefi}
\end{equation}

\subsection{Super chain relations}
\label{subsec:SuperChainRelations}
A handy tool in calculating supergraphs are the super chain relations.
They allow the reduction of the super-convolution of two two-point functions to a single one and are the superspace equivalents of the bosonic- and fermionic relations in sections \ref{subsec:ScalarChainRelations} and \ref{sec:TheGeneralizedFermionicPropagatorAsLatticeWeight}, respectively.
We compactly display the relations graphically for both superspaces, the three-dimensional $\mathcal{N} = 2$ one and the four-dimensional $\mathcal{N} = 1$ one.
At first, we consider the convolution of two generalized superpropagators \eqref{eq:gen_superWeight_graphical}.
We find their anti-chiral chain rule\footnote{Note that we can lift the restriction to chiral external superspace points by action with the operators $\e^{\I \theta_1 \gamma^\mu \bar{\theta}_1 \partial_{1,\mu}}$ and $\e^{\I \theta_2 \gamma^\mu \bar{\theta}_2 \partial_{2,\mu}}$. Accordingly, the chiral chain rule can be lifted by similar operators.}
\begin{subequations}
\begin{align}
\left.
\int \dd^D x_0\, \dd^2\bar{\theta}_0
\frac{1}{\left[x_{1\bar{0}}^2 \right]^{u_1}}
\frac{1}{\left[x_{2\bar{0}}^2 \right]^{u_2}}
\right|_{\substack{\theta_0 = 0 \\ \bar{\theta}_{1,2}=0}}
=
\frac{\theta_{12}^2}{\left[x_{12}^2 \right]^{u_1+u_2 - \frac{D}{2} + 1}} 
\cdot 
\left\lbrace 
\begin{matrix}
 4 \, r_0 ( 2 - u_1 - u_2 , u_1 , u_2 ) \\
- 4 \, r_0 ( 3 - u_1 - u_2 , u_1 , u_2 )
\end{matrix}
\right. 
~, \\
\adjincludegraphics[valign=c,scale=0.8]{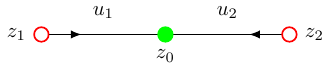} 
= 
\adjincludegraphics[valign=c,scale=0.8]{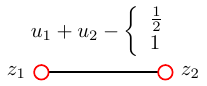}
\cdot 
\left\lbrace 
\begin{matrix}
4 \, r_0 ( 2 - u_1 - u_2 , u_1 , u_2 ) \\
- 4 \, r_0 ( 3 - u_1 - u_2 , u_1 , u_2 )
\end{matrix}
\right. 
 ~,
\end{align}
\label{eq:ChainRelAntiChiral}%
\end{subequations}
and its chiral counterpart
\begin{subequations}
\begin{align}
\left.
\int \dd^D x_0\, \dd^2\theta_0
\frac{1}{\left[x_{0\bar{1}}^2 \right]^{u_1}}
\frac{1}{\left[x_{0\bar{2}}^2 \right]^{u_2}}
\right|_{\substack{\bar{\theta}_0 = 0 \\ \theta_{1,2}=0}}
=
\frac{\bar{\theta}_{12}^2}{\left[x_{12}^2 \right]^{u_1+u_2 - \frac{D}{2} + 1}}
\cdot 
\left\lbrace 
\begin{matrix}
4 \, r_0 ( 2 - u_1 - u_2 , u_1 , u_2 ) \\
- 4 \, r_0 ( 3 - u_1 - u_2 , u_1 , u_2 )
\end{matrix}
\right. 
 ~, \\
\adjincludegraphics[valign=c,scale=0.8]{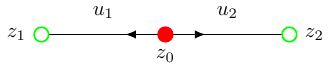} 
= 
\adjincludegraphics[valign=c,scale=0.8]{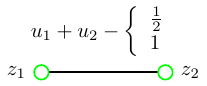} 
\cdot 
\left\lbrace 
\begin{matrix}
4 \, r_0 ( 2 - u_1 - u_2 , u_1 , u_2 ) \\
- 4 \, r_0 ( 3 - u_1 - u_2 , u_1 , u_2 )
\end{matrix}
\right. 
 ~.
\end{align}
\label{eq:ChainRelChiral}%
\end{subequations}
Here, the upper case corresponds to the chain relation in $D = 3$ $\mathcal{N} = 2$ superspace, while the lower case is related to $D = 4$ $\mathcal{N} = 1$ superspace.
Respectively, the spinor conventions of appendix \ref{appsec:TheThreeDimensionalN2Superspace} and \ref{appsec:TheFourDimensionalN1Superspace} have to be used, and their difference is the reason for the respective minus sign in the chain relations.
The chain relations \eqref{eq:ChainRelAntiChiral} and \eqref{eq:ChainRelChiral} are proven in appendix \ref{subsec:SuperChainRule}.
A comment on the color-coding is in place:
here and in the following we will denote internal, integrated, chiral (anti-chiral) vertices by a filled red (green) dot, which indicate the integration over the bosonic subspace in addition to the chiral (anti-chiral) fermionic subspace.
Note that the Gra\ss mann delta functions in \eqref{eq:ChainRelAntiChiral} and \eqref{eq:ChainRelChiral} annihilate the part of the fermionic integration whose chirality is opposite to the one of the vertex at hand.
Hence, propagators \eqref{eq:gen_superWeight_graphical} always connect the chiral and anti-chiral subspaces of superspace, while R-charged two-point functions \eqref{eq:FeynmanRules_AuxTwoPointFctns} connect the subspaces homogeneously.
When an external point in a super Feynman diagram is expected to be integrated by a chiral (anti-chiral) vertex in a later step according to the Feynman vertex rules of the QFT under examination, we denote this by a un-filled red (green) circle.

There exists a critical limiting case of \eqref{eq:ChainRelAntiChiral} and \eqref{eq:ChainRelChiral}, which is when $u_1$ and $u_2$ are chosen to add up to $D - 1$.
Then, the r.\,h.\,s.\ also becomes a delta function on the bosonic space, c.\,f.\ \eqref{eq:Delta_bosonic}.
The resulting relations are 
\begin{subequations}
\begin{align}
\lim_{\varepsilon \rightarrow 0}
\adjincludegraphics[valign=c,scale=0.8]{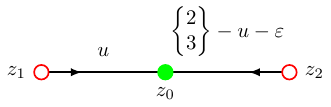} 
&= 
\adjincludegraphics[valign=c,scale=0.8]{figures/superFN/basics/deltadist/chiral_delta/DeltaDist.pdf}
\cdot 
\left\lbrace 
\begin{matrix}
4 \,\pi^{3} \cdot a_0 ( u )\, a_0 ( 2 - u ) \\
- 4\,\pi^{4} \cdot a_0 ( u )\, a_0 ( 3 - u )
\end{matrix}
\right.
 ~,
\label{eq:ResolutionOfUnity_antichiral}\\
\lim_{\varepsilon \rightarrow 0}
\adjincludegraphics[valign=c,scale=0.8]{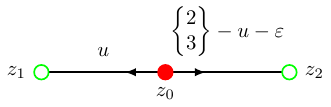} 
&= 
\adjincludegraphics[valign=c,scale=0.8]{figures/superFN/basics/deltadist/antichiral_delta/DeltaDist.pdf}
\cdot 
\left\lbrace 
\begin{matrix}
4 \,\pi^{3} \cdot a_0 ( u )\, a_0 ( 2 - u ) \\
- 4\,\pi^{4} \cdot a_0 ( u )\, a_0 ( 3 - u )
\end{matrix}
\right.
~. \label{eq:ResolutionOfUnity_chiral}
\end{align}
\label{eq:ResolutionOfUnity}%
\end{subequations}
Furthermore, we consider convolutions of a superpropagator \eqref{eq:gen_superWeight_graphical} with an R-charged two-point function \eqref{eq:FeynmanRules_AuxTwoPointFctns}.
Starting with the anti-chiral version, we find 
\begin{subequations}
\begin{align}
&\left.
\int \dd^D x_0\, \dd^2\bar{\theta}_0
\frac{1}{\left[x_{1\bar{0}}^2 \right]^{u_1}}
\frac{\bar{\theta}_{02}^2}{\left[x_{20}^2 \right]^{u_2}}
\right|_{\substack{\theta_{0,2} = 0 \\ \bar{\theta}_{1}=0}}
=
r_0 ( D - u_1 - u_2 , u_1 , u_2 )\;
\frac{1}{\left[x_{1\bar{2}}^2 \right]^{u_1+u_2 - \frac{D}{2}}} ~, \\
&\adjincludegraphics[valign=c,scale=0.8]{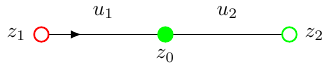} 
= 
\adjincludegraphics[valign=c,scale=0.8]{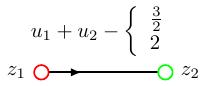} 
\cdot 
\left\lbrace 
\begin{matrix}
r_0 ( 3 - u_1 - u_2 , u_1 , u_2 ) \\
r_0 ( 4 - u_1 - u_2 , u_1 , u_2 )
\end{matrix}
\right. 
 ~,
\end{align}
\label{eq:ChainRelAntiChiral_Aux}%
\end{subequations}
and its chiral counterpart
\begin{subequations}
\begin{align}
&\left.
\int \dd^D x_0\, \dd^2\theta_0
\frac{1}{\left[x_{0\bar{1}}^2 \right]^{u_1}}
\frac{\theta_{02}^2}{\left[x_{20}^2 \right]^{u_2}}
\right|_{\substack{\bar{\theta}_{0,2} = 0 \\ \theta_{1}=0}}
=
r_0 ( D - u_1 - u_2 , u_1 , u_2 )\;
\frac{1}{\left[x_{2\bar{1}}^2 \right]^{u_1+u_2 - \frac{D}{2}}} ~, \\
& \adjincludegraphics[valign=c,scale=0.8]{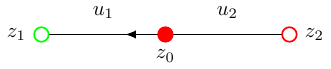} 
= 
\adjincludegraphics[valign=c,scale=0.8]{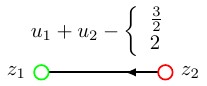} 
\cdot 
\left\lbrace 
\begin{matrix}
r_0 ( 3 - u_1 - u_2 , u_1 , u_2 ) \\
r_0 ( 4 - u_1 - u_2 , u_1 , u_2 )
\end{matrix}
\right. 
 ~.
\end{align}
\label{eq:ChainRelChiral_Aux}%
\end{subequations}
The proof is straightforward: after writing the superpropagator via a differential shift operator, we can perform the bosonic spacetime integral with the bosonic chain relation \eqref{eq:ChainRuleScalar} and the fermionic space integral is trivially annihilated by the Gra\ss mann delta function $\theta^2_{ij}$ or $\bar{\theta}^2_{ij}$.

\subsection{Super x-unity relation}
The super x-unity relation is the pivotal auxiliary relation for calculating the critical coupling of supersymmetric QFTs in chapter \ref{chpt:SupersymmetricDoubleScaledDeformations}.
It allows us to find the inverse of the graph-building row-matrix of the super vacuum graphs and to apply the method of inversion relations.
The derivation of the super x-unity relation relies on the super chain relations in section \ref{subsec:SuperChainRelations}.

Let us make two observations before deriving the super x-unity relation. 
\begin{itemize}
\item 
Consider a three-spiked star integral, where the weights add up to $D-1$, and where we take one superpropagator weight to zero.
Then the propagator disappears in the limit, and we obtain the integral relation \eqref{eq:ResolutionOfUnity_antichiral}.
This yields the relation
\begin{equation}
\lim_{\varepsilon\rightarrow 0}
\adjincludegraphics[valign=c,scale=1]{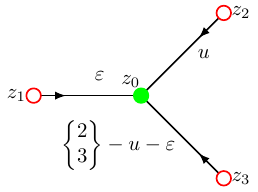} 
=
\left\lbrace 
\begin{matrix}
4\,\pi^3  a_0(u)\, a_0(2 - u) \\
- 4\,\pi^4  a_0(u)\, a_0(3 - u) 
\end{matrix}
\right\rbrace
\cdot ~
\adjincludegraphics[valign=c,scale=1]{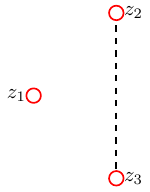}  ~.
\label{eq:3ptFctnWithOneVanishingParameter}
\end{equation}

\item
We obtain another helpful relation when integrating one external point of a three-spiked star integral over the chiral subspace of superspace, in the case where the propagator weights add up to $D - 1$: 
\begin{equation}
\adjincludegraphics[valign=c,scale=1]{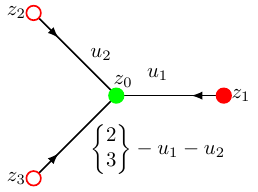} 
=
\left\lbrace 
\begin{matrix}
4\,\pi^3  a_0(u)\, a_0(2 - u) \\
- 4\,\pi^4  a_0(u)\, a_0(3 - u) 
\end{matrix}
\right\rbrace
\cdot ~
\adjincludegraphics[valign=c,scale=1]{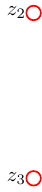} 
\label{eq:3ptFctnOnePointIntegrated}
\end{equation}
The derivation of the relation \eqref{eq:3ptFctnOnePointIntegrated} is shown in detail in appendix \ref{subsec:auxRelForXUnity}.
Therein, the use of the bosonic STR \eqref{eq:CFTUniqueness_Scalar} is crucial, which is a direct imprint of integrability.
We observe that the whole expression reduces to a factor independent of $u_2$.
\end{itemize}
With the help of the relations \eqref{eq:ResolutionOfUnity_chiral}, \eqref{eq:3ptFctnWithOneVanishingParameter} and \eqref{eq:3ptFctnOnePointIntegrated} we can follow the steps of \eqref{eq:XUnity_Scalar_derivation} to derive the super x-unity relations for the $D=3$ $\mathcal{N} = 2$ (upper case) and $D = 4$ $\mathcal{N} = 1$ superspace (lower case) and we find, respectively,
\begin{subequations}
\begin{align}
\adjincludegraphics[valign=c,scale=1]{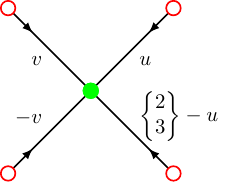} 
=
\left\lbrace 
\begin{matrix}
4\,\pi^3  a_0(u)\, a_0(2 - u) \\
- 4\,\pi^4  a_0(u)\, a_0(3 - u) 
\end{matrix}
\right\rbrace
\cdot \hspace{0.5cm}
\adjincludegraphics[valign=c,scale=1]{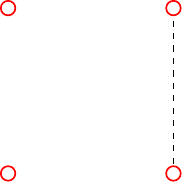} ~, \label{eq:SuperXUnity_green}\\
\adjincludegraphics[valign=c,scale=1]{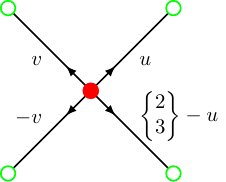} 
=
\left\lbrace 
\begin{matrix}
4\,\pi^3  a_0(u)\, a_0(2 - u) \\
- 4\,\pi^4  a_0(u)\, a_0(3 - u) 
\end{matrix}
\right\rbrace
\cdot \hspace{0.5cm}
\adjincludegraphics[valign=c,scale=1]{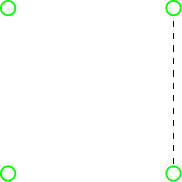} 
 ~.\label{eq:SuperXUnity_red}
\end{align}\label{eq:SuperXUnity}% 
\end{subequations}
The last equation is the chiral counterpart of the x-unity derived above and it is shown the same way by using the relations with inverted chirality.

\subsection{Superconformal star integral and a super star-triangle relation?}
\label{subsec:SuperconfStarIntegral}
To construct commuting transfer matrices, a STR is sufficient. 
However, for superspace propagators, it seems to be missing.
It prevents us from building an integrable model from the superspace propagators.
This raises the question of how integrability is present in superspace vacuum diagrams.
We argue that superspace vacuum graphs still hide an integrable structure, since the super x-unity relations \eqref{eq:SuperXUnity} still required the bosonic STR \eqref{eq:CFTUniqueness_Scalar}.
However, the explicit form of a superspace STR is still unclear.

We can think of the naive superspace generalization where the star side of the STR consists of three superspace propagators of the form \eqref{eq:gen_superWeight_graphical} meet in an integrated chiral or anti-chiral vertex.
In four-dimensional $\mathcal{N} = 1$ superspace, there is Osborn's star integral \cite{Osborn:1998qu,Dolan:2000uw}.
It reads
\begin{equation}
\begin{split}
& \I\int \dd^4x_0\; \dd^2\theta_0\, \dd^2 \bar{\theta}_0\; \delta^{(2)}(\theta_0)\;
\frac{1}{\left[x_{1\bar{0}}^2 \right]^{u_1}}
\frac{1}{\left[x_{2\bar{0}}^2 \right]^{u_2}}
\frac{1}{\left[x_{3\bar{0}}^2 \right]^{u_3}}\\
& \stackrel{u_1+u_2+u_3=3}{=}
-4\,
r_0 (u_1, u_2, u_3)\;
\frac
{
\left(\theta_{12}\theta_{13}\right) x_{23,+}^2 +
\left(\theta_{23}\theta_{21}\right) x_{31,+}^2 +
\left(\theta_{31}\theta_{32}\right) x_{12,+}^2 
}
{
[ x_{12,+}^2 ]^{2-u_3}
[ x_{23,+}^2 ]^{2-u_1}
[ x_{31,+}^2 ]^{2-u_2}
} ~.
\end{split}
\label{eq:Osborn_original}
\end{equation}
Here, the abbreviations $x_{ij,+}^\mu := x_{i,+}^\mu - x_{j,+}^\mu$ are used.
Note the similarity of \eqref{eq:Osborn_original} with a star-triangle relation (STR), see e.\,g.\ \cite{Chicherin:2012yn,DEramo:1971hnd,Symanzik:1972wj}.
Generally, a STR is a relation between a subgraph in the shape of a three-spiked star and a triangle, modulo some factor, which depends on the model under investigation.
However, the r.h.s. of \eqref{eq:Osborn_original} is not quite of the form of a triangle built from superpropagators of the type \eqref{eq:gen_superPropagator_graphical}, due to the non-factorizing numerator.
Remarkably, despite this shortcoming, in the below chapter \ref{chpt:SupersymmetricDoubleScaledDeformations} we will provide exciting evidence for the model's integrability by demonstrating that Zamolodchikov's method of inversions \cite{Zamolodchikov:1980mb}, see also \cite{Kade:2023xet}, may nevertheless be successfully applied.

\subsection{Spinning generalized superpropagators}
\label{eq:SpinningGeneralizedSuperpropagators}
Similar to the fermionic generalized propagators in section \eqref{sec:TheGeneralizedFermionicPropagatorAsLatticeWeight}, we can generalize the superpropagator by acting with $S$ derivatives on it.
The same can be done with the R-charged two-point functions \eqref{eq:FeynmanRules_AuxTwoPointFctns}.
We require the derivatives to be contracted with a traceless and symmetric tensor and then find 
\begin{subequations}
\begin{align}
\adjincludegraphics[valign=c,scale=1]{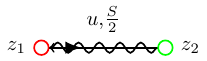}
:= &
\frac{\Gamma ( u-\frac{S}{2} ) }{\Gamma ( u+\frac{S}{2} ) } 
\frac{\partial_1^{\mu_1} \cdots \partial_1^{\mu_S}}{(-2)^S}
\adjincludegraphics[valign=c,scale=1]{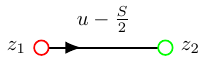} 
=
\frac{1}{\left[x_{1\bar{2}}^2 \right]^{u}}
\prod_{i=0}^S
\frac{x_{1\bar{2}}^{\mu_i}}{|x_{1\bar{2}}|} ~, \label{eq:SpinningSuperpropagators_prop}\\
\adjincludegraphics[valign=c,scale=1]{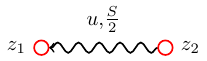}
:= &
\frac{\Gamma ( u-\frac{S}{2} ) }{\Gamma ( u+\frac{S}{2} ) } 
\frac{\partial_1^{\mu_1} \cdots \partial_1^{\mu_S}}{(-2)^S}
\adjincludegraphics[valign=c,scale=1]{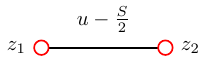} 
=
\frac{ \theta_{12}^2 }{\left[x_{12}^2 \right]^{u}}
\prod_{i=0}^S
\frac{x_{12}^{\mu_i}}{|x_{12}|} ~.
\label{eq:SpinningSuperpropagators_aux}
\end{align}
\label{eq:SpinningSuperpropagators}%
\end{subequations} 
Of course, a similar expression is obtained for the anti-chiral R-charged two-point function, but we focus here on those relations that are necessary for the diagonalization of the superfishnet graph-builder in section \ref{subsec:ZeroMagnonCase}.
From the definitions of the spinning superweights \eqref{eq:SpinningSuperpropagators_prop}, we observe that we can merge a spinning $S \neq 0$ with a non-spinning $S=0$ superweight, whenever they start and end at the same superspace points.
The resulting spinning superpropagator carries the sum of the two merged spectral parameters.\\
With the spinning super-two-point functions, we can generalize some of the three-dimensional $\mathcal{N} = 2$ super chain relations in section \ref{subsec:SuperChainRelations}.
We find 
\begin{subequations}
\begin{align}
\adjincludegraphics[valign=c,scale=1]{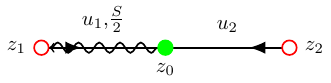} 
&=
4\,
r_\frac{S}{2} (u_2 , u_1 , 2 - u_1 - u_2)
\adjincludegraphics[valign=c,scale=1]{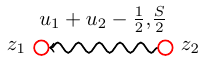} ~, \\
\adjincludegraphics[valign=c,scale=1]{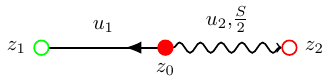} 
&=
r_\frac{S}{2} (u_1 , u_2 , 3 - u_1 - u_2)
\adjincludegraphics[valign=c,scale=1]{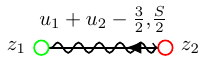} ~.
\end{align}
\label{eq:Superchainrelations_spinning}%
\end{subequations}
Notice that the only difference to the non-spinning super chain relations \ref{subsec:SuperChainRelations} is that the index of the $r_\ell$-factor \eqref{eq:Factor_rEll} captures the dependence on $S$.
The proof of \eqref{eq:Superchainrelations_spinning} is easily obtained by letting the differentials of \eqref{eq:SpinningSuperpropagators} act on the non-spinning super chain relations \ref{subsec:SuperChainRelations} and rearranging the gamma functions in the $r_\ell$-factor.

The (super) propagators of this chapter can be thought of as weights in the bulk of a Feynman (super) graph.
Before taking the insights of this chapter into practice in chapters \ref{sec:IntegrableVacuumDiagramsAndTheCriticalCoupling} and \eqref{chpt:SupersymmetricDoubleScaledDeformations}, we present another class of lattice weights in the next chapter, which relate to the concept of boundary integrability and are therefore called boundary weights.

%%%----------------------------------------------------------------------------------------
%%%	THESIS CONTENT - PART 4 - Non-supersymmetric fishnet theories and its relatives
%%%----------------------------------------------------------------------------------------
\chapter{Boundary integrability in Feynman graphs}
\label{chpt:BoundaryIntegrabilityInFeynmanGraphs}

Establishing the integrability of a model frequently requires finding a suitable R-matrix that solves the Yang-Baxter equation (YBE).
We may then introduce the concept of spectral lines intersecting at an R-matrix.
If possible and as a first step, one postpones the question about boundary conditions and considers several spectral lines living on a doubly periodic torus.
This is instrumental for the construction of the transfer matrix:
a closed spectral line, with intersecting further spectral lines, and wrapping around one cycle of a torus.
The YBE implies the commutativity of the product of two transfer matrices by the train track argument; see \eqref{eq:6V_CommutingRows}.
Graphically, the ``product'' means that the intersecting spectral lines of the two transfer matrices are suitably contracted.
Commuting transfer matrices are a sufficient condition for integrability, since expanding the transfer matrix in its spectral parameter (graphically, the difference of slopes of the intersecting spectral lines) gives infinitely many conserved charges.

Aside from the doubly periodic boundary condition, one can imagine introducing one, or even multiple, boundaries for the spectral lines.
Some examples of such bounded spaces are the semi-infinite cylinder with one boundary, the finite cylinder with two boundaries and one periodic direction, and the box as introduced for the six-vertex model, see section \ref{sec:TheBoxBoundaryCondition}.
Consequently, one has to determine what happens when a spectral line hits a boundary.
Commonly, the spectral lines are assumed to be reflected by the boundary, or similarly the boundary serves as a source or sink for a pair of spectral lines.
The behavior is described by another object, the so-called K-matrix, which graphically sits at the point where the spectral lines bounce, spawn, or vanish.

Along a non-periodic direction, the construction of a periodic transfer matrix is impossible.
Hence, one might wonder how to construct commuting objects and how to maintain integrability.
In the case of two boundaries on opposite sides, Sklyanin showed that commuting double-row transfer matrices can be built \cite{Sklyanin:1988yz}.
The commutativity is a consequence of the YBE and a novel relation, the boundary Yang-Baxter equation (bYBE).
Thereby, the bYBE has to be fulfilled by the K-matrix, while the R-matrix in the bulk is constrained by the requirement that the YBE holds.

In this chapter, we will introduce the notion of boundary integrability for generalized Feynman graphs.
First, we introduce reflected medial lines that have to be compatible with bulk integrability.
We use the map \eqref{eq:Map_FeynmanMedial} to translate the graphical picture of reflecting medial lines satisfying a bYBE to Feynman diagrams.
Thus, the K-matrix is associated with the so-called \textit{boundary weights} and we find two distinct versions of them.
Furthermore, we show that the bYBE gives rise to two boundary star-triangle relations (bSTR).
Based on this, we continue with deriving the fused K-matrix in analogy to \eqref{eq:Rmatrix_Scalar_Fused} and we show that it solves a bYBE as well.
Recently, a similar K-matrix was derived from algebraic considerations of $\SL{2, \mathbb{C}}$ \cite{antonenko2024reflection1,antonenko2024reflectionoperatorhypergeometryii}, here we present a graphical approach.

\section{The K-matrix and the boundary weights}
\label{eq:TheKmatrixAndTheBoundaryWeights}
We start by assigning a K-matrix to a medial line that bounces off a boundary. 
First, we consider the case of a boundary to the left. 
Of course, there is also a K-matrix for possible other boundaries and for every boundary that can be a different solution of the bYBE.
In contrast to the source/sink formulation in section \ref{sec:TheBoxBoundaryCondition} for the six-vertex model, we restrict ourselves to the case where the medial lines are reflected at the boundary.
Then the faces of the medial graph correspond to spacetime points, and since we do not introduce a shading yet, we obtain the factorized K-matrix analog of \eqref{eq:Rmatrix_facetype_WWbar},
\begin{equation}
K^{y_1}_{y_2 x}(p)
=
V_{p}(x)
V_{p}(y_1 , y_2)
=
\adjincludegraphics[valign=c,scale=0.9]{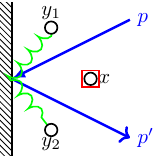} ~.
\label{eq:Kmatrix_facetype}
\end{equation}
We introduced some new graphical notations.
To obtain two independent Feynman graphs after shading the medial graph, we have two different boundary weights $V_{p}(x)$ and $V_{p}(y_1 , y_2)$.
They depend on one (red) and two (green) points, and we call them of types \RNum{1} and \RNum{2}, respectively.
We denote them by drawing a box around the spacetime point in the argument or by connecting the two spacetime points by a curly line.
Importantly, the boundary weights $V_{p}(x)$ and $V_{p}(y_1 , y_2)$ are fundamentally different and cannot be related by a crossing transformation, as is the case for the bulk weights, which build up the R-matrix \eqref{eq:Rmatrix_facetype_WWbar}.
Furthermore, we denote the reflected medial line by a different arrowhead. 
Its spectral parameter $p'$ has to be related to the original one, $p$, and it turns out that setting $p' = -p$ is a useful choice to avoid additional factors in the bYBEs.
Furthermore, the solutions we find below require this relation.
However, in the following, we keep using the prime-notation $p' = - p$, since it allows for a concise notation.

The reflected medial lines have to satisfy the STR as well to preserve bulk integrability.
In chapter \ref{chpt:AuxiliaryRelations} the canonical directions of the medial lines were up and to the left, here we potentially will also encounter reflected lines, which point down or to the right.
Therefore we require the YBEs 
\begin{equation}
\begin{array}{ccc}
\adjincludegraphics[valign=c,scale=0.8]{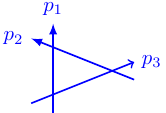} 
=
\adjincludegraphics[valign=c,scale=0.8]{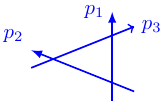}, &
\adjincludegraphics[valign=c,scale=0.8]{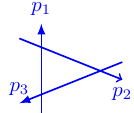} 
=
\adjincludegraphics[valign=c,scale=0.8]{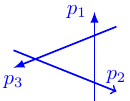}, &
\adjincludegraphics[valign=c,scale=0.8]{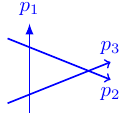} 
=
\adjincludegraphics[valign=c,scale=0.8]{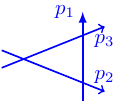},
\end{array}
\label{eq:YBE_Scalar_reflected}
\end{equation}
to hold.
In turn, we have to extend the relation between generalized propagators and medial lines in \eqref{eq:GenPropagatorWWeight_medial} to intersections involving reflected medial lines.
In accordance with the uniqueness relation \eqref{eq:CFTUniqueness_Scalar_Diagram} as a YBE in the sense of \eqref{eq:YBE_Scalar_reflected}, we have additionally to \eqref{eq:GenPropagatorWWeight_medial}
\begin{subequations}
\begin{align}
W_u (x_{12})
\stackrel{u = q - p}{=}
\adjincludegraphics[valign=c,scale=1]{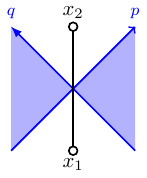} 
\quad \mathrm{and} \quad
\bar{W}_u (x_{12})
\stackrel{u = q - p}{=}
\adjincludegraphics[valign=c,scale=1]{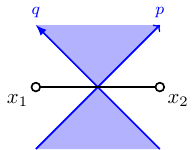} 
 ~, \\
W_u (x_{12})
\stackrel{u = p - q}{=}
\adjincludegraphics[valign=c,scale=1]{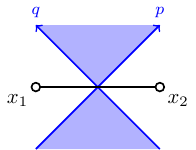} 
\quad \mathrm{and} \quad
\bar{W}_u (x_{12})
\stackrel{u = p - q}{=}
\adjincludegraphics[valign=c,scale=1]{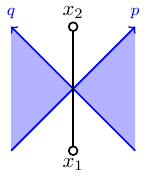} ~.
\end{align}\label{eq:GenPropagatorWWeight_medial_reflected}%
\end{subequations}
In summary, whenever a medial line intersects with a reflected one, we have an exponent of the generalized propagator related to the \textit{left minus right} spectral parameter.
Furthermore, we exchange the crossed weight with the ordinary one in this case.
If two reflected medial lines intersect, the exponent is related to a difference of the same order as if two ordinary medial lines meet, \textit{right minus left}.
In addition, the crossed weights are, as usual, \eqref{eq:GenPropagatorWWeight_medial}, related to a propagator passing in between the spectral lines.

The bYBE, expressed graphically, is the relation
\begin{equation}
\adjincludegraphics[valign=c,scale=0.9]{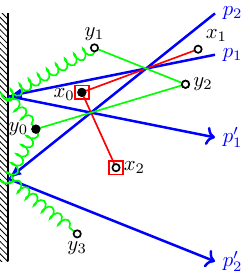} 
\qquad = \qquad
\adjincludegraphics[valign=c,scale=0.9]{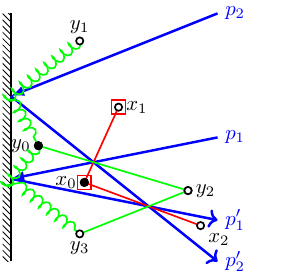} ~.
\label{eq:bYBE_facetype_Diagram}
\end{equation} 
It is the boundary analog of \eqref{eq:YBE_facetype_Diagram2} and since the both objects, the R-matrix and the K-matrix, factorize, we have two independent, but dual, generalized Feynman graphs $\mathscr{F}$ and $\mathscr{F}^*$ in \eqref{eq:bYBE_facetype_Diagram}, drawn in red and green.
Algebraically, the diagram \eqref{eq:bYBE_facetype_Diagram} encodes the equation
\begin{equation}
\begin{split}
&\int \dd^D x_0 \, \dd^D y_0 ~
R_{x_1 y_2}^{x_0 y_1} ( p_{12} )
K_{y_0 x_0}^{y_1} ( p_1 )
R_{y_0 x_0}^{y_2 x_2} ( p_{1'2} )
K_{y_3 x_2}^{y_0} ( p_2 ) \\
=
&\int \dd^D x_0 \, \dd^D y_0 ~
K_{y_0 x_1}^{y_1} ( p_2 )
R_{y_0 x_1}^{y_2 x_0} ( p_{2'1} )
K_{y_3 x_0}^{y_0} ( p_1 )
R_{x_0 y_2}^{x_2 y_3} ( p_{2'1'} ) ~,
\end{split}
\end{equation}
which is a face-type bYBE.
With the help of intertwining vectors \cite{Isaev:2022mrc}, we can turn it into the familiar vertex-type form due to Sklyanin \cite{Sklyanin:1988yz}.
We formally expressed the solution \eqref{eq:Kmatrix_facetype} in terms of the boundary weights $V_{p}(x)$ and $V_{p}(y_1 , y_2)$.
Consequently, we wish to determine the boundary weights, and for this purpose, we first derive the necessary relations for them from \eqref{eq:bYBE_facetype_Diagram}.

Hence, due to the fundamental difference of the boundary weights in the K-matrix \eqref{eq:Kmatrix_facetype}, we obtain two different bSTRs.
One corresponds to the generalized Feynman graph and the other to its dual.
This is in contrast to the bulk YBE \eqref{eq:YBE_facetype_Diagram}, which implies only the STR \label{eq:YBE_facetype_Diagram} to hold for the bulk weights on the Feynman graph and its dual.
The two bSTRs are 
\begin{subequations}
\begin{align}
\begin{split}
\adjincludegraphics[valign=c,scale=1]{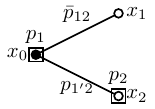} 
~ & = ~
\adjincludegraphics[valign=c,scale=1]{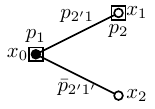} ~, \\
V_{p_2}(x_2) 
\int d^D x_0 \;
W_{\bar{p}_{12}} (x_{10})
V_{p_1}(x_0) &
W_{p_{1'2}} (x_{20})  \\
~ = ~
V_{p_2}(x_1) &
\int d^D x_0 \;
W_{p_{2'1}} (x_{10}) 
V_{p_1}(x_0)
W_{\bar{p}_{2'1'}} (x_{20}) ~, 
\end{split} \label{eq:bSTR_type1}\\
\begin{split}
\adjincludegraphics[valign=c,scale=1]{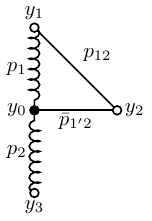} 
~ & = ~
\adjincludegraphics[valign=c,scale=1]{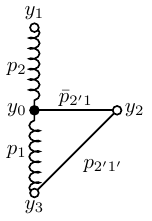} ~, \\
W_{p_{12}} (y_{12})  
\int d^D y_0 \;
V_{p_1}(y_1 , y_0)
W_{\bar{p}_{1'2}} (y_{20}) &
V_{p_2}(y_0 , y_3) \\
~ = ~ 
W_{p_{2'1'}} (y_{32}) 
\int d^D y_0 & \;
V_{p_2}(y_1 , y_0)
W_{\bar{p}_{2'1}} (y_{20}) 
V_{p_1}(y_0 , y_3) ~.
\end{split} \label{eq:bSTR_type2}
\end{align}
\label{eq:bSTR_type1And2}%
\end{subequations}
They separately only depend on one type of boundary weight, $V_{p}(x)$ and $V_{p}(y_1 , y_2)$, respectively.
Therefore, we have to find solutions to the bSTR which we can use to build integrable graphs with a integrability-preserving boundary.
Here and in the following, we use the abbreviations $p_{i'j} = p_i' - p_j$, $p_{ij'} = p_i - p_j'$ and $p_{i'j'} = p_i' - p_j'$.
In the field of statistical physics similar bSTR relations where obtained also for the chiral Potts model, among others \cite{Zhou:1997vkj,Zhou:1998xxr}.

Before finding solutions to the bSTRs \eqref{eq:bSTR_type1And2}, we fuse a pair of medial lines.
This gives us a shading-preserving K-matrix $\mathbb{K}$, in analogy to the R-matrix $\mathbb{R}$ in \eqref{eq:Rmatrix_Scalar_Fused}.
However, the reflection of a medial line is not covariantly transforming under inversion of the shading, since the two types of boundary weights are fundamentally different.
Therefore, we make a choice and construct the K-matrix $\mathbb{K}$ related to the boundary weight of type \RNum{1}.
Accordingly, the shaded medial- and the Feynman graph satisfy a fused boundary YBE that we denote by $\mathbb{bYBE}$,
\begin{equation}
\adjincludegraphics[valign=c,scale=1]{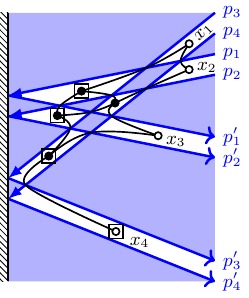} 
\qquad = \qquad
\adjincludegraphics[valign=c,scale=1]{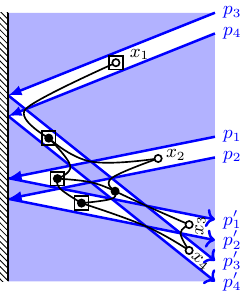} ~.
\label{eq:bYBE_fused_type1}
\end{equation}
It can be proven graphically by using the STR \eqref{eq:STR_Scalar_medial} and the red version of \eqref{eq:bYBE_facetype_Diagram} multiple times for various medial lines, which help to shift the $(p_1 ,p_2)$-pair to the bottom.
We will show \eqref{eq:bYBE_fused_type1} on the Feynman graph level, using the STR \eqref{eq:CFTUniqueness_Scalar_Diagram} and the bSTR \eqref{eq:bSTR_type1}.
We start with the Feynman graph on the l.\,h.\,s.\ of \eqref{eq:bYBE_fused_type1} and perform the steps
\begin{equation}
\begin{split}
\adjincludegraphics[valign=c,scale=0.6]{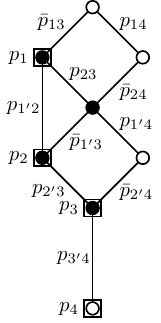} 
\stackrel{\eqref{eq:CFTUniqueness_Scalar_Diagram}}{\longrightarrow}
\adjincludegraphics[valign=c,scale=0.6]{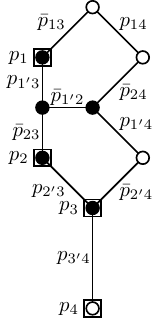} 
& \stackrel{\eqref{eq:CFTUniqueness_Scalar_Diagram}}{\longrightarrow}
\adjincludegraphics[valign=c,scale=0.6]{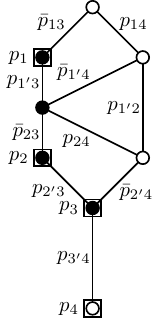}
\stackrel{\eqref{eq:bSTR_type1}}{\longrightarrow}
\adjincludegraphics[valign=c,scale=0.6]{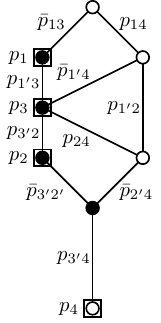} \\
\stackrel{\eqref{eq:bSTR_type1}}{\longrightarrow}
\adjincludegraphics[valign=c,scale=0.6]{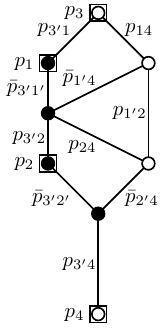}
\stackrel{\eqref{eq:CFTUniqueness_Scalar_Diagram}}{\longrightarrow}
\adjincludegraphics[valign=c,scale=0.6]{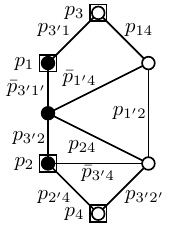} 
\stackrel{\eqref{eq:CFTUniqueness_Scalar_Diagram}}{\longrightarrow}  
& \adjincludegraphics[valign=c,scale=0.6]{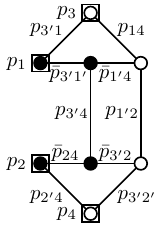} 
\stackrel{\eqref{eq:CFTUniqueness_Scalar_Diagram}}{\longrightarrow} 
\adjincludegraphics[valign=c,scale=0.6]{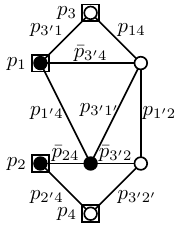} 
\stackrel{\eqref{eq:CFTUniqueness_Scalar_Diagram}}{\longrightarrow}   
\adjincludegraphics[valign=c,scale=0.6]{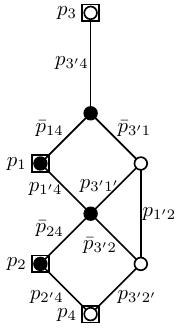} \\
\stackrel{\eqref{eq:bSTR_type1}}{\longrightarrow}  
\adjincludegraphics[valign=c,scale=0.6]{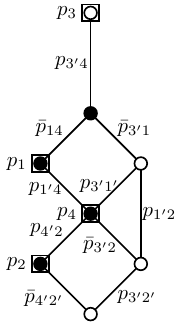} 
\stackrel{\eqref{eq:bSTR_type1}}{\longrightarrow} 
\adjincludegraphics[valign=c,scale=0.6]{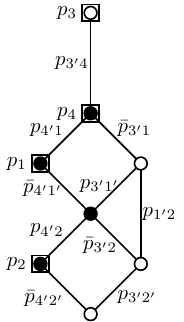} 
& \stackrel{\eqref{eq:CFTUniqueness_Scalar_Diagram}}{\longrightarrow} 
\adjincludegraphics[valign=c,scale=0.6]{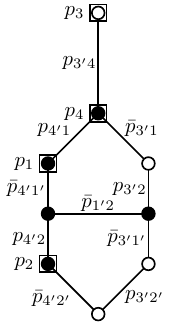} 
\stackrel{\eqref{eq:CFTUniqueness_Scalar_Diagram}}{\longrightarrow} 
\adjincludegraphics[valign=c,scale=0.6]{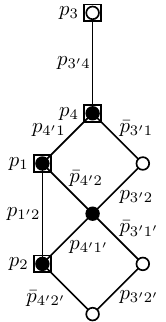} ~,
\end{split}\label{eq:bYBE_fused_type1_proof}
\end{equation}
such that we find the r.\,h.\,s.\ of \eqref{eq:bYBE_fused_type1}.
As in the case of the $\mathbb{YBE}$, the steps \eqref{eq:bYBE_fused_type1_proof} produce an overall factor.
It consists of all the $r_0$-factors that are produced every time we use the STR \eqref{eq:CFTUniqueness_Scalar_Diagram}.
In the end they cancel, since the overall factor is 
\begin{equation}
\frac{
r_0 ( \bar{p}_{1'2}, \bar{p}_{24}, p_{1'4} ) \,
r_0 ( \bar{p}_{2'4}, \bar{p}_{3'2'}, p_{3'4} ) \,
r_0 ( \bar{p}_{3'1'}, \bar{p}_{1'4}, p_{3'4} ) \,
r_0 ( \bar{p}_{1'2}, \bar{p}_{4'1'}, p_{4'2} ) \,
}{
r_0 ( \bar{p}_{1'2}, \bar{p}_{23}, p_{1'3} ) \,
r_0 ( \bar{p}_{3'2}, \bar{p}_{24}, p_{3'4} ) \,
r_0 ( \bar{p}_{14}, \bar{p}_{3'1}, p_{3'4} ) \,
r_0 ( \bar{p}_{1'2}, \bar{p}_{3'1'}, p_{3'2} ) \,
}
~ \stackrel{p' = - p}{=} ~
1 ~.
\end{equation}
To show that this really collapses to one, we cancel $a_0$-factors in the $r_0$-abbreviations \eqref{eq:Factor_rEll} in the nominator and denominator, use the the relation $a_0 (v) a_0 (\bar{v}) = 1$ (c.\,f.\ \eqref{eq:Factor_aEll}) and explicitly use the parametrization $p' = - p$.

Within the $\mathbb{bYBE}$, we find the fused K-matrix $\mathbb{K}$ related to the boundary weights of type \RNum{1},
\begin{equation}
\mathbb{K}^{x_1}_{x_2} (p_1, p_2)
=
\adjincludegraphics[valign=c,scale=1]{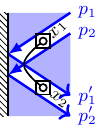}
=
V_{p_1} (x_1)
W_{p_{1'2}} (x_{12})
V_{p_2} (x_2)
~.
\label{eq:Kmatrix_fused}
\end{equation}
Recall that the reflected medial lines must, in any case, carry a spectral parameter related to the unreflected one, hence, the K-matrix \eqref{eq:Kmatrix_fused} only depends on two spectral parameters.

So far, we have paved the way for a notion of boundary integrability that yields a K-matrix \eqref{eq:Kmatrix_fused} for the construction of boundary integrable Feynman graphs.
However, we still have to find a concrete solution of the type \RNum{1} boundary weight $V_p (x)$ for the bSTR \eqref{eq:bSTR_type1}.
We present two solutions and recall that $p' = - p$:
\begin{itemize}
\item 
The \textit{trivial boundary weight} is $V_{p_i} (x) = 1$ or graphically $\adjincludegraphics[valign=c,scale=0.8]{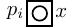} 
= 1$.
It solves the bSTR \eqref{eq:bSTR_type1}, because
\begin{equation}
\adjincludegraphics[valign=c,scale=1]{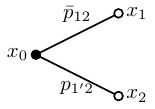}
\stackrel{\eqref{eq:ChainRuleScalar}}{=}
r_0 (\bar{p}_{1'1} , \bar{p}_{12} , p_{1'2})
\cdot
\adjincludegraphics[valign=c,scale=1]{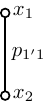}
\stackrel{\eqref{eq:ChainRuleScalar}}{=}
\adjincludegraphics[valign=c,scale=1]{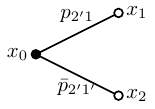} ~.
\label{eq:bweight_trivial_proof}
\end{equation}
It can be seen by starting on each side of \eqref{eq:bSTR_type1} separately and applying the chain relation \eqref{eq:ChainRuleScalar} to find the middle expression in \eqref{eq:bweight_trivial_proof}.
Note that $r_0 (\bar{p}_{1'1} , \bar{p}_{12} , p_{1'2}) = r_0 (\bar{p}_{1'1} , \bar{p}_{2'1'} , p_{2'1})$ holds.

\item
The \textit{canonical boundary weight} is another solution of the type \RNum{1} bSTR \eqref{eq:bSTR_type1}.
It is $V_{p_i} (x_j) = W_{\bar{p}_{i'i}} (x_{\mathrm{L}j})$ or graphically $\adjincludegraphics[valign=c,scale=0.8]{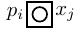} 
= \adjincludegraphics[valign=c,scale=0.8]{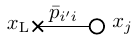} 
$.
The boundary weight depends on an external point $x_\mathrm{L}$, where the subscript refers to the position of the boundary on the left.
It has to be understood as a global parameter, common to all left-hand side boundary weights in some generalized Feynman graph.
We denote it by a cross instead of an unfilled circle to stress its special relation to the left boundary.
The bSTR is satisfied with the canonical boundary weight, 
\begin{equation}
\adjincludegraphics[valign=c,scale=0.8]{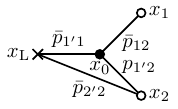}
\stackrel{\eqref{eq:CFTUniqueness_Scalar_Diagram}}{=}
r_0 (\bar{p}_{1'1} , \bar{p}_{12} , p_{1'2})
\cdot
\adjincludegraphics[valign=c,scale=0.8]{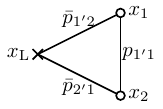}
\stackrel{\eqref{eq:CFTUniqueness_Scalar_Diagram}}{=}
\adjincludegraphics[valign=c,scale=0.8]{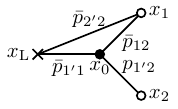} ~,
\label{eq:bweight_canonic_proof}
\end{equation}
thanks to the STR \eqref{eq:CFTUniqueness_Scalar_Diagram}. 
This explains its name; the bulk STR implies the bSTR for the particular choice of boundary weight.
For this weight the K-matrix $\mathbb{K}$ \eqref{eq:Kmatrix_fused} has the form of a triangle,
\begin{equation}
\mathbb{K}^{x_1}_{x_2} (p_1, p_2 ; x_\mathrm{L})
=
W_{\bar{p}_{1'1}} (x_{1\mathrm{L}})
W_{p_{1'2}} (x_{12})
W_{\bar{p}_{2'2}} (x_{2\mathrm{L}})
=
\adjincludegraphics[valign=c,scale=1]{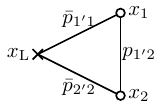} ~.
\label{eq:Kmatrix_fused_canonical}
\end{equation}
\end{itemize}
It would be very interesting to continue the classification of solutions to the bSTR \eqref{eq:bSTR_type1} coming from integrable generalized Feynman graphs.
Usually, the ``boundaries'' of a Feynman graph are external spacetime points, which have their origin in considering QFT correlation functions of local operators.
The canonical boundary weight can be employed to model generalized Feynman graphs that contribute to these quantities.
Thus, each boundary is associated with one external point, which is called $x_\mathrm{L}$ above.
Inserting Wilson lines into the correlation functions could yield Feynman diagrams, which admit a generalization in the exponents of the propagators \cite{Gromov:2021ahm}.
If the propagators connecting the Wilson line and the bulk of a Feynman diagram have a suitable generalization, which satisfies the bSTR, the whole toolbox of boundary integrability could be applied to the calculation of correlation functions invloving these objects.

\section{K-matrix of a boundary on the right hand side}
Next, we study the bYBE, $\mathbb{bYBE}$ and K-matrices on a boundary on the right.
The methods are very similar to the previous left-boundary case.
However, requiring the medial lines to be reflected going upward changes the exponents one assigns to the generalized propagators.
Aside from this small difference, the derivation of the K-matrix $K$, the bYBE factorizing into the two types of bSTRs, the doubling of the medial lines for consistent shading, the $\mathbb{bYBE}$ and the K-matrix $\mathbb{K}$ is very similar to the left-boundary case.
Therefore, we present these results concisely for the right boundary.

We consider the reflection of a medial line on a boundary on the right and describe the process by the K-matrix
\begin{equation}
K^{\mathrm{R} y_1}_{~ y_2 x}(p)
=
V^\mathrm{R}_{p}(x)
V^\mathrm{R}_{p}(y_1 , y_2)
=
\adjincludegraphics[valign=c,scale=0.8]{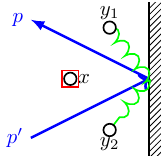} ~.
\label{eq:Kmatrix_facetype_right}
\end{equation}
The K-matrices and boundary weights carry the superscript $\mathrm{R}$ to make clear that they correspond to the right boundary.
In the last section, consistently, we should have indicated the correspondence of objects to the left boundary by a superscript $\mathrm{L}$, but we did not for the sake of conciseness.
Their representation in Feynman graphs is the same as for the left boundary \eqref{eq:Kmatrix_facetype}.
We denote the boundary weights of type \RNum{1}, $V^\mathrm{R}_{p}(x)$, as a box around the spacetime point $x$ and those of type \RNum{2}, $V^\mathrm{R}_{p}(y_1 , y_2)$, as a wiggled line that connects the points $y_1$ and $y_2$.
For the sake of integrability, we require also the right K-matrix $K^\mathrm{R}$ to satisfy a bYBE, which corresponds to the picture
\begin{equation}
\adjincludegraphics[valign=c,scale=0.8]{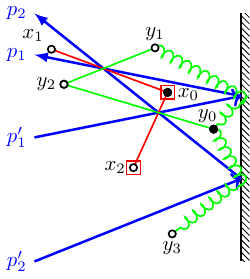} 
\qquad = \qquad
\adjincludegraphics[valign=c,scale=0.8]{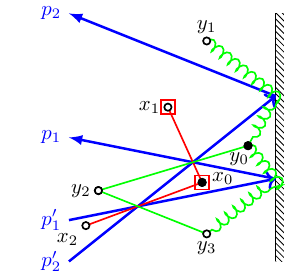} ~.
\label{eq:bYBE_facetype_right_Diagram}
\end{equation}
In terms of generalized Feynman graphs, we find two equations ensuring integrability as well.
They are the right hand side bSTRs,
\begin{subequations}
{\allowdisplaybreaks
\begin{align}
\begin{split}
\adjincludegraphics[valign=c,scale=0.8]{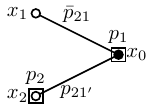} 
~ & = ~
\adjincludegraphics[valign=c,scale=0.8]{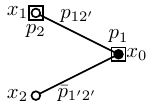} ~, \\
V^\mathrm{R}_{p_2}(x_2) 
\int d^D x_0 \;
W_{\bar{p}_{21}} (x_{10})
V^\mathrm{R}_{p_1}(x_0) &
W_{p_{21'}} (x_{20})  \\
~ = ~
V^\mathrm{R}_{p_2}(x_1) &
\int d^D x_0 \;
W_{p_{12'}} (x_{10}) 
V^\mathrm{R}_{p_1}(x_0)
W_{\bar{p}_{1'2'}} (x_{20}) ~, 
\end{split} \label{eq:bSTR_right_type1}\\
\begin{split}
\adjincludegraphics[valign=c,scale=0.8]{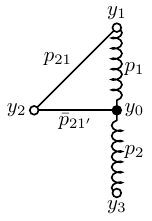}
~ & = ~
\adjincludegraphics[valign=c,scale=0.8]{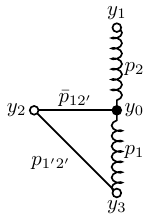} ~, \\
W_{p_{21}} (y_{12})  
\int d^D y_0 \;
V^\mathrm{R}_{p_1}(y_1 , y_0)
W_{\bar{p}_{21'}} (y_{20}) &
V^\mathrm{R}_{p_2}(y_0 , y_3) \\
~ = ~ 
W_{p_{1'2'}} (y_{32}) 
\int d^D y_0 & \;
V^\mathrm{R}_{p_2}(y_1 , y_0)
W_{\bar{p}_{12'}} (y_{20}) 
V^\mathrm{R}_{p_1}(y_0 , y_3) ~,
\end{split} \label{eq:bSTR_right_type2}
\end{align}
\label{eq:bSTR_right_type1And2}%
}%
\end{subequations}
of type \RNum{1} and \RNum{2}, respectively.
It is important to note that the solutions of the bSTRs on the left and the right do not have to be related in any way.
We are free to choose a particular solution on the left, e.\,g.\ the trivial one, and a solution of a different type on the right boundary.
Speaking of solutions, we focus again on the boundary weight of type \RNum{1} and \eqref{eq:bSTR_right_type1}.
We find again the trivial one, $V^\mathrm{R}_{p}(x) = 1$, and a canonical one.
However, the latter is slightly different from the left boundary solution, due to the different propagator exponents of \eqref{eq:bSTR_right_type1} compared to \eqref{eq:bSTR_type1}.
It reads
\begin{equation}
V^\mathrm{R}_{p_i} (x_j) 
=
\adjincludegraphics[valign=c,scale=0.8]{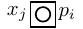} 
=
W_{\bar{p}_{ii'}} (x_{\mathrm{R}j})
= 
\adjincludegraphics[valign=c,scale=0.8]{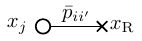}  
\end{equation}
and proving that both solutions, the trivial one and the canonical one, satisfy the bSTR \eqref{eq:bSTR_right_type1} works in the same way as for the left boundary, \eqref{eq:bweight_trivial_proof} and \eqref{eq:bweight_canonic_proof}, respectively.
Note that the introduction of an additional boundary also implies that the canonical solution depends on a different external point $x_\mathrm{R}$ than the left-boundary canonical solution, which depends on $x_\mathrm{L}$.

Moreover, requiring a K-matrix with consistent shading of the medial graph leads to staggering two K-matrices $K^\mathrm{R}$ into the K-matrix $\mathbb{K}^\mathrm{R}$, which is 
\begin{equation}
\mathbb{K}^{\mathrm{R} x_1}_{~\; x_2} (p_1, p_2)
=
\adjincludegraphics[valign=c,scale=1]{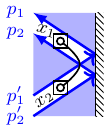}
=
V^\mathrm{R}_{p_1} (x_1)
W_{p_{21'}} (x_{12})
V^\mathrm{R}_{p_2} (x_2) 
\stackrel{\mathrm{canonical}}{\longrightarrow}
\adjincludegraphics[valign=c,scale=1]{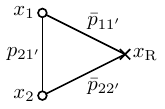} 
~,
\label{eq:Kmatrix_fused_right}
\end{equation}
and we show the K-matrix with the canonical solution of the bSTR.
It satisfies a staggered version of the bYBE on the right boundary, the right $\mathbb{bYBE}$
\begin{equation}
\adjincludegraphics[valign=c,scale=0.8]{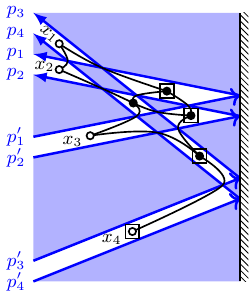} 
\qquad = \qquad
\adjincludegraphics[valign=c,scale=0.8]{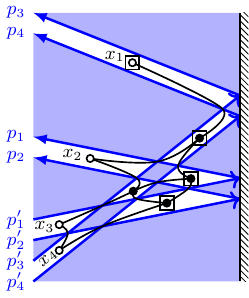} ~.
\label{eq:bYBE_fused_right_type1}
\end{equation}
The proof works analogously to the left-boundary version \eqref{eq:bYBE_fused_type1} by performing similar steps to \eqref{eq:bYBE_fused_type1_proof}.

\section{The double-row transfer matrix}
The two $\mathbb{bYBE}$s \eqref{eq:bYBE_fused_type1} and \eqref{eq:bYBE_fused_right_type1} for the left and right boundary, respectively, together with the bulk $\mathbb{YBE}$ \eqref{eq:YBE_Scalar_Fused} and the unitarity relations \eqref{eq:Unitarity_Rmatrix_scalar_fused} are instrumental for the construction of commuting double-row transfer matrices.
This was first shown in \cite{Sklyanin:1988yz}.
In the presence of boundaries, the double-row transfer matrix replaces the periodic transfer matrices \eqref{eq:TransferMatix_periodic} presented below as the generating object for conserved charges.
In the case of integrable, scalar generalized Feynman graphs, we can graphically represent the inhomogeneous, staggered, double-row transfer matrix as 
\begin{equation}
\mathbb{T}_N (p_a, p_b, \lbrace p_i \rbrace)
=
\adjincludegraphics[valign=c,scale=1.3]{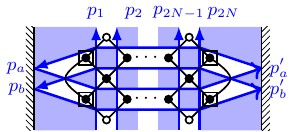} ~.
\label{eq:TransferMatrix_QFT_doubleRow_staggered}
\end{equation}
Two such transfer matrices with identical vertical medial lines and distinct horizontal medial lines commute, which can be shown by performing the same steps as for the six-vertex model in \eqref{eq:6V_CommutingRows} with one pair of medial lines here replace a single line there.
The steps using bYBEs, YBE and unitarity should be adapted to the staggered relations $\mathbb{bYBE}$ on the left \eqref{eq:bYBE_fused_type1}, $\mathbb{bYBE}$ on the right \eqref{eq:bYBE_fused_right_type1}, $\mathbb{YBE}$ \eqref{eq:YBE_Scalar_Fused} and unitarity \eqref{eq:Unitarity_Rmatrix_scalar_fused}.
The factors from unitarity cancel and hence we can show that
\begin{equation}
\com{
\mathbb{T}_N (p_a, p_b, \lbrace p_i \rbrace)
}{
\mathbb{T}_N (q_a, q_b, \lbrace p_i \rbrace)
}
=
0 ~.
\label{eq:TransferMatrix_QFT_doubleRow_staggered_commutation}
\end{equation}
The product of two transfer matrices is understood as a convolution over $N$ spacetime points on the top of one factor and the bottom of the other one.
To remember this, we use the abbreviation $\circ$ for the product.

We observe that the staggered double-row transfer matrix \eqref{eq:TransferMatrix_QFT_doubleRow_staggered} can be factorized: we can disentangle the two horizontal medial lines $p_a$ and $p_b$ by $2N$-fold application of the STR \eqref{eq:STR_Scalar_medial} from one boundary to the other.
This way, we obtain
\begin{equation}
\mathbb{T}_N (p_a, p_b, \lbrace p_i \rbrace)
=
\left[
\prod_{i=1}^N 
\frac{
a_0 (\bar{p}_{i+1,a'}) 
a_0 (p_{i+1,b}) 
}{
a_0 (\bar{p}_{ia'}) 
a_0 (p_{ib})
}
\right]
\cdot
T_N (p_a, \lbrace p_i \rbrace)
\circ
T_N (p_b, \lbrace p_i \rbrace) ~,
\end{equation}
and the two factors are double-row transfer matrices themselves, however, not staggered anymore.
They are represented by
\begin{equation}
T_N (p_a, \lbrace p_i \rbrace)
=
\adjincludegraphics[valign=c,scale=1.5]{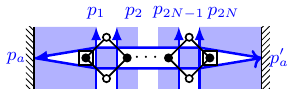} 
\label{eq:TransferMatrix_QFT_doubleRow}
\end{equation}
and they are the fundamental commuting object, which can be shown again in analogy to the six-vertex model \eqref{eq:6V_CommutingRows}, now by employing the bYBEs \eqref{eq:bYBE_facetype_Diagram} and \eqref{eq:bYBE_facetype_right_Diagram}, the STR \eqref{eq:STR_Scalar_medial} and unitarity \eqref{eq:Unitarity_Scalar_medial}.
We find 
\begin{equation}
\com{
T_N (p_a, \lbrace p_i \rbrace)
}{
T_N (q_a, \lbrace p_i \rbrace)
}
=
0 ~,
\label{eq:TransferMatrix_QFT_doubleRow_commutation}
\end{equation}
which of course implies \eqref{eq:TransferMatrix_QFT_doubleRow_staggered_commutation}.

In addition to the periodic transfer matrix \eqref{eq:TransferMatix_periodic}, the double-row transfer matrix \eqref{eq:TransferMatrix_QFT_doubleRow} allows us to construct other classes of integrable generalized Feynman graphs.
The specification of the boundary weights has to be adapted to the quantity under investigation, which admits an expansion in Feynman graphs generated by the double-row transfer matrix after tuning the spectral parameters to a particular value.
Performing this procedure with the canonical solutions on the left and right boundaries, periodic identification of a stack of $M$ double row transfer matrices gives two-point functions depending on $x_\mathrm{L}$ and $x_\mathrm{R}$.
Once we fix the vertical and horizontal spectral parameters, $\lbrace p_i \rbrace$ and $\lbrace p_m \rbrace$, respectively, the remaining question is which QFT produces the Feynman graphs in the perturbative expansion of two-point functions.
A promising candidate for future research is the checkerboard theory \cite{Alfimov:2023vev}.

In the next chapter, we restrict our study to bulk integrability where the match to non-supersymmetric QFTs may be established and facilitate the calculation of exact results for the critical coupling, anomalous dimensions and OPE coefficient.
Similarly, we hope that the introduction of boundary integrability can lead to exact results of more observables in the future.

%%%----------------------------------------------------------------------------------------
%%%	THESIS CONTENT - PART 5 - Boundary integrability in Feynman graphs
%%%----------------------------------------------------------------------------------------
\chapter{Non-supersymmetric fishnet theories and their relatives}
\label{chpt:NonSusyFishnetTheoriesAndItsRelatives}

In this chapter it is shown how the double-scaling limit of the $\gamma$-deformation of $\mathcal{N} = 4$ SYM yields multiple simplified and integrable QFTs \cite{Gurdogan:2015csr}.
Supersymmetry is generically broken and we obtain a family of integrable fishnet theories.
They consist of the $\chi$-CFT, which is also called dynamical fishnet theory \cite{Kazakov:2018gcy}, the $\chi_0$-theory and the bi-scalar fishnet theory.
Here, we explicitly show their derivation starting from the $\mathcal{N} = 1$ superspace formulation of $\mathcal{N} = 4$ SYM and present a detailed review of the bi-scalar fishnet theory.
We consider its double-trace beta functions \cite{Sieg:2016vap,Grabner:2017pgm,Korchemsky:2018hnb}, and a family of operators, whose scaling dimension can be determined exactly, as well as an all-loop result for a four-point correlation function \cite{Grabner:2017pgm,Gromov:2018hut}. 

Moreover, we review Zamolodchikov's computation of the critical coupling of the bi-scalar fishnet theory, the fishnet-deformation of ABJM, and a six-dimensional theory with hexagonal vacuum graphs \cite{Zamolodchikov:1980mb}.
We present the generalization of his techniques, the method of inversion relations, to compute the critical couplings of fermionic theories as well.
We apply it to find the results for the brick wall theory and the three-dimensional fermionic fishnet theory \cite{Kade:2023xet}.

\section{The bi-scalar fishnet theory}
\label{sec:TheBiScalarFishnetTheory}
We review the derivation of bi-scalar fishnet theory from $\mathcal{N} = 4$ SYM and its quantization by regular Feynman graphs with R-matrix graph builders in the planar limit.
Some of the theory's couplings run, but we show that they have a fixed point, which we determine perturbatively.
Working at the fixed point, we can use the methods of CFTs. 
Furthermore, we present the integrability-based computations of the critical coupling and exact scaling dimensions of various operators.

\subsection[Double-scaling limit of $\gamma$-deformed $\mathcal{N} = 4$ SYM]{\boldmath Double-scaling limit of $\gamma$-deformed $\mathcal{N} = 4$ SYM}
\label{susec:DoubleScalingLimitOfGammaDeformedSYM}
We present the derivation of the bi-scalar fishnet theory, starting from the $\mathcal{N} = 1$ superspace action of $\mathcal{N} = 4$ SYM, then performing the $\gamma$-deformation and consecutively taking a double-scaling limit in the 't Hooft coupling and the deformation parameter.
We present the actions of the $\chi$-, $\chi_0$-CFT and the bi-scalar fishnet theory with its Feynman rules.
 
\subsubsection{\boldmath The $\mathcal{N} = 4$ SYM action from a superspace formulation}
Integrability is an important property of four-dimensional $\mathcal{N} = 4$ SYM in the planar limit.
The theory is a supersymmetric gauge theory and we take the gauge group to be $\SU{\mathrm{N}}$, where the number of colors is $\mathrm{N}$ and in the limit $\mathrm{N} \rightarrow \infty$ planar diagrams dominate.
The supercharges generate the R-symmetry group $\SU{4} \cong \SO{6}$.
Therefore, the action of $\mathcal{N} = 4$ SYM can be formulated in four-dimensional $\mathcal{N} = 1$ superspace, where the R-symmetry subgroup $\U{1}$ is manifest.
It reads (see e.\,g.\ \cite{Penati:2001sv,Fokken:2013aea})
\begin{equation}
\begin{split}
S
= &
\int \dd^4 x\; \dd^2\theta\, \dd^2 \bar{\theta}\;
\sum_{i=1}^{3} \,
\mathrm{tr} \left(
\e^{-g \mathcal{V}} \Phi_i^\dagger \e^{g \mathcal{V}} \Phi_i
\right)
+
\frac{1}{2g^2}
\int \dd^4 x\; \dd^2\theta\;
\mathrm{tr} \left(
W^\alpha W_\alpha
\right)\\
& +
\frac{\I g}{\sqrt{2}}
\int \dd^4 x\; \dd^2\theta\;
\mathrm{tr} \left(
\Phi_1 \com{\Phi_2}{\Phi_3}
\right)
+
\frac{\I g}{\sqrt{2}}
\int \dd^4 x\; \dd^2\bar{\theta}\;
\mathrm{tr} \left(
\Phi^\dagger_1 \com{\Phi^\dagger_2}{\Phi^\dagger_3}
\right) ~.
\end{split}
\label{eq:SYM_Action}
\end{equation}
Its field content is a real vector superfield $\mathcal{V}$ and three chiral superfields $\Phi_i$, all four in the adjoint representation of the gauge group $\SU{\mathrm{N}}$, i.\ e.\ $\mathcal{V} = \mathcal{V}_A T^A$ and $\Phi_i = \Phi_{i,A} T^A$ with the adjoint generators $T^A$ ($A = 1, ..., \mathrm{N}^2 - 1$).
They satisfy
\begin{equation}
\begin{array}{ccc}
\com{T^A}{T^B} = \I f^{ABC} T_C ~, &
\mathrm{tr}\left( T^A T^B \right) = \delta^{AB} ~, &
( T^A )^a_b \left( T_A \right)^c_d = \delta^a_d \delta^c_b - \frac{1}{\mathrm{N}} \delta^a_b \delta^c_d ~.
\end{array}
\label{eq:Adjoint_generators}
\end{equation}
The supersymmetric field strength is $W_\alpha = \frac{\I}{4} \bar{D}^2 \left( \e^{-g \mathcal{V}} D_\alpha \e^{g \mathcal{V}} \right)$, with the covariant superderivatives defined in \eqref{eq:SuperCovariantDerivatives}.

We will show briefly how to obtain the component action of \eqref{eq:SYM_Action}.
First, the real vector superfield in Wess-Zumino gauge admits the expansion 
\begin{equation}
\mathcal{V}
=
- \theta \sigma^\mu \bar{\theta} A_\mu
+ \I \theta^2 \bar{\theta} \bar{\chi}
- \I \bar{\theta}^2 \theta \chi
+ \frac{1}{2} \theta^2 \bar{\theta}^2 \mathrm{D}
\label{eq:VectorSuperfield}
\end{equation}
with the gauge vectorfield $A_\mu$, the gaugino $\chi$ and the auxiliary scalar $\mathrm{D}$.
By the vanishing of cubic Gra\ss mann orders, we have $\mathcal{V}^2 = - \frac{1}{2} \theta^2 \bar{\theta}^2 A_\mu A^\mu$ and $\mathcal{V}^3 = 0$.
This means the exponentials containing $\mathcal{V}$ in the action \eqref{eq:SYM_Action} truncate after the second order.
Expanding the supersymmetric field strength gives the expression 
\begin{equation}
W_\alpha
=
\frac{\I g}{4} \bar{D}^2 D_\alpha \mathcal{V}
-
\frac{\I g^2}{8}
\bar{D}^2 \com{\mathcal{V}}{D_\alpha \mathcal{V}}
\end{equation}
whose explicit component evaluation is lengthy.
It is handy to keep in mind that in the action the bilinear of $W$ will be integrated only over chiral subspace, such that one can drop the $\bar{\theta}$-dependent terms eventually.
We find 
\begin{equation}
W_\alpha \vert_{\bar{\theta} = 0}
=
g \chi_\alpha 
+ \, \I\, g\, \theta_\alpha \mathrm{D}
+ \frac{1}{2} \, g \, \sigma^\mu_{\alpha \dot{\alpha}} \bar{\sigma}^{\nu , \dot{\alpha} \beta} \theta_\beta F_{\mu\nu}
+ \, \I \, g \, \theta^2 \sigma^\mu_{\alpha \dot{\alpha}} \mathcal{D}_\mu \bar{\chi}^{\dot{\alpha}}
\end{equation}
with the field strength $F_{\mu\nu} = \partial_\mu A_\nu - \partial_\nu A_\mu + \frac{\I g}{2} \com{A_\mu}{A_\nu}$ and the gauge covariant derivative $\mathcal{D}_\mu = \partial_\mu + \frac{\I g}{2} \com{A_\mu}{\cdot}$.
Therefore, the gauge part of the action \eqref{eq:SYM_Action} is in components
\begin{equation}
S_\mathrm{g}
=
\frac{1}{2g^2}
\int \dd^4 x\; \dd^2\theta\;
\mathrm{tr} \left(
W^\alpha W_\alpha
\right)
=
\int \dd^4 x\;
\mathrm{tr} \left[
- \I \bar{\chi} \bar{\sigma}^\mu \mathcal{D}_\mu \chi
- \frac{1}{4} F_{\mu\nu} F^{\mu\nu}
- \frac{\I}{8} F_{\mu\nu} F^{*,\mu\nu}
+ \frac{1}{2} \mathrm{D}^2
\right] ~.
\label{eq:SYM_gauge_components}
\end{equation}
We find the kinetic term for the gaugino, the kinetic term for the non-Abelian field strength as well as the quadratic term of the auxiliary scalar.
The third term is the contraction of the field strength with its Hodge-dual, which is a total derivative and thus topological.

Next, we expand the first term in \eqref{eq:SYM_Action}, containing the canonical K\"{a}hler potential, in its components to obtain 
\begin{equation}
\begin{split}
S_\mathrm{K}
&=
\int \dd^4 x\; \dd^2\theta \dd^2 \bar{\theta}\; 
\sum_{i=1}^{3} \,
\mathrm{tr} \left(
\e^{-g \mathcal{V}} \Phi_i^\dagger \e^{g \mathcal{V}} \Phi_i
\right) \\
&=
\int \dd^4 x\;
\sum_{i=1}^{3} \,
\mathrm{tr} \left[ 
- \mathcal{D}_\mu \phi_i^\dagger \mathcal{D}^\mu \phi_i 
- \I \bar{\psi}_i \bar{\sigma}^\mu \mathcal{D}_\mu \psi_i
+ \frac{1}{2} g \mathrm{D} \com{\phi_i^\dagger}{\phi_i}
+ F_i^\dagger F_i
\right. \\
& \hspace{3cm} \left.
+ \frac{\I g}{\sqrt{2}}
\left(
\bar{\psi}_i \bar{\chi} \phi_i
+ \phi_i^\dagger \chi \psi_i
- \bar{\chi} \bar{\psi}_i \phi_i
- \chi \phi_i^\dagger \psi_i
\right)
\right] ~.
\end{split}
\label{eq:SYM_kaehler_components}
\end{equation}
At this point, we can integrate out the auxiliary field $\mathrm{D}$ by eliminating it through its equation of motion that reads
\begin{equation}
\mathrm{D} 
= 
\mathrm{D}_A T^A 
=
\frac{g}{2}
\sum_{i=1}^3
\com{\phi_i}{\phi_i^\dagger} ~. 
\end{equation}
Since it is on-shell expressed by a commutator, potential double-trace terms by the last relation in \eqref{eq:Adjoint_generators} vanish.
This will be not the case anymore in the deformed theory \cite{Fokken:2013aea}, as we will see below in section \ref{subsec:ComponentAction_SBW}.
Note that, once replaced in $S_\mathrm{K}$ and $S_\mathrm{g}$, the field will on-shell generate quartic scalar interaction terms in the component action.
So does the superpotential in \eqref{eq:SYM_Action}, after integrating out the auxiliary fields.
It solely contains (anti) chiral superfields, thus, the expansion \eqref{eq:Superfield_Components} gives the component action of the superpotential 
\begin{equation}
\begin{split}
&S_\mathrm{pot}
=
\frac{\I g}{\sqrt{2}}
\int \dd^4 x\; \dd^2\theta\;
\mathrm{tr} \left(
\Phi_1 \com{\Phi_2}{\Phi_3}
\right)\\
&=
\frac{\I g}{\sqrt{2}}
\int \dd^4 x\;
\mathrm{tr} \Bigl(
\phi_1 \com{\phi_2}{F_3} +
\phi_1 \com{F_2}{\phi_3} +
F_1 \com{\phi_2}{\phi_3} -
\phi_1 \com{\psi_2}{\psi_3} -
\psi_1 \com{\phi_2}{\psi_3} -
\psi_1 \com{\psi_2}{\phi_3}
\Bigr)
\end{split}
\label{eq:SYM_superpot_componentsWithF}
\end{equation}
and similarly for the hermitian conjugated superpotential action $\bar{S}_{\mathrm{pot}}$ (replace fields by their daggered/barred partners).
The auxiliary fields are integrated out by their on-shell expression obtained by their equations of motion
\begin{equation}
\begin{aligned}[c]
F_1 = \frac{\I g}{\sqrt{2}} \com{\phi_3^\dagger}{\phi_2^\dagger} ~,\\
F^\dagger_1 = \frac{\I g}{\sqrt{2}} \com{\phi_3}{\phi_2} ~,
\end{aligned}
\qquad
\begin{aligned}[c]
F_2 = \frac{\I g}{\sqrt{2}} \com{\phi_1^\dagger}{\phi_3^\dagger} ~,\\
F^\dagger_2 = \frac{\I g}{\sqrt{2}} \com{\phi_1}{\phi_3} ~,
\end{aligned}
\qquad
\begin{aligned}[c]
F_3 = \frac{\I g}{\sqrt{2}} \com{\phi_2^\dagger}{\phi_1^\dagger} ~,\\
F^\dagger_3 = \frac{\I g}{\sqrt{2}} \com{\phi_2}{\phi_1} ~.
\end{aligned}
\label{eq:Auxiliary_SYM_F}
\end{equation}
Again, this procedure potentially can produce double-trace terms, but, as it was the case for the auxiliary field $\mathrm{D}$, replacing the fields $F_i$ by the commutators \eqref{eq:Auxiliary_SYM_F} and using \eqref{eq:Adjoint_generators} would give the trace of a commutator, which is zero.
Note that the sesquilinear terms of $F_i$ in $S_\mathrm{K}$ yield additional quartic scalar interactions.
So does the superpotential \eqref{eq:SYM_superpot_componentsWithF} after the replacements \eqref{eq:Auxiliary_SYM_F}, it is 
\begin{equation}
S_\mathrm{pot}
=
\int \dd^4 x\;
\mathrm{tr} \left[
\frac{g^2}{2}
\sum_{i<j} \com{\phi_i}{\phi_j} \com{\phi_i^\dagger}{\phi_j^\dagger}
-
\frac{\I g}{\sqrt{2}}
\varepsilon^{ijk}
\psi_i \phi_j \psi_k
\right]
\end{equation}
and the hermitian conjugated superpotential is again obtained by the replacement of fields by their daggered/barred partners.
We see that the first term is the same in both, $S_\mathrm{pot}$ and $\bar{S}_{\mathrm{pot}}$.

Before reassembling the component action, we have a closer look on the quartic scalar interaction terms in the Lagrange density.
When we deform the theory in section \ref{subsubsec:TheGammaDeformationOfN4SYM}, it is useful to collect interaction terms with zero charge under R-symmetry subgroups.
The quartic terms in the Lagrangian are 
\begin{equation}
\mathrm{tr}\left[
\frac{g^2}{2}
\sum_{i<j} 
\com{\phi_i}{\phi_j} \com{\phi_i^\dagger}{\phi_j^\dagger}
-
\frac{g^2}{8}
\sum_{i,j=1}^3
\com{\phi_i}{\phi_i^\dagger} \com{\phi_j}{\phi_j^\dagger}
\right] ~,
\label{eq:SYM_components_quartic_1}
\end{equation}
where the first term comes from the $F_i^\dagger F_i$-terms and from the chiral and anti-chiral superpotential, which yields the factor $-\frac{g^2}{2} + 2\cdot \frac{g^2}{2} = \frac{g^2}{2}$.
The second terms is due to the terms from the auxiliary field, $\frac{1}{2} g \mathrm{D} \com{\phi_i^\dagger}{\phi_i}$ and $\frac{1}{2} \mathrm{D}^2$, yielding the factor $- \frac{g^2}{4} + \frac{g^2}{8} = -\frac{g^2}{8}$.
We split the second sum in \eqref{eq:SYM_components_quartic_1} into the diagonal and off-diagonal parts, $i=j$, $i<j$ and $j<i$, respectively.
Obeying the symmetry properties of permutation of the indices, we can be combine it with the first term in \eqref{eq:SYM_components_quartic_1} and we find the expression
\begin{equation}
\frac{g^2}{4}
\mathrm{tr}\left[
\sum_{i<j} 
\left(
2 \phi_i \phi_j \phi_i^\dagger \phi_j^\dagger +
2 \phi_j \phi_i \phi_j^\dagger \phi_i^\dagger -
\lbrace \phi_i^\dagger , \phi_i \rbrace
\lbrace \phi_j^\dagger , \phi_j \rbrace
\right)
+
\sum_{i=1}^3
\left(
2 \phi_i \phi_i \phi_i^\dagger \phi_i^\dagger -
\frac{1}{2}
\lbrace \phi_i^\dagger , \phi_i \rbrace^2
\right)
\right] ~.
\label{eq:SYM_components_quartic_2}
\end{equation}
Since the anti-commutator-term is symmetric under $i \leftrightarrow j$, we can combine the two sums and obtain the quartic terms 
\begin{equation}
\frac{g^2}{2}
\sum_{i,j=1}^3
\mathrm{tr}\left[
\phi_i \phi_j \phi_i^\dagger \phi_j^\dagger 
-
\frac{1}{4}
\lbrace \phi_i^\dagger , \phi_i \rbrace
\lbrace \phi_j^\dagger , \phi_j \rbrace
\right] ~.
\label{eq:SYM_components_quartic_3}
\end{equation}
The anti-commutator-term has the handy property that the fields $\phi_i$ and $\phi_i^\dagger$ are next to each other within the trace, which implies that they are unaltered by R-charge deformations.

Finally, we perform two rescalings: first, for convenience, we redefine $g\rightarrow - \sqrt{2}\; g$ and second, for the sake of the 't Hooft expansion, we rescale all fields by $\sqrt{\mathrm{N}}$ and introduce the 't Hooft coupling $\lambda := g \cdot \sqrt{\mathrm{N}}$. 
Effectively, the rescaling multiplies the action by the rank of the gauge group $\mathrm{N}$, whereas the coupling $g$ gets replaced by $\lambda$, i.\,e.\ $S(g) \rightarrow \mathrm{N} \cdot S(\lambda)$.
We find the component action of $\mathcal{N} = 4$ SYM to take the form
\begin{equation}
\begin{split}
S
=
\mathrm{N}
\int \dd^4 x\;
\mathrm{tr} & \left[ 
- \mathcal{D}_\mu \phi_i^\dagger \mathcal{D}^\mu \phi^i 
- \I \bar{\psi}_i \bar{\sigma}^\mu \mathcal{D}_\mu \psi^i
- \I \bar{\chi} \bar{\sigma}^\mu \mathcal{D}_\mu \chi
- \frac{1}{4} F_{\mu\nu} F^{\mu\nu}
- \frac{\I}{8} F_{\mu\nu} F^{*,\mu\nu} \right. \\
& ~  
- \I\lambda
\left(
\bar{\chi} \phi^i \bar{\psi}_i 
+ \psi^i \phi_i^\dagger \chi
- \bar{\psi}_i \phi^i \bar{\chi}
- \chi \phi_i^\dagger \psi^i
-
\varepsilon^{ijk}
\psi_i \phi_j \psi_k
-
\varepsilon^{ijk}
\bar{\psi}_i \phi_j^\dagger \bar{\psi}_k
\right)  \\
& ~ + \left.
\lambda^2
\left(
\phi^i \phi^j \phi_i^\dagger \phi_j^\dagger 
-
\frac{1}{4}
\lbrace \phi_i^\dagger , \phi^i \rbrace
\lbrace \phi_j^\dagger , \phi^j \rbrace
\right)
\right] ~.
\end{split}
\label{eq:SYM_Action_components}
\end{equation}
We changed to the Einstein sum-convention for the indices $i,j,k = 1,2,3$.
The action \eqref{eq:SYM_Action_components} will be studied in the planar limit, where the coupling $g$ goes to zero, while the rank of the gauge group, $\mathrm{N}$, is taken to infinity.
This is the planar limit and in summary we have
\begin{equation}
\begin{array}{ccc}
g \rightarrow 0 ~, &
\mathrm{N} \rightarrow \infty ~, &
\lambda^2 = g^2 \cdot \mathrm{N} ~ \mathrm{finite.}
\end{array}
\label{eq:SYM_PlanarLimit}
\end{equation}
Let us have a closer look at the component action \eqref{eq:SYM_Action_components}.
The degrees of freedom are the gauge vector field $A_\mu$, three complex scalars $\phi_i$, three chiral fermions $\psi_i$ and the chiral gaugino $\chi$.
All of them are in the adjoint representation of the gauge group $\SU{\mathrm{N}}$, see \eqref{eq:Adjoint_generators}, and they are represented by $\mathrm{N}\times\mathrm{N}$ matrices.
Due to the overall factor of $\mathrm{N}$ in the action \eqref{eq:SYM_Action_components}, we can determine the overall scaling of $\mathrm{N}$ of a particular Feynman graph in a perturbation expansion after quantization.
Interaction vertices contribute with a factor $N^{b_0}$, where $b_0$ is the number of vertices of the graph.
A single propagator contributes $\frac{1}{\mathrm{N}}$ to a Feynman graph, such that the overall propagator contribution is $\mathrm{N}^{-b_1}$, where $b_1$ is the number of edges in the Feynman graph.
Lastly, since the fields are matrix valued, contracting interaction terms with propagators potentially yield traces of multiple fields.
Every trace is of the order $\mathrm{N}$.
To find the total number of such traces, the so-called \textit{fat graph} notation, which we introduce for the fishnet theory in \eqref{eq:Fishnet_biscalar_Feynmanrules}.
It allows to identify every face of the Feynman graph with a trace.
Therefore, the faces of the Feynman graph contribute $\mathrm{N}^{b_2}$.
Putting all the factors together, we see that the Feynman graphs can be organized according to their Euler characteristic (not to be confused with the gaugino)
\begin{equation}
\chi = b_0 - b_1 + b_2
\label{eq:EulerCharacteristic}
\end{equation}
with overall factor $\mathrm{N}^\chi$. 
This means the diagrams with the lowest Euler characteristic dominate the perturbation expansion in the planar limit and we can neglect higher contributions.

Moreover, the action of $\mathcal{N} = 4$ SYM \eqref{eq:SYM_Action_components} has many symmetries.
First, the action is invariant under the $\mathcal{N} = 4$ super Poincar\'{e} algebra, which is generated by translations, Lorentz transformation and four supercharges.
The $\mathcal{N} = 1$ action \eqref{eq:SYM_Action} makes the invariance under one of them, by grouping the fields into the chiral \eqref{eq:Superfield_Components_antichiral} and anti-chiral superfields \eqref{eq:Superfield_Components_antichiral}, as well as the real vector superfield \eqref{eq:VectorSuperfield}.
Second, the action \eqref{eq:SYM_Action_components} does not contain masses and all interaction terms are marginal (i.\,e.\ the coupling is dimensionless).
This implies that the theory is classically conformal invariant.
And even after quantization, $\mathcal{N} = 4$ SYM maintains this property \cite{Sohnius:1981sn}, because the contributions of divergent loop diagrams cancel each other, when summing over all possible fields, which can run in the diagrams' loops \cite{Minahan:2010js}.
Hence, no masses or quantum corrections to the coupling are generated, and the beta function vanishes.
Third, for the superalgebra to close, $\mathcal{N} = 4$ SYM is invariant under a global $\SU{4} \cong \SO{6}$ R-symmetry.
It acts as the fundamental representation $\mathbf{4}$ on the fermions $(\psi_i, \chi)$ and as the $\mathbf{6}$ on the six scalars, which are the real- and imaginary part of the three complex scalars $\phi_i$.
We can use the global R-symmetry transformations to rotate the scalars into a diagonal basis of the three Cartan generators of the $\mathfrak{su} (4)_\mathrm{R}$.
Doing so, we can obtain the frame used in the literature \cite{Fokken:2013aea}, and the corresponding charges are displayed in table \ref{tab:SYM_Rcharges}.
Altogether, the fields of $\mathcal{N} = 4$ SYM describe representations of the $\mathcal{N} = 4$ superconformal algebra $\mathfrak{psu}(2,2\vert 4)$.

\begin{table}[]
\centering
\begin{tabular}{c|llll|lll|l}
 & $\psi_1$ & $\psi_2$ & $\psi_3$ & $\chi$ & $\phi_1$ & $\phi_2$ & $\phi_3$ & $A_\mu$\\ \hline
$q_1$ & $\phantom{-}\frac{1\mathstrut}{2}$ & $-\frac{1}{2}$ & $-\frac{1}{2}$ & $\phantom{-}\frac{1}{2}$ & $1$ & $0$ & $0$ & $0$\\
$q_2$ & $-\frac{1\mathstrut}{2}$ & $\phantom{-}\frac{1}{2}$ & $-\frac{1}{2}$ & $\phantom{-}\frac{1}{2}$ & $0$ & $1$ & $0$ & $0$\\
$q_3$ & $-\frac{1\mathstrut}{2}$ & $-\frac{1}{2}$ & $\phantom{-}\frac{1}{2}$ & $\phantom{-}\frac{1}{2}$ & $0$ & $0$ & $1$ & $0$
\end{tabular}
\caption{The charges of the component fields of $\mathcal{N} = 4$ SYM under the Cartan-generators of the R-symmetry $\mathfrak{su}(4)_\mathrm{R}$. The conjugated fields transform with the negative charge under the Cartan generators.}
\label{tab:SYM_Rcharges}
\end{table}

\subsubsection{The $\gamma$-deformation of $\mathcal{N} = 4$ SYM}
\label{subsubsec:TheGammaDeformationOfN4SYM}
Starting from the component action of $\mathcal{N} = 4$ SYM \eqref{eq:SYM_Action_components}, we will perform the so-called $\gamma$-deformation \cite{Frolov:2005iq,Fokken:2013aea}.
It deforms the pointwise product of fields to a non-commutative \textit{star product}.
In the case of the $\gamma$-deformation, the star product is meant to deform the R-symmetry by introducing a phase which depends on the R-charge of the factors.
For two fields $\varphi_1$ and $\varphi_2$ with $\mathfrak{su}(4)$-charges $q^{\varphi_1}$ and $q^{\varphi_2}$, respectively, it is of the form 
\begin{equation}
\varphi_1 \cdot \varphi_2
\rightarrow
\varphi_1 \star \varphi_2
=
\e^{- \frac{\I}{2} \varepsilon^{ijk} \gamma_i\, q_j^{\varphi_1} q_k^{\varphi_2} } \varphi_1 \cdot \varphi_2
\label{eq:StarProduct}
\end{equation}
and depends on three complex parameters $\gamma_1$, $\gamma_2$ and $\gamma_3$.
We observe that we recover the undeformed, regular point-wise product when we set all three parameters to zero.
Next, we replace the products in all terms of the $\mathcal{N} = 4$ SYM action \eqref{eq:SYM_Action_components} by the star product.
Its definition \eqref{eq:StarProduct} involves two factors, but the action also contains interactions, which have more factors. 
For them we use the useful relation
\begin{equation}
\varphi_1 \star \cdots \star \varphi_p
=
\e^{- \frac{\I}{2} \sum_{m < n} \varepsilon^{ijk} \gamma_i\, q_j^{\varphi_m} q_k^{\varphi_n} } \varphi_1 \cdots \varphi_p ~,
\label{eq:StarProduct_multipleFactors}
\end{equation}
which follows from \eqref{eq:StarProduct} if we recall that the Cartan-charges of a product combine additively.

We find the $\gamma$-deformation of $\mathcal{N} = 4$ SYM \cite{Frolov:2005iq,Fokken:2013aea} with the action
\begin{equation}
\begin{split}
S
=
\mathrm{N}
\int \dd^4 x\;
\mathrm{tr} & \left[ 
- \mathcal{D}_\mu \phi_i^\dagger \mathcal{D}^\mu \phi^i 
- \I \bar{\psi}_i \bar{\sigma}^\mu \mathcal{D}_\mu \psi^i
- \I \bar{\chi} \bar{\sigma}^\mu \mathcal{D}_\mu \chi
- \frac{1}{4} F_{\mu\nu} F^{\mu\nu}
- \frac{\I}{8} F_{\mu\nu} F^{*,\mu\nu} \right. \\
& ~  
- \I\lambda
\left(
\e^{\frac{\I}{2} \gamma_i^-} \bar{\chi} \phi^i \bar{\psi}_i 
+ \e^{- \frac{\I}{2} \gamma_i^-} \psi^i \phi_i^\dagger \chi
- \e^{- \frac{\I}{2} \gamma_i^-} \bar{\psi}_i \phi^i \bar{\chi}
- \e^{\frac{\I}{2} \gamma_i^-} \chi \phi_i^\dagger \psi^i
\right. \\
& ~~~ - \left.
\e^{\frac{\I}{2} \varepsilon_{ijk} \gamma_j^+}
\varepsilon^{ijk}
\psi_i \phi_j \psi_k
-
\e^{\frac{\I}{2} \varepsilon_{ijk} \gamma_j^+}
\varepsilon^{ijk}
\bar{\psi}_i \phi_j^\dagger \bar{\psi}_k
\right)  \\
& ~ + \left.
\lambda^2
\left(
\e^{-\I \varepsilon_{ijk} \gamma_k}
\phi^i \phi^j \phi_i^\dagger \phi_j^\dagger
-
\frac{1}{4}
\lbrace \phi_i^\dagger , \phi^i \rbrace
\lbrace \phi_j^\dagger , \phi^j \rbrace 
\right)
\right] ~.
\end{split}
\label{eq:SYM_GammaDeformation}
\end{equation}
Indices, which might appear more than two times are understood to be implicitly summed over.
We observe that for the different interaction terms, different combinations of the deformation parameters enter the action.
To write them compactly we use the abbreviations \cite{Gurdogan:2015csr}
\begin{equation}
\begin{array}{ccc}
\gamma_1^\pm 
= 
- \frac{\gamma_3 \pm \gamma_2}{2} ~, &
\gamma_2^\pm 
= 
- \frac{\gamma_1 \pm \gamma_3}{2} ~, &
\gamma_3^\pm 
= 
- \frac{\gamma_2 \pm \gamma_1}{2} ~.
\end{array}
\end{equation}

\subsubsection{The double-scaling limit and fishnet theories}
\label{subsubsec:TheDoubleScalingLimitAndFishnetTheories}
For the sake of simplifying the action of the $\gamma$-deformation of $\mathcal{N} = 4$ SYM dramatically, while preserving integrability, in \cite{Gurdogan:2015csr} the double-scaling limit is proposed.
Conceptually similar to the planar limit, the double-scaling limit assumes the combinations $q_i := \e^{- \frac{\I}{2} \gamma_i}$ (not to be confused with the Cartan-charges) to tend to infinity at the same rate, and, in parallel, the 't Hooft coupling $\lambda$ is taken to zero.
These limits are carefully related to keep the three quantities $\xi_i := q_i \cdot \lambda$ at a finite value, i.\,e.\
\begin{equation}
\begin{array}{ccc}
q_i \rightarrow \infty ~, &
\lambda \rightarrow 0 ~, &
\mathrm{while}~ \xi_i = q_i \cdot \lambda ~ \mathrm{finite.}
\end{array}
\label{eq:DoubleScalingLimit_SYM}
\end{equation}
In the $\gamma$-deformed action \eqref{eq:SYM_Action_components}, we find deformed interaction terms with an overall factor $\lambda$ (by the gauge covariant derivative), $\lambda \sqrt{q_k / q_j}$ for the Yukawa couplings with the gaugino, $\lambda \sqrt{q_i q_k}$ for Yukawa terms with an even permutation $(i,j,k)$ of $(1,2,3)$, $\lambda / \sqrt{q_i q_k}$ for the odd permutations, and $\lambda^2 q_i^2$, $\lambda^2 / q_i^2$ and $\lambda^2$ for the quartic scalar interactions.
In the limit \eqref{eq:DoubleScalingLimit_SYM}, only the even-permutation Yukawa terms and some quartic terms do not go to zero.
Additionally, we see that the gauge fields $A_\mu$ and the gaugino $\chi$ decouple from the theory, since all their interaction terms with the scalars and the remaining fermions vanish.
Hence, we neglect their contributions in the limit \eqref{eq:DoubleScalingLimit_SYM}, which are their kinetic terms and interactions among themselves.
Finally, we are left with the double-scaled $\gamma$-deformation of $\mathcal{N} = 4$ SYM,
\begin{equation}
\begin{split}
& S^\chi
=
\mathrm{N}
\int \dd^4 x\;
\mathrm{tr} \left[ 
- 
\sum_{i = 1}^3
\partial_\mu \phi_i^\dagger \partial^\mu \phi_i
- 
\sum_{i = 1}^3
\I \bar{\psi}_i \bar{\sigma}^\mu \partial_\mu \psi_i
\right. \\
& + \I \sqrt{\xi_1 \xi_2} \left(
\psi_2 \phi_3 \psi_1
+ 
\bar{\psi}_2 \phi_3^\dagger \bar{\psi}_1
\right) 
+\I \sqrt{\xi_1 \xi_3} \left(
\psi_1 \phi_2 \psi_3
+
\bar{\psi}_1 \phi_2^\dagger \bar{\psi}_3
\right) 
+\I \sqrt{\xi_2 \xi_3} \left(
\psi_3 \phi_1 \psi_2
+
\bar{\psi}_3 \phi_1^\dagger \bar{\psi}_2
\right)  \\
& + 
\left.
\xi_1^2
\phi_2 \phi_3 \phi_2^\dagger \phi_3^\dagger
+
\xi_2^2
\phi_3 \phi_1 \phi_3^\dagger \phi_1^\dagger
+
\xi_3^2
\phi_1 \phi_2 \phi_1^\dagger \phi_2^\dagger
\right] ~.
\end{split}
\label{eq:Chi_SYM_action}
\end{equation}
This theory depends on the three parameters $\xi_1$, $\xi_2$ and $\xi_3$.
It is known as $\chi$-CFT or dynamical fishnet theory, because the interaction vertices only allow for Feynman graphs, which resemble fishing nets.
Crucially, the double-scaling limit \eqref{eq:DoubleScalingLimit_SYM} renders the theory \eqref{eq:Chi_SYM_action} non-Hermitian. 
The reason is that the hermitian conjugated interaction terms vanished in the limit.
Commonly, when aiming for phenomenological applications, non-Hermitian theories yield negative-norm states, which spoil causality.
However, here, the motivation to study theories such as \eqref{eq:Chi_SYM_action} is conceptual, since the simplified Lagrangian allows us to study the integrable structure in more detail \cite{Gurdogan:2015csr,Caetano:2016ydc,Kazakov:2018gcy}.
Yet, the many interaction terms and the many different fields make it very tedious to actually consider all the Feynman diagrams for an interesting quantity.
This motivates further simplifications of the dynamical fishnet theory \eqref{eq:Chi_SYM_action}.

The dynamical fishnet theory is the starting point for the construction of many theories which are obtained by tuning the $\xi_i$'s to particular values.
For example, we will show in section \ref{subsec:ComponentAction_SBW} that for the case $\xi_1 = \xi_2 = \xi_3$, the theory \eqref{eq:Chi_SYM_action} is $\mathcal{N} = 1$ supersymmetric.
We show therein that superspace techniques for Feynman diagrams help to grasp the many possible Feynman graphs produced by the theory \eqref{eq:Chi_SYM_action}. 
For a general value of $\xi_i$' s, the supersymmetry is broken.
Another special case, which simplifies the action even more, is the case where one $\xi_i$ is set to zero.
Thereby, four of the Yukawa interactions and one quartic interaction drop out of the action \eqref{eq:Chi_SYM_action} and we obtain the so-called $\chi_0$-CFT
\begin{equation}
\begin{split}
S^{\chi_0}
=
\mathrm{N}
\int \dd^4 x\;
\mathrm{tr} & 
\left[ 
- 
\sum_{i = 1}^3
\partial_\mu \phi_i^\dagger \partial^\mu \phi^i 
- 
\sum_{i = 2}^3
\I \bar{\psi}_i \bar{\sigma}^\mu \partial_\mu \psi^i
\right. \\
& \left. ~ + \I \sqrt{\xi_2 \xi_3} 
\left(
\psi_3 \phi_1 \psi_2
+
\bar{\psi}_3 \phi_1^\dagger \bar{\psi}_2
\right) + 
\xi_2^2
\phi_3 \phi_1 \phi_3^\dagger \phi_1^\dagger
+
\xi_3^2
\phi_1 \phi_2 \phi_1^\dagger \phi_2^\dagger
\right] ~,
\end{split}
\label{eq:Chi0_SYM_action}
\end{equation}
where we set $\xi_1 = 0$ in \eqref{eq:Chi_SYM_action} and dropped the decoupled kinetic term of $\psi_1$.
We can continue the idea of turning off couplings by setting $\xi_2$ to zero as well.
This gives a theory
\begin{equation}
S^\mathrm{FN}
=
\mathrm{N}
\int \dd^4 x\;
\mathrm{tr} 
\left[ 
- 
\sum_{i = 1}^2
\partial_\mu \phi_i^\dagger \partial^\mu \phi^i 
+
\xi^2
\phi_1 \phi_2 \phi_1^\dagger \phi_2^\dagger
\right] ~,
\label{eq:Fishnet_biscalar_action}
\end{equation}
of two complex scalars, which are coupled through a single quartic, non-Hermitian interaction.
Furthermore, we rename $\xi := \xi_3$.
This is the bi-scalar fishnet theory \cite{Gurdogan:2015csr}.

The Feynman rules can be read off the action \eqref{eq:Fishnet_biscalar_action}.
We recall that the fields in the bi-scalar fishnet theory are in the adjoint representation of the former gauge group $\SU{\mathrm{N}}$, that is, they are matrix-valued $(\phi_i)^a_b = \phi_{i,A} (T^A)^a_b$.
Hence, we can represent the Feynman diagrams as fatgraphs, which facilitates the determination of the order of a graph in $\mathrm{N}$ by its Euler characteristic \eqref{eq:EulerCharacteristic} in the planar limit $\mathrm{N}\rightarrow \infty$.
In the fatgraph formulation, each index of the field $a,b,...$ is related to a \textit{color line}.
This means that a propagator is blown up to two color lines and the trace-structure of interaction is reflected in the wiring of the color lines at a vertex.
The Feynman rules of \eqref{eq:Fishnet_biscalar_action} are
\begin{equation}
\begin{array}{ccc}
\adjincludegraphics[valign=c,scale=0.9]{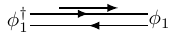} = \frac{1}{\mathrm{N}} \frac{1}{k^2}~, &
\adjincludegraphics[valign=c,scale=0.9]{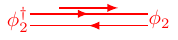} = \frac{1}{\mathrm{N}} \frac{1}{k^2}~, &
\adjincludegraphics[valign=c,scale=0.9]{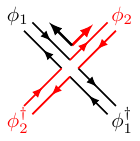} = \I\, \mathrm{N} (4\pi)^2 \xi^2 ~.
\end{array}
\label{eq:Fishnet_biscalar_Feynmanrules}
\end{equation}
The momentum-space propagator in four Euclidean dimensions is derived before \eqref{eq:GenPropagatorGWeight}.
Whenever it is informative, we draw the $\phi_1$ propagators in black and the $\phi_2$ propagators in red.
Note that the vertex is \textit{chiral}, not in a particle physics sense, but in the way it is oriented on a surface.
This is symbolized by the small black and red arrows next to the vertex in \eqref{eq:Fishnet_biscalar_Feynmanrules}.
They should remind us that the Hermitian-conjugated interaction is missing such that we are only allowed to rotate the vertex for the construction of Feynman graphs, but we must not flip it.

\subsection{Double-trace terms, renormalization and conformal fixed-points}
\label{subsec:DoubleTraceTermsRenormalizationAndConformalFixedPoint}
One immediate question arises when quantizing the bi-scalar fishnet theory \eqref{eq:Fishnet_biscalar_action}:
do quantum corrections, in the form of Feynman graphs with loops, generate a mass term for the scalars and does the coupling $\xi^2$ get corrections?
In both cases, the answer is negative.
In the planar limit, it is not possible to draw Feynman graphs for correlation functions with a single-trace structure.
For example, one-loop bubbles would require the existence of the Hermitian conjugate counterpart of the chiral vertex in \eqref{eq:Fishnet_biscalar_action}.
Therefore, the coupling $\xi^2$ does not receive radiative corrections \cite{Kazakov:2018ugh}.
The same argument also holds for diagrams, which would yield mass corrections to the propagators in the planar limit.
To illustrate this, we show the non-existing diagrams
\begin{equation}
\begin{array}{cc}
\adjincludegraphics[valign=c,scale=0.6]{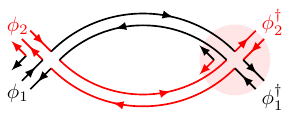} ~, &
\adjincludegraphics[valign=c,scale=0.6]{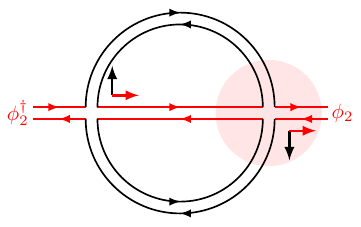}  ~,
\end{array}
\label{eq:Fishnet_missingXiRenormalization}
\end{equation}
where the light-red shading signals the non-existing Hermitian-conjugate vertex.
The left diagram would contribute to radiative corrections of the coupling $\xi^2$ and the right one would generate a mass. 

However, there exist diagrams which are in accordance with the Feynman rules \eqref{eq:Fishnet_biscalar_Feynmanrules} and which are divergent \cite{Sieg:2016vap}.
Consider the diagrams 
\begin{equation}
\begin{array}{cccc}
\adjincludegraphics[valign=c,scale=0.6]{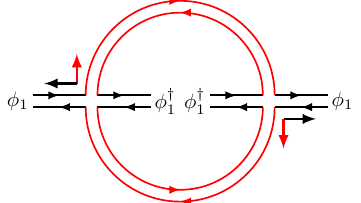} ~, &
\adjincludegraphics[valign=c,scale=0.6]{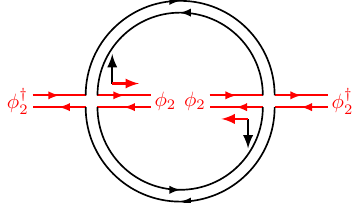} ~, &
\adjincludegraphics[valign=c,scale=0.6]{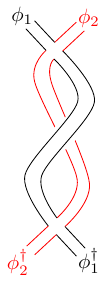} ~, &
\adjincludegraphics[valign=c,scale=0.6]{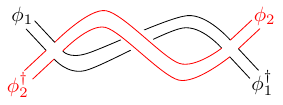} ~.
\end{array}
\label{eq:Fishnet_doubleTraceDiagrams}
\end{equation}
They are one-loop diagrams and proportional to $\xi^4$.
For the sake of simplicity, we do neither show the color line directions, nor the chirality indicators in the last two diagrams.
Following the color lines in each of them, we observe that a pair of them is connecting only two external points with each other.
Hence, for the sake of renormalization, we have to add double-trace counter terms to the action of the bi-scalar fishnet theory \eqref{eq:Fishnet_biscalar_action}.
They are, corresponding to the diagrams \eqref{eq:Fishnet_doubleTraceDiagrams},
\begin{subequations}
\begin{align}
\mathcal{L}_{\mathrm{dt}} 
~=~ +~ & \alpha_1^2 ~ \mathrm{tr}\left(\phi_1\phi_1\right) \mathrm{tr}\left(\phi_1^\dagger\phi_1^\dagger\right)
~+~ \tilde{\alpha}_1 ^2 ~ \mathrm{tr}\left(\phi_2\phi_2\right) \mathrm{tr}\left(\phi_2^\dagger\phi_2^\dagger\right) \label{eq:Fishnet_DoubleTraceTerms_sameflavor} \\
~-~ & \alpha_2 ^2 ~ \mathrm{tr}\left(\phi_1\phi_2\right) \mathrm{tr}\left(\phi_1^\dagger\phi_2^\dagger\right)
~-~ \alpha_3 ^2 ~ \mathrm{tr}\left(\phi_1\phi_2^\dagger\right) \mathrm{tr}\left(\phi_2\phi_1^\dagger\right) ~. \label{eq:Fishnet_DoubleTraceTerms_mixedflavor}
\end{align} \label{eq:Fishnet_DoubleTraceTerms}%\left( 4\pi\right)^2
\end{subequations}
The double-trace couplings $\alpha_1^2$, $\tilde{\alpha}_1^2$, $\alpha_2^2$ and $\alpha_3^2$ are counter terms, which are running.
Double-trace counter terms are already present in the $\gamma$-deformation \eqref{eq:SYM_GammaDeformation} and consequently also in the bi-scalar fishnet theory \cite{Sieg:2016vap,Fokken:2013aea}
To be able to draw all the contributing diagrams for the calculation of the beta-function of these couplings, we introduce the Feynman rules for the double-trace couplings, 
\begin{subequations}
{\allowdisplaybreaks
\begin{align}
&\adjincludegraphics[valign=c,scale=0.7]{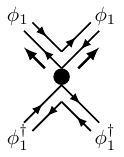} 
  ~\sim~ \I\cdot \left( 4\pi \right)^2  \alpha_1^2 ~,
&&\adjincludegraphics[valign=c,scale=0.7]{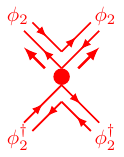} ~\sim~ \I\cdot \left( 4\pi \right)^2  \tilde{\alpha}_1^2 ~, \\
&\adjincludegraphics[valign=c,scale=0.7]{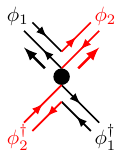} ~\sim~ \I\cdot \left( 4\pi \right)^2  \alpha_2^2 ~,
&&\adjincludegraphics[valign=c,scale=0.7]{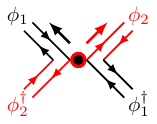} ~\sim~ \I\cdot \left( 4\pi \right)^2  \alpha_3^2 ~.
\label{eq:FeynmanVerticesA2A3}
\end{align}%
}%
\end{subequations}
The graphical representation should resemble joints, since within one of the two traces of the double-trace couplings one can permute the fields freely.

For the sake of conformal symmetry, the double-trace couplings have to be tuned to a value where their beta functions vanish.
This is the conformal fixed-point.
We start with the determination of the conformal fixed-points of the couplings \eqref{eq:Fishnet_DoubleTraceTerms_mixedflavor}.
The divergences, which have to be renormalized by the couplings $\alpha_2^2$ and $\alpha_3^2$, are a chain of bubble integrals.
In the case of $\alpha_3^2$, they look like
\begin{equation}
\adjincludegraphics[valign=c,scale=1]{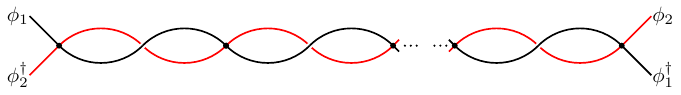} ~,
\label{eq:Fishnet_alpha23_RenDiags}
\end{equation}
where we use the skeleton graph (i.\,e.\ the ordinary Feynman graph, in contrast to the fatgraph) for a concise graphic.
Each vertex could contribute the coupling $\xi^2$ or the coupling $\alpha_3^2$.
In momentum space, the diagrams of type \eqref{eq:Fishnet_alpha23_RenDiags} are proportional to the integral of the four-dimensional scalar massless bubble, raised to the power of the number of loops.
This is the length of the diagram \eqref{eq:Fishnet_alpha23_RenDiags}.
Furthermore, a momentum-space bubble diagram is proportional to the scalar chain relation in \eqref{eq:ChainRuleScalar}.
In dimensional regularization with $D = 4 - 2\varepsilon$, a bubble diagram is therefore proportional to $r_0 (2\varepsilon ,2-2\varepsilon, 2-2\varepsilon)$.
The divergent part of the factor is $\Gamma (2\varepsilon)$ in the limit $\varepsilon \rightarrow 0$ and for a $L$-loop diagram, which is the product of $L$ bubbles, it is $\Gamma (2\varepsilon)^L$.
Therefore, we only have to do the combinatorial sum over the $\xi^2$ and $\alpha_3^2$ couplings.
Hence, a $L$-loop diagram of the form \eqref{eq:Fishnet_alpha23_RenDiags} is proportional to $\Gamma (\varepsilon)^L (\xi^2 - \alpha_3^2)^{L+1}$.
This immediately implies that no correction is generated if $\xi^2 - \alpha_3^2 = 0$.
Hence, the all-loop conformal fixed-point is $\alpha_{3,*}^2 = \xi^2$.
The same argument applies to the coupling $\alpha_{2}^2$, and the fixed-point is $\alpha_{2,*}^2 = \xi^2$ as well \cite{Grabner:2017pgm}.

The beta function for the couplings \eqref{eq:Fishnet_DoubleTraceTerms_sameflavor} is shown in the example of $\alpha_{1}^2$ in appendix \ref{sec:RenormalizationOfDoubleTraceCouplings} up to three loops.
Unfortunately, the fixed-point is only known perturbatively and for weak coupling $\xi^2$ it is the result \eqref{eq:Alpha1FixedPointExpansion3L}.
We repeat it here,
\begin{equation}
\alpha_{1*}^2 
~=~ 
\pm \dfrac{\I}{2}\xi^2 -\dfrac{\xi^4}{2} \mp \dfrac{3\I}{4}\xi^6 + \mathcal{O}(\xi^8) ~.
\label{eq:Fishnet_Alpha1FixedPointExpansion3L_maintext}
\end{equation}
An analog calculation yields the same fixed point for $\tilde{\alpha}_{1}^2$, with the above expansion.
In the literature, it is even determined up to seven loops \cite{Grabner:2017pgm,Korchemsky:2018hnb}.

\subsection{Exact all-loop anomalous dimensions and correlation function}
\label{subsec:ExactAnomalousDimensionBiScalarFN}
Traditionally, integrability manifests itself in $\mathcal{N} = 4$ SYM in the anomalous dimensions of single-trace operators, see e.\,g.\ \cite{Beisert:2010jr}.
These quantities can be calculated by solving the spectral problem of spin chain Hamiltonians, which are constructed from perturbative data.
However, instead of in a perturbative Hamiltonian, the integrable properties of the theories come in the form of STRs \eqref{eq:STR_Scalar_medial} and \eqref{eq:STR_Fermion_medial}, which are YBEs on the medial lattice.
Especially the chain relations \eqref{eq:ChainRuleScalar} and \eqref{eq:ChainRuleScalarFermion}, the STR's special case, is enabling the computation alongside the theory's conformal symmetry.
Here, we present the exact computation of the anomalous dimension of the length-two single-trace operator $\mathcal{O} = \mathrm{tr} ( \phi_1 \phi_1 )$ and its spinning generalization, where derivatives may act on the fields.
Furthermore, we compute the exact four-point function, where these operators are exchanged in the operator product expansion (OPE) \cite{Gromov:2018hut}.

Instead of calculating the two-point function perturbatively, which is done to three loops in appendix \ref{sec:PerturbativeCalculationAnomalousDimension}, we follow the strategy of \cite{Grabner:2017pgm,Kazakov:2018qbr,Gromov:2018hut,Kazakov:2018gcy}.
It consists of considering the four-point function, where the operator $\mathcal{O} = \mathrm{tr} ( \phi_1 \phi_1 )$ is exchanged in the OPE expansion.
It is crucial that its Feynman diagrams admit a ladder structure in terms of a repetitive graph-building operator, such that one can sum up the perturbative expansion in a geometric series.
Next, the knowledge of a suitable eigenfunction is necessary to diagonalize the graph-building operators and to express the correlation function in a spectral integral representation.
It develops poles for each exchanged operator and we can read off the operators' exact scaling dimension from their position.
In coordinate-space Feynman diagrams and in planar limit, the four-point correlation function has the perturbative expansion
\begin{equation}
\begin{split}
\left\langle 
\mathrm{tr}
\left[
\phi_1 (x_1)
\phi_1 (x_2)
\right]
\mathrm{tr}
\left[
\phi_1^\dagger (x_3)
\phi_1^\dagger (x_4)
\right]
\right\rangle 
=
\adjincludegraphics[valign=c,scale=1]{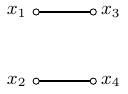} 
+
\xi^4
\adjincludegraphics[valign=c,scale=1]{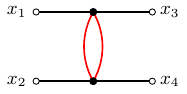} \\
+
\xi^8
\adjincludegraphics[valign=c,scale=1]{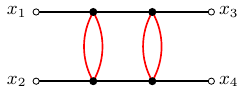}
+
\xi^{12}
\adjincludegraphics[valign=c,scale=1]{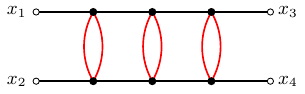}\\
+
\cdots
+ 
(x_3 \leftrightarrow x_4)
+
\mathcal{O} (\alpha_1^2) ~.
\end{split}
\label{eq:Fishnet_0MagnonPertExpansion}
\end{equation}
We see that its structure is very regular.
Increasing the loop order just adds a new, red, vertically running $\phi_2$-bubble to the diagram.
Of course, there are suppressed double-trace terms.
They ensure the finiteness and the conformal properties of \eqref{eq:Fishnet_0MagnonPertExpansion}, when tuned to the fixed-point \eqref{eq:Fishnet_Alpha1FixedPointExpansion3L_maintext}, but they do not alter the poles of exchanged operators.
The graphs appearing in \eqref{eq:Fishnet_0MagnonPertExpansion} are referred to as ``zero-magnon'' graphs \cite{Gromov:2018hut}.
The name corresponds to closed vertical subgraphs made up of red $\phi_2$ propagators, which run in circles and do not contract with the operators at the external points of the diagrams.
They are also called mirror-magnons \cite{Basso:2018agi,Basso:2019xay}.

The perturbative expansion consists of the convolution of two different kinds of integral operators
\begin{equation}
\begin{aligned}[c]
\mathbb{H}
=
\adjincludegraphics[valign=c,scale=1]{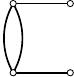} ~,
\end{aligned}
\qquad \qquad
\begin{aligned}[c]
\mathbb{P}
=
\adjincludegraphics[valign=c,scale=1]{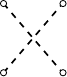} ~,
\end{aligned}
\label{eq:Fishnet_0MagnonBuilders}
\end{equation}
which we will also call graph-building operators.
We define the product $\circ$ between two integral kernels as the convolution in two spacetime points, so that we can use the integral operators to build the diagrams in \eqref{eq:Fishnet_0MagnonPertExpansion}.
The correlation function can be resumed by usage of the geometric series and we have the expression 
\begin{equation}
\begin{split}
& \left\langle 
\mathrm{tr}
\left[
\phi_1 (x_1)
\phi_1 (x_2)
\right]
\mathrm{tr}
\left[
\phi_1^\dagger (x_3)
\phi_1^\dagger (x_4)
\right]
\right\rangle  \\
&=
\adjincludegraphics[valign=c,scale=1]{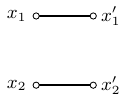} 
\circ
\left[
1
+
\xi^4
\mathbb{H}
+
\xi^8
\mathbb{H} \circ \mathbb{H}
+
\xi^{12}
\mathbb{H} \circ \mathbb{H} \circ \mathbb{H}
+ 
\cdots
\right]
\left[
1 + \mathbb{P}
\right]
+
\mathcal{O} (\alpha_1^2) \\
& =
x_{12}^4
\mathbb{H}
\circ
\left[
\frac{1 + \mathbb{P}}{1 - \xi^4 \mathbb{H}}
\right]
+
\mathcal{O} (\alpha_1^2) ~.
\end{split}
\label{eq:Fishnet_0MagnonPertExpansion_2}
\end{equation}
We observe that the permutation operator $\mathbb{P}$ accounts for the terms in \eqref{eq:Fishnet_0MagnonPertExpansion}, where $x_3$ and $x_4$ are exchanged.
The excess propagators on the left can be expressed by the graph-building operator, which is multiplied by $x_{12}^4$ to eliminate the vertical propagators.

\subsubsection{Conformal triangles}
In order to evaluate the expression \eqref{eq:Fishnet_0MagnonPertExpansion_2}, we take inspiration from quantum mechanics and insert a complete set of eigenvectors into \eqref{eq:Fishnet_0MagnonPertExpansion_2} and diagonalize the operators.
In \cite{Dobrev:1977qv,Polyakov:1970xd,Fradkin:1978pp} such a two-point completeness relation was shown to exist and it reads
\begin{equation}
\delta^{(D)}\left( x_{13} \right) \cdot
\delta^{(D)}\left( x_{24} \right)
=
\frac{1}{\left[ x_{12}^2\right]^{D - \Delta_1 - \Delta_2}}
\sumint{\Delta,S} ~~~~~~
\int \dd^D x_0 ~
\ket{\bar{\Omega}^{\Delta,S}_{\mu_1 \cdots \mu_S} {\scriptstyle (x_1, x_2 ; x_0) }} \;
\bra{\Omega_{\Delta,S}^{\mu_1 \cdots \mu_S} {\scriptstyle (x_3, x_4 ; x_0) }}
\label{eq:Bosonic_CompletenessRelation}
\end{equation}
The integration/summation measure is defined as\footnote{
In the following, we use the abbreviations \cite{Dobrev:1977qv,Grabner:2017pgm,Gromov:2018hut} 
\begin{subequations}
\begin{align}
c_1(\Delta , S)
&=
\frac{2^{S+1} S!}{\pi^{-\frac{3D}{2} - 1}}
\frac{(\Delta - 1) ( D- \Delta)}{(\frac{D}{2} + S -1)^2 - (\Delta - \frac{D}{2})^2}
\frac{a_0(\Delta) a_0 (D - \Delta)}{\Gamma (\frac{D}{2} + S)} \\
c_2(\Delta , S)
&=
2 \pi ^{D+1} (-1)^S S!
\frac{ 
\Gamma ( \Delta -\frac{D}{2} )
\Gamma ( S + \Delta - 1 ) 
\Gamma ( \frac{D + S - \Delta + \Delta_1 - \Delta_2 }{2} ) 
\Gamma ( \frac{D + S - \Delta - \Delta_1 + \Delta_2 }{2} )
}{
\Gamma ( \Delta -1 )
\Gamma ( \frac{D}{2}+S ) 
\Gamma ( D + S - \Delta ) 
\Gamma ( \frac{ S + \Delta + \Delta_1 - \Delta_2 }{2} ) 
\Gamma ( \frac{ S + \Delta - \Delta_1 + \Delta_2 }{2} )
}
\end{align} \label{eq:Coeff_ConformalBlocks_nonsusy}%
\end{subequations}
and the conformal blocks in four dimensions are defined as \cite{Grabner:2017pgm,Gromov:2018hut,Kazakov:2018gcy} 
\begin{subequations}
\begin{align}
& g_{\Delta,S} (r_1 , r_2)
=
(-1)^S
\frac{
z\, z^*
}{
z - z^*
}
\left[
h (\Delta + S, z) h(\Delta - S - 2, z^*)
-
h (\Delta + S, z^*) h(\Delta - S - 2, z)
\right] \\
& \mathrm{with} ~~~
h (t, z ) 
:=
z^\frac{t}{2} ~
_2 F_1
\left(
\frac{t - \Delta_1 + \Delta_2}{2},
\frac{t + \Delta_3 - \Delta_4}{2},
t, z
\right) ~,
\end{align}
\end{subequations}
and $_2 F_1$ denotes the hypergeometric function.
The four-dimensional cross-ratios $r_1$ and $r_2$ are defined as 
\begin{equation}
r_1 
= 
z \, z^*
=
\frac{x_{12}^2 x_{34}^2}{x_{13}^2 x_{24}^2}
~~~~ \mathrm{and} ~~~~
r_2
= 
(1 - z) (1 - z^*)
=
\frac{x_{14}^2 x_{23}^2}{x_{13}^2 x_{24}^2} ~.
\end{equation}
\label{footnote:ConformalTrianglesAbbreviations}
} 
\cite{Grabner:2017pgm,Gromov:2018hut,Kazakov:2018gcy}
\begin{equation}
\sumint{\Delta,S} ~~~~~~
=
\frac{\I}{2}
\sum_{S=0}^\infty
(-1)^{S+1}
\int_{\frac{D}{2}}^{\frac{D}{2} + \I\infty}
\frac{\dd \Delta}{c_1(\Delta , S)}
\end{equation}
and it captures the integrating/summing over the Cartan labels of the conformal algebra in $D$ dimensions.
The factor $a_0$ is given in \eqref{eq:Factor_aEll} and is a ratio of two gamma functions.
The eigenvectors are so-called \textit{conformal triangles} that furnish a non-compact representation of the conformal group and they read
\begin{equation}
\Omega_{\Delta,S}^{\mu_1 \cdots \mu_S} {\scriptstyle (x_1, x_2 ; x_0)}
=
\frac{
\prod_{i=0}^S
\left[
\frac{2 x_{02}^{\mu_i} }{x_{02}^2}
-
\frac{2 x_{01}^{\mu_i} }{x_{01}^2}
\right]
}{
\left[
x_{12}^2
\right]^{\frac{\Delta_1 + \Delta_2 - (\Delta - S)}{2}}
\left[
x_{10}^2
\right]^{\frac{\Delta_1 - \Delta_2 + (\Delta - S)}{2}}
\left[
x_{20}^2
\right]^{\frac{- \Delta_1 + \Delta_2 + (\Delta - S)}{2}}
}
\label{eq:Conformal_Triangles}
\end{equation}
Their form is fixed by conformal symmetry and they are symmetric and traceless in the Lorentz indices $\mu_1 \cdots \mu_S$.
They equal the three-point function of the operators $\mathcal{O}_1$, $\mathcal{O}_2$ and $\mathcal{O}_{\Delta ,S}$ and the completeness relation \eqref{eq:Bosonic_CompletenessRelation} describes the exchange of the operators $\mathcal{O}_{\Delta ,S}$.
In our case at hand, we have the scaling dimensions $\Delta_1 = \Delta_2 = 1$ for the operators $\mathcal{O}_1 = \mathcal{O}_2 = \phi_1$ and the fishnet theory is in four dimensions, $D=4$.
Diagonalizing the operators \eqref{eq:Fishnet_0MagnonBuilders} on these three-point functions is unnecessarily hard, and we simplify $\Omega_{\Delta,S}^{\mu_1 \cdots \mu_S} {\scriptstyle (x_1, x_2 ; x_0)}$ by sending $x_0\rightarrow \infty$.
This gives the two-point function
\begin{equation}
\Omega_{\Delta,S}^{\mu_1 \cdots \mu_S} {\scriptstyle (x_1, x_2 ; x_0)}
\stackrel{x_0 \rightarrow \infty}{\sim}
\Psi_{\frac{\Delta_1 + \Delta_2 - \Delta}{2},\frac{S}{2}}^{\mu_1 \cdots \mu_S} (x_1 , x_2)
:= W_u^{\frac{S}{2}, \mu_1 \cdots \mu_S} (x_{12})
=
\adjincludegraphics[valign=c,scale=1]{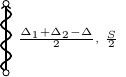} ~,
\label{eq:Fishnet_0Magnon_Omega_scaling}
\end{equation}
which is a generalized higher-spin weight \eqref{eq:GenFermionicPropagatorWWeight_higherSpin}.

Furthermore, the conformal triangles can be expressed in terms of conformal blocks $g_{\Delta,S}(r_1, r_2)$ \cite{Kazakov:2018gcy} and in four dimensions the relation is exactly the spacetime integral in \eqref{eq:Bosonic_CompletenessRelation},
\begin{equation}
\begin{split}
\int & \dd^D x_0 ~ 
\ket{\bar{\Omega}^{\Delta,S}_{\mu_1 \cdots \mu_S} {\scriptstyle (x_1, x_2 ; x_0) }} \;
\bra{\Omega_{\Delta,S}^{\mu_1 \cdots \mu_S} {\scriptstyle (x_3, x_4 ; x_0) }} \\
& =
\left(
\frac{1}{x_{12}^2 x_{34}^2}
\right)^{\frac{\Delta_1 + \Delta_2}{2}}
\left(
\frac{x_{24}^2}{ x_{13}^2}
\right)^{\frac{\Delta_1 - \Delta_2}{2}}
\left[
\frac{c_1(\Delta, S)}{c_2 (\Delta, S)}
g_{\Delta , S} (r_1 , r_2)
+
\frac{c_1(\Delta^*, S)}{c_2 (\Delta^*, S)}
g_{\Delta^* , S} (r_1 , r_2)
\right] ~.
\end{split}
\label{eq:Conformal_blocks4D}
\end{equation}
The abbreviations $c_1 (\Delta , S)$, $c_2 (\Delta ,S)$ and the conformal blocks are defined in \eqref{eq:Coeff_ConformalBlocks_nonsusy} in the footnote.

\subsubsection{Eigenvalues and correlation function}
We can decompose the correlation function \eqref{eq:Fishnet_0MagnonPertExpansion_2} into two parts 
\begin{equation}
\left\langle 
\mathrm{tr}
\left[
\phi_1 (x_1)
\phi_1 (x_2)
\right]
\mathrm{tr}
\left[
\phi_1^\dagger (x_3)
\phi_1^\dagger (x_4)
\right]
\right\rangle
=
G(x_1 , x_2 \vert x_3 , x_4)
+
G(x_1 , x_2 \vert x_3 , x_4) \circ \mathbb{P}
\end{equation}
and focus on the calculation of $G(x_1 , x_2 \vert x_3 , x_4)$, since the second term is obtained by interchanging $x_3 \leftrightarrow x_4$.
When we insert the completeness relation \eqref{eq:Bosonic_CompletenessRelation} into \eqref{eq:Fishnet_0MagnonPertExpansion_2}, we can replace the graph-building operator $\mathbb{H}$ by its eigenvalue $\bra{\Omega}\circ \mathbb{H} = \bra{\Omega}\cdot E (\Delta , S)$.
Using \eqref{eq:Conformal_blocks4D}, we obtain the integral representation \cite{Grabner:2017pgm,Gromov:2018hut}
\begin{equation}
\begin{split}
G(x_1 , x_2 \vert x_3 , x_4)
=
\frac{1}{x_{12}^2 x_{34}^2}
\cdot
\frac{\I}{2}
\sum_{S=0}^\infty
(-1)^{S+1}
\int_{2 - \I\infty}^{2 + \I\infty}
\frac{ \dd \Delta}{c_2(\Delta , S)}
\frac{E(\Delta , S)}{1 - \xi^4 E(\Delta ,S)} \;
g_{\Delta , S} (r_1 , r_2) ~.
\end{split}
\label{eq:FishnetCorrelationFunction_SpectralForm}
\end{equation}
The integral was completed to a infinite line into the imaginary direction, by anticipating the property $E(4 - \Delta ,S) = E(\Delta ,S)$, which we show below.
The second line in \eqref{eq:FishnetCorrelationFunction_SpectralForm} will be denoted by $\mathcal{G} (r_1 ,r _2)$ and its integral can be evaluated by the residue theorem.
Because spurious poles are absent \cite{Gromov:2018hut}, we find 
\begin{equation}
\mathcal{G} (r_1 ,r _2) 
=
\sum_{S, \Delta}
C_{\Delta , S}
\,
g_{\Delta , S} (r_1 , r_2)
=
\sum_{S, \Delta}
(-1)^{S+1}
\pi\,
\underset{\Delta}{\mathrm{Res}}
\left[
\frac{1}{c_2 (\Delta , S)}
\frac{E(\Delta , S)}{1 - \xi^4 E(\Delta ,S)}
\right]
\,
g_{\Delta , S} (r_1 , r_2)
~,
\label{eq:CPWexpansion_4D}
\end{equation}
which has the form of a conformal partial wave expansion and when we know the values $\Delta$ of the poles, we can calculate the residues and read off the OPE coefficient $C_{\Delta , S}$.

We can calculate the eigenvalue of the operator $\mathbb{H}$, when acting on the simplified eigenfunction \eqref{eq:Fishnet_0Magnon_Omega_scaling}.
After merging the vertical propagators, the spinning chain relation \eqref{eq:ChainRuleScalarFermion} is crucial and allows to derive the eigenvalue, 
\begin{equation}
\begin{split}
\bra{\Psi_{u,\frac{S}{2}}} \circ \mathbb{H}
=&
\adjincludegraphics[valign=c,scale=1]{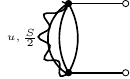} 
=
\adjincludegraphics[valign=c,scale=1]{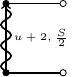}
=
r_{\frac{S}{2}} (1, u + 2 , 1 - u )
\cdot
\adjincludegraphics[valign=c,scale=1]{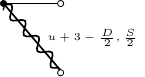} \\
=&
r_{\frac{S}{2}} (1, u + 2 , 1 - u )\, r_{\frac{S}{2}} (1, u + 1 , 2 - u )
\cdot
\adjincludegraphics[valign=c,scale=1]{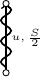}\\
=& ~
\mathrm{E} (u,\tfrac{S}{2}) 
\cdot 
\bra{\Psi_{u,\tfrac{S}{2}}} ~.
\end{split}
\label{eq:Fishnet_0MagnonPreDiag}
\end{equation}
Hence, the eigenvalue $\mathrm{E} (u) $ is a combination of gamma functions, by the factor \eqref{eq:Factor_rEll}.
According to \eqref{eq:Fishnet_0Magnon_Omega_scaling}, we should use a specific value for $u$, which involves the scaling dimension of the exchanged operator $\Delta$ and the dimensions of the external operators $\Delta_1 = \Delta_2 = 1$.
We define the parameterized eigenvalue as $E (\Delta, S) = \mathrm{E} (\frac{2 - \Delta}{2},\frac{S}{2})$ and find that the gamma functions are reduced to a factor due to their functional relation $\Gamma (z + 1) = z \cdot \Gamma (z)$.
Thus, the eigenvalue of $\mathbb{H}$ when acting on the eigenfunction \eqref{eq:Fishnet_0Magnon_Omega_scaling} is
\begin{equation}
E (\Delta,S)
=
\frac{
16 \pi^4
}{
(- \Delta + S + 2) (-\Delta + S + 4) (\Delta + S - 2) (\Delta + S)
} ~.
\end{equation}
We can now solve the pole condition of \eqref{eq:CPWexpansion_4D}, which is $1 = \xi^4 E (\Delta,S)$, for the position of the poles, which coincides with the scaling dimensions of the exchanged operators in \eqref{eq:FishnetCorrelationFunction_SpectralForm}.
They read
\begin{subequations}
\begin{align}
\Delta_2
&=
2
\pm
\sqrt{
(S + 1)^2 + 1 
-
2 \sqrt{ (S + 1)^2 + 4\pi^4 \xi^4}
} ~, \\
\Delta_4
&=
2
\pm
\sqrt{
(S + 1)^2 + 1 
+
2 \sqrt{ (S + 1)^2 + 4\pi^4 \xi^4}
} 
\end{align}\label{eq:Fishnet_ScalingDimensions}%
\end{subequations}
and correspond to operators, which have the classical scaling dimensions $\Delta_2 \rightarrow 2$ and $\Delta_4 \rightarrow 4$ when $\xi^2 \rightarrow 0$.
The operators have the form $\mathrm{tr}\left[ \phi_1 \partial^S \phi_1 \right]$ and $\mathrm{tr}\left[ \square \phi_1 \partial^S \phi_1 \right] + ...$ and the dots denote operators where the derivatives are unevenly distributed on the two fields.
The derivatives should be contracted with a symmetric traceless tensor.
In particular, we can obtain the exact scaling dimension of the length-two operator $\mathrm{tr}\left[ \phi_1 \phi_1 \right]$ by setting $S=0$ and expanding in $\xi$.
After the rescaling $\xi \rightarrow \xi / \pi$, the seven-loop result is 
\begin{equation}
\Delta 
=
2
-
2 i \xi ^2
+
i \xi ^6
-
\frac{7 i \xi ^{10}}{4}
+
\frac{33 i \xi ^{14}}{8}
+
\mathcal{O}\left(\xi ^{18}\right)
\label{eq:Fishnet_ScalingDimensions_S0}
\end{equation}
and we obtain the three-loop result \eqref{eq:Fishnet_ResultAnomalousDimPerturbative} of the perturbative calculation in appendix \ref{sec:PerturbativeCalculationAnomalousDimension}.
In section \ref{sec:ExactAnomalousDimensionsFrom4PtCorrelationFunctions}, we derive the scaling dimension of similar operators by the superspace generalization of the method presented here.

Finally, we determine the last missing piece for the exact correlation function \eqref{eq:Fishnet_0MagnonPertExpansion_2}, the residues in \eqref{eq:CPWexpansion_4D}.
We obtain the OPE coefficient \cite{Gromov:2018hut}
\begin{equation}
C_{\Delta , S}
=
\frac{
4^{3-\Delta } (S+1) 
}{
(S^2+2 S-2-\Delta ^2+4 \Delta) 
}
\frac{
\Gamma (\frac{S-\Delta +5}{2}) 
\Gamma (\frac{S+\Delta }{2})
}{
\Gamma (\frac{S-\Delta +4}{2} ) 
\Gamma (\frac{S+\Delta -1}{2} )
} ~,
\end{equation}
and we can obtain the perturbative expansion by expanding in the coupling, where $\Delta$ is implicitly dependent on $\xi^2$ by \eqref{eq:Fishnet_ScalingDimensions}.
In \cite{Gromov:2018hut}, other correlation functions and scaling dimensions of the bi-scalar fishnet theory are also calculated exactly in the presented way.

\section{Integrable vacuum diagrams and the critical coupling}
\label{sec:IntegrableVacuumDiagramsAndTheCriticalCoupling}
Remarkably, from the results of the previous section, we see that integrability, in the form of the chain relation \eqref{eq:ChainRuleScalar} or its spinning generalization \eqref{eq:ChainRuleScalarFermion}, allows for exact determination of the scaling dimensions of single-trace operators.
However, so far, chain relations have only been used as a tool to enable these computations.
In this section, we conceptually follow another way: based on the integrable propagator weights from chapter \ref{chpt:AuxiliaryRelations}, we construct a partition function in the spirit of statistical physics.
Therein, a convenient choice of boundary conditions are doubly-periodic ones; the model lives on a torus.
We adapt these boundary conditions and find that the model of propagator weights describes an integrable generalization of vacuum Feynman diagrams.
One standard calculation, as presented in the case of the eight-vertex model in section \ref{subsec:FreeEnergyInThermoDynLimit_8V}, is the derivation of the thermodynamic limit of the model's partition function.
In our case of vacuum Feynman diagrams, the thermodynamic limit corresponds to the critical coupling of the theory \cite{Zamolodchikov:1980mb,Bazhanov:2016ajm}.
That is, the radius of divergence of the free energy of the theory which produces the vacuum graphs under examination when the sum of all vacuum graphs is considered as a function of the coupling.
The eminent question is which theories allow for the vacuum graphs describing such an integrable propagator model.
In the previous section \eqref{sec:TheBiScalarFishnetTheory}, we present such a theory as a double-scaling limit of $\gamma$-deformed $\mathcal{N} = 4$ SYM, the bi-scalar fishnet theory.
But there exist also other theories, which might not even come from prominent examples of integrable QFTs.
As an instance, we will present the brick wall theory and a three-dimensional fishnet theory that contain fermions and we calculate their critical coupling \cite{Kade:2023xet}.

\subsection{Integrable vacuum diagrams and the generalized partition function}
\label{subsec:IntegrableVacuumDiagramsFromTheGeneralizedPartitionFunction}
The fundamental building block of an integrable lattice model is the R-matrix.
In the model made from generalized Feynman propagators, the R-matrix is \eqref{eq:Rmatrix_Fermion_Fused}.
We use it to construct a periodic, inhomogeneous transfer matrix, by intersecting a pair of scalar horizontal medial lines with $N$ pairs of fermionic vertical ones.
On the level of the Feynman graphs, this yields a chain of rhombuses,
\begin{equation}
\begin{split}
\mathbb{T}_{N} (p_a , p_b \vert \lbrace \mathbf{p}_j \rbrace)
&= 
\mathrm{tr}
\left[
\mathbb{R}( p_a, p_b \vert \mathbf{p}_1 ,\mathbf{p}_2 )
\mathbb{R}( p_a, p_b \vert \mathbf{p}_3 , \mathbf{p}_4 )
\cdots
\mathbb{R}( p_a, p_b \vert \mathbf{p}_{2N-1} , \mathbf{p}_{2N} ) 
\right] \\
&=
\mathbb{R}^{x_1' y_2}_{x_1 y_1}( p_a, p_b \vert \mathbf{p}_1 ,\mathbf{p}_2 )
\mathbb{R}^{x_2' y_3}_{x_2 y_2}( p_a, p_b \vert \mathbf{p}_3 ,\mathbf{p}_4 )
\cdots
\mathbb{R}^{x_N' y_1}_{x_N y_{N}}( p_a, p_b \vert \mathbf{p}_{2N-1} ,\mathbf{p}_{2N} )\\
&=
\adjincludegraphics[valign=c,scale=1]{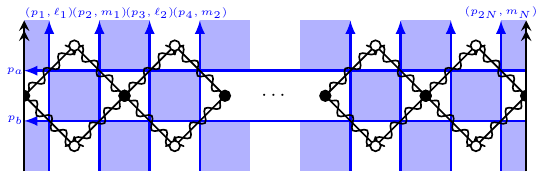} ~.
\end{split}
\label{eq:TransferMatix_periodic}
\end{equation}
The spectral parameters for the vertical lines contain the spin label.
In the second line, we explicitly show the spacetime arguments of the $N$ R-matrices. 
We used the integration convention that we have to integrate over spacetime coordinates, which appear upstairs and downstairs.
Here, the integrated coordinates are $y_1, ..., y_N$.
For the spectral parameters, we have $\mathbf{p}_j = (p_j ,\ell_j)$ for $j$ odd and $\mathbf{p}_j = (p_j ,m_j)$ for $j$ even.
%Comparing the expression with the R-matrix \eqref{eq:Rmatrix_Fermion_Fused}, we deduce that not all spectral parameters are independent.
%There are $N$ constraints, $p_a + p_b + p_{2i-1} + p_{2i} = D$ and we can solve them for example for the odd spectral parameters.
%Then we are left with the $N+2$ free parameters $p_a$, $p_b$ and $p_i$ for $i=1,..., N$.
The periodicity in \eqref{eq:TransferMatix_periodic} due to the trace is indicated by the left and right boundary and the double-arrows indicate that they should be identified.
Respectively, the integration point on the left and right boundary, $y_1$, is the same and also the generalized Feynman graph is periodically identified.
The transfer matrix commutes with another transfer matrix for different values of $p_a$ and $p_b$ and we have 
\begin{equation}
\mathbb{T}_{N} (p_a , p_b \vert \lbrace \mathbf{p}_j \rbrace)
\circ 
\mathbb{T}_{N} (q_a , q_b \vert \lbrace \mathbf{p}_j \rbrace)
=
\mathbb{T}_{N} (q_a , q_b \vert \lbrace \mathbf{p}_j \rbrace)
\circ 
\mathbb{T}_{N} (p_a , p_b \vert \lbrace \mathbf{p}_j \rbrace) ~.
\label{eq:CommutativityTransferMatrix_nonsusy}
\end{equation}
The product $\circ$ means that we stack the two factors on top of each other and convolute them by integrating over their common spacetime points.
The proof is carried out easily on the medial lattice, we just have to repeat the train track argument from \eqref{eq:8V_periodicTransferMatrix_comm} with pairs of medial lines.
The fermionic version of unitarity \eqref{eq:Unitarity_Rmatrix_scalar_fused} and the $\mathbb{YBE}$ \eqref{eq:YBE_Scalar_Fused} are sufficient to do so. 
As usual, the commutativity of the transfer matrices implies the existence of conserved charges, which are the coefficients in an expansion of the transfer matrix in the spectral parameters.
When we switch to the Feynman graph picture, we can read off the propagator exponents from \eqref{eq:TransferMatix_periodic} via \eqref{eq:GenFermionicPropagatorWWeight_medial}.
We drop the shading and the medial lines and find the transfer matrix as a generalized Feynman graph,
\begin{equation}
\mathbb{T}_{N} (p_a , p_b \vert \lbrace \mathbf{p}_j \rbrace)
=
\adjincludegraphics[valign=c,scale=1]{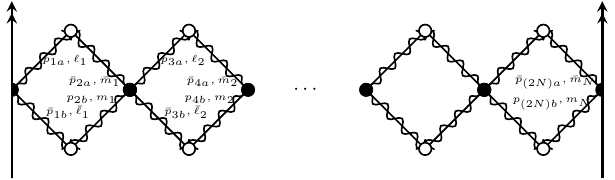} ~.
\label{eq:TransferMatix_periodic_Feynman}
\end{equation}
In order to build a toroidal partition function, we stack $M$ transfer matrices on top of each other, using the product $\circ$ and eventually convolute the top and bottom of this stack.
The last operation can be understood as a trace, since the stack of transfer matrices has two ``indices'', the top and bottom row of external spacetime points with $N$ external vertices each.
Contracting the indices by convolution means to take the trace, and we obtain the toroidal partition function
\begin{equation}
Z_{MN}
(\lbrace q_i \rbrace_{i=1,...,2M} \vert \lbrace \mathbf{p}_j \rbrace_{j=1,...,2N}  )
=
\mathrm{Tr}
\left[
\prod_{I=1}^{M}
\mathbb{T}_{N} (q_{2I - 1} , q_{2I} \vert \lbrace \mathbf{p}_j \rbrace)
\right] ~.
\end{equation}
We called the spectral parameters of the horizontal lines $q_i$ and the spectral parameters of the vertical ones $p_j$.
Thus, the spectral parameters are so far inhomogeneous in both directions.
We restrict ourselves to the vertically homogeneous case, which is $q_{2I - 1} \equiv p_a$ and $q_{2I} \equiv p_b$ for all $I=1,..,M$.
Furthermore, we complete the transition to the Feynman graph formulation, by changing the parametrization in terms of the spectral parameters of the medial lines, to a parametrization in terms of the values of the exponent of the propagators in the transfer matrix \eqref{eq:TransferMatix_periodic_Feynman}.
We define 
\begin{equation}
U_J
:=
\begin{pmatrix}
\mathbf{u}^+_J & \mathbf{v}^+_J \\
\mathbf{u}^-_J & \mathbf{v}^-_J
\end{pmatrix}
=
\begin{pmatrix}
( u^+_J , \ell_J ) & ( v^+_J , \bar{m}_J ) \\
( u^-_J , \bar{\ell}_J ) & ( v^-_J , m_J )
\end{pmatrix}
:=
\begin{pmatrix}
( p_{(2J-1)a} , \ell_{J} ) & ( \bar{p}_{(2J)a} , \bar{m}_{J} ) \\
( \bar{p}_{(2J-1)b} , \bar{\ell}_{J} ) & ( p_{(2J)b} , m_{J} )
\end{pmatrix}
\label{eq:Parametermatrix_U}
\end{equation}
for $J = 1, ... , N$ and the matrix $U_J$ contains all the information of the propagators in the R-matrix at site $J$.
We recall that the four spectral parameters at each R-matrix are not independent: the sum of them has to be equal to $D$, which is $u^+_J + u^-_J + v^+_J + v^-_J = D$, as we see from the parametrization of the medial lines \eqref{eq:TransferMatix_periodic_Feynman}.
Note that our convention to express the differences between spectral parameters of medial lines is e.\,g.\ $p_{(2J-1)a} = p_{2J-1} - p_a$ and the bracket in the subscript helps to specify the first term.
The transfer matrix \eqref{eq:TransferMatix_periodic_Feynman} in this parametrization is 
\begin{equation}
\mathbb{T}_{N} 
\left(
\left\lbrace
U_J
\right\rbrace
\right)
=
\adjincludegraphics[valign=c,scale=0.9]{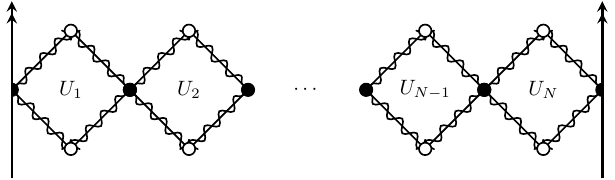} ~.
\label{eq:TransferMatix_periodic_Feynman_2}
\end{equation}
From the transfer matrix \eqref{eq:TransferMatix_periodic_Feynman_2}, we can build the vertically homogeneous, but horizontally inhomogeneous partition function
\begin{equation}
Z_{MN}
\left(
\left\lbrace
U_J
\right\rbrace
\right)
=
\mathrm{Tr}
\left[
\mathbb{T}_{N} 
\left(
\left\lbrace
U_J
\right\rbrace
\right)^M
\right] ~.
\label{eq:ZMN_Inhomogeneous_vacuumgraph}
\end{equation}
The partition function $Z_{MN}$ consists of vacuum Feynman graphs with generalized propagator exponents, which possess a toroidal topology.
Therefore we call them generalized vacuum graphs.
By keeping the spectral parameters in form of the parameter matrix \eqref{eq:Parametermatrix_U} generic, we are able to employ the toolbox of integrability.
Specifically we will calculate the thermodynamic limit of \eqref{eq:ZMN_Inhomogeneous_vacuumgraph} in section \ref{subsec:ThermodynamicLimitOfVacuumDiagramsAndCriticalCoupling}.
A critical comment is in place: vacuum diagrams in field theory are generally proportional to the spacetime volume, and thus, the overall free energy is usually infrared divergent. 
Since we are only interested in the free energy density, this divergence is best fixed by leaving one of the points in the vacuum graphs unintegrated, which fixes the zero mode.
By this regularization, our results for the thermodynamic limit will be well-defined.

However, we first want to understand which physical vacuum graphs of which QFTs are generalized by \eqref{eq:ZMN_Inhomogeneous_vacuumgraph}.
Therefore, we turn to the question of how to choose the values of $U_J$ to construct graphs of physical theories.
First, we need to normalize the transfer matrix \eqref{eq:TransferMatix_periodic_Feynman_2} because it is built from the propagator weights \eqref{eq:GenPropagatorWWeight} and \eqref{eq:GenFermionicPropagatorWWeight} instead of the physical propagators \eqref{eq:GenPropagatorGWeight} and \eqref{eq:GenFermionicPropagatorGWeight}, respectively.
When specifying the exponents and spin labels of the generalized propagators in \eqref{eq:TransferMatix_periodic_Feynman_2}, we obtain one of the three different physical objects\footnote{Note that there are more choices than the four presented ones, if one considers theories with fields of higher spin.}.
\begin{itemize}
\item
Scalar propagator in $D$ dimensions.
If we set the spin label $\ell_i$ or $m_i$ to zero, we obtain the scalar propagator weights $W_u(x)$ from \eqref{eq:GenPropagatorWWeight}.
To obtain a scalar propagator in $D$ dimensions, $G_v(x)$, we have to adjust for the factor that appears in \eqref{eq:GenPropagatorGWeight} and is defined in \eqref{eq:Factor_cEll}.
Let us repeat their relation here,
\begin{equation}
G_{\frac{D}{2} - u} (x)
=
c_0(\bar{u}) \;
W_{u} (x)
~~~~
\mathrm{with}
~~~~
c_\ell (v) 
=
- (-1)^v
2^{\frac{D}{2} - 2v}
a_\ell (v) ~.
\label{eq:ScalarPropagator_WvsG}
\end{equation}
Furthermore, to find the scalar propagator of a local QFT in $D$ dimensions, we have to tune the exponent to $u=\frac{D-2}{2}$.

\item
Fermionic propagator in $D$ dimensions.
Comparing the propagator weight $W_{u, \alpha \dot{\beta}}^{\frac{1}{2}}(x) $ from \eqref{eq:GenFermionicPropagatorWWeight} to the fermionic propagator in $D$ dimensions, we find a match up to the factor $c_\ell (v)$ and a shift in the exponent.
We have
\begin{equation}
G_{\frac{D}{2}- u -\frac{1}{2} ,\alpha \dot{\beta}} (x)
=
c_{\frac{1}{2}} (\bar{u}) \;
W_{u,\alpha \dot{\beta}}^{\frac{1}{2}} (x)
\label{eq:FermionicPropagator_WvsG}
\end{equation}
and the factor $c_\ell (u)$ is defined in \eqref{eq:Factor_cEll} and repeated in \eqref{eq:ScalarPropagator_WvsG}.
At the point $u=\frac{D-1}{2}$, we obtain the fermionic propagator of an undeformed $D$-dimensional QFT.

\item
$D$-dimensional delta function.
If we multiply the scalar generalized propagator (i.\,e.\ $\ell_i = 0$ or $m_i = 0$) by $\pi^{-\frac{D}{2}} a_0 (\varepsilon)$, tune the exponent to $u=\frac{D}{2} - \varepsilon $ and take eventually the limit $\varepsilon \rightarrow 0$ we obtain a delta function based on \eqref{eq:Delta_bosonic}.
Within a vacuum graph, the delta function can increase the number of edges that are adjacent to a vertex.
It implies that graphs $Z_{MN} \left( \left\lbrace U_J \right\rbrace \right)$ generalize not only vacuum graphs with quartic vertices.

\item
No two-point function at all.
If we set the spin labels and the exponent to zero, we annihilate the generalized propagator $W^0_0 (x) = 1$ and detach the connection between the two vertices.
This can drastically change the topology of the vacuum graph.
\end{itemize}
We aim for one of the four choices at each generalized propagator in the vacuum graph \eqref{eq:ZMN_Inhomogeneous_vacuumgraph}, which we do by specifying the value of the $N$ matrices $\lbrace U_J \rbrace$ and fixing the dimension $D$.
We present a selection of theories and the corresponding vacuum graphs according to \eqref{eq:ZMN_Inhomogeneous_vacuumgraph}.

\subsubsection{Integrable QFTs and their vacuum graphs}
Here, we consider the case of homogeneous vacuum graphs in both directions on the torus.
Since \eqref{eq:ZMN_Inhomogeneous_vacuumgraph} is already homogeneous along the vertical direction, setting $U_J = U$ for all $J=1, ... , N$ and some $U$ yields also horizontal homogeneity.

First, the vacuum graphs of the fishnet theory \eqref{eq:Fishnet_biscalar_action} can be obtained by a specific choice of the parameters $U$.
We choose one propagator of the R-matrices in \eqref{eq:TransferMatix_periodic_Feynman_2} to vanish, the one on the opposite side to become a delta function, and the remaining two should become scalar propagators in four dimensions, i.\,e.\ $D=4$.
Thus, we choose a parameter matrix
\begin{equation}
U^\mathrm{FN}
=
\begin{pmatrix}
( 0 , 0 ) & ( 1 , 0 ) \\
( 1 , 0 ) & ( 2 - \varepsilon , 0 )
\end{pmatrix} ~.
\label{eq:ParameterMatrix_FN}
\end{equation}
Note that, in the limit $\varepsilon \rightarrow 0$, the spectral parameters add up to the dimension of the spacetime, $0 + 1 + 1 + 2 - \varepsilon \rightarrow 4$, which means that the corresponding row matrix is indeed a transfer matrix evaluated at a special value.
Additionally, we multiply each R-matrix in \eqref{eq:TransferMatix_periodic_Feynman_2} with $\pi^{-2} a_0 (\varepsilon)$, such that, in the limit $\varepsilon \rightarrow 0 $, they have the shape
\begin{equation}
\lim_{\varepsilon \rightarrow 0}
\pi^{-2} a_0 (\varepsilon)
\cdot
\adjincludegraphics[valign=c,scale=1]{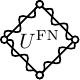} 
=
\adjincludegraphics[valign=c,scale=1]{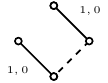} ~. 
\label{eq:Graphbuilder_Fishnet}
\end{equation}
In principle, by \eqref{eq:ScalarPropagator_WvsG}, we also would have to multiply the factors $c_0 (1) $ but they evaluate to one.
The rhombus \eqref{eq:Graphbuilder_Fishnet} is indeed a graph-builder for vacuum graphs of the bi-scalar fishnet theory, as we argue next.
We can chain \eqref{eq:Graphbuilder_Fishnet} via \eqref{eq:TransferMatix_periodic_Feynman_2} into a periodic transfer matrix
\begin{equation}
\begin{split}
\hat{\mathbb{T}}_{N}^\mathrm{FN}
:=
\lim_{\varepsilon \rightarrow 0}
\left[
\pi^{- 2} a_0 (\varepsilon)
\right]^N
\mathbb{T}_{N} 
\left(
\left\lbrace
U^\mathrm{FN}
\right\rbrace
\right)
&=
\adjincludegraphics[valign=c,scale=0.5]{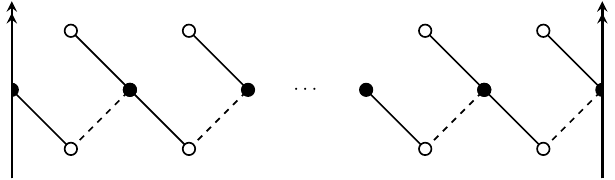} \\
&=
\adjincludegraphics[valign=c,scale=0.5]{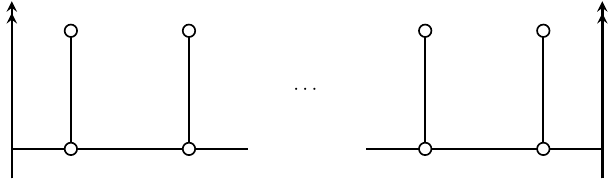} ~.
\end{split}
\label{eq:TransferMatrix_fishnet}
\end{equation}
Here and in the following, we denote by a hat over the transfer matrix or the partition function the physical quantities appearing in a particular QFT.
To obtain the second line of \eqref{eq:TransferMatrix_fishnet}, the delta function can be integrated by joining the two vertices it connects, and it annihilates one internal integration each.
Eventually, we stack the fishnet transfer matrices \eqref{eq:TransferMatrix_fishnet} together to form a torus, according to \eqref{eq:ZMN_Inhomogeneous_vacuumgraph}, evaluated at $U_J = U^\mathrm{FN}$.
The vacuum graph for $M=3$ and $N=4$ looks like
\begin{equation}
\hat{Z}_{34}^\mathrm{FN}
:=
\mathrm{Tr}
\left[
\left(
\hat{\mathbb{T}}_{4}^\mathrm{FN}
\right)^3
\right]
=
\adjincludegraphics[valign=c,scale=0.9]{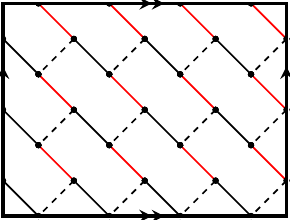} 
=
\adjincludegraphics[valign=c,scale=0.9]{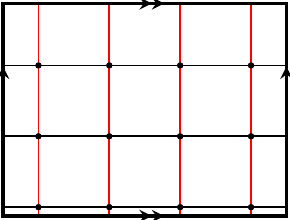} ~.
\label{eq:VacuumGraph_Fishnet}
\end{equation}
We clearly see the fishing-net pattern appearing, which gives the theory \eqref{eq:Fishnet_biscalar_action} its name. 
The red propagators are representing the $\phi_2$ scalar field, and the vacuum graphs are the only ones allowed\footnote{For $M=2$ or $N=2$, the fishnet vacuum graphs resemble periodically identified graphs from the calculation of the anomalous dimension in \eqref{eq:Fishnet_0MagnonPertExpansion}.
These graphs contain pairs of internal vertices that are connected by two propagators. 
This subgraph is UV-divergent and requires other vacuum diagrams with double-trace interactions \eqref{eq:Fishnet_DoubleTraceTerms} to renormalize the divergence. In fact, in the planar limit, double-trace diagrams contribute to $M=2$ or $N=2$.} for $M,N > 2$, according to the Feynman rules \eqref{eq:Fishnet_biscalar_Feynmanrules}.
The vacuum graph \eqref{eq:VacuumGraph_Fishnet} has a toroidal topology, which is indicated by the arrows on the boundary: opposite sides have the same number of arrows and should be identified.
This procedure yields a torus.

A second example of a theory, which admits vacuum diagrams generalized by \eqref{eq:ZMN_Inhomogeneous_vacuumgraph}, is a double-scaled $\gamma$-deformation of the three-dimensional ABJM theory.
ABJM theory, with a special focus on the double-scaling limit of its $\beta$-deformation, is studied in detail in section \ref{sec:DoubleScaledBetaDeformationOfABJM}.
Similar to the bi-scalar fishnet theory \eqref{eq:Fishnet_biscalar_action}, the double-scaled $\gamma$-deformation of ABJM theory contains three scalars in the adjoint representation of the gauge group $\U{\mathrm{N}} \otimes \U{\mathrm{N}}$, which are coupled in the sextic interaction term \cite{Caetano:2016ydc}
\begin{equation}
\mathcal{L}^\mathrm{Tri}
=
-
\mathrm{N}
\cdot
\sum_{i=1,2,4}
\mathrm{tr}
\left[
\partial_\mu Y_i^\dagger
\partial^\mu Y^i
\right]
+
\mathrm{N}
\cdot
\xi
\cdot
\mathrm{tr}
\left[
Y^1 Y_4^\dagger Y^2 Y_1^\dagger Y^4 Y_2^\dagger
\right] ~.
\label{eq:ABJM_fishnet_interationterm}
\end{equation}
The interaction is chiral, since its Hermitian conjugate counterpart is not included in the theory.
In the planar limit, also this theory's vacuum diagrams are dominated by toroidal topology and one finds them to be of triangular form
\begin{equation}
\adjincludegraphics[valign=c,scale=0.9]{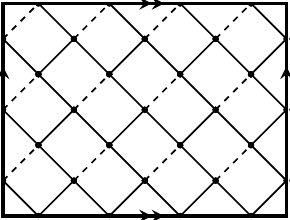} 
=
\adjincludegraphics[valign=c,scale=0.9]{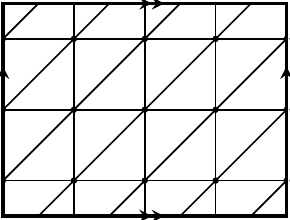} ~.
\label{eq:ABJM_fishnet_vacuumgraphs}
\end{equation}
As the l.\,h.\,s.\ indicates, the vacuum graphs of this theory are special cases of the generalized partition function \eqref{eq:ZMN_Inhomogeneous_vacuumgraph}.
To complete the matching, we have to specify the parameter matrix \eqref{eq:Parametermatrix_U} to $U_J = U^\mathrm{Tri}$.
The theory \eqref{eq:ABJM_fishnet_interationterm} only contains scalars, which means that the spin labels $\ell_J$ and $m_J$ are zero and we choose $u_J^- = v_J^+ = v_J^- = \frac{1}{2}$.
Furthermore, a scalar in $D=3$ dimensions comes with a factor $c_0(1) = \sqrt{\pi /2}$, according to \eqref{eq:ScalarPropagator_WvsG}.
The remaining generalized propagator, corresponding to $u_J^+$, is set to produce a delta function.
In three dimensions, this means $u_J^+ = \frac{3}{2} - \varepsilon$ and we have to multiply a factor $\pi^{- \frac{3}{2}} a_0 (\varepsilon)$ before taking the limit $\varepsilon \rightarrow 0$.
In summary, this justifies the choice
\begin{equation}
U^\mathrm{Tri}
=
\begin{pmatrix}
( \frac{3}{2} - \varepsilon , 0 ) & ( \frac{1}{2} , 0 ) \\[1pt]
( \frac{1}{2} , 0 ) & ( \frac{1}{2} , 0 )
\end{pmatrix} 
~~~~
\mathrm{with}
~~~~
\lim_{\varepsilon \rightarrow 0}
\pi^{-\frac{3}{2}} a_0 (\varepsilon)
\cdot
\adjincludegraphics[valign=c,scale=1]{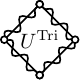} 
=
\adjincludegraphics[valign=c,scale=1]{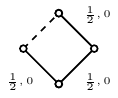}
~.
\end{equation}
Based on the graph-builder, we can construct the special cases of the transfer matrix \eqref{eq:TransferMatix_periodic_Feynman_2}
\begin{equation}
\begin{split}
\hat{\mathbb{T}}_{N}^\mathrm{Tri}
: & =
\lim_{\varepsilon \rightarrow 0}
\left[
c_0(1)^3
\pi^{- \frac{3}{2}} a_0 (\varepsilon)
\right]^N
\mathbb{T}_{N} 
\left(
\left\lbrace
U^\mathrm{Tri}
\right\rbrace
\right) \\
&=
c_0(1)^{3N} \cdot
\adjincludegraphics[valign=c,scale=0.5]{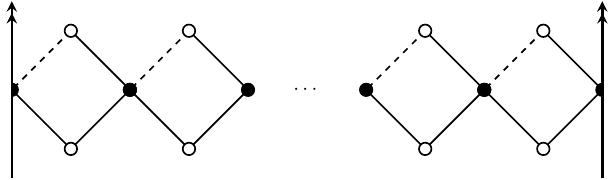} 
=
c_0(1)^{3N} \cdot
\adjincludegraphics[valign=c,scale=0.5]{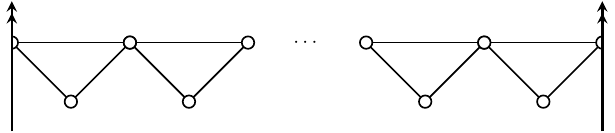} ~.
\end{split}
\label{eq:TransferMatrix_Tri}
\end{equation}
Stacking them on top of each other and taking the trace yields the vacuum graphs of the theory \eqref{eq:ABJM_fishnet_interationterm}.
For example, the graph shown in \eqref{eq:ABJM_fishnet_vacuumgraphs} is obtained from $\hat{Z}_{34}^\mathrm{Tri} :=\mathrm{Tr} [ ( \hat{\mathbb{T}}_{4}^\mathrm{Tri} )^3 ]$.

Another example of a theory admitting integrable vacuum diagrams given by Zamolodchikov \cite{Zamolodchikov:1980mb} is proposed and studied in \cite{Mamroud:2017uyz,Kade:2023xet} and is a theory of three complex scalars in six dimensions, $D=6$.
They are understood as $\mathrm{N} \times \mathrm{N}$ matrices in the adjoint representation of some gauge group of an unknown mother theory.
They are coupled by
\begin{equation}
\mathcal{L}^\mathrm{Hex} 
=
\mathrm{N}
\cdot
\sum_{i=1}^3
\mathrm{tr}
\left[
\partial^\mu \phi_i^\dagger
\partial_\mu \phi_i
\right]
+
\mathrm{N}
\cdot
\rho
\cdot
\mathrm{tr}
\left[
\phi_1 \phi_2 \phi_3
+
\phi_1^\dagger \phi_2^\dagger \phi_3^\dagger
\right] ~,
\label{eq:SexticScalarTheory_interactionterm}
\end{equation}
and the interaction terms are chiral again, since Hermitian-conjugation would shuffle the indices of the scalars to terms not contained in \eqref{eq:SexticScalarTheory_interactionterm}.
We can convince ourselves that the vacuum diagrams produced by the interactions \eqref{eq:SexticScalarTheory_interactionterm} are of brick wall type. 
Qualitatively, the graphs resemble a honeycomb with hexagonal faces.
These vacuum diagrams are illustrated by the example
\begin{equation}
\hat{Z}_{34}^\mathrm{Hex} 
:=
\mathrm{Tr} [ ( \hat{\mathbb{T}}_{4}^\mathrm{Hex} )^3 ]
=
(-2)^{3\cdot 3 \cdot 4} \cdot
c_0 (1)^{3\cdot 3 \cdot 4} \cdot
\adjincludegraphics[valign=c,scale=0.9]{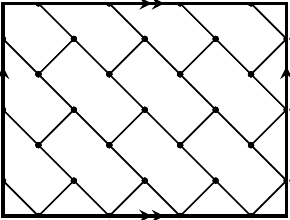}  ~.
\label{eq:SexticScalarTheory_vacuumgraphs}
\end{equation}
In order to specify the general vacuum graphs \eqref{eq:ZMN_Inhomogeneous_vacuumgraph} and transfer matrix \eqref{eq:TransferMatix_periodic_Feynman_2}, we must realize that the brick wall topology is due to the switching off of one side of the rhombuses in the transfer matrix.
Hence, we choose the north-west propagator and therefore set $u_J^+ = 0$.
The other propagators should become scalars in six dimensions, which implies $u_J^- = v_J^+ = v_J^- = 2$.
Note that the vertices are marginal, since $u_J^+ + u_J^- + v_J^+ + v_J^- = 6$
Naturally, since the theory only contains scalars, the spin labels are zero.
In six dimensions, we find that the factor $c_0 (1) = 2$ by \eqref{eq:ScalarPropagator_WvsG} and the factor $-2$ for each bosonic propagator, coming from the normalization of the kinetic term of \eqref{eq:SexticScalarTheory_interactionterm}, should be respected.
After all, we find the graph-builder 
\begin{equation}
U^\mathrm{Hex}
=
\begin{pmatrix}
( 0 , 0 ) & ( 2 , 0 ) \\
( 2 , 0 ) & ( 2 , 0 )
\end{pmatrix} 
~~~~
\mathrm{with}
~~~~
\adjincludegraphics[valign=c,scale=1]{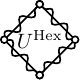} 
=
\adjincludegraphics[valign=c,scale=1]{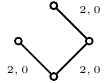}
~.
\end{equation}
This implies a row matrix
\begin{equation}
\hat{\mathbb{T}}_{N}^\mathrm{Hex}
:=
\left(-2 c_0 (1)\right)^{3N}
\mathbb{T}_{N} 
\left(
\left\lbrace
U^\mathrm{Hex}
\right\rbrace
\right)
=
\left(-2 c_0 (1)\right)^{3N}
\cdot
\adjincludegraphics[valign=c,scale=0.5]{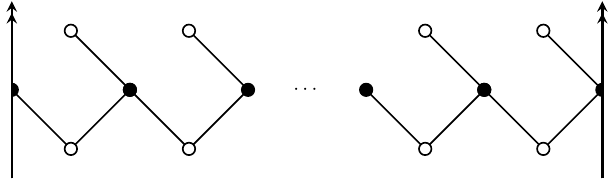} ~,
\label{eq:TransferMatrix_Six}
\end{equation}
which can build up the vacuum graphs \eqref{eq:SexticScalarTheory_vacuumgraphs} via the relation $\hat{Z}_{MN}^\mathrm{Hex} :=\mathrm{Tr} [ ( \hat{\mathbb{T}}_{N}^\mathrm{Hex} )^M ]$.

Next to the scalar vacuum graphs, we can also consider graphs, which contain fermions.
One of these theories is the so-called brick wall theory, originally proposed in \cite{Pittelli:2019ceq} and studied in \cite{Kade:2023xet}.
Conceptually, it can be motivated by the $\chi_0$-theory \eqref{eq:Chi0_SYM_action} simply by considering its Yukawa interaction vertices and eliminating the fields $\phi_2$ and $\phi_3$ from the theory.
However, this operation leaves the realm of integrable deformations of $\mathcal{N} = 4$ SYM, yet the brick wall theory was obtained as a double-scaling limit of another theory, namely a $\mathcal{N}=2$ SCFT \cite{Pittelli:2019ceq}.
We present the Lagrangian of the brick wall theory,
\begin{equation}
\begin{split}
&\mathcal{L}^{\mathrm{BW}} 
~=~
\mathrm{N}  \cdot \mathrm{tr}\left[ 
- \frac{1}{2} \partial^\mu \phi^\dagger \partial_\mu \phi 
+ \I \sum_{k=1}^2 \bar{\psi}_k \slashed{\partial} \psi_{k}
+ 
\rho 
%\I \sqrt{\xi_2 \xi_3}  
\cdot 
\left(
\psi_1\phi \psi_2 + \bar{\psi}_1\phi^\dagger \bar{\psi}_2
%\psi_3\phi_1 \psi_2 + \bar{\psi}_3\phi_1^\dagger \bar{\psi}_2
\right)
\right] ~.
\end{split}
\label{eq:BrickWall_Lagrangian}
\end{equation}
Its vertices are still chiral, since the Hermitian conjugates of the Yukawa interaction terms, for which the fields are ordered in reverse, are missing.  
This chirality, in combination with the fact that the two vertices have to alternate when following a fermion line, results in the typical ``brick wall'' structure of the generated Feynman diagrams.
As an example we present a vacuum graph of the theory,
\begin{equation}
\hat{Z}_{34}^\mathrm{BW}
=
(-1)^{2\cdot 3 \cdot 4}
(2)^{3 \cdot 4}
c_\frac{1}{2} (\tfrac{3}{2})^{2\cdot 3 \cdot 4}
\cdot
\adjincludegraphics[valign=c,scale=0.9]{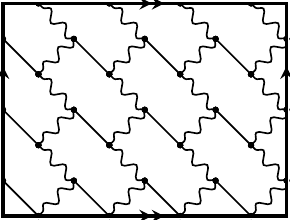} ~.
\label{eq:BrickWall_vacuumgraphs}
\end{equation}
We see that it resembles the vacuum graph \eqref{eq:SexticScalarTheory_vacuumgraphs} of the hexagonal theory. 
The only difference is that the brick wall theory lives in a four-dimensional spacetime, which allows for the marginal, cubic Yukawa vertices.
In addition to the $c_\ell$ factors, we equipped the normalized brick wall graph in \eqref{eq:BrickWall_vacuumgraphs} with the factor $2$ for each bosonic propagator and the factor $-1$ for each fermionic propagator, due to the noncanonical kinetic terms in \eqref{eq:BrickWall_Lagrangian} compared to \eqref{eq:ScalarUndefKinTerm} and \eqref{eq:FermionicUndefKinTerm}, respectively.
The parameter matrix and the graph-builder of the brick wall theory are 
\begin{equation}
U^\mathrm{BW}
=
\begin{pmatrix}
( 0 , 0 ) & ( \frac{3}{2} , \bar{\frac{1}{2}} ) \\[1pt]
( 1 , 0 ) & ( \frac{3}{2} , \frac{1}{2} )
\end{pmatrix} 
~~~~
\mathrm{with}
~~~~
\adjincludegraphics[valign=c,scale=1]{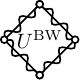} 
=
\adjincludegraphics[valign=c,scale=1]{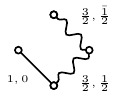} 
~.
\label{eq:ParameterMatrix_BW_graphbuilder}
\end{equation}
Since $c_\frac{1}{2} (\frac{3}{2}) = -2\I$ by \eqref{eq:FermionicPropagator_WvsG}, we find that the row matrix, which builds up the graph \eqref{eq:BrickWall_vacuumgraphs}, is obtained from the generalized transfer matrix \eqref{eq:TransferMatix_periodic_Feynman_2} as
\begin{equation}
\hat{\mathbb{T}}_{N}^\mathrm{BW}
:=
\left( 
2
c_\frac{1}{2} (\tfrac{3}{2})^2
\right)^N
\mathbb{T}_{N} 
\left(
\left\lbrace
U^\mathrm{BW}
\right\rbrace
\right)
=
2^N
c_\frac{1}{2} (\tfrac{3}{2})^{2N}
\cdot
\adjincludegraphics[valign=c,scale=0.5]{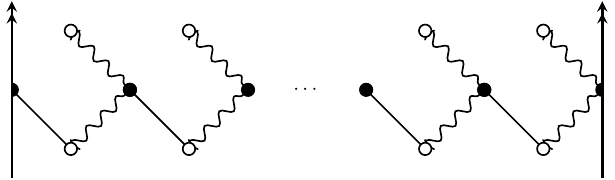} 
~,
\label{eq:TransferMatrix_BW}
\end{equation}

The last non-supersymmetric theory whose vacuum diagrams we want to study is a fermionic version of the fishnet theory \eqref{eq:Fishnet_biscalar_action}, which lives in three spacetime dimensions.
Therefore, marginal quartic couplings in the fields between two bosons and two fermions are possible.
The $D=3$ Lagrangian reads \cite{Kade:2023xet}
\begin{equation}
\begin{split}
&\mathcal{L}^{\mathrm{FFN}}
~=~
-\mathrm{N}  \cdot \mathrm{tr}\left[ 
\sum_{j=1}^2 \partial^\mu Y_j^\dagger \partial_\mu Y_j^{\phantom{\dagger}}
+ \I \sum_{k=1}^2 \Psi^\dagger_k \slashed{\partial} \Psi_{k}^{\phantom{\dagger}}
+
\rho 
%\I \sqrt{\xi_2 \xi_3}  
\cdot 
\left(
Y_1^{\phantom{\dagger}} Y_2^\dagger \Psi_1^{\phantom{\dagger}} \Psi_2^\dagger
-Y_1^\dagger Y_2^{\phantom{\dagger}} \Psi_1^\dagger \Psi_2^{\phantom{\dagger}}
\right)
\right] ~ .
\end{split}
\label{eq:FermionicFN_Lagrangian}
\end{equation}
The theory is related to ABJM in the same way as the brick wall theory \eqref{eq:BrickWall_Lagrangian} is related to $\mathcal{N} = 4$ SYM: 
after performing a double-scaling limit of the $\gamma$-deformation, set one fixed effective coupling $\xi_i$ to zero (this gives the $\chi_0$-theory \eqref{eq:Chi0_SYM_action}) and we drop purely scalar interaction terms and only keep the mixed fermion-boson ones.
The last step leaves the realm of integrable deformations and is rather an ad hoc construction.
The chiral interactions in \eqref{eq:FermionicFN_Lagrangian} are three-dimensional versions of Yukawa interactions and they only allow very regular square-lattice shaped vacuum diagrams 
\begin{equation}
\hat{Z}_{34}^\mathrm{FFN}
=
c_\frac{1}{2} (1)^{2\cdot 3 \cdot 4}
c_0 (\tfrac{1}{2})^{2\cdot 3 \cdot 4}
\cdot
\adjincludegraphics[valign=c,scale=0.9]{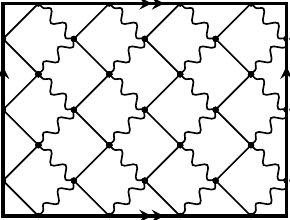} ~.
\label{eq:BrickWall_vacuumgraphs}
\end{equation}
Also these vacuum diagrams fit into the general class of graphs generated by the transfer matrix \eqref{eq:TransferMatix_periodic_Feynman_2}.
The parameter matrix and the graph-builder, which are required, are
\begin{equation}
U^\mathrm{FFN}
=
\begin{pmatrix}
( \frac{1}{2} , 0 ) & ( 1 , \bar{\frac{1}{2}} ) \\[1pt]
( \frac{1}{2} , 0 ) & ( 1 , \frac{1}{2} )
\end{pmatrix} 
~~~~
\mathrm{with}
~~~~
\adjincludegraphics[valign=c,scale=1]{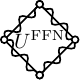} 
=
\adjincludegraphics[valign=c,scale=1]{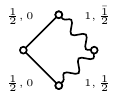} 
~.
\end{equation}
By taking the additional factors $c_{\frac{1}{2}} (1) = - \I \sqrt{\pi /2}$ and $c_{0} (\frac{1}{2}) = \sqrt{\pi /2}$ from \eqref{eq:ScalarPropagator_WvsG} and \eqref{eq:FermionicPropagator_WvsG}, we find the row matrix 
\begin{equation}
\hat{\mathbb{T}}_{N}^\mathrm{FFN}
:=
\left(
c_0 (1)^2 
c_\frac{1}{2} (\tfrac{1}{2})^2
\right)^N
\mathbb{T}_{N} 
\left(
\left\lbrace
U^\mathrm{FFN}
\right\rbrace
\right)
=
\left(
c_0 (1)
c_\frac{1}{2} (\tfrac{1}{2})
\right)^{2N}
\cdot
\adjincludegraphics[valign=c,scale=0.5]{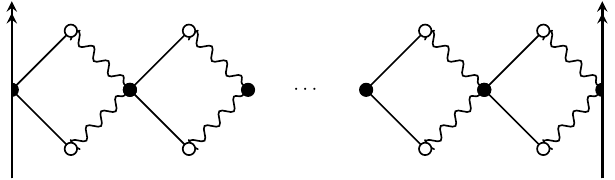} 
~,
\label{eq:TransferMatrix_FFN}
\end{equation}

\subsection{Thermodynamic limit of vacuum diagrams and critical coupling}
\label{subsec:ThermodynamicLimitOfVacuumDiagramsAndCriticalCoupling}
The partition function \eqref{eq:ZMN_Inhomogeneous_vacuumgraph} has an interesting thermodynamic limit, where $M,N \rightarrow \infty$.
In statistical physics, this limit is of importance, since it allows us to compute macroscopic quantities of the model, such as the entropy of ice in the case of the six-vertex model, see section \ref{subsec:FreeEnergyInThermoDynLimit_8V}.

In the context of vacuum graphs of QFTs, the thermodynamic limit relates to the critical coupling of the theory in question.
The term ``critical coupling'' has multiple meanings in QFT, but here it describes a particular value of the coupling, which is the radius of convergence of the free energy of the QFT.
We can see this by the perturbation expansion of the free energy in vacuum diagrams,
\begin{equation}
\hat{Z}
~=~
\sum_{M,N=1}^\infty
\hat{Z}^{\mathrm{QFT}}_{MN}
\xi^{b\cdot MN} ~ .
\label{eq:VacuumFreeEnergyQFT}
\end{equation}
The general QFT we study is dependent on a single coupling $\xi$ and its vacuum diagrams are denoted by $\hat{Z}^{\mathrm{QFT}}_{MN}$, which are \eqref{eq:ZMN_Inhomogeneous_vacuumgraph} evaluated at particular $U_J$'s and decorated with factors from \eqref{eq:ScalarPropagator_WvsG} or \eqref{eq:FermionicPropagator_WvsG}.
The integers $M$ and $N$ count the number of vertical and horizontal pairs of medial lines, which are directly related to the number of propagator cycles that wrap the toroidal graphs by the map \eqref{eq:Map_FeynmanMedial}.
The number $b= \frac{\mathrm{\# ~ vertices}}{MN}$ counts the ratio between vertices and pairwise intersections of medial lines in the theory's vacuum graphs.

Viewing \eqref{eq:VacuumFreeEnergyQFT} as a double series, the radius of divergence is related to the limit
\begin{equation}
\hat{K}^{\mathrm{QFT}}
~:=~
\lim_{M,N\rightarrow \infty} 
\left|
\hat{Z}^{\mathrm{QFT}}_{MN}
\right|^{\frac{1}{MN}} ~ .
\label{eq:K_CriticalCouplingQFT}
\end{equation}
The critical coupling, which is the value where \eqref{eq:VacuumFreeEnergyQFT} diverges, is then 
\begin{equation}
\xi_\mathrm{cr}
~ = ~
\frac{1}{\left(\hat{K}^{\mathrm{QFT}}\right)^b} ~.
\end{equation}
This is the quantity we want to determine exactly for the instances of integrable, non-supersymmetric QFTs presented in the previous section.
The scalar theories are the $D=4$ bi-scalar fishnet theory \eqref{eq:Fishnet_biscalar_action}, the $D=3$ triangular fishnet theory from ABJM \eqref{eq:ABJM_fishnet_interationterm}, and the $D=6$ hexagonal scalar theory \eqref{eq:SexticScalarTheory_interactionterm}.
The thermodynamic limit of their graphs was studied in the seminal paper \cite{Zamolodchikov:1980mb}, and we give a detailed review on its calculation in the following using the method of inversion relations.
Fermionic theories are the $D=4$ brick wall model \eqref{eq:BrickWall_Lagrangian} with hexagonal graphs and the $D=3$ fermionic fishnet theory \eqref{eq:FermionicFN_Lagrangian} with vacuum graphs made from square faces.
For them, the critical coupling is obtained in \cite{Kade:2023xet}, also by the method of inversion relations, which we present next.

\subsubsection{Method of inversion relations}
The method of inversion relation is rooted in the generalization of the partition function/vacuum diagram to \eqref{eq:ZMN_Inhomogeneous_vacuumgraph}, by introducing horizontally inhomogeneous spectral parameters.
They are the propagators' exponents and spin-labels and we collectively denote them by the parameter matrix \eqref{eq:Parametermatrix_U}.
The dependence on the spectral parameters carries over to the thermodynamic limit,
\begin{equation}
K 
(\lbrace U_J \rbrace)
~:=~
\lim_{M,N\rightarrow \infty} 
\left|
Z_{MN}
\left(
\left\lbrace
U_J
\right\rbrace
\right)
\right|^{\frac{1}{MN}} ~ .
\label{eq:K_CriticalCouplingQFT}
\end{equation}
Next, we use the fact that the generalized vacuum graphs $Z_{MN}\left( \left\lbrace U_J \right\rbrace \right)$ are expressed as a trace of a stack of $M$ generalized transfer matrices \eqref{eq:TransferMatix_periodic_Feynman_2}.
This implies that the trace can be rewritten as a sum of all the eigenvalues of the transfer matrix, which we denote by $\Lambda_{N,i} \left( \left\lbrace U_J \right\rbrace \right)$.
Thus, $i$ labels the eigenvalues. 
We find that the maximal eigenvalue is dominating the thermodynamic limit \eqref{eq:K_CriticalCouplingQFT},
\begin{equation}
\begin{split}
K 
(\lbrace U_J \rbrace)
& =
\lim_{M,N\rightarrow \infty} 
\left|
\Lambda_{N,\mathrm{max}}
\left(
\left\lbrace
U_J
\right\rbrace
\right)^M
\left[
1
+
\sum_{i \neq \mathrm{max}}
\left(
\frac{
\Lambda_{N,i}
\left(
\left\lbrace
U_J
\right\rbrace
\right)
}{
\Lambda_{N,\mathrm{max}}
\left(
\left\lbrace
U_J
\right\rbrace
\right)
}
\right)^M
\right]
\right|^{\frac{1}{MN}} \\
& = 
\lim_{M,N\rightarrow \infty} 
\left|
\Lambda_{N,\mathrm{max}}
\left(
\left\lbrace
U_J
\right\rbrace
\right)
\right|^{\frac{1}{N}}
~ ,
\end{split}
\label{eq:K_CriticalCouplingQFT_ev}
\end{equation}
because the second term in the first line is negligible in the $M \rightarrow \infty$.

Thus, in order to determine $K (\lbrace U_J \rbrace)$, we derive functional equations for the eigenvalues of the generalized transfer matrix \eqref{eq:TransferMatix_periodic_Feynman_2}.
We are able to do so by constructing different representations of the inverse of the generalized transfer matrix.
Therefore, the x-unity relation \eqref{eq:XUnity_Fermion} is crucial.
We use it to annihilate the product of two generalized transfer matrices at different values of the spectral parameters.
In detail the annihilation works as 
\begin{subequations}
\begin{align}
\mathbb{T}&_{N} 
\left(
\lbrace U_J \rbrace
\right)
\circ
\mathbb{T}_{N}
\left(
\lbrace U_J' \rbrace
\right)
=
\adjincludegraphics[valign=c,scale=0.5]{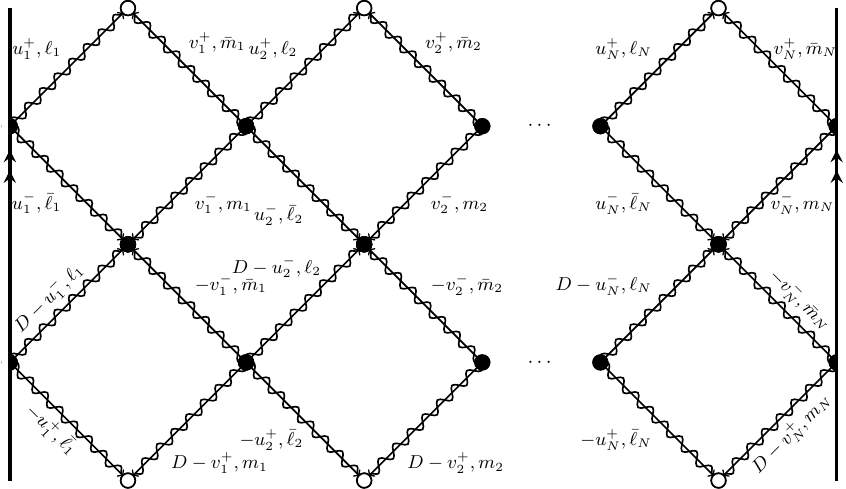} 
\label{eq:TransferMatrixAnnihilation1}\\
%%%%%%%%%%
~&\stackrel{\eqref{eq:XUnity_Fermion}}{=}~
\adjincludegraphics[valign=c,scale=0.5]{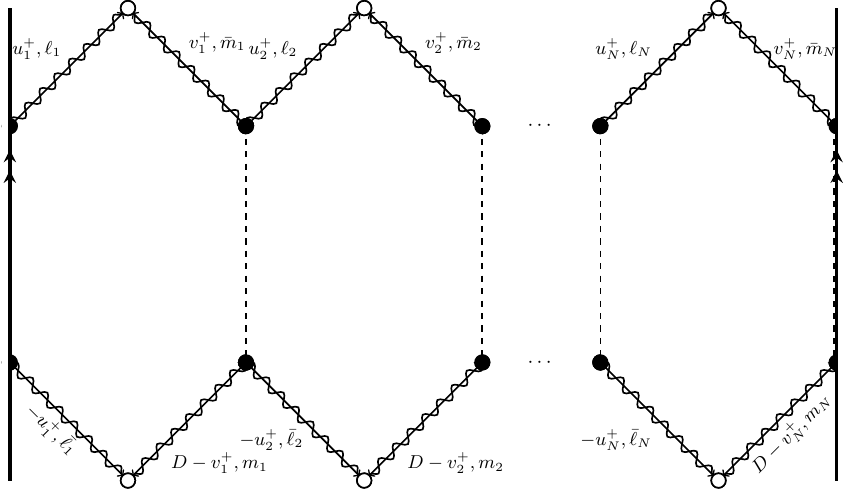} 
\cdot 
\prod_{J=1}^N
A_{\ell_J} ( u^-_J )
\cdot \mathbb{I}^{(\ell_J)} \mathbb{I}^{(m_J)}
\\
%%%%%%%%%%
~&\stackrel{\phantom{\eqref{eq:XUnity_Fermion}}}{=}~
\adjincludegraphics[valign=c,scale=0.5]{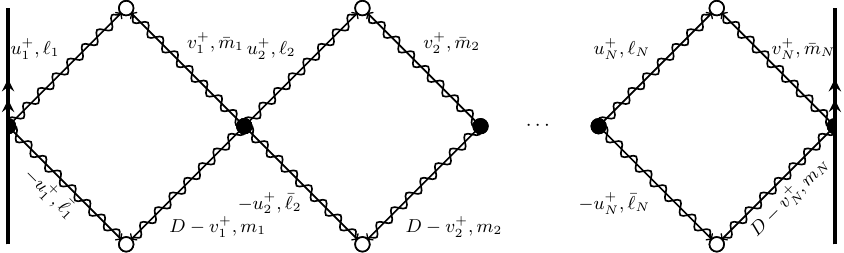}	
\cdot 
\prod_{J=1}^N
A_{\ell_J} ( u^-_J )
\cdot \mathbb{I}^{(\ell_J)} \mathbb{I}^{(m_J)}
\\
%%%%%%%%%%
~&\stackrel{\eqref{eq:XUnity_Fermion}}{=}~
\adjincludegraphics[valign=c,scale=0.5]{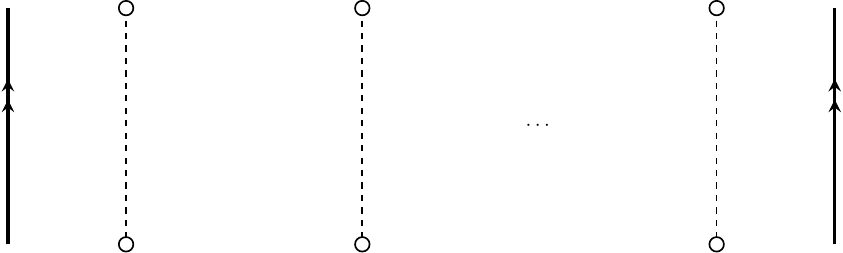}
\cdot
\prod_{J=1}^N
A_{\ell_J} ( u^-_J ) \,
A_{m_J} ( v^+_J )
\cdot \mathbb{I}^{(\ell_J)} \mathbb{I}^{(m_J)}
\\
%%%%%%%%%%
~&\stackrel{\phantom{\eqref{eq:XUnity_Fermion}}}{=}
\mathbb{1}_N \cdot
\prod_{J=1}^N
A_{\ell_J} ( u^-_J )
A_{m_J} ( v^+_J )
\cdot \mathbb{I}^{(\ell_J)} \mathbb{I}^{(m_J)}
\label{eq:TransferMatrixAnnihilationEnd}\; 
\end{align}\label{eq:TransferMatrixAnnihilation}%
\end{subequations}
and we use the shorthand $\mathbb{1}_N$ in \eqref{eq:TransferMatrixAnnihilationEnd} for the $N$ delta functions and 
\begin{equation}
A_{\ell} ( u )
:=
\pi^{D} \;
a_{\ell} ( u )\, a_{\ell} ( D - u )
\label{eq:Factor_AEll}
\end{equation}
for the collective dependence on one single spectral parameter of the final expression \eqref{eq:TransferMatrixAnnihilationEnd} .
The argument of the two transfer matrices are the spectral parameter matrices 
\begin{equation}
\begin{array}{ccc}
U_J
=
\begin{pmatrix} 
( u^+_J , \ell_J ) & ( v^+_J , \bar{m}_J ) \\
( u^-_J , \bar{\ell}_J ) & ( v^-_J , m_J ) 
\end{pmatrix} &
\mathrm{and} &
U_{J, \mathrm{inv} }
=
\begin{pmatrix} 
( D - u^-_J , \ell_J ) & ( - v^-_J , \bar{m}_J ) \\
( -u^+_J , \bar{\ell}_J ) & ( D - v^+_J , m_J )
\end{pmatrix} ~.
\end{array}
\end{equation}
We observe from \eqref{eq:TransferMatrixAnnihilation} that multiplying the transfer matrix with spectral parameters $\lbrace U_J \rbrace$ by another one with parameters $\lbrace U_{J, \mathrm{inv} } \rbrace$, leaves us with a unit kernel of $N$ sites, $\mathbb{1}_N$ times a factor.
The factor is a combinations of gamma functions, via the definitions \eqref{eq:Factor_AEll} and $a_\ell$ in \eqref{eq:Factor_aEll}, depending on the arguments $\mathbf{u}^-_J$ and $\mathbf{v}^+_J$.
At first, it seems mysterious that the factor depends only on these specific two spectral parameters, but by choosing different parameter matrices $\lbrace U_{J, \mathrm{inv} } \rbrace$, we can also access the other parameters $\mathbf{u}^+_J$ and $\mathbf{v}^-_J$.
Hence, we also find three other representations of the inverse of $\mathbb{T}_N ( \lbrace U_J \rbrace )$, which are
\begin{subequations}
\begin{align}
& \mathbb{T}_{N}
\left(
\left\lbrace
\left(
\begin{smallmatrix}
( - u^-_J , \ell_J ) & ( D - v^-_J , \bar{m}_J ) \\
( D - u^+_J , \bar{\ell}_J ) & ( - v^+_J , m_J )
\end{smallmatrix}
\right)
\right\rbrace
\right)
\cdot
\frac{1}{
\prod_{J=1}^N
A_{\ell_J} ( u^+_J )
A_{m_J} ( v^-_J )
} ~, \\
& \mathbb{T}_{N}
\left(
\left\lbrace
\left(
\begin{smallmatrix}
( D - u^-_J , \ell_J ) & ( - v^-_J , \bar{m}_J ) \\
( D - u^+_J , \bar{\ell}_J ) & ( - v^+_J , m_J )
\end{smallmatrix}
\right)
\right\rbrace
\right)
\cdot
\frac{1}{
\prod_{J=1}^N
A_{\ell_J} ( u^-_J )
A_{\ell_J} ( u^+_J )
} ~, \\
& \mathbb{T}_{N}
\left(
\left\lbrace
\left(
\begin{smallmatrix}
( - u^-_J , \ell_J ) & ( D - v^-_J , \bar{m}_J ) \\
( - u^+_J , \bar{\ell}_J ) & ( D - v^+_J , m_J )
\end{smallmatrix}
\right)
\right\rbrace
\right)
\cdot
\frac{1}{
\prod_{J=1}^N
A_{m_J} ( v^-_J )
A_{m_J} ( v^+_J )
} ~.
\end{align}
\end{subequations}
The property, that shifts in the spectral parameters yield an inverse up to a factor, implies functional relations for the eigenvalues, which carry over to the thermodynamic limit \eqref{eq:K_CriticalCouplingQFT_ev}.
We find four functional relations for the thermodynamic limit of the form 
\begin{equation}
K
\left(
\left\lbrace
U_J
\right\rbrace
\right)
\cdot
K
\left(
\left\lbrace
U_{J, \mathrm{inv}}
\right\rbrace
\right)
=
F 
\left(
\left\lbrace
U_J
\right\rbrace
\right)
\end{equation}
with four different options for $U_{J, \mathrm{inv}}$ and $F (\lbrace U_J \rbrace)$,
\begin{subequations}
\begin{align}
& U_{J, \mathrm{inv}}
=
\left(
\begin{smallmatrix}
( D - u^-_J , \ell_J ) & ( - v^-_J , \bar{m}_J ) \\
( -u^+_J , \bar{\ell}_J ) & ( D - v^+_J , m_J )
\end{smallmatrix}
\right)
~~~ \mathrm{with} ~~~
F(\lbrace U_J \rbrace)
&& =
\lim_{N \rightarrow \infty}
\left|
\prod_{J=1}^N
A_{\ell_J} ( u^-_J )
A_{m_J} ( v^+_J )
\right|^{\frac{1}{N}} ~, 
\\
& U_{J, \mathrm{inv}}
=
\left(
\begin{smallmatrix}
( - u^-_J , \ell_J ) & ( D - v^-_J , \bar{m}_J ) \\
( D - u^+_J , \bar{\ell}_J ) & ( - v^+_J , m_J )
\end{smallmatrix}
\right)
~~~ \mathrm{with} ~~~
F(\lbrace U_J \rbrace)
&& =
\lim_{N \rightarrow \infty}
\left|
\prod_{J=1}^N
A_{\ell_J} ( u^+_J )
A_{m_J} ( v^-_J )
\right|^{\frac{1}{N}} ~, 
\\
& U_{J, \mathrm{inv}}
=
\left(
\begin{smallmatrix}
( D - u^-_J , \ell_J ) & ( - v^-_J , \bar{m}_J ) \\
( D - u^+_J , \bar{\ell}_J ) & ( - v^+_J , m_J )
\end{smallmatrix}
\right)
~~~ \mathrm{with} ~~~
F(\lbrace U_J \rbrace)
&& =
\lim_{N \rightarrow \infty}
\left|
\prod_{J=1}^N
A_{\ell_J} ( u^-_J )
A_{\ell_J} ( u^+_J ) 
\right|^{\frac{1}{N}} ~, 
\\
& U_{J, \mathrm{inv}}
=
\left(
\begin{smallmatrix}
( - u^-_J , \ell_J ) & ( D - v^-_J , \bar{m}_J ) \\
( - u^+_J , \bar{\ell}_J ) & ( D - v^+_J , m_J )
\end{smallmatrix}
\right)
~~~ \mathrm{with} ~~~
F(\lbrace U_J \rbrace)
&& =
\lim_{N \rightarrow \infty}
\left|
\prod_{J=1}^N
A_{m_J} ( v^-_J )
A_{m_J} ( v^+_J ) 
\right|^{\frac{1}{N}} ~.
\end{align}\label{eq:K_ThermodynamicLimit_FunctionalRelation}%
\end{subequations}
In order to find a minimal solution, we make an ansatz for the thermodynamic limit $K$, which reflects the assumption that the dependence of its four arguments factorizes,
\begin{equation}
K
\left(
\left\lbrace
\left(
\begin{smallmatrix}
( u^-_J , \ell_J ) & ( v^-_J , \bar{m}_J ) \\
( u^+_J , \bar{\ell}_J ) & ( v^+_J , m_J )
\end{smallmatrix}
\right)
\right\rbrace
\right)
=
\lim_{N \rightarrow \infty}
\left|
\prod_{J=1}^N
\kappa_{\ell_J} (u^+_J)
\kappa_{\ell_J} (u^-_J)
\kappa_{m_J} (v^+_J)
\kappa_{m_J} (v^-_J)
\right|^{\frac{1}{N}} ~.
\label{eq:K_ThermodynamicLimit_Ansatz}
\end{equation}
Making this ansatz associates to each generalized propagator in the transfer matrix a function $\kappa_{\ell} (u)$, which is its contribution in the thermodynamic limit.
To have a non-vanishing impact, the particular value $\mathbf{u}$ as to appear in a number of propagators, which scales like $N$. 
Plugging the ansatz \eqref{eq:K_ThermodynamicLimit_Ansatz} back into \eqref{eq:K_ThermodynamicLimit_FunctionalRelation} and find that the function $\kappa_\ell ( u )$ has to satisfy the compact functional relations 
\begin{subequations}
\begin{align}
\kappa_\ell (u) \kappa_\ell (-u) ~&=~ 1  ~, \label{eq:InversionRelationsKappa1}\\
\kappa_\ell ( u ) \kappa_\ell (D - u) ~&=~ A_{\ell} ( u ) ~=~ \pi^D a_\ell ( u ) a_\ell (D - u)  ~. \label{eq:InversionRelationsKappa2}
\end{align}\label{eq:InversionRelationsKappa}%
\end{subequations}
The functional relation are called inversion relations and in this form \cite{Kade:2023xet}, they are a spinning generalization of the scalar case, which is already presented in Zamolodchikov's work \cite{Zamolodchikov:1980mb}.
The inversion relations \eqref{eq:InversionRelationsKappa} alone do not determine the function $\kappa_\ell (u)$ uniquely, rather they have many solutions, which correspond to different eigenvalues $\Lambda_{N,i}\left(\left\lbrace U_J\right\rbrace\right)$ of the transfer matrix.
Thanks to the commutativity of transfer matrices \eqref{eq:CommutativityTransferMatrix_nonsusy}, the eigenvalues are entire functions of the spectral parameters \cite{Baxter:1982xp}.
To find the solution corresponding to the maximal eigenvalue, as required by \eqref{eq:K_CriticalCouplingQFT_ev}, we follow Stroganov's hypothesis \cite{StroganovInvRelLatticeModels}, which states that the searched solution is analytic in the so-called physical strip $\left[ 0 , \frac{D}{2} \right)$ as a function of $u$.
Along the lines of \cite{Bombardelli:2016scq,Shankar:1977cm,Zamolodchikov:1976uc}, and presented in detail in appendix \ref{sec:SolvingInversionRelations}, we can construct the solution 
\begin{equation}
\kappa_\ell (u) 
~=~
\pi^u
\frac{\Gamma (\frac{D}{2} - u + \ell)}{\Gamma (\frac{D}{2} + \ell)}
\prod_{k=1}^\infty
\frac{\Gamma (D k + \frac{D}{2} - u + \ell) \, \Gamma (D k + u + \ell) \, \Gamma (D k - \frac{D}{2} + \ell)}
{\Gamma (D k - \frac{D}{2} + u + \ell) \, \Gamma (D k - u + \ell) \, \Gamma (D k + \frac{D}{2} + \ell)} ~.
\label{eq:Kappagenfinalresult}
\end{equation}
In the following, we evaluate the solution \eqref{eq:Kappagenfinalresult} for different values of $u$, $\ell$ and $D$.
For the values under consideration, the infinite product truncates at some $k$ and higher factors cancel.
Technically, the evaluation can be carried out by taking the logarithm of \eqref{eq:Kappagenfinalresult} and using the integral representation of the log-gamma function, but in most cases the evaluation with a computer, e.\,g.\ \cite{Mathematica}, is more practical.
Two important special values for generic $D$ are $\kappa_\ell (0) = 1$ and $\lim_{\varepsilon \rightarrow 0} \pi^{-\frac{D}{2}} a_0 (\varepsilon) \kappa_0 (\frac{D}{2} - \varepsilon) = 1$.
They correspond to a generalized propagator with exponent zero, which is one, and the delta function limit of the scalar generalized propagator \eqref{eq:Delta_bosonic}, respectively.
We notice that a vanishing propagator and a delta function do not contribute to the thermodynamic limit of a vacuum graph.
Furthermore, from the delta function limit, we can infer that $\kappa_0 (u)$ has a pole at $u = \frac{D}{2} = \eta$, which is exactly at the upper end of the physical strip, at the value of the crossing parameter.

\subsubsection{The critical coupling of integrable QFTs}
Having established the thermodynamic limit of the generalized vacuum diagrams in the form of \eqref{eq:K_ThermodynamicLimit_Ansatz} with the solutions \eqref{eq:Kappagenfinalresult}, we are now in the position to evaluate the thermodynamic limit at the special points of the spectral parameters to find the critical coupling of the QFTs presented in section \ref{subsec:IntegrableVacuumDiagramsFromTheGeneralizedPartitionFunction}.
All of these vacuum graphs have a thermodynamic limit, where the spectral parameters are chose homogeneously along the transfer matrix, $\forall J= 1,...,N : ~ U_J = U$.
In this homogeneous case, the thermodynamic limit \eqref{eq:K_ThermodynamicLimit_Ansatz} reduces to 
\begin{equation}
K (U)
=
K
\left(
\begin{smallmatrix}
( u^- , \ell ) & ( v^- , \bar{m} ) \\
( u^+ , \bar{\ell} ) & ( v^+ , m )
\end{smallmatrix}
\right)
=
\left|
\kappa_{\ell} (u^+)
\kappa_{\ell} (u^-)
\kappa_{m} (v^+)
\kappa_{m} (v^-)
\right| ~.
\label{eq:K_ThermodynamicLimit_Ansatz_homogeneous}
\end{equation}
For the normalized thermodynamic limit \eqref{eq:K_CriticalCouplingQFT}, where the factors from \eqref{eq:ScalarPropagator_WvsG} and \eqref{eq:FermionicPropagator_WvsG} are respected, we find the following thermodynamic limits 
\begin{subequations}
\begin{align}
\hat{K}^\mathrm{FN}
& =
\left|
\lim_{\varepsilon \rightarrow 0}
\pi^{-2} a_0 (\varepsilon)
\kappa_0 (2 - \varepsilon)
\kappa_0 (0)
\kappa_0 (1)^2
\right| 
&& \stackrel{D=4}{=}
\frac{\pi^3}{16}
\frac{\Gamma \left(\frac{1}{4}\right)^2}{\Gamma \left(\frac{3}{4}\right)^2} ~, \\
\hat{K}^\mathrm{Tri}
& =
\left|
\sqrt{\tfrac{\pi}{2}}^3
\cdot
\lim_{\varepsilon \rightarrow 0}
\pi^{-\frac{3}{2}} a_0 (\varepsilon)
\kappa_0 (\tfrac{3}{2} - \varepsilon)
\kappa_0 (1)^3
\right| 
&& \stackrel{D=3}{=}
\frac{\pi ^3}{54 \sqrt{2}}
\frac{\Gamma \left(\frac{1}{6}\right)^3}{\Gamma \left(\frac{2}{3}\right)^3} ~, \\
\hat{K}^\mathrm{Hex}
& =
\left|
(-2)^3 
2^3 
\cdot
\kappa_0 (0)
\kappa_0 (2)^3
\right| 
&& \stackrel{D=6}{=}
\frac{8 \pi ^{15/2}}{27}
\frac{\Gamma \left(\frac{1}{3}\right)^3}{\Gamma \left(\frac{5}{6}\right)^3} ~, \\
\hat{K}^\mathrm{BW}
& =
\left|
2\cdot
( -2\I )^2
\cdot
\kappa_0 (0)
\kappa_0 (1)
\kappa_{\frac{1}{2}} (\tfrac{3}{2})^2
\right| 
&& \stackrel{D=4}{=}
\frac{\pi ^{11/2}}{2}
\frac{\Gamma \left(\frac{1}{4}\right)}{\Gamma \left(\frac{3}{4}\right)} ~, \\
\hat{K}^\mathrm{FFN}
& =
\left|
\sqrt{\tfrac{\pi}{2}}^2
\left(
-\I \sqrt{\tfrac{\pi}{2}}
\right)^2
\cdot
\kappa_0 (\tfrac{1}{2})^2
\kappa_{\frac{1}{2}} (1)^2
\right| 
&& \stackrel{D=3}{=}
\frac{\pi^5}{27}
\frac{\Gamma \left(\frac{1}{6}\right)^2}{\Gamma \left(\frac{2}{3}\right)^2} ~.
\end{align}
\end{subequations}
The limits for the bosonic theories were first obtained by Zamolodchikov \cite{Zamolodchikov:1980mb} and later confirmed by  TBA-methods \cite{Basso:2018agi,Basso:2019xay}.
The results for the theories with fermions were first derived in \cite{Kade:2023xet}.
Thereby, the normalization of the propagators is modified, which is the reason for the difference in the factors in front of the gamma functions here and in the literature, see the appendix of \cite{Kade:2023xet}.
Finally, we obtain the critical couplings of all five theories by taking the power of couplings in units of $N$ into account.
We find the results
\begin{subequations}
\begin{align}
& (\xi^\mathrm{FN} )^2_\mathrm{cr}
\stackrel{b=1}{=}
\frac{32}{\pi \, \Gamma (\tfrac{1}{4})^{4} }
=
\frac{2}{\pi^4 \, \eta (\I)^4 } ~,
&& \rho_\mathrm{cr}^\mathrm{BW}
\stackrel{b=2}{=}
\frac{8}{\pi^9 \, \Gamma (\tfrac{1}{4})^{4}} 
=
\frac{1}{2 \pi^{12} \, \eta (\I)^4 } ~, \\
& \xi_\mathrm{cr}^\mathrm{Tri}
\stackrel{b=1}{=}
\frac{144\, 3^{1/4}}{\pi ^{3/4} \Gamma (\frac{1}{6})^{9/2}}
=
\frac{3^{3/4}}{\sqrt{2} \pi ^{9/2} | \eta ( \e^{\frac{i \pi }{3}} ) | ^6} ~,
&& \rho_\mathrm{cr}^\mathrm{FFN}
\stackrel{b=2}{=}
\frac{1944\, 2^{1/3}}{\pi ^7 \Gamma ( \frac{1}{6} )^6}
=
\frac{27}{ 2^{8/3} \pi ^{12} | \eta ( \e^{\frac{i \pi }{3}} ) | ^8} ~, \\
& \rho_\mathrm{cr}^\mathrm{Hex}
\stackrel{b=2}{=}
\frac{2187 \sqrt{3}}{2 \pi ^{21/2} \Gamma ( \frac{1}{6} )^9} 
=
\frac{81 \sqrt{3}}{1024 \pi ^{18} | \eta ( \e^{\frac{i \pi }{3}} )| ^{12}} ~. &&
\end{align}\label{eq:ResultsCriticalCoupling_NonSusy}%
\end{subequations}
The final form of the results is obtained by using several relations between gamma function values.
Notable ones are $\Gamma (\tfrac{1}{4}) \Gamma (\tfrac{3}{4}) = \pi \sqrt{2}$, $\Gamma (\tfrac{1}{3}) \Gamma (\tfrac{2}{3}) = 2\pi / \sqrt{3}$, $\Gamma (\tfrac{1}{6}) \Gamma (\tfrac{2}{3}) = 2^{2/3} \sqrt{\pi}\, \Gamma (\tfrac{1}{3})$ and $\Gamma (\tfrac{5}{6}) \Gamma (\tfrac{1}{3}) = 2^{1/3} \sqrt{\pi}\, \Gamma (\tfrac{2}{3})$.
Furthermore, we have expressed our results in terms of special values of the Dedekind eta function, which obey the relations $2\, \pi^{3/4}\, \eta (\I) = \Gamma (\tfrac{1}{4})$ and $24\, \pi^{5/2} | \eta ( \e^{\I\pi /3} ) |^4 = \Gamma (\tfrac{1}{6})^3$.
It is remarkable that the critical coupling of all theories, which were considered here, are proportional to values of $\eta (\tau)$ at $\tau = \I$ and $\tau = \e^{\I\pi /3}$.
The complex number $\tau$ is known as the \textit{modulus}, on which $\SL{2, \mathbb{Z}}$ modular transformations act.
The modulus can be interpreted as a complex structure modulus of a torus and the above values are exactly the two singularities of its fundamental domain, when it is viewed as an orbifold.
This raises the question if the numerical coincidence has a physical explanation.
The AdS/CFT-correspondence could give an answer:
in the limit of strong coupling, the vacuum diagrams become dense and are dual to a string worldsheet.
In the planar limit, the worldsheet has the topology of a torus, which might have the form set by the modulus $\tau$.
The free energy of the 2D worldsheet CFT can then be calculated by the partition functions of bosons and fermions on a torus, and the results in the literature \cite{DiFrancesco:1997nk} famously involve Dedekind eta functions as well.
We leave it for future explorations to make the duality explicit.

\subsubsection{Inhomogeneous vacuum diagrams}
\label{subsubsec:InhomogeneousVacuumDiagrams}
In \eqref{eq:K_ThermodynamicLimit_Ansatz} and \eqref{eq:Kappagenfinalresult} we find the thermodynamic limit of inhomogeneous vacuum graphs, which allow for $N$ different parameter matrices \eqref{eq:Parametermatrix_U}.
The QFTs presented in this section all have homogeneous vacuum graphs, as we have seen above.
But inhomogeneous vacuum graphs are also imaginable:
we notice that combining two interaction vertices of the type of the bi-scalar fishnet theory \eqref{eq:Fishnet_biscalar_action} with the interactions of the brick wall theory \eqref{eq:BrickWall_Lagrangian} gives the $\chi_0$-CFT \eqref{eq:Chi0_SYM_action}, where the flavor indices have to be modified.
Therefore, we can think of the vacuum graphs of the $\chi_0$-CFT as the sum of all possible arrangements of $U^\mathrm{FN}$ in \eqref{eq:ParameterMatrix_FN} and $U^\mathrm{BW}$ in \eqref{eq:ParameterMatrix_BW_graphbuilder} within the generalized vacuum diagram $Z_{MN} (\lbrace U_J \rbrace)$ \eqref{eq:ZMN_Inhomogeneous_vacuumgraph}.
However, the sum over all possible arrangements adds an additional sum to the free energy \eqref{eq:VacuumFreeEnergyQFT} and the radius of convergence of the expansion of the free energy would no longer correspond to \eqref{eq:K_CriticalCouplingQFT}.
In section \eqref{sec:VacuumGraphsThermoLimit_susy}, we circumvent this problem by considering supersymmetric vacuum graphs for a special case of the $\chi$-CFT \eqref{eq:Chi_SYM_action}.

Nevertheless, inhomogeneous vacuum graphs can be used to resolve a subtlety, we swept under the carpet so far.
We exemplify this with the vacuum graphs of the bi-scalar fishnet theory; see e.\,g.\ \eqref{eq:VacuumGraph_Fishnet}.
There are additional vacuum diagrams produced by the theory, which are of the form 
\begin{equation}
\adjincludegraphics[valign=c,scale=1]{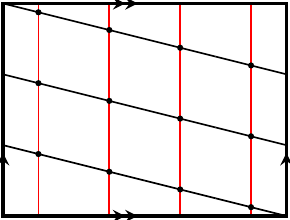} 
=
\adjincludegraphics[valign=c,scale=1]{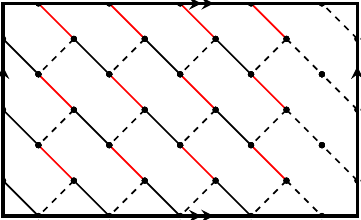} 
\label{eq:Twisted_FN_diagram}
\end{equation}
and where the black propagators wrap around the horizontal cycle of the torus in a skewed way.
We call it a twist, and in the diagram \eqref{eq:Twisted_FN_diagram} it is one, since the black propagator is lifted by one site after completing one horizontal cycle.
Consequently, the twist can be any integer from zero to $M-1$, since twist-$M$ diagrams are identical with twist-$0$ ones.
Diagrams, which are twisted around the vertical cycle, are equivalent to diagrams with horizontal twist but zero vertical twist.

We can show that the thermodynamic limit of the twisted vacuum diagrams is the same as that of the untwisted one.
In \eqref{eq:Twisted_FN_diagram}, we see that a twist of one unit can be implemented by adding a column of unit integral operators of the form of a R-matrix at zero spectral parameter, c.\,f.\ \eqref{eq:Rmatrix_spectralParameterZero}.
This is a permutation operator $\mathbb{P}$ an it has the parameter matrix 
\begin{equation}
U'
=
\begin{pmatrix}
( 0 , 0 ) & ( \frac{D}{2} - \varepsilon , 0 ) \\
( \frac{D}{2} - \varepsilon , 0 ) & ( 0 , 0 )
\end{pmatrix} ~,
\label{eq:ParameterMatrix_FN}
\end{equation}
and we have to multiply the factors $\pi^{ - \frac{D}{2}} a_0 (\varepsilon)$ twice, to make sure the propagators turn into delta functions in the limit $\varepsilon \rightarrow 0$, see \eqref{eq:Delta_bosonic}.
For fishnet theory, we have $D=4$, but the argument of negligible twists also applies to any other $D$-dimensional theory.
For higher twist-$P$, we have to multiply $P$ columns of this kind.
Thus, the twist-$P$ vacuum diagram is built from transfer matrices \eqref{eq:TransferMatix_periodic_Feynman_2} with spectral parameter matrices $\lbrace U \rbrace_{J = 1, ... , N} \cup \lbrace U' \rbrace_{J = N + 1, ... , N + P}$.
Plugging them into the solution \eqref{eq:K_ThermodynamicLimit_Ansatz} and normalizing it to the actual number of sites, which are not $\mathbb{P}$, gives
\begin{equation}
K (U , U')
=
K(U)
\cdot 
K(U')
=
K (U)
\cdot
\lim_{N \rightarrow \infty}
\lim_{\varepsilon \rightarrow 0}
\left|
\pi^{ - D} a_0 (\varepsilon)^2
\kappa_0 (\tfrac{D}{2} - \varepsilon )^2
\kappa_0 (0)
\right|^{\frac{P}{N}} ~.
\end{equation}
When we recall that, for generic $D$, we have $\kappa_\ell (0) = 1$ and $\lim_{\varepsilon \rightarrow 0} \pi^{-\frac{D}{2}} a_0 (\varepsilon) \kappa_0 (\frac{D}{2} - \varepsilon) = 1$, we find that $K(U') = 1$.
Therefore, the twisted thermodynamic limit coincides with the untwisted limit, $K (U , U') = K(U)$.
This justifies why we neglected the twisted vacuum diagrams in the first place.

The next chapter presents a different approach to inhomogeneous vacuum diagrams by organizing them in a supersymmetry covariant manner into vacuum supergraphs.
This implies that many ordinary, inhomogeneous vacuum diagrams combine to form a homogeneous superspace vacuum graph.
While this works neatly in supersymmetric theories, it is interesting to think of the supergraph methods as an effective technique to tackle also isolated, non-supersymmetric, inhomogeneous vacuum graphs.

%%%----------------------------------------------------------------------------------------
%%%	THESIS CONTENT - PART 6 - Supersymmetric double-scaled deformations
%%%----------------------------------------------------------------------------------------
\chapter{Supersymmetric double-scaled deformations}
\label{chpt:SupersymmetricDoubleScaledDeformations}
The previous chapter shows that exact results make bi-scalar fishnet theory a remarkable toy model for the study of integrable Feynman diagrams.
In the end, however, it was constructed to learn more about integrability in full-fledged $\mathcal{N} = 4$ SYM \cite{Gurdogan:2015csr}.
Therefore, translating the methods of bi-scalar fishnet theory to less radical deformed theories like the $\chi_0$-CFT, $\chi$-CFT and eventually $\mathcal{N} = 4$ SYM must be the goal, see section \ref{susec:DoubleScalingLimitOfGammaDeformedSYM}.
In section \ref{subsubsec:InhomogeneousVacuumDiagrams} we comment about the difficulties on this path, which are the sum over many interaction vertices and many vacuum diagrams at a particular order in the coupling.
Nevertheless, in the case of the $\chi$-CFT, it was shown that a careful study of all possible contributing diagrams can give the exact anomalous dimension of a length-two operator and its associated four-point function \cite{Kazakov:2018gcy}.

Here, we present a more economic framework to employ the methods of integrable Feynman diagrams, notably the chain relations, based on Feynman diagrams in superspace.
In the case of supersymmetric double-scaled deformations, we can use a superspace formulation of the action of the theory to expand observables in supergraphs.
Therein, propagators are replaced by propagators of chiral superfields, which depend on the fermionic coordinates of superspace as well as on the ordinary bosonic ones.
Vertices, where one conventionally integrates over the whole spacetime, are generalized to superspace integrals that integrate over spacetime and the chiral or anti-chiral subspace of the fermionic coordinates.
In section \ref{sec:TheGeneralizedSuperspacePropagatorAsLatticeWeight}, we study the interpretation of superpropagators as weights in the sense of lattice models and find many helpful superspace relations, which presumably encode integrability on the level of the supergraphs.

In this chapter, these relations will be applied to the supergraphs of the double-scaled $\beta$-deformation of the ABJM theory and $\mathcal{N} = 4$ SYM since both remain supersymmetric even after the deformation.
We find that the supergraphs are very much restricted to regular patterns, similar to the ordinary Feynman graphs of bi-scalar fishnet theory and the hexagonal theory \eqref{eq:SexticScalarTheory_interactionterm}, see \eqref{eq:VacuumGraph_Fishnet} and \eqref{eq:SexticScalarTheory_vacuumgraphs}.
The two double-scaled $\beta$-deformations of ABJM theory and $\mathcal{N} = 4$ SYM are consequently coined the \textit{superfishnet} theory and the \textit{super brick wall} theory, respectively.
Their supergraphs furnish a vast amount of ordinary diagrams from the scalar and fermionic components of the chiral superfield.
Amazingly, the super chain relations from section \ref{sec:TheGeneralizedSuperspacePropagatorAsLatticeWeight} do not require us to unpack the component diagrams, and the exact computations of scaling dimensions, four-point functions and critical couplings can be uplifted to superspace.
We believe that our superspace approach paves the way for the expansion of the fishnet program closer to ABJM theory and $\mathcal{N}=4$ super Yang-Mills theory.

Therefore, we present the derivation of the superspace action of the superfishnet and super brick wall theory and their appealing superspace diagrams in sections \ref{sec:DoubleScaledBetaDeformationOfABJM} and \ref{sec:TheSuperBrickWallTheory_sec}, respectively, below.
Furthermore, we can compute the exact scaling dimensions of families of length-two and length-four operators and an exact OPE coefficient in the superfishnet theory.
Conceptually, we follow the bi-scalar fishnet calculation of section \ref{subsec:ExactAnomalousDimensionBiScalarFN}.
In the case of the super brick wall theory, we can exactly determine the scaling dimension of a length-two operator, which reproduces a result from the literature \cite{Kazakov:2018gcy} and supports the validity of the superspace approach.
Finally, we compute the critical coupling of the vacuum supergraphs of both theories in section \ref{sec:VacuumGraphsThermoLimit_susy}, by using the super x-unity relation \eqref{eq:SuperXUnity}.

\section[The double-scaled $\beta$-deformation of ABJM]{The double-scaled $\boldsymbol{\beta}$-deformation of ABJM}
\label{sec:DoubleScaledBetaDeformationOfABJM}
We present the $\mathcal{N}=2$ superspace formulation of three-dimensional, $\mathcal{N}=6$ ABJM theory \cite{Aharony:2008ug,Benna:2008zy,Leoni:2010tb,Bianchi:2012cq}, following closely \cite{Benna:2008zy}, and show the reduction to the superfishnet theory \cite{Kade:2024lkc} by combining R-symmetry deformations with double-scaling limits.
Double-scaled deformations of the R-symmetry of the ABJM theory, supersymmetric and non-supersymmetric, were studied in \cite{Caetano:2016ydc}.
Here, we focus on the $\beta$-deformation, which preserves $\mathcal{N}=2$ supersymmetry, which allows us to perform the deformation at the level of the superspace action \cite{Imeroni:2008cr} and which leaves the superspace intact.
Moreover, we take a double-scaling limit of the 't Hooft coupling and the deformation parameter, which renders the theory non-unitary, but leaves us with a set of very restricted, quartic interaction terms.
We present the super Feynman rules for the resulting $\mathcal{N}=2$ superfishnet theory.
Additionally, we present the calculation of exact anomalous dimensions of various operators in this theory, which is possible due to the regularity of the supergraphs.
This section follows \cite{Kade:2024lkc} very closely, but the exact results for anomalous dimensions of spinning zero-magnon operators and the OPE coefficient are new.
All calculations in three-dimensional $\mathcal{N} = 2$ superspace employ the spinor conventions of \cite{Benna:2008zy}, which are collected in appendix \ref{appsec:TheThreeDimensionalN2Superspace}.

\subsection[ABJM $\U{\mathrm{N}} \otimes \U{\mathrm{N}}$ gauge theory]{\boldmath ABJM $\U{\mathrm{N}} \otimes \U{\mathrm{N}}$ gauge theory}
The ABJM theory is an $\mathcal{N}=6$ superconformal Chern-Simons theory of the gauge group $\U{\mathrm{N}} \otimes \U{\mathrm{N}}$ with levels $k$ and $-k$ and bi-fundamental matter, which couples to the gauge fields, and a superpotential.
Its global superconformal algebra is $\mathfrak{osp}(6 \vert 4)$.
Schematically, we may write the action as \cite{Benna:2008zy}
\begin{equation}
S_\mathrm{ABJM} = -\I\, \tfrac{k}{\lambda} \cdot S_\mathrm{CS} + S_\mathrm{mat} + \tfrac{\lambda}{k} \cdot S_\mathrm{pot} ~.
\label{eq:Action_ABJM}
\end{equation}
It depends on the Chern-Simons level $k$ and the 't Hooft coupling $\lambda = \frac{\mathrm{N}}{k}$.
The action can be conveniently formulated in a three-dimensional $\mathcal{N}=2$ superspace.
We present its different terms along with the corresponding field content, following the notation of \cite{Benna:2008zy}.

The Chern-Simons part $S_\mathrm{CS}$ is given by 
\begin{equation}
S_\mathrm{CS} \left[ \mathcal{V} , \hat{\mathcal{V}} \right]
=
\int \dd^3 x \; \dd^2 \theta \, \dd^2 \bar{\theta}
\int_0^1 \dd t ~
\mathrm{tr}
\left[
\mathcal{V} \bar{D}^\alpha 
\left(
\e^{t\mathcal{V}} D_\alpha \e^{-t\mathcal{V}}
\right)
-
\hat{\mathcal{V}} \bar{D}^\alpha 
\left(
\e^{t\hat{\mathcal{V}}} D_\alpha \e^{-t\hat{\mathcal{V}}}
\right)
\right] ~.
\end{equation}
Its degrees of freedom are two gauge vector superfields $\mathcal{V}$ and $\hat{\mathcal{V}}$, they correspond to the representations $\left( \mathbf{N}\otimes \bar{\mathbf{N}} , \mathbf{1} \right)$ and $\left( \mathbf{1} , \mathbf{N}\otimes \bar{\mathbf{N}} \right)$ of the gauge group $\U{\mathrm{N}} \otimes \U{\mathrm{N}}$, respectively.
The supersymmetric covariant derivatives $D_\alpha$ and $\bar{D}_\alpha$ are defined in \eqref{eq:SuperCovariantDerivatives}.
The vector superfield contains the gauge field $A_\mu$ and admits an expansion in Gra\ss mann components, similar to its four dimensional counterpart in \eqref{eq:VectorSuperfield},
\begin{equation}
\mathcal{V}
=
2\I\, \theta \bar{\theta}\, \sigma (x) 
+ 2\, \theta \gamma^\mu \bar{\theta} \, A_\mu (x) 
+ \sqrt{2} \I\, \theta^2\, \bar{\theta}\bar{\chi} (x)
- \sqrt{2} \I\, \bar{\theta}^2\, \theta \chi (x)
+ \theta^2\, \bar{\theta}^2\, \mathrm{D} (x) ~.
\end{equation}
Analogously, $\hat{\mathcal{V}}$ contains $\hat{A}_\mu$.
The remaining component-fields are auxiliary scalars ($\sigma$, $\mathrm{D}$ and their hatted versions) and auxiliary fermions ($\chi$, $\bar{\chi}$ and their hatted versions).

The matter part $S_\mathrm{mat}$ of the ABJM theory is 
\begin{equation}
S_\mathrm{mat} \left[ \mathcal{Z}, \mathcal{W}, \bar{\mathcal{Z}}, \bar{\mathcal{W}} , \mathcal{V}, \hat{\mathcal{V}} \right]
=
\int \dd^3 x \; \dd^2 \theta \, \dd^2 \bar{\theta} ~
\mathrm{tr}
\left[
- \bar{\mathcal{Z}}_A \e^{-\mathcal{V}} \mathcal{Z}^A \e^{\hat{\mathcal{V}}}
- \bar{\mathcal{W}}^A \e^{-\hat{\mathcal{V}}} \mathcal{W}_A \e^{\mathcal{V}}
\right] ~.
\end{equation}
It contains the chiral superfields $\mathcal{Z}$ and $\mathcal{W}$, as well as the anti-chiral superfields $\bar{\mathcal{Z}}$ and $\bar{\mathcal{W}}$ as matter degrees of freedom.
Furthermore, it includes the coupling of the matter fields to the gauge fields $\mathcal{V}$ and $\hat{\mathcal{V}}$.
The chiral superfields $\mathcal{Z}$ and $\mathcal{W}$ transform in the representations $(\mathbf{2},\mathbf{1})$ and $(\mathbf{1},\bar{\mathbf{2}})$ of the global symmetry group $\SU{2}\otimes \SU{2}$, respectively.
We denote the corresponding indices by capital Latin letters.  
Additionally, they transform in the bi-fundamental representations $(\mathbf{N}, \bar{\mathbf{N}})$ and $(\bar{\mathbf{N}}, \mathbf{N})$ of the gauge group $\U{\mathrm{N}} \otimes \U{\mathrm{N}}$, respectively.
The anti-chiral superfields $\bar{\mathcal{Z}}$ and $\bar{\mathcal{W}}$ transform accordingly in the representations $(\bar{\mathbf{2}},\mathbf{1})$ and $(\mathbf{1}, \mathbf{2})$ of the global symmetry group and in the representations $(\bar{\mathbf{N}}, \mathbf{N})$ and $(\mathbf{N}, \bar{\mathbf{N}})$ of the gauge group, respectively.
The expansions of the chiral superfields in Gra\ss mann components are
\begin{subequations}
\begin{align}
\mathcal{Z}^A 
&= Z^A (x_+) + \sqrt{2}\, \theta \zeta^A (x_+) + \theta^2\, F^A (x_+) ~,\\
\mathcal{W}_A
&= W_A (x_+) + \sqrt{2}\, \theta \omega_A (x_+) + \theta^2\, G_A (x_+) ~,
\end{align}
\label{eq:ComponentExpansion_ChiralSuperfields}%
\end{subequations}
and the ones of the anti-chiral superfields are 
\begin{subequations}
\begin{align}
\bar{\mathcal{Z}}_A 
&= Z^\dagger_A (x_-) - \sqrt{2}\, \bar{\theta} \bar{\zeta}_A (x_-) - \bar{\theta}^2\, F^\dagger_A (x_-) ~,\\
\bar{\mathcal{W}}^A
&= W^{\dagger A} (x_-) - \sqrt{2}\, \bar{\theta} \bar{\omega}^{A} (x_-) - \bar{\theta}^2 G^{\dagger A} (x_-) ~.
\end{align}
\label{eq:ComponentExpansion_AntiChiralSuperfields}%
\end{subequations}
The bosons $F^A$, $G_A$, $F^\dagger_A$ and $G^{\dagger A}$ are auxiliary fields and we recall the shorthand notation $x^\mu_\pm = x^\mu \pm \I \theta \gamma^\mu \bar{\theta}$.
Note that \eqref{eq:ComponentExpansion_AntiChiralSuperfields} is in accordance with the conventions for three-dimensional superspace in \eqref{eq:Superfield_Components_3DN2} and the different signs in comparison to the four-dimensional case in \eqref{eq:Superfield_Components} are due to the difference in conventions \eqref{eq:ThetaSquare_ThetaCube} and \eqref{eq:ThetaSquare_ThetaCube_D3N2} from \cite{Benna:2008zy}.

The matter superfields couple in a quartic interaction in the superpotential part of the action
\begin{equation}
\begin{split}
S_\mathrm{pot} \left[ \mathcal{Z}, \mathcal{W}, \bar{\mathcal{Z}}, \bar{\mathcal{W}} \right]
=
& \int \dd^3 x \; \dd^2 \theta ~
\tfrac{1}{4}\, \varepsilon_{AC}\varepsilon^{BD}\;
\mathrm{tr}
\left[
\mathcal{Z}^A \mathcal{W}_B \mathcal{Z}^C \mathcal{W}_D
\right]\\
& +
\int \dd^3 x \; \dd^2 \bar{\theta} ~
\tfrac{1}{4}\, \varepsilon^{AC}\varepsilon_{BD}\;
\mathrm{tr}
\left[
\bar{\mathcal{Z}}_A \bar{\mathcal{W}}^B \bar{\mathcal{Z}}_C \bar{\mathcal{W}}^D
\right] ~.
\end{split}
\label{eq:Superpotential_ABJM}
\end{equation}
Here we denote the antisymmetric tensor of $\SU{2}$ by $\varepsilon$ with the conventions $\varepsilon^{12} = \varepsilon_{21} = +1$.
The superpotential action and the other parts of \eqref{eq:Action_ABJM} are symmetric under a global $\SU{2} \otimes \SU{2}$, as well as under a global $\U{1}_\mathrm{R}$ symmetry, which is the manifest part of the R-symmetry in a three-dimensional $\mathcal{N}=2$ superspace action.
The action of $\U{1}_\mathrm{R}$ on chiral superfields is shown in \eqref{eq:U1_Rsymmetry_transformation} and it is the same as in four-dimensional $\mathcal{N} = 1$ superspace.
For convenience, we recall the action here:
when acting on the fermionic coordinates as $\theta \rightarrow \e^{\I \alpha}\theta$ and $\bar{\theta} \rightarrow \e^{-\I \alpha} \bar{\theta}$, the matter fields transform as
\begin{subequations}
\begin{align}
\mathcal{Z}^A &\rightarrow \e^{-\frac{\I}{2} \alpha}\, \mathcal{Z}^A (x,\e^{\I \alpha}\theta ,\e^{-\I \alpha} \bar{\theta}) ~,
&&\mathcal{W}_A \rightarrow \e^{-\frac{\I}{2} \alpha}\, \mathcal{W}_A (x,\e^{\I \alpha}\theta ,\e^{-\I \alpha} \bar{\theta}) ~,  \\
\bar{\mathcal{Z}}_A &\rightarrow \e^{\frac{\I}{2} \alpha}\, \bar{\mathcal{Z}}_A (x,\e^{\I \alpha}\theta ,\e^{-\I \alpha} \bar{\theta}) ~,
&&\bar{\mathcal{W}}^A \rightarrow \e^{\frac{\I}{2} \alpha}\, \bar{\mathcal{W}}^A (x,\e^{\I \alpha}\theta ,\e^{-\I \alpha} \bar{\theta}) ~.
\end{align}
\label{eq:U1_intakt_transformation}%
\end{subequations}
The global symmetry $\SU{2} \otimes \SU{2}$ enhances the manifest R-symmetry $\U{1}_\mathrm{R}$ to the R-symmetry $\SU{4}_\mathrm{R}$ of $\mathcal{N}=6$ ABJM theory \eqref{eq:Action_ABJM}.
To summarize, we display the charges of the degrees of freedom under the three Cartan-generators of $\mathfrak{su}(4)_\mathrm{R}$ in table \ref{tab:Rcharges}.

\begin{table}[]
\centering
\begin{tabular}{lc|llll|llll|ll}
 &  & $\mathcal{Z}^1$ & $\mathcal{Z}^2$ & $\mathcal{W}_1$ & $\mathcal{W}_2$ & $\bar{\mathcal{Z}}_1$ & $\bar{\mathcal{Z}}_2$ & $\bar{\mathcal{W}}^1$ & $\bar{\mathcal{W}}^2$ & $\mathcal{V}$ & $\hat{\mathcal{V}}$ \\ \hline
$\mathfrak{su}(2)$ & $q^1$ & $\phantom{-}\frac{1}{2}$ & $-\frac{1}{2}$ & $\phantom{-}0$ & $\phantom{-}0$ & $-\frac{1}{2}$ & $\phantom{-}\frac{1}{2}$ & $\phantom{-}0$ & $\phantom{-}0$ & $0$ & $0$ \\
$\mathfrak{su}(2)$ & $q^2$ & $\phantom{-}0$ & $\phantom{-}0$ & $-\frac{1}{2}$ & $\phantom{-}\frac{1}{2}$ & $\phantom{-}0$ & $\phantom{-}0$ & $\phantom{-}\frac{1}{2}$ & $-\frac{1}{2}$ & $0$ & $0$ \\
$\mathfrak{u}(1)$ & $R = q^3$ & $\phantom{-}\frac{1}{2}$ & $\phantom{-}\frac{1}{2}$ & $\phantom{-}\frac{1}{2}$ & $\phantom{-}\frac{1}{2}$ & $-\frac{1}{2}$ & $-\frac{1}{2}$ & $-\frac{1}{2}$ & $-\frac{1}{2}$ & $0$ & $0$
\end{tabular}
\caption{The charges of the superfields in  the ABJM theory under the Cartan-generators of the subalgebras of the R-symmetry $\mathfrak{su}(4)_\mathrm{R}$. The $\beta$-deformation breaks the R-symmetry, only keeping the $\mathfrak{u}(1)$ component intact.}
\label{tab:Rcharges}
\end{table}

In order to establish the genus expansion in the planar large-$\mathrm{N}$ limit, we find it convenient to rescale the (anti-)chiral superfields as $(\mathcal{Z}, \mathcal{W}, \bar{\mathcal{Z}}, \bar{\mathcal{W}}) \rightarrow \sqrt{\mathrm{N}} \cdot (\mathcal{Z}, \mathcal{W}, \bar{\mathcal{Z}}, \bar{\mathcal{W}})$ and the gauge vector superfields as $(\mathcal{V}, \hat{\mathcal{V}}) \rightarrow \lambda \cdot (\mathcal{V}, \hat{\mathcal{V}})$.
This way, the ABJM action \eqref{eq:Action_ABJM} turns into 
\begin{equation}
S'_\mathrm{ABJM}
=
- \I\, \tfrac{k}{\lambda} \cdot S_\mathrm{CS} \left[ \lambda \mathcal{V} ,\lambda \hat{\mathcal{V}} \right]
+ \mathrm{N} \cdot S_\mathrm{mat}\left[ \mathcal{Z}, \mathcal{W}, \bar{\mathcal{Z}}, \bar{\mathcal{W}}, \lambda \mathcal{V} ,\lambda \hat{\mathcal{V}} \right]
+ \mathrm{N}\lambda^2 \cdot S_\mathrm{pot} \left[ \mathcal{Z}, \mathcal{W}, \bar{\mathcal{Z}}, \bar{\mathcal{W}} \right] ~.
\label{eq:Action_ABJM_prime}
\end{equation}
The action $S'_\mathrm{ABJM}$ will be subject to the $\beta$-deformation and a double-scaling limit, eventually yielding the superfishnet theory.

\subsection[$\beta$-Deformation and double-scaling limit]{\boldmath $\beta$-Deformation and double-scaling limit}
In analogy to the derivation of the four-dimensional $\chi$-CFT in section \ref{susec:DoubleScalingLimitOfGammaDeformedSYM}, we obtain the superfishnet theory from the action \eqref{eq:Action_ABJM_prime} by a two-step procedure.
It consists of a $\beta$-deformation of the $\SU{4}_\mathrm{R}$ R-symmetry and a consecutive double-scaling limit of the 't Hooft coupling $\lambda$ and the deformation parameter.
The $\beta$-deformation of the ABJM theory on the level of the superpotential was described in \cite{Imeroni:2008cr} and on the level of the component fields in \cite{Caetano:2016ydc,He:2013hxd,Chen:2016geo}.
It consists of twisting the two $\U{1}$ Cartan-subgroups in the $\SU{2} \otimes \SU{2}$ subgroup of the global R-symmetry group $\SU{4}_\mathrm{R}$.
Therefore, the $\U{1}_\mathrm{R}$-factor corresponding to the transformations \eqref{eq:U1_intakt_transformation} stays intact and will be the residual R-symmetry of the $\beta$-deformed theory, which hence is $\mathcal{N}=2$ supersymmetric.
This is important, since it allows us to keep our $\mathcal{N}=2$ superspace formalism untouched by the deformation.

In the superspace action $S'_\mathrm{ABJM}$ in \eqref{eq:Action_ABJM_prime}, the deformation is implemented by replacing ordinary products of superfields by the star product \eqref{eq:StarProduct}.
Instead of the general deformation in all three Cartan charges of $\SU{4}_\mathrm{R}$, as it is the case for the $\gamma$-deformation, we consider the diagonal case $(\gamma_1 , \gamma_2 , \gamma_3) = (-\beta , -\beta , -\beta)$.
Naturally, we denote the deformation parameter of the $\beta$-deformation as $\beta$.
In this case, the star product of multiple factors \eqref{eq:StarProduct_multipleFactors} takes the form 
\begin{equation}
\Phi_1 \cdots \Phi_p
~ \rightarrow ~
\e^{-\frac{\I}{2} \sum^p_{m > n} \beta \cdot \varepsilon_{ij}\, q^i_{\Phi_m} q^j_{\Phi_n}}\;
\Phi_1 \cdots \Phi_p ~.
\label{eq:BetaDeformation_StarProduct}
\end{equation}
Here, the antisymmetric $\varepsilon_{ij}$ has the component $\varepsilon_{12} = 1$ and the charges $q^1_\Phi$ and $q^2_\Phi$ correspond to the charges of the field $\Phi$ under the Cartan-generators of the $\SU{2} \otimes \SU{2}$ subgroup.
They are listed in the first two rows of table \ref{tab:Rcharges}.
Since the gauge vector superfields do not transform under the R-symmetry, we find the Chern-Simons part of the action \eqref{eq:Action_ABJM_prime} unaltered by the $\beta$-deformation $S_\mathrm{CS} \rightarrow S_\mathrm{CS}$.
The matter superfields $\mathcal{Z}$ and $\bar{\mathcal{Z}}$, as well as $\mathcal{W}$ and $\bar{\mathcal{W}}$ have opposite charges under the R-symmetry, respectively, resulting in an unchanged matter term of the action \eqref{eq:Action_ABJM_prime}, $S_\mathrm{mat} \rightarrow S_\mathrm{mat}$.
The only term affected by the deformation in the action \eqref{eq:Action_ABJM_prime} is the superpotential part $S_\mathrm{pot}$.
The holomorphic and anti-holomorphic superpotentials \eqref{eq:Superpotential_ABJM} get deformed as
\begin{subequations}
\begin{align}
\mathrm{tr} 
\left[
\mathcal{Z}^1 \mathcal{W}_2 \mathcal{Z}^2 \mathcal{W}_1
-
\mathcal{Z}^1 \mathcal{W}_1 \mathcal{Z}^2 \mathcal{W}_2
\right]
\rightarrow
\mathrm{tr} 
\left[
q \cdot \mathcal{Z}^1 \mathcal{W}_2 \mathcal{Z}^2 \mathcal{W}_1
-
q^{-1} \cdot \mathcal{Z}^1 \mathcal{W}_1 \mathcal{Z}^2 \mathcal{W}_2
\right] ~,\\
\mathrm{tr} 
\left[
\bar{\mathcal{Z}}_1 \bar{\mathcal{W}}^2 \bar{\mathcal{Z}}_2 \bar{\mathcal{W}}^1
-
\bar{\mathcal{Z}}_1 \bar{\mathcal{W}}^1 \bar{\mathcal{Z}}_2 \bar{\mathcal{W}}^2
\right]
\rightarrow
\mathrm{tr} 
\left[
q \cdot \bar{\mathcal{Z}}_1 \bar{\mathcal{W}}^2 \bar{\mathcal{Z}}_2 \bar{\mathcal{W}}^1
-
q^{-1} \cdot \bar{\mathcal{Z}}_1 \bar{\mathcal{W}}^1 \bar{\mathcal{Z}}_2 \bar{\mathcal{W}}^2
\right] ~,
\end{align}
\end{subequations}
respectively \cite{Imeroni:2008cr}.
We use the abbreviation $q = \e^{\frac{\I}{4} \beta}$, not to be confused with the charges.
To summarize, we find the action of the $\beta$-deformation of $S'_\mathrm{ABJM}$ to be
\begin{equation}
\begin{split}
S'_{\beta ,\, \mathrm{ABJM}}
=
& - \I\, \tfrac{k}{\lambda} \cdot S_\mathrm{CS} \left[ \lambda \mathcal{V} ,\lambda \hat{\mathcal{V}} \right]
+ \mathrm{N} \cdot S_\mathrm{mat}\left[ \mathcal{Z}, \mathcal{W}, \bar{\mathcal{Z}}, \bar{\mathcal{W}}, \lambda \mathcal{V} ,\lambda \hat{\mathcal{V}} \right]\\
& + \mathrm{N}\lambda^2 
\int \dd^3 x \; \dd^2 \theta ~
\tfrac{1}{2}\,
\mathrm{tr} 
\left[
q \cdot \mathcal{Z}^1 \mathcal{W}_2 \mathcal{Z}^2 \mathcal{W}_1
-
q^{-1} \cdot \mathcal{Z}^1 \mathcal{W}_1 \mathcal{Z}^2 \mathcal{W}_2
\right]\\
& + \mathrm{N}\lambda^2 
\int \dd^3 x \; \dd^2 \bar{\theta} ~
\tfrac{1}{2}\,
\mathrm{tr} 
\left[
q \cdot \bar{\mathcal{Z}}_1 \bar{\mathcal{W}}^2 \bar{\mathcal{Z}}_2 \bar{\mathcal{W}}^1
-
q^{-1} \cdot \bar{\mathcal{Z}}_1 \bar{\mathcal{W}}^1 \bar{\mathcal{Z}}_2 \bar{\mathcal{W}}^2
\right] ~.
\end{split}
\label{eq:Action_ABJM_prime_betaDef}
\end{equation}

The second step consists of taking the double-scaling limit of the 't Hooft coupling $\lambda$ and the deformation parameter $q$ of the $\beta$-deformed ABJM theory \eqref{eq:Action_ABJM_prime_betaDef} \cite{Caetano:2016ydc}.
We consider the limit $\lambda \rightarrow 0$ and $q \rightarrow \infty$ (or equivalently $\beta \rightarrow - \I \infty$), while requiring the product $2\xi := q \cdot \lambda^2$ to stay finite.
The double-scaling procedure has two consequences.
Firstly, due to the limit $\lambda \rightarrow 0$, the matter part of the action $S'_{\beta ,\, \mathrm{ABJM}}$ in \eqref{eq:Action_ABJM_prime_betaDef} loses its dependency on the gauge fields $\mathcal{V}$ and $\hat{\mathcal{V}}$.
We find $S_\mathrm{mat} \rightarrow S_\mathrm{mat}\left[ \mathcal{Z}, \mathcal{W}, \bar{\mathcal{Z}}, \bar{\mathcal{W}}, 0 ,0 \right]$.
Hence, the gauge vector superfields decouple.
We focus on the remainder of the action, which contains solely the matter superfields, discarding the Chern-Simons action $S_\mathrm{CS}$ in the following.
Secondly, the terms proportional to $q^{-1}$ will vanish in the double-scaling limit because $q^{-1} \rightarrow 0$, and there is no factor to compensate.
However, the terms proportional to $q$ combine with the prefactor $\lambda^2$ to the new coupling $\xi$, which we require to be finite.
We could also have conversely considered the limit $q \rightarrow 0$, while keeping the product $q^{-1}\lambda^2$ fixed.
In this way, we would keep the other two terms in the superpotential part.
However, the obtained theory would be the hermitian conjugate of the double-scaled theory with $q \rightarrow \infty$ and $\xi$ fixed.

Finally, we find the action of the superfishnet theory \cite{Kade:2024lkc}
\begin{equation}
\begin{split}
S_{\mathrm{SFN}}
~=~
&\mathrm{N}
\int \dd^3 x \; \dd^2 \theta \, \dd^2 \bar{\theta} ~
\mathrm{tr}
\left[
- \bar{\mathcal{Z}}_A \mathcal{Z}^A
- \bar{\mathcal{W}}^A \mathcal{W}_A
\right]\\
& + \mathrm{N}\xi 
\int \dd^3 x \; \dd^2 \theta ~
\mathrm{tr} 
\left[
\mathcal{Z}^1 \mathcal{W}_2 \mathcal{Z}^2 \mathcal{W}_1
\right]
+ \mathrm{N}\xi 
\int \dd^3 x \; \dd^2 \bar{\theta} ~
\mathrm{tr} 
\left[
\bar{\mathcal{Z}}_1 \bar{\mathcal{W}}^2 \bar{\mathcal{Z}}_2 \bar{\mathcal{W}}^1
\right]
\end{split}
\label{eq:Action_superfishnet_Z_W}
\end{equation}
for the double-scaled $\beta$-deformation of ABJM theory.
It is a three-dimensional $\mathcal{N}=2$ supersymmetric field theory of four chiral and four anti-chiral superfields.
The double-scaling limit broke unitarity; however, similarly to the double-scaling limits of $\mathcal{N}=4$ SYM in section \ref{susec:DoubleScalingLimitOfGammaDeformedSYM}, this sacrifice will lead to very regular Feynman supergraphs.
Since the $\beta$-deformation breaks the global $\SU{2} \otimes \SU{2}$, the symmetry algebra is reduced to the superconformal $\mathfrak{osp}(2 \vert 4)$, see e.\,g.\ \cite{Frappat:1996pb}.

\subsection{Component action}
Let us derive the superfishnet action \eqref{eq:Action_superfishnet_Z_W} in bosonic spacetime.
The strategy is to replace the superfields in \eqref{eq:Action_superfishnet_Z_W} by their expansions in Gra\ss mann components \eqref{eq:ComponentExpansion_ChiralSuperfields} and \eqref{eq:ComponentExpansion_AntiChiralSuperfields}.
Then we can eliminate the remaining auxiliary fields $F^A$, $G_A$, $F_A^\dagger$ and $G^{\dagger A}$ by using their equations of motion\footnote{We have used identities like $\theta \varphi \cdot \theta \psi = - \frac{1}{2} \theta^2 \varphi \psi$ and $\bar{\theta} \bar{\varphi} \cdot \bar{\theta} \bar{\psi} = - \frac{1}{2} \bar{\theta}^2 \bar{\varphi} \bar{\psi}$, see appendix \ref{appsec:TheThreeDimensionalN2Superspace} for our conventions.}
\begin{equation}
\begin{aligned}[c]
F^1 &= \xi \, W^{\dagger 2} Z_2^\dagger W^{\dagger 1} ~,\\
F^2 &= \xi \, W^{\dagger 1} Z_1^\dagger W^{\dagger 2} ~,\\
\end{aligned}
\quad
\begin{aligned}[c]
G_1 &= \xi \, Z_1^{\dagger} W^{\dagger 2} Z_2^{\dagger} ~,\\
G_2 &= \xi \, Z_2^{\dagger} W^{\dagger 1} Z_1^{\dagger} ~,\\
\end{aligned}
\quad
\begin{aligned}[c]
F_1^\dagger &= - \xi \, W_2 Z^2 W_1 ~,\\
F_2^\dagger &= - \xi \, W_1 Z^1 W_2 ~,\\
\end{aligned}
\quad
\begin{aligned}[c]
G^{\dagger 1} &= - \xi \, Z^1 W_2 Z^2 ~,\\
G^{\dagger 2} &= - \xi \, Z^2 W_1 Z^1 ~.\\
\end{aligned}
\end{equation}
Since the gauge group is $\U{\mathrm{N}} \otimes \U{\mathrm{N}}$, we do not find additional double-trace terms, in contrast to the super brick wall theory in section \ref{subsec:ComponentAction_SBW}.
Eventually, we find the superfishnet action formulated on bosonic spacetime to be
\begin{equation}
\begin{split}
& S_\mathrm{SFN}
=
\mathrm{N}
\int \dd^3 x ~
\mathrm{tr}
\left\lbrace
Z_A^\dagger \square Z^A
+ W^{\dagger A} \square W_A
+ \I\, \bar{\zeta}_A^{\alpha} \gamma^\mu_{\alpha \beta} \partial_\mu \zeta^{A \beta}
+ \I\, \bar{\omega}^{A \alpha} \gamma^\mu_{\alpha \beta} \partial_\mu \omega_A^{\beta}
\right. \\
&+ \xi^2
\left[
Z^1 W_2 Z^2 Z_1^\dagger W^{\dagger 2} Z_2^\dagger +
W_1 Z^1 W_2 W^{\dagger 1} Z_1^\dagger W^{\dagger 2}
+
Z^2 W_1 Z^1 Z_2^\dagger W^{\dagger 1} Z_1^\dagger +
W_2 Z^2 W_1 W^{\dagger 2} Z_2^\dagger W^{\dagger 1}
\right] \\
&- \xi \;
\left[
\zeta^1 \omega_2 Z^2 W_1 +
\omega_2 \zeta^2 W_1 Z^1 + 
\zeta^2 \omega_1 Z^1 W_2 -
\bar{\omega}^{1} \bar{\zeta}_1 W^{\dagger 2} Z_2^\dagger 
\right] \\
&- \xi \;
\left[
\bar{\zeta}_1 \bar{\omega}^{2} Z_2^\dagger W^{\dagger 1} +
\bar{\omega}^{2} \bar{\zeta}_2 W^{\dagger 1} Z_1^\dagger +
\bar{\zeta}_2 \bar{\omega}^{1} Z_1^\dagger W^{\dagger 2} -
\omega_1 \zeta^1 W_2 Z^2 
\right] \\
&- \xi \;
\left.
\left[
\zeta^1 W_2 \zeta^2 W_1 +
\omega_2 Z^2 \omega_1 Z^1 +
\bar{\zeta}_1 W^{\dagger 2} \bar{\zeta}_2 W^{\dagger 1} +
\bar{\omega}^{2} Z_2^\dagger \bar{\omega}^{1} Z_1^\dagger 
\right] ~
\right\rbrace ~.
\end{split}
\end{equation}
Furthermore, we can follow \cite{Benna:2008zy} and compose the matter fields into a $\SU{4}_\mathrm{R}$ fundamental representation (although it is broken to $\U{1}_\mathrm{R}$),
\begin{equation}
Y^M
=
\begin{psmallmatrix}
Z^1 \\
Z^2 \\
W^{\dagger 1} \\
W^{\dagger 2}
\end{psmallmatrix} , \hspace{0.5cm}
Y_M^\dagger
=
\begin{psmallmatrix}
Z_1^\dagger \\
Z_2^\dagger \\
W_1 \\
W_2
\end{psmallmatrix} , \hspace{0.5cm}
\Psi_M
=
\e^{-\frac{\I \pi}{4}}
\begin{psmallmatrix}
- \zeta^2 \\
\zeta^1 \\
\I\, \bar{\omega}^{2} \\
- \I\, \bar{\omega}^{1}
\end{psmallmatrix} , \hspace{0.5cm}
\bar{\Psi}^{M}
=
\e^{\frac{\I \pi}{4}}
\begin{psmallmatrix}
- \bar{\zeta}_2 \\
\bar{\zeta}_1 \\
- \I\, \omega _2 \\
\I\, \omega_1
\end{psmallmatrix}.
\end{equation}
The superfishnet action in terms of the fields $Y^M$, $Y_M^\dagger$, $\Psi_M$ and $\bar{\Psi}^{M}$ reads
\begin{equation}
\begin{split}
& S_\mathrm{SFN}
=
\mathrm{N}
\int \dd^3 x ~
\mathrm{tr}
\left\lbrace
Y_M^\dagger \square Y^M
+ \I\, \bar{\Psi}^{M \alpha} \gamma^\mu_{\alpha \beta} \partial_\mu \Psi_M^{\beta}
\right. \\
&+ \xi^2
\left[
Y^1 Y_4^\dagger Y^2 Y_1^\dagger Y^4 Y_2^\dagger +
Y^1 Y_4^\dagger Y^3 Y_1^\dagger Y^4 Y_3^\dagger
+
Y^2 Y_3^\dagger Y^1 Y_2^\dagger Y^3 Y_1^\dagger +
Y^2 Y_3^\dagger Y^4 Y_2^\dagger Y^3 Y_4^\dagger
\right] \\
&- \I \xi \;
\left[
\Psi_2 \bar{\Psi}^{3} Y^2 Y_3^\dagger -
\bar{\Psi}^{3} \Psi_1 Y_3^\dagger Y^1 +
\Psi_1 \bar{\Psi}^{4} Y^1 Y_4^\dagger -
\Psi_4 \bar{\Psi}^{2} Y^4 Y_2^\dagger 
\right] \\
&+ \I \xi \;
\left[
\bar{\Psi}^{2} \Psi_3 Y_2^\dagger Y^3 -
\Psi_3 \bar{\Psi}^{1} Y^3 Y_1^\dagger  +
\bar{\Psi}^{1} \Psi_4 Y_1^\dagger Y^4 -
\bar{\Psi}^{4} \Psi_2 Y_4^\dagger Y^2
\right] \\
&+ \I \xi \;
\left.
\left[
\Psi_2 Y_4^\dagger \Psi_1 Y_3^\dagger +
\bar{\Psi}^{3} Y^2 \bar{\Psi}^{4} Y^1 -
\bar{\Psi}^{2} Y^4 \bar{\Psi}^{1} Y^3 -
\Psi_3 Y_2^\dagger \Psi_4 Y_1^\dagger 
\right] ~
\right\rbrace ~.
\end{split}
\label{eq:Action_superfishnet_Y_Psi}
\end{equation}
The action contains sextic, scalar interactions and three-dimensional versions of the Yukawa coupling that are quadratic in the fermions and quadratic in the bosons.
We find that the sextic scalar interaction are of the type of the ABJM fishnet theory, presented in \eqref{eq:ABJM_fishnet_interationterm}.
It makes sense, since the stronger $\gamma$-deformation, which yields to the fishnet theories discards all but one of the purely scalar interaction terms.
The elegant packing of the many interaction terms of \eqref{eq:Action_superfishnet_Y_Psi} into the superfield action \eqref{eq:Action_superfishnet_Z_W} motivates us to study the superfishnet theory in a supersymmetry-covariant way, namely in the diagrammatics of Feynman supergraphs.

Another $\beta$-deformation was obtained in \cite{Caetano:2016ydc}, which coincides with \eqref{eq:Action_superfishnet_Y_Psi} up to the $\SU{4}_\mathrm{R}$-indices.
The difference is due to different bases of the $\mathfrak{su}(4)_\mathrm{R}$ Cartan-torus, c.\,f.\ table \ref{tab:Rcharges} and table 1 in \cite{Caetano:2016ydc}.
The $\beta$-deformations here and in \cite{Caetano:2016ydc} were taken into different directions in the $\SU{2} \otimes \SU{2}$ group.

\subsection{The Superfishnet theory}
\label{sec:SuperfishnetTheory}
We derived the superspace action of the superfishnet theory  in section \ref{sec:DoubleScaledBetaDeformationOfABJM} and found it to be \eqref{eq:Action_superfishnet_Z_W}.
In this section we will adopt a more suitable notation, present the generalized superfishnet theory, and derive the supersymmetric Feynman rules for this theory.

In order to keep track of the different fields more efficiently, we redefine the chiral superfields as $(\mathcal{Z}^1 , \mathcal{W}_2 , \mathcal{Z}^2 , \mathcal{W}_1) = \Phi_i$ and the anti-chiral ones as $(\bar{\mathcal{Z}}_1 , \bar{\mathcal{W}}^2 , \bar{\mathcal{Z}}_2 , \bar{\mathcal{W}}^1) = \Phi_i^\dagger$.
We will refer to the index $i$ as flavor.
Additionally, we introduce Gra\ss mann-squares to extend the superspace integral of the superpotential over the full chiral and anti-chiral superspace.
Finally, the superfishnet theory takes the form
\begin{equation}
\begin{split}
S_{\mathrm{SFN}}
~=~
&\mathrm{N}
\int \dd^3 x \; \dd^2 \theta \, \dd^2 \bar{\theta} ~\;
\mathrm{tr}
\left[
- \sum_{i=1}^4 \Phi_i^\dagger \Phi_i^{\phantom{\dagger}}
+ \xi \cdot \bar{\theta}^2 \, \Phi_1 \Phi_2 \Phi_3 \Phi_4
+ \xi \cdot \theta^2 \, \Phi_1^\dagger \Phi_2^\dagger \Phi_3^\dagger \Phi_4^\dagger
\right] ~.
\end{split}
\label{eq:Action_superfishnet_Phi}
\end{equation}
Despite the decoupling of the gauge degrees of freedom, the superfields are still matrices in the bi-fundamental representation of the gauge group $\U{\mathrm{N}} \otimes \U{\mathrm{N}}$.
Due to the prefactor $\mathrm{N}$ in the superfishnet action \eqref{eq:Action_superfishnet_Phi}, the supergraphs admit a genus expansion, of which we study the toroidal order that dominates in the limit $\mathrm{N} \rightarrow \infty$.

In close analogy to \cite{Kade:2024ucz}, we propose a non-local generalization \cite{Kimura:2016irk} of the superfishnet theory, where the K\"{a}hler potential is modified by fractional derivatives,
\begin{equation}
\begin{split}
S_{\mathrm{SFN},\boldsymbol{\omega}}
= \;
&\mathrm{N}
\int \dd^3 x \; \dd^2 \theta \, \dd^2 \bar{\theta} ~\;
\mathrm{tr}
\left[
- \sum_{i=1}^4 \Phi_i^\dagger \square^{\omega_i} \Phi_i^{\phantom{\dagger}}
+ \xi \cdot \bar{\theta}^2 \, \Phi_1 \Phi_2 \Phi_3 \Phi_4
+ \xi \cdot \theta^2 \, \Phi_1^\dagger \Phi_2^\dagger \Phi_3^\dagger \Phi_4^\dagger
\right] ~.
\end{split}
\label{eq:Action_genSuperfishnet_Phi}
\end{equation}
Dimensional analysis shows that the mass dimensions of the superfields get deformed to $\left[\Phi_i\right] = [\Phi^\dagger_i ] = \frac{1}{2} - \omega_i$.
Demanding a marginal interaction implies the constraint $\sum_{i=1}^4 \omega_i = 0$ for the four deformation parameters $\omega_i$. 
The form of the action \eqref{eq:Action_genSuperfishnet_Phi} is strikingly similar to the so-called checkerboard theory \cite{Alfimov:2023vev}.
Instead of bosonic fields and spacetime integration, the generalized superfishnet action contains superfields as degrees of freedom and integration over superspace.
Lastly, we note that the generalized superfishnet reduces to the ordinary superfishnet theory by setting all deformation parameters to zero, $\omega_i = 0$.

Note that the $\boldsymbol{\omega}$-deformed action \eqref{eq:Action_genSuperfishnet_Phi} is formally still supersymmetric.
Indeed, the kinetic part is still a D-term, expressed as a full superspace integral, such that the usual argument applies.
Defining $f(z)$ to be an arbitrary function of the supercoordinate $z = (x,\theta, \bar{\theta})$, we have  
\begin{equation}
0
\stackrel{\mathrm{!}}{=}
\delta_\varepsilon
\int \dd^3 x\; \dd^2\theta \dd^2 \bar{\theta} f(z)
=
\int \dd^3 x\; \dd^2\theta \dd^2 \bar{\theta}\; \delta_\varepsilon f(z)
\end{equation}
with 
\begin{equation}
\delta_\varepsilon f(z)
=
\left(
\varepsilon^\alpha Q_\alpha + \bar{\varepsilon}_{\dot{\alpha}} \bar{Q}^{\dot{\alpha}}
\right)
f(z)
=
\left(
\varepsilon^\alpha \partial_\alpha + \bar{\varepsilon}_{\dot{\alpha}} \bar{\partial}^{\dot{\alpha}}
\right)
f(z)
+ \mathrm{total~derivative}.
\end{equation}
Finally, the superspace integral over a Gra\ss mann derivative of $f(z)$ vanishes, 
\begin{equation}
\int \dd^3 x\; \dd^2\theta \dd^2 \bar{\theta}\; \partial_\alpha f(z)
=
0
=
\int \dd^3 x\; \dd^2\theta \dd^2 \bar{\theta}\; \bar{\partial}^{\dot{\alpha}} f(z) ~ ,
\end{equation}
since otherwise $f(z)$ would need to have a $\theta^3 \bar{\theta}^2$ (l.\,h.\,s.), respectively  $\theta^2 \bar{\theta}^3$ (r.\,h.\,s.), component.
The superpotential is unaltered and therefore still supersymmetric as well.
The supersymmetry of \eqref{eq:Action_genSuperfishnet_Phi} is also consistent with a formal counting of the degrees of freedom, since the latter are in total unchanged due to the constraint $\sum_{i=1}^4 \omega_i = 0$.

We read off the vertex Feynman rules from the superfishnet action \eqref{eq:Action_superfishnet_Phi} or the generalization \eqref{eq:Action_genSuperfishnet_Phi} alike.
After Wick-rotating, they are
\begin{equation}
\adjincludegraphics[valign=c,scale=0.6]{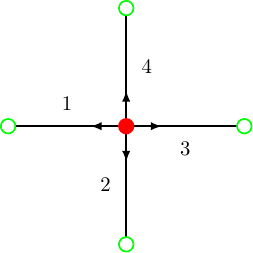}
\sim
\mathrm{N}\, \xi \cdot \I \int \dd^3x\; \dd^2\theta\,\dd^2\bar{\theta} \;\delta^{(2)}(\bar{\theta}) ~, ~
\adjincludegraphics[valign=c,scale=0.6]{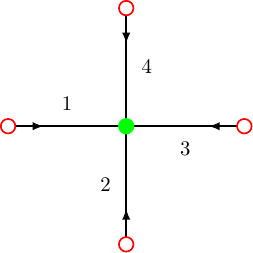}
\sim
\mathrm{N}\, \xi \cdot \I \int \dd^3x\; \dd^2\theta\,\dd^2\bar{\theta} \;\delta^{(2)}(\theta) ~.
\label{eq:FeynmanRules_vertices}
\end{equation}
Firstly, note that the labels in \eqref{eq:FeynmanRules_vertices} are the flavors of the superfields, which meet at the respective vertex.
They should not be confused with the spectral parameter of the superpropagator.
Due to the non-unitarity of the superfishnet theory \eqref{eq:Action_superfishnet_Phi}, the two vertices cannot be mapped to each other by hermitian conjugation.
Respecting the order of the flavor in each vertex reduces the number of allowed supergraphs and leads to the regular fishnet pattern.
Secondly, similar to superpropagators, we will neglect the factor $\mathrm{N}\xi$ in our calculations for now.
In the end, we reinstate them for the diagram under consideration.

Lastly, we comment on possible double-trace interaction terms.
The only options allowed by supersymmetry would be of the form $\mathrm{tr} [ \Phi_1 \Phi_2 ] \mathrm{tr} [ \Phi_3 \Phi_4 ]$.
The supergraphs, which contribute to the radiative corrections, look similar to long chains, as in the case of the bi-scalar fishnet theory.
However, in contrast to fishnet theory, we find the supergraph versions of \eqref{eq:Fishnet_alpha23_RenDiags} to be finite, and we leave it for future work to determine the corresponding four-point function.
Thus, the double-trace terms do not have to be added.

\subsection{Exact all-loop anomalous dimensions and correlation function}
\label{sec:ExactAnomalousDimensionsFrom4PtCorrelationFunctions}
This section is devoted to the study of four-point functions in the undeformed $\boldsymbol{\omega} = 0$ superfishnet theory that admit a perturbative expansion in regular, fishnet Feynman supergraphs in the large-$\mathrm{N}$ limit.
We follow the strategy presented in section \ref{subsec:ExactAnomalousDimensionBiScalarFN} \cite{Grabner:2017pgm,Kazakov:2018qbr,Gromov:2018hut,Kazakov:2018gcy} to extract exact scaling dimensions out of the four-point correlators:
The correlation functions under investigation can be built up by repetitive application of a specific supergraph-building operator.
The operator can be resummed in a geometric series and diagonalized on a set of eigenfunctions.
The resummed series has poles that correspond to exchanged operators.
We identify the position of the poles and may access the anomalous dimension of specific operators.
They are the ones where we can construct the corresponding eigenfunctions of the graph-building operator.

In \cite{Chang:2021fmd} a class of three-dimensional $\mathcal{N}=2$ eigenfunctions were derived, and we can even access the exact correlation function.
For another class, however, we are unaware of the eigenfunctions' precise form, yet we can construct some of them in a particular limit. 
In both cases, the limit is enough to compute the eigenvalue of the correlator's graph-builder and to calculate the anomalous dimension of some operators.
In the so-called zero-magnon case in section \ref{subsec:ZeroMagnonCase}, we give a more detailed outline of the procedure to be concise in the subsequent two-magnon case in section \ref{subsec:TwoMagnonCase} and the super brick wall computation in section \ref{sec:ExactAnomalousDimensions_SBW}.

\subsubsection{Zero-magnon case}
\label{subsec:ZeroMagnonCase}
The zero-magnon case is the four-point correlation function $
\langle \mathrm{tr} [\Phi_1 (z_1)\Phi_3^\dagger (z_2)] \mathrm{tr}[\Phi_1^\dagger (z_3)\Phi_3 (z_4)] \rangle $ in the large-$\mathrm{N}$ limit.
Alternatively, we could also have chosen fields with interchanged flavors $(\mathrm{1},\mathrm{3}) \leftrightarrow (\mathrm{2},\mathrm{4})$, where the calculation is essentially unchanged.
The Feynman supergraphs are of toroidal order in the genus expansion, that is, they scale as $\mathrm{N}^0$.
Due to the restricting flavor-ordering in the vertices \eqref{eq:FeynmanRules_vertices}, the diagrams have a very regular, ladder-like structure,
\begin{equation}
\begin{split}
\left\langle 
\mathrm{tr}
\left[
\Phi_1 (z_1)
\Phi_3^\dagger (z_2)
\right]
\mathrm{tr}
\left[
\Phi_1^\dagger (z_3)
\Phi_3 (z_4)
\right]
\right\rangle 
=
\adjincludegraphics[valign=c,scale=1]{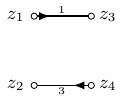} 
+
\xi^2
\adjincludegraphics[valign=c,scale=1]{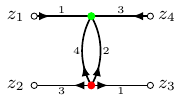} \\
+
\xi^4
\adjincludegraphics[valign=c,scale=1]{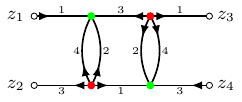} 
+
\xi^6
\adjincludegraphics[valign=c,scale=1]{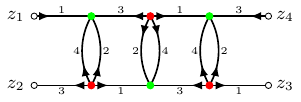} 
+
%\adjincludegraphics[valign=c,scale=1]{figures/superFN/2ptfctn/zeromagnon/diags/4ptfunction_4.pdf} 
%+ 
\cdots ~.
\end{split}
\label{eq:0MagnonPertExpansion}
\end{equation}
Note that in the odd terms, the superpoints $z_3$ and $z_4$ are interchanged with respect to the even terms.
The diagrammatics is similar to bi-scalar fishnet four-point functions in \eqref{eq:Fishnet_0MagnonPertExpansion}, however, here we are considering supergraphs.

We can identify a graph-building operator $\mathbb{H}$ that, together with its conjugate $\bar{\mathbb{H}}$ and the permutation operator of two points $\mathbb{P}$, build up the graphs of \eqref{eq:0MagnonPertExpansion}.
They are
\begin{equation}
\begin{aligned}[c]
\mathbb{H}
=
\adjincludegraphics[valign=c,scale=1]{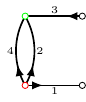} ~,
\end{aligned}
\qquad\qquad
\begin{aligned}[c]
\bar{\mathbb{H}}
=
\adjincludegraphics[valign=c,scale=1]{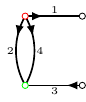} ~,
\end{aligned}
\begin{aligned}[c]
\qquad \qquad
\mathbb{P}
=
\adjincludegraphics[valign=c,scale=1]{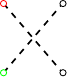} ~,
\end{aligned}
\label{eq:0MagnonGraphBuilder}
\end{equation}
and we can use them to rewrite the correlation function \eqref{eq:0MagnonPertExpansion} formally as
\begin{equation}
\begin{split}
&\left\langle 
\mathrm{tr}
\left[
\Phi_1 (z_1)
\Phi_3^\dagger (z_2)
\right]
\mathrm{tr}
\left[
\Phi_1^\dagger (z_3)
\Phi_3 (z_4)
\right]
\right\rangle \\
&=
\adjincludegraphics[valign=c,scale=1]{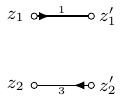} 
\circ
\left[
1
+
\xi^2
\mathbb{H} \circ \mathbb{P}
+
\xi^4
\mathbb{H} \circ \bar{\mathbb{H}}
+
\xi^6
\mathbb{H} \circ \bar{\mathbb{H}} \circ \mathbb{H} \circ \mathbb{P}
+
%\adjincludegraphics[valign=c,scale=1]{figures/superFN/2ptfctn/zeromagnon/diags/4ptfunction_4.pdf} 
%+ 
\cdots
\right] \\
&=
x_{1\bar{2}}^2\, \bar{\mathbb{H}}
\circ
\left[
\sum_{n = 0}^\infty 
(\xi^4 \mathbb{H} \circ \bar{\mathbb{H}} )^n
\right]
\left[
1 + \xi^2 \mathbb{H} \circ \mathbb{P}
\right] 
=
x_{1\bar{2}}^2\, \bar{\mathbb{H}}
\circ
\frac{
1 + \xi^2 \mathbb{H} \circ \mathbb{P}
}{
1 - \xi^4 \mathbb{H} \circ \bar{\mathbb{H}}
}
 ~.
\end{split}
\label{eq:0MagnonPertExpansion_2}
\end{equation}
The symbol $\circ$ denotes the super-convolution of chiral or anti-chiral superspace points. 
Its precise meaning can be inferred from the supergraph expansion \eqref{eq:0MagnonPertExpansion}.
We observe that only the combinations $\mathbb{H} \circ \bar{\mathbb{H}}$ and $\mathbb{H} \circ \mathbb{P}$ of the graph-builders \eqref{eq:0MagnonGraphBuilder} appear in \eqref{eq:0MagnonPertExpansion_2}.
Therefore, we would like to diagonalize them over a complete set of eigenfunctions.
In the non-supersymmetric case, we encountered a similar set of eigenfunctions, the conformal triangles \eqref{eq:Conformal_Triangles}.
We assume the existence of a superspace generalization of their construction, which means that there exists a superconformal three-point function
\begin{equation}
\Omega_{\Delta,S,R} {\scriptstyle (z_1, z_2 ; z_0)}
=
\left\langle
\mathrm{tr}
\left[ 
\mathcal{O}_1  (z_1)
\mathcal{O}_2  (z_2)
\right]
\mathcal{O}_{\Delta,S,R} (z_0)
\right\rangle  ~,
\label{eq:ConformalTriangle3pt}
\end{equation}
and the Cartan labels $\Delta$, $S$ and $R$ of the superconformal algebra in three dimensions.
The labels correspond to the scaling dimension, the spin and the R-charge of the operator $\mathcal{O}_{\Delta,S,R}$, respectively.
The functions $\Omega_{\Delta,S,R}$ should form a complete basis in the vector space of a non-compact representation of the three-dimensional $\mathcal{N}=2$ superconformal group $\mathrm{OSP}(2 \vert 4)$.
Accordingly, we formally write the completeness relation of the eigenfunctions as 
\begin{equation}
\mathbb{1}
=
\delta^{(7)}\left( z_{13} \right) \cdot
\delta^{(7)}\left( z_{24} \right)
=
\sumint{\Delta,S} ~~~~~~
\int \dd^7 z_0 ~
\ket{
\bar{\Omega}_{\Delta,S,R} {\scriptstyle (z_1, z_2 ; z_0) }
} \;
\bra{
\Omega_{\Delta,S,R} {\scriptstyle (z_3, z_4 ; z_0) }
}
\label{eq:CompletenessRelation}
\end{equation}
and we can insert the resolution of unity into supergraphs like \eqref{eq:0MagnonPertExpansion_2} and replace the graph-builders with their eigenvalues.
We introduce the shorthand for the chiral and anti-chiral superspace integration measure $\dd^5 z = \dd^3 x\; \dd^2 \theta\, \dd^2 \bar{\theta} ~ \bar{\theta}^2$ and $\dd^5 \bar{z} = \dd^3 x\; \dd^2 \theta\, \dd^2 \bar{\theta} ~ \theta^2$, respectively, and $\dd^7 z = \dd^3 x\; \dd^2 \theta\, \dd^2 \bar{\theta}$ for the full superspace measure for a concise notation.

For the zero-magnon case at hand, we require the superconformal triangle $\Omega_{\Delta,S,R} {\scriptstyle (z_1, z_2 ; z_0) }$ to be the three-point function of $\mathcal{O}_1 = \Phi_1^\dagger$, $\mathcal{O}_2 = \Phi_3$, both with scaling dimension $\Delta_\Phi = \frac{1}{2}$, and the exchanged operator $\mathcal{O}_{\Delta,S,0}$, which must have zero R-charge.
For this case, the superconformal triangles were found in \cite{Chang:2021fmd}, and they read
\begin{equation}
\Omega_{\Delta,S,0} {\scriptstyle (z_1, z_2 ; z_0)}
=
\frac{
C_{\Phi_1^\dagger \Phi_3 \mathcal{O}}}
{
\left[x_{2\bar{1}}^2 \right]^{\Delta_\Phi - \frac{\Delta}{2} + \frac{S}{2}} 
\left[x_{0\bar{1}}^2 \right]^{\frac{\Delta}{2} + \frac{S}{2}}
\left[x_{2\bar{0}}^2 \right]^{\frac{\Delta}{2} + \frac{S}{2}}
} 
\left[
X_{3,-}^{\mu_1}
\cdots
X_{3,-}^{\mu_S}
-
\mathrm{traces}
\right]
~.
\end{equation}
Thereby, $X_{3,-}^{\mu} = X_{3}^{\mu} - \I \Theta_3 \gamma^\mu \bar{\Theta}_3$ contains the supersymmetric three-point structures $X_{3}^{\mu}$ and $\Theta_3$, which are given in \cite{Chang:2021fmd}.
The traces are all possible contractions of $X_{3,-}^{\mu_1} \cdots X_{3,-}^{\mu_S}$ with the metric to render the superconformal triangle traceless and symmetric.
Furthermore, the conformal triangles admit an expansion in three-dimensional $\mathcal{N} = 2$ superconformal blocks $\mathsf{g}_{\Delta , S} (r_1, r_2 )$ \cite{Chang:2021fmd},
\begin{equation}
\begin{split}
\int & \dd^7 z_0 ~
\ket{
\bar{\Omega}_{\Delta,S,0} {\scriptstyle (z_1, z_2 ; z_0) }
} \;
\bra{
\Omega_{\Delta,S,0} {\scriptstyle (z_3, z_4 ; z_0) }
} \\
& =
\left(
\frac{1}{
x_{1\bar{2}}^2
x_{4\bar{3}}^2
}
\right)^{\frac{1}{2}}
\cdot
\left[
\frac{\mathsf{c}_1(\Delta, S)}{\mathsf{c}_2 (\Delta, S)}
\mathsf{g}_{\Delta , S} (r_1 , r_2)
+
\frac{\mathsf{c}_1(\Delta^*, S)}{\mathsf{c}_2 (\Delta^*, S)}
\mathsf{g}_{\Delta^* , S} (r_1 , r_2)
\right] ~,
\end{split}
\label{eq:Conformal_blocks3DSusy}
\end{equation}
which is the superspace analog\footnote{
The superspace coefficients were found in \cite{Chang:2021fmd} as 
\begin{subequations}
\begin{align}
\mathsf{c}_1(\Delta, S)
& =
16 \pi ^3
\frac{
(-1)^{2 S} (\Delta -S-1)^2 (\Delta +S)
}{
(2 \Delta -1) (2 S+1) (-\Delta +S+1)
}
\; \mathrm{tan} (\pi  \Delta )
\\
\mathsf{c}_2(\Delta, S)
& =
\pi ^3 (-1)^S 
2^{2 \Delta -S+2} 
\frac{
\Gamma (S+1) 
\Gamma (\Delta -\frac{1}{2}) 
\Gamma (\frac{S-\Delta + 3}{2}) 
\Gamma (\frac{S+\Delta +1}{2})
}{
\Gamma (S+\frac{3}{2}) 
\Gamma (\Delta ) 
\Gamma (\frac{S-\Delta + 2}{2}) 
\Gamma (\frac{S + \Delta}{2})
} ~,
\end{align}
\end{subequations}
and the superconformal blocks $\mathsf{g}_{\Delta , S} (r_1 , r_2)$ may be found in \cite{Bobev:2015jxa}.
The cross-ratios $r_1$ and $r_2$ can be related to the four external points by using superconformal symmetry to set $x_1 = (0,0,0)$, $x_2 = (\frac{r_1 + r_2}{2},\frac{r_1 - r_2}{2\I},0)$, $x_3 = (1,0,0)$ and $x_4 = \infty$ for the bosonic coordinates and $\bar{\theta}_1 = \theta_2 = \theta_3 = \bar{\theta}_4 = 0$ for the fermionic ones.
} 
of \eqref{eq:Conformal_blocks4D}.

We can denote the eigenvalues of the zero-magnon graph-building kernels in \eqref{eq:0MagnonPertExpansion_2} as
\begin{subequations}
\begin{align}
\int 
\dd^5 \bar{z}_1
\dd^5 z_2 ~
%\dd^3x_1\, \dd^2\bar{\theta}_1\, % \dd^2\theta_1\, \theta_1^2\;
%\dd^3x_2\, \dd^2\theta_2 \, % \dd^2\bar{\theta}_2\, \bar{\theta}_2^2\;
\Omega_{\Delta,S,0} {\scriptstyle (z_1, z_2 ; z_0)}
\left[ \mathbb{H} \circ \bar{\mathbb{H}} \right] {\scriptstyle (z_1, z_2 ; z_3, z_4) }
&=
E_{0} (\Delta, S)^2
\cdot \Omega_{\Delta,S,0} {\scriptstyle (z_3, z_4 ; z_0)} ~, \\
\int 
\dd^5 \bar{z}_1
\dd^5 z_2 ~
\Omega_{\Delta,S,0} {\scriptstyle (z_1, z_2 ; z_0)}
\left[ \mathbb{H} \circ \mathbb{P} \right] {\scriptstyle (z_1, z_2 ; z_3, z_4) }
&=
E_{0} (\Delta, S)
\cdot \Omega_{\Delta,S,0} {\scriptstyle (z_3, z_4 ; z_0)} ~.
\end{align}
\label{eq:0Magnon_ev_Omega}%
\end{subequations}
The subscript of $E_{0} (\Delta, S)$ signals that it corresponds to the zero-magnon case.
In \eqref{eq:Fishnet_0Magnon_Omega_scaling}, we see that the computation of the eigenvalue can be facilitated by considering the limit $x_0 \rightarrow \infty$ and in the superconformal case at hand, we additionally set $\theta_0, \bar{\theta}_0 =0$, which can be inverted by superconformal transformations.
The asymptotic expression for the corresponding eigenfunction gives
\begin{equation}
\Omega_{\Delta,S,0} (z_1, z_2 ; z_0)
\stackrel{\substack{x_0 \rightarrow \infty \\ \theta_0, \bar{\theta}_0 = 0 }}{\sim}
\Psi_{\frac{1}{2}-\frac{\Delta}{2} , \frac{S}{2}} (z_1 , z_2)
:=
\adjincludegraphics[valign=c,scale=1]{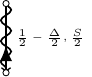} ~.
\label{eq:0Magnon_Omega_scaling}
\end{equation}
We notice that $\Psi_{u , \frac{S}{2}}$ has the form of a generalized spinning superpropagator \eqref{eq:SpinningSuperpropagators_prop}.
The calculation of the eigenvalue $E_{0} (\Delta, S)$ on $\Psi_{\frac{1}{2}-\frac{\Delta}{2} , \frac{S}{2}} (z_1 , z_2)$ of $\mathbb{H} \circ \bar{\mathbb{H}}$ and $\mathbb{H} \circ \mathbb{P}$ starts by making the following observation:
the generalized zero-magnon graph building operator $\mathbb{H}_{\boldsymbol{\omega}}$, which corresponds to the generalized superfishnet theory\footnote{Turning to the non-locally deformed theory \eqref{eq:Action_genSuperfishnet_Phi} is not necessary at this point, but we find it curious that the resulting eigenvalue admits only in the undeformed case the form of an inverse of an polynomial in $\Delta$, see \eqref{eq:0Magnon_ev_spinless}.} \eqref{eq:Action_genSuperfishnet_Phi}, has a valuable action on the zero-R-charge eigenfunction from \eqref{eq:0Magnon_Omega_scaling}.
It is not an eigenfunction, however, after merging the vertical superpropagators we can use the spinning chain rules \eqref{eq:SpinningSuperpropagators} of section \ref{eq:SpinningGeneralizedSuperpropagators} to obtain 
\begin{equation}
\begin{split}
&
\bra{ \Psi_{u , \frac{S}{2}} } \circ \mathbb{H}_{\boldsymbol{\omega}}
=
\adjincludegraphics[valign=c,scale=1.2]{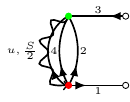} 
=
\adjincludegraphics[valign=c,scale=1.2]{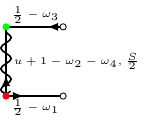}
\cdot
\prod_{i = 1}^4
c_0 (1 +\omega_i) \\
=&
4\, 
r_{\frac{S}{2}} ( \tfrac{1}{2} - \omega_3 , u + 1 - \omega_2 - \omega_4 , \tfrac{1}{2} - u - \omega_1 ) (-1)^S \cdot
\adjincludegraphics[valign=c,scale=1.2]{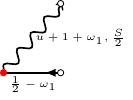}
\cdot 
\prod_{i = 1}^4
c_0 (1 +\omega_i) \\
=&
4\, (-1)^S
r_{\frac{S}{2}} ( \tfrac{1}{2} - \omega_3 , u + 1 - \omega_2 - \omega_4 , \tfrac{1}{2} - u - \omega_1 )
r_{\frac{S}{2}} ( \tfrac{1}{2} - \omega_1 , u + 1 + \omega_1, \tfrac{3}{2} - u  )
 \cdot
\adjincludegraphics[valign=c,scale=1]{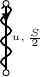}
\cdot 
\prod_{i = 1}^4
c_0 (1 +\omega_i) \\
=& : ~
\bra{\Psi_{u , \frac{S}{2}}^\dagger}
\cdot
(-1)^S \,
\mathrm{E}_{0,\boldsymbol{\omega}} (u,\tfrac{S}{2}) ~.
\end{split}
\label{eq:0MagnonPreDiag}
\end{equation}
Here, the first diagram is equipped with the flavor indices of the superfields, whose propagators make up the generalized graph building operator $\mathbb{H}_{\boldsymbol{\omega}}$.
We recall the restriction on the deformation parameters $\omega_1 + \omega_2 + \omega_3 + \omega_4 = 0$ and the definition of the factor $c_0 (v)$ in \eqref{eq:Factor_cEll}.
The factor $\mathrm{E}_{0,\boldsymbol{\omega}} (u , \tfrac{S}{2})$ is a combination of gamma functions involving the deformation parameters $\omega_i$ and the label of the eigenfunction $u$.
After using their functional relation, we find
\begin{equation}
\mathrm{E}_{0,\boldsymbol{\omega}} (u , \tfrac{S}{2})
=
\pi ^3
\frac{
\Gamma (\frac{S}{2}+u) 
\Gamma (\frac{1}{2}-\omega_2) 
\Gamma (\frac{1}{2}-\omega_4) 
\Gamma (\frac{S}{2}-u+\omega_2+\omega_4+\frac{1}{2})
}{
\Gamma (\frac{S}{2}- u +\frac{3}{2}) 
\Gamma (\omega_2+1) 
\Gamma (\omega_4+1) 
\Gamma (\frac{S}{2}+u-\omega_2-\omega_4+1)
}
 ~.
\end{equation}
Note that $\mathrm{E}_{0,\boldsymbol{\omega}} (u , \tfrac{S}{2})$ only depends on $\omega_2$ and $\omega_4$.
From \eqref{eq:0MagnonPreDiag}, and its hermitian conjugate 
\begin{equation}
\bra{\Psi_{u , \frac{S}{2}}^\dagger} \circ \bar{\mathbb{H}}_{\boldsymbol{\omega}} = \bra{\Psi_{u , \frac{S}{2}}} \cdot (-1)^S  \mathrm{E}_{0,\boldsymbol{\omega}} (u , \tfrac{S}{2}) ~,
\end{equation}
we find that $\mathbb{H}_{\boldsymbol{\omega}} \circ \bar{\mathbb{H}}_{\boldsymbol{\omega}}$ has the eigenfunction $\Psi_{u , \frac{S}{2}}$.
Furthermore, we observe that the permutation operator $\mathbb{P}$, which permutes start- and end-point, maps $\Psi_u$ to its hermitian conjugate $\Psi_u^\dagger$ times $(-1)^S$.
We can combine it with $\mathbb{H}_{\boldsymbol{\omega}}$ and find the eigenvalue equations
\begin{subequations}
\begin{align}
\bra{ \Psi_{u , \frac{S}{2}} } \circ
\left( \mathbb{H}_{\boldsymbol{\omega}} \circ \bar{\mathbb{H}}_{\boldsymbol{\omega}} \right)
&=
\bra{ \Psi_{u , \frac{S}{2}} } \cdot
\mathrm{E}_{0,\boldsymbol{\omega}} (u , \tfrac{S}{2})^2 ~, \\
\bra{ \Psi_{u , \frac{S}{2}} } \circ
\left( \mathbb{H}_{\boldsymbol{\omega}} \circ \mathbb{P} \right)
&=
\bra{ \Psi_{u , \frac{S}{2}} } \cdot
\mathrm{E}_{0,\boldsymbol{\omega}} (u , \tfrac{S}{2}) ~.
\end{align}
\end{subequations}
Setting the deformation parameters to zero, $\boldsymbol{\omega} = 0$, and substituting $u = \frac{1}{2} - \frac{\Delta}{2}$ gives the zero-magnon eigenvalue of the superfishnet theory \eqref{eq:Action_superfishnet_Phi}
\begin{equation}
E_{0} (\Delta , S)
=
\mathrm{E}_{0,\boldsymbol{\omega}=0} (\tfrac{1}{2} - \tfrac{\Delta}{2} , \tfrac{S}{2})
=
\frac{4 \pi ^4}{(1 + S - \Delta )( S + \Delta ) } ~.
\label{eq:0Magnon_ev_spinless}
\end{equation}
We observe that the eigenvalue is invariant under a transformation $E_{0} (1 - \Delta , S) = E_{0} (\Delta , S)$.
Now we can return to the four-point function \eqref{eq:0MagnonPertExpansion_2} and insert unity in the form of the completeness relation \eqref{eq:CompletenessRelation}.
Then the action of $\mathbb{H} \circ \bar{\mathbb{H}}$ and $\mathbb{H} \circ \mathbb{P}$ on the eigenfunctions will give the eigenvalues by \eqref{eq:0Magnon_ev_Omega}.
The decomposition into superconformal blocks \eqref{eq:Conformal_blocks3DSusy} gives a similar spectral decomposition as in \cite{Chang:2021fmd} and we obtain the superspace analog of \eqref{eq:FishnetCorrelationFunction_SpectralForm} 
\begin{equation}
\begin{split}
&\left\langle 
\mathrm{tr}
\left[
\Phi_1 (z_1)
\Phi_3^\dagger (z_2)
\right]
\mathrm{tr}
\left[
\Phi_1^\dagger (z_3)
\Phi_3 (z_4)
\right]
\right\rangle \\
&=
\sum_{S=0}^\infty
(-1)^{S+1}
\int_{\frac{1}{2}}^{\frac{1}{2} + \I \infty}
\frac{\I\, \dd \Delta}{2\pi\,\mathsf{c}_1(\Delta, S)}
\frac{E_{0} (\Delta, S)}{1 - \xi^2 E_{0} (\Delta, S)}
\int \dd^7 z_0 ~
\ket{
\bar{\Omega}_{\Delta,S,0} {\scriptstyle (z_1, z_2 ; z_0) }
} \;
\bra{
\Omega_{\Delta,S,0} {\scriptstyle (z_3, z_4 ; z_0) }
} \\
& =
\left(
\frac{1}{
x_{1\bar{2}}^2
x_{4\bar{3}}^2
}
\right)^{\frac{1}{2}}
\sum_{S, \Delta}
(-1)^S \,
\underset{\Delta}{\mathrm{Res}}
\left[
\frac{1}{\mathsf{c}_2 (\Delta , S)}
\frac{E_0(\Delta , S)}{1 - \xi^2 E_0(\Delta ,S)}
\right]
\,
\mathsf{g}_{\Delta , S} (r_1 , r_2) \\
&=
\left(
\frac{1}{
x_{1\bar{2}}^2
x_{4\bar{3}}^2
}
\right)^{\frac{1}{2}}
\sum_{S, \Delta}
\mathsf{C}_{\Delta ,S}
\,
\mathsf{g}_{\Delta , S} (r_1 , r_2)
 ~.
\end{split}
\label{eq:0MagnonPertExpansion_3}
\end{equation}
Here, we use $E_{0} (1 - \Delta , S) = E_{0} (\Delta , S)$ to extend the contour integral to the infinite line in the imaginary direction.
Using the residue theorem in combination with the assumption that spurious poles are absent gives an expression for the OPE coefficients $\mathsf{C}_{\Delta ,S}$.
Therefore, we have to determine the positions of the poles in the $\Delta$-integrant, which boils down to solve $1 = \xi^2 E_0(\Delta ,S)$ for $\Delta$.
Additionally, this signals the exchange of the operator $\mathcal{O}_{\Delta,S,0}$ and we obtain their exact scaling dimensions 
\begin{equation}
\Delta
=
1 
+
\frac{1}{2}
\left(
-1
\pm
2
\sqrt{
(S+\tfrac{1}{2})^2 - 4 \pi^4 \xi^2
}
\right) ~.
\label{eq:0Magnon_result_scalingDim_spinning}
\end{equation}
The upper expression is the scaling dimension of the operators $\mathrm{tr}[ \Phi_1 \partial^S \Phi_3^\dagger ]$ with classical dimension $\Delta \rightarrow \Delta_0 = 1 + S$ in the limit $\xi \rightarrow 0$.
It is remarkable to find a closed form of the scaling dimension at any value of the coupling.
In view of the AdS/CFT-correspondence, the result \eqref{eq:0Magnon_result_scalingDim_spinning} bridges between the weak-coupling expansion of the double-scaled $\beta$-deformation of ABJM theory and the strong-coupling expansion of string masses.
We observe that the form is a single square-root in contrast to the double-square-root of the bi-scalar fishnet theory \eqref{eq:Fishnet_ScalingDimensions}.

In particular, we obtain the exact scaling dimension of the operator $\mathrm{tr} [ \Phi_1 \Phi_3^\dagger ]$ by the solution corresponding to $\Delta_0 = 1$ with $S=0$, which is \cite{Kade:2024lkc}
\begin{equation}
\Delta\vert_{S=0}
=
1 + \gamma
=
1 + \frac{1}{2} \left( -1 + \sqrt{ 1 + 16 \pi ^4 \xi^2}\right) ~.
\label{eq:0Magnon_result_scalingDim}
\end{equation}
Therefore, we can confirm the result for the dispersion relation of the length-2 operator, which was obtained by Bethe ansatz in \cite{Caetano:2016ydc} and by re-summation of ladder diagrams in \cite{Bak:2009tq} for the undeformed ABJM theory.

Finally, we are able to compute the residues and eventually the OPE coefficient from \eqref{eq:0MagnonPertExpansion_3}.
We find the new result
\begin{equation}
\mathsf{C}_{\Delta ,S}
=
- 
2^{S-1-2 \Delta}
\pi
\frac{
\Gamma (S+\frac{3}{2})
\Gamma (\Delta )  
\Gamma (\frac{S-\Delta +2}{2} ) 
\Gamma (\frac{S+\Delta }{2} )
}{
\Gamma (S+1) 
\Gamma (\Delta +\frac{1}{2} ) 
\Gamma (\frac{S-\Delta +3}{2} ) 
\Gamma (\frac{S+\Delta +1}{2} )
} ~,
\label{eq:Superfishnet_exactOPEcoefficient}
\end{equation}
which is exact and yields the all-loop result for the correlation function by \eqref{eq:0MagnonPertExpansion_3}.
Expanding in $\xi$, where the dependence is hidden in $\Delta$ by \eqref{eq:0Magnon_result_scalingDim_spinning}, gives its perturbative corrections.

\subsubsection{Two-Magnon case}
\label{subsec:TwoMagnonCase}
After the successful diagonalization of the graph-builders appearing in the zero-magnon case in section \ref{subsec:ZeroMagnonCase}, we consider the supergraph expansion of another correlation function, which is
\begin{equation}
\begin{split}
& \left\langle 
\mathrm{tr}
\left[
\Phi_2 (z_1)
\Phi_1 (z_1)
\Phi_2 (z_2)
\Phi_1 (z_2)
\right]
\mathrm{tr}
\left[
\Phi_1^\dagger (z_3)
\Phi_2^\dagger (z_3)
\Phi_1^\dagger (z_4)
\Phi_2^\dagger (z_4)
\right]
\right\rangle \\
= &
\adjincludegraphics[valign=c,scale=1]{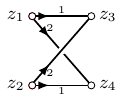}
+
\xi^4
\adjincludegraphics[valign=c,scale=1]{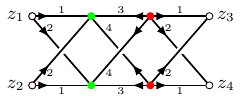}
+
\xi^8
\adjincludegraphics[valign=c,scale=1]{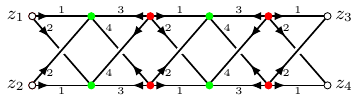} \\
& + \cdots + (z_3 \leftrightarrow z_4) ~.
\end{split}
\label{eq:2MagnonPertExpansion}
\end{equation}
We call the study of this correlation function the two-magnon case because there are two strands of propagators composed by alternating flavor-two and flavor-four superpropagators.
They wind around the cylindrical supergraphs, but unlike in the zero-magnon case \eqref{eq:0MagnonPertExpansion}, they are not closed, but rather start and end in the external operators of the correlation function.
At this point, one might wonder if there is a one-magnon case, much like in bi-scalar fishnet theory \cite{Gromov:2018hut}.
However, the alternating chirality of the vertices in the horizontal direction rules out the corresponding supergraphs.
Furthermore, we could have considered the correlation function with flavors $(\mathrm{1} , \mathrm{2}) \leftrightarrow (\mathrm{3} , \mathrm{4})$ exchanged, which would give an equivalent derivation of the scaling dimensions.

In \eqref{eq:2MagnonPertExpansion}, we can identify the graph-builders
\begin{equation}
\begin{aligned}[c]
\mathbb{H}
=
\adjincludegraphics[valign=c,scale=1]{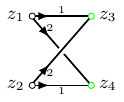} ~,
\end{aligned}
\qquad\qquad
\begin{aligned}[c]
\bar{\mathbb{H}}
=
\adjincludegraphics[valign=c,scale=1]{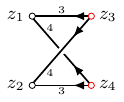} ~,
\end{aligned}
\begin{aligned}[c]
\qquad \qquad
\mathbb{P}
=
\adjincludegraphics[valign=c,scale=1]{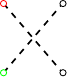} ~,
\end{aligned}
\label{eq:2MagnonGraphBuilder}
\end{equation}
and rewrite the correlation function \eqref{eq:2MagnonPertExpansion} as
\begin{equation}
 \left\langle 
\mathrm{tr}
\left[
\Phi_2 (z_1)
\Phi_1 (z_1)
\Phi_2 (z_2)
\Phi_1 (z_2)
\right]
\mathrm{tr}
\left[
\Phi_1^\dagger (z_3)
\Phi_2^\dagger (z_3)
\Phi_1^\dagger (z_4)
\Phi_2^\dagger (z_4)
\right]
\right\rangle
=
\frac{\mathbb{H}
\circ
(1 + \mathbb{P})}{1 - \xi^4 \mathbb{H} \circ \bar{\mathbb{H}} } ~.
\label{eq:2MagnonPertExpansion_2}
\end{equation}
Unfortunately, we are unaware of the precise form of the conformal triangle \eqref{eq:ConformalTriangle3pt} for the operators $\mathcal{O}_1 = \mathcal{O}_2 = \Phi_4 \Phi_3$.
However, since both operators are charged under the R-symmetry \eqref{eq:U1_intakt_transformation} with charge $R_{1} = R_{2} = \frac{1}{2} + \frac{1}{2} = 1$, the superconformal triangle corresponds to an operator $\mathcal{O}_{\Delta , S , R}$ with non-trivial R-charge $R = -2$.
We want to employ the same trick as in the zero-magnon case \eqref{eq:0Magnon_Omega_scaling} and amputate the superspace point $z_0$ where the operator $\mathcal{O}_{\Delta , S , R}$ sits.
Therefore, we again consider the limit $x_0 \rightarrow \infty$ and set the fermionic coordinates to zero $\theta_0 , \bar{\theta}_0 = 0$.
The two-point function that we should obtain will carry R-charge $-2$, hence, regarding \eqref{eq:FeynmanRules_AuxTwoPointFctns}, we assume the eigenfunction to have the form
\begin{equation}
\Omega_{\Delta,0,-2} (z_1, z_2 ; z_0)
\stackrel{\substack{ x_0 \rightarrow \infty \\ \theta_0, \bar{\theta}_0 = 0 }}{\sim}
\Psi_{\frac{1}{2}-\frac{\Delta}{2}} (z_1 , z_2)
:=
\frac{\theta_{12}^2}{\left[x_{12}^2 \right]^{\frac{1}{2}-\frac{\Delta}{2}}} 
= ~
\adjincludegraphics[valign=c,scale=1]{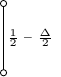} 
\label{eq:2Magnon_Omega_scaling}
\end{equation}
for the spinless case $S=0$.
We may anticipate the form of the spinning eigenfunction, it has the form of the spinning auxiliary two-point function \eqref{eq:SpinningSuperpropagators_aux}.
However, we are unable to find the eigenvalue in the case $S>0$.
To find the exponent of $\Psi_{u}$, we had to use the more general relation for its scaling dimension $\Delta_{\Psi_{u}} = 1 - 2u = \Delta - \Delta_1 - \Delta_2 + R_1 + R_2$ also considering the R-charges of the operators in $\Omega_{\Delta,0,R} (z_1, z_2 ; z_0)$.

Next, diagonalization of the graph-builder $\mathbb{H}$ on the eigenfunction \eqref{eq:2Magnon_Omega_scaling} may be carried out by explicitly doing the superspace integration.
The eigenvalue of the two-magnon graph-builder can be determined by reducing the superspace integral to the three-dimensional bosonic kite integral \cite{Kotikov:2024yok}.
Since the kite integral is more complicated in the case of arbitrary propagator powers \cite{Grozin:2012xi}, we proceed with the calculation in the case of the undeformed superfishnet theory \eqref{eq:Action_superfishnet_Phi}.
Acting with the graph-builder $\mathbb{H}$ of \eqref{eq:2MagnonGraphBuilder} on the eigenfunction, we find
\begin{equation}
\begin{split}
\Psi_u \circ \mathbb{H}
&=
\adjincludegraphics[valign=c,scale=1]{figures/superFN/2ptfctn/twomagnon/diagonalization_left/Builder_H_ev.pdf}
=
-\left[
\int 
\dd^5 z_0 \; \dd^5 z_{0'} ~
\frac{c_0(1)}{\left[x_{0\bar{1}}^2 \right]^{\frac{1}{2}}}
\frac{c_0(1)}{\left[x_{0'\bar{1}}^2 \right]^{\frac{1}{2}}} 
\frac{\theta_{00'}^2}{\left[x_{00'}^2 \right]^{u}} 
\frac{c_0(1)}{\left[x_{0\bar{2}}^2 \right]^{\frac{1}{2}}}
\frac{c_0(1)}{\left[x_{0'\bar{2}}^2 \right]^{\frac{1}{2}}} 
\right]_{\theta_{1,2} = 0} \\
&=
-c_0(1)^4
\int 
\dd^2 \theta_0 ~
\e^{2\I \theta_0 \gamma^\mu \bar{\theta}_{12} \partial_{1,\mu}}
\int \dd^3 x_0 \; \dd^3 x_{0'}
\frac{1}{\left[x_{10}^2 \right]^{\frac{1}{2}}}
\frac{1}{\left[x_{10'}^2 \right]^{\frac{1}{2}}} 
\frac{1}{\left[x_{00'}^2 \right]^{u}} 
\frac{1}{\left[x_{20}^2 \right]^{\frac{1}{2}}}
\frac{1}{\left[x_{20'}^2 \right]^{\frac{1}{2}}} \\
&=
c_0(1)^4 \cdot
\bar{\theta}_{12}^2 \square_1 \;
\mathsf{kite}^{(3)}
( x_{12}^2, u )  ~.
\end{split}
\label{eq:2Magnon_ev_1}
\end{equation}
Here, we have evaluated the Gra\ss mann delta function $\theta_{00'}^2$ by the $x_{0'}$-integration, which allows us to factor out the common $\theta_0$-dependent exponential with the differential operators.
The integral over bosonic space is the three-dimensional kite diagram, which can be determined as $\mathsf{kite}^{(3)}( x_{12}^2, u ) = \frac{I^{(3)}(u)}{\left[ x^2 \right]^{u-1}}$ \cite{Kotikov:2024yok}, with the function
\begin{equation}
I^{(3)}(u)
=
\frac{1}{\pi^2 \left( \frac{3}{2} - u \right) \left( u-1 \right) }
\frac{1}{32 \pi^2}
\int_1^\infty
\dd s ~
\frac{ s^{\frac{1}{2} - u} + s^{ - 2 + u } }{\sqrt{1 + s}}\;
\mathrm{log}
\left[
\frac{\sqrt{1 + s} + 1}{\sqrt{1 + s} - 1}
\right] ~.
\label{eq:2Magnon_Ifactor_intrepr}
\end{equation}
The integral representation can be expressed in terms of the hypergeometric $_3 F_2$-function at unit argument; the corresponding expression for the eigenvalue is presented in \eqref{eq:2Magnon_ev_3F2}.
The eigenvalue, corresponding to the eigenvalue equation $\Psi_u \circ \mathbb{H} = \mathrm{E}_2 (u) \cdot \Psi_u^\dagger$, is obtained after performing the derivatives in \eqref{eq:2Magnon_ev_1} and reads
\begin{equation}
\mathrm{E}_2 (u)
=
-4 c_0(1)^4 \cdot
\left(
u - 1 
\right)
\left(
\tfrac{3}{2} - u
\right)
I^{(3)} (u) ~.
\label{eq:2Magnon_ev_mathrmE}
\end{equation}
 
\begin{figure}[h]
\includegraphics[scale=0.7]{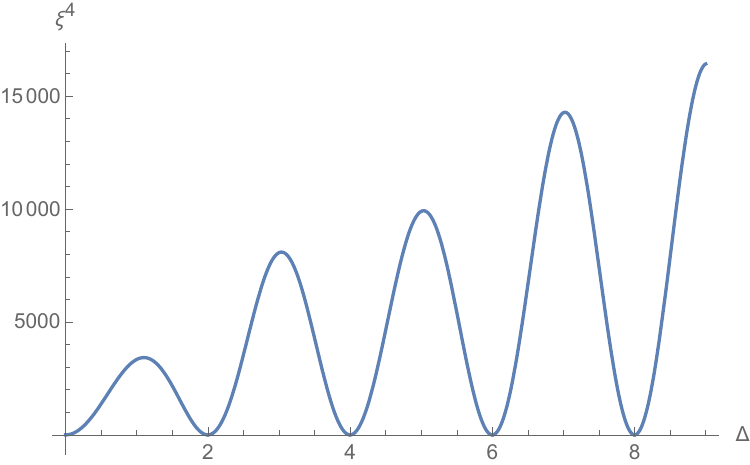} 
\centering
\caption{The plot shows the function $\xi^4 = E_2 (\Delta)^{-2}$ for $\Delta \in [0,9]$. We observe that the zeros $\Delta_0$ occur at $0$, $2$, $4$, $6$, $8$, ... . Unfortunately, the approximations in \eqref{eq:DeltaApproximatinos} are of too low order to be accurate in the shown range of the coupling. Each operator's scaling dimension will collide with the neighboring operators at a local maximum for increasing coupling. The critical value for the coupling increases with the classical dimension of the operators.}
\label{fig:Plot_xiofDelta}
\end{figure}

The calculation presented here works analogously for the conjugated graph-builder $\bar{\mathbb{H}}$ with the same eigenvalue, $\Psi_{u}^\dagger \circ \bar{\mathbb{H}} = \mathrm{E}_2 (u) \cdot \Psi_u$.
Hence, we obtain the eigenvalue equation
\begin{equation}
\Psi_u \circ 
\left(\mathbb{H} \circ \bar{\mathbb{H}}\right) 
= 
\mathrm{E}_2 (u)^2 \cdot \Psi_u
\end{equation}
where the eigenvalue upon action on the eigenfunction is $E_2 (\Delta) = \mathrm{E}_2 (\tfrac{1}{2}-\tfrac{\Delta}{2})$ and from \eqref{eq:2Magnon_Ifactor_intrepr} and \eqref{eq:2Magnon_ev_mathrmE} we find
\begin{equation}
E_2 (\Delta)
=
\frac{\csc (\pi ( \frac{\Delta }{2} + 1 ))\, \Gamma ( \frac{\Delta}{2} + 1 )}
{32 \sqrt{\pi }\, \Gamma ( \frac{\Delta}{2} + \frac{3}{2} )}
-
\frac{\, _3F_2\left(1 , 1 , \frac{\Delta}{2} + \frac{3}{2} ; \frac{\Delta}{2} + 2 , \frac{\Delta}{2} + \frac{5}{2} ; 1 \right)}
{16 \pi ^2 \left( \Delta + 3 \right) \left(\frac{\Delta }{2}+1\right)} ~.
\label{eq:2Magnon_ev_3F2}
\end{equation}
We observe that the action of the graph-builder $\mathbb{P}$ is trivial, since $\Psi_u$ is symmetric under exchange of the external points, $\Psi_u \circ \mathbb{P} = \Psi_u$.
The condition of \eqref{eq:2MagnonPertExpansion_2} for having a pole relates the coupling with the square of $E_2 (\Delta)$ and reads
\begin{equation}
1 
=
\xi^4 
E_2 (\Delta)^2 ~.
\label{eq:2Magnon_ImplicitScalingDims}
\end{equation}
In the weak-coupling limit, the function $E_2 (\Delta)^2$ has to have poles at the values of classical scaling dimensions $\Delta_0$ of the exchanged operators.
We deduce them at even, positive integers, including zero, $\Delta_0 \in 2\mathbb{N}_0$, see figure \ref{fig:Plot_xiofDelta}.
Contrary to the zero-magnon case, we cannot invert the function $E_2 (\Delta)^{-2}$ to access the anomalous dimension of the exchanged operator $\Delta$ analytically.
However, we can expand the function $E_2 (\Delta)^{-2}$ around the zeros $\Delta_0$ and invert the series perturbatively, where we obtain two solutions for each branch, related by $\xi^2 \rightarrow - \xi^2$.
The perturbative inversion can be performed to an arbitrary order in the coupling.
For the lowest three scaling dimensions, up to eight loops and after rescaling $\xi \rightarrow \pi \cdot \xi$ we find
\begin{subequations}
{\allowdisplaybreaks
\begin{align}
\begin{split}
\Delta^{(2)}
~=~ &
2 
\pm \frac{\xi^2}{12} 
- \frac{\xi^4}{576} 
\pm \frac{18 \pi ^2-97}{248832} \xi^6 \\
& +\frac{ 3803 - 2268 \zeta_3 + 18 \pi ^2 (\log (4096)-19)}{35831808} \xi^8
+\mathcal{O}(\xi^{10}) ~,
\end{split} \\
\begin{split}
\Delta^{(4)}
~=~  &
4
\pm\frac{4 \xi^2}{15}
-\frac{\xi^4}{7200}
\pm \frac{28800 \pi ^2-191191}{777600000} \xi^6 \\
& \frac{ 191678057 - 145152000 \zeta_3 + 28800 \pi ^2 (480 \log (2)-421)}{5598720000000} \xi^8 
+\mathcal{O}(\xi^{10}) ~,
\end{split} \\
\begin{split}
\Delta^{(6)}
~=~ &
6
\pm \frac{2 \xi^2}{35}
-\frac{4 \xi^4}{2940}
\pm \frac{ 352800 \pi ^2 - 2244421}{15126300000} \xi^6 \\
& +\frac{2972114029 - 2074464000 \zeta_3 + 117600 \pi ^2 (1680 \log (2)-1801)}{148237740000000} \xi^8
+\mathcal{O}(\xi^{10}) .
\end{split}
\end{align}
\label{eq:DeltaApproximatinos}%
}%
\end{subequations}
The scaling dimension $\Delta^{(2)}$ corresponds to the exchanged operator $\mathrm{tr}\left[ \Phi_1^\dagger \Phi_2^\dagger \Phi_1^\dagger \Phi_2^\dagger \right]$, which matches with the classical values of the charges $\Delta^{(2)} = 2$ and $R = -2$.
The other solutions with higher classical scaling dimensions correspond to insertions of bosonic space derivatives, like e.\,g.\ $\square \; \mathrm{tr}\left[ \Phi_1^\dagger \Phi_2^\dagger \Phi_1^\dagger \Phi_2^\dagger \right]$.

\section{The super brick wall theory}
\label{sec:TheSuperBrickWallTheory_sec}
The super brick wall theory is the double-scaled $\beta$-deformation of $\mathcal{N} = 4$ SYM theory.
The $\beta$-deformation is a special case of the $\gamma$-deformation, which is presented in section \ref{susec:DoubleScalingLimitOfGammaDeformedSYM}, where all three deformation parameters have the same value, i.\,e.\ $\gamma_1 = \gamma_2 = \gamma_3 = \beta$.
Therefore, the deformation preserves one supercharge and $\beta$-deformed SYM is $\mathcal{N} = 1$ supersymmetric.
It implies that the deformed theory can be written as a superspace action like $\mathcal{N} = 4$ SYM and the latter does not have to be decomposed in the spacetime component action \eqref{eq:SYM_Action_components} prior to the deformation, as it was the case for the $\gamma$-deformation at generic $\gamma_i$.
Hence, the plan of this section is to start with the $\mathcal{N} = 1$ superspace formulation of $\mathcal{N} = 4$ SYM and to perform the $\beta$-deformation in a $\mathcal{N} = 1$ supersymmetry covariant manner, which results in the superspace action of the super brick wall theory.
Nevertheless, we will expand the superspace action in components as well, since it yields some of the double-trace terms, which we had to include in the case of the bi-scalar fishnet theory by quantum corrections \eqref{eq:Fishnet_DoubleTraceTerms_mixedflavor}, automatically.
Furthermore, we present a non-local deformation of the super brick wall theory.
In addition, we use superspace integral relations, such as the chain relations from section \ref{subsec:SuperChainRelations}, to reproduce the result for the length-two operator $\mathrm{tr} \left( \Phi_1^2 \right) $ from the literature on dynamical fishnet theory \cite{Kazakov:2018gcy}.  

We copy the $\mathcal{N} = 1$ superspace action of $\mathcal{N} = 4$ SYM from \eqref{eq:SYM_Action},
\begin{equation}
\begin{split}
S
= &
\int \dd^4 x\; \dd^2\theta\, \dd^2 \bar{\theta}\;
\sum_{i=1}^{3} \,
\mathrm{tr} \left(
\e^{-g \mathcal{V}} \Phi_i^\dagger \e^{g \mathcal{V}} \Phi_i
\right)
+
\frac{1}{4g^2}
\int \dd^4 x\; \dd^2\theta\;
\mathrm{tr} \left(
W^\alpha W_\alpha
\right)\\
& +
\I g
\int \dd^4 x\; \dd^2\theta\;
\mathrm{tr} \left(
\Phi_1 \com{\Phi_2}{\Phi_3}
\right)
+
\I g
\int \dd^4 x\; \dd^2\bar{\theta}\;
\mathrm{tr} \left(
\Phi^\dagger_1 \com{\Phi^\dagger_2}{\Phi^\dagger_3}
\right) ~,
\end{split}
\label{eq:SYM_Action_gRescaled}
\end{equation}
where we have rescaled coupling $g\rightarrow \sqrt{2} g$.
In the spirit of section \ref{susec:DoubleScalingLimitOfGammaDeformedSYM}, we perform the $\beta$-deformation of the R-symmetry $\SU{4}$ of \eqref{eq:SYM_Action_gRescaled} by replacing the ordinary products of fields with the star product $\Phi_i \star \Phi_j := \e^{\frac{\I}{2}\; \mathrm{det}\left(\boldsymbol{\gamma} \vert \mathbf{q}_i \vert \mathbf{q}_j\right)} \Phi_i \Phi_j$, c.\,f.\ \eqref{eq:StarProduct} and \eqref{eq:StarProduct_multipleFactors}, and setting the three deformation parameters to the same value $\boldsymbol{\gamma} = (\beta, \beta, \beta)$ \cite{Imeroni:2008cr}.
The $\mathbf{q}_i$ are the charges of the chiral superfields $\Phi_i$ under the Cartan elements of the R-symmetry $\SU{4}$.
They are $\mathbf{q}_1 = \left( \frac{1}{2}, -\frac{1}{2}, -\frac{1}{2} \right)$, $\mathbf{q}_2 = \left( -\frac{1}{2}, \frac{1}{2}, -\frac{1}{2} \right)$, and $\mathbf{q}_3 = \left( -\frac{1}{2}, -\frac{1}{2}, \frac{1}{2} \right)$ for the three chiral superfields and zero for the gauge field \cite{Jin:2012np}.
The superfields share the R-charges with their component fields, but note that in table \ref{tab:SYM_Rcharges}, we use a different R-symmetry basis for the scalar fields $\phi_i$.
This deformation leaves us with the superpotential of the $\beta$-deformation of $\mathcal{N}=4$ SYM \cite{Lunin:2005jy,Jin:2012np,Fokken:2013mza}
\begin{equation}
\I g \int \dd^4 x \; \dd^2 \theta ~ \mathrm{tr}\left[ q\; \Phi_1 \Phi_2 \Phi_3 - q^{-1} \Phi_1 \Phi_3 \Phi_2 \right] + \mathrm{h.c.} ~,
\label{eq:MarginalDeformation}
\end{equation}
where we introduced the abbreviation $q = \e^{\frac{\I}{2}\beta}$.

Next, we perform the planar limit \eqref{eq:SYM_PlanarLimit}, by sending $g\rightarrow 0$ and $\mathrm{N} \rightarrow \infty$, while keeping the 't Hooft coupling $\lambda = g \sqrt{\mathrm{N}}$ fixed.
After appropriate rescalings of the fields by $\sqrt{\mathrm{N}}$, as indicated above \eqref{eq:SYM_Action_components}, superspace Feynman graphs in double-line notation with the smallest genus will dominate.
In a second step, we perform the double-scaling limit consisting of $\lambda \rightarrow 0$ and $\beta \rightarrow -\I \infty \Rightarrow q\rightarrow \infty$, while the product $\xi := \lambda\cdot q$ remains finite. 
After this operation the gauge fields decouple and two out of the four terms of the deformation \eqref{eq:MarginalDeformation} vanish. 
We consider the obtained theory in the planar limit $\mathrm{N}\rightarrow \infty$, where the leading order is described by toroidal double-line Feynman graphs.
We obtain the concise action
\begin{equation}
\begin{split}
S ~=~
& S_\mathrm{kin} + S_\mathrm{int}\\
~=~
& \mathrm{N}\int \dd^4 x\; \dd^2\theta \dd^2 \bar{\theta}
\left\lbrace 
\sum_{i=1}^{3} 
\mathrm{tr} \left[ 
\Phi_i^\dagger \Phi_i
\right]
+
\I \xi \cdot \bar{\theta}^2\,
\mathrm{tr}\left[ \Phi_1 \Phi_2 \Phi_3 \right]
+
\I \xi \cdot \theta^2\,
\mathrm{tr}\left[ \Phi_1^\dagger \Phi_2^\dagger \Phi_3^\dagger \right]
\right\rbrace ~,
\end{split}
\label{eq:DSbetaDeformation}
\end{equation}
where the squares $\theta^2$ and $\bar{\theta}^2$ in general act as delta function in the fermionic coordinates, see appendix \ref{sec:BerezinIntegral_D4N1}.
Accordingly, we can read off the two Feynman supergraph vertices in Minkowski space
\begin{equation}
\adjincludegraphics[valign=c,scale=0.5]{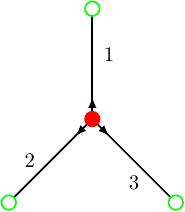}
\sim
- \mathrm{N}\, \xi \int \dd^4x\; \dd^2\theta\,\dd^2\bar{\theta} \;\delta^{(2)}(\bar{\theta}) ~, ~~~
\adjincludegraphics[valign=c,scale=0.5]{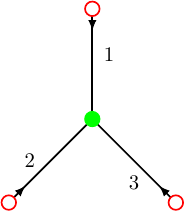}
\sim
- \mathrm{N}\, \xi \int \dd^4x\; \dd^2\theta\,\dd^2\bar{\theta} \;\delta^{(2)}(\theta) ~,
\label{eq:betaDef_FeynmanRules_vertices}
\end{equation}
after performing a Wick rotation.

\subsection{Component action}
\label{subsec:ComponentAction_SBW}
Let us now make contact with the component field action of the $\mathcal{N} = 1 $ double-scaled $\beta$-deformation of $\mathcal{N}=4$ SYM, which is the $\chi$-CFT in \eqref{eq:Chi_SYM_action} with the couplings identified, i.\,e.\ $\xi_1 = \xi_2 = \xi_3 = \xi$.
We may expand \eqref{eq:DSbetaDeformation} in Gra\ss mann components.
The kinetic terms from the canonical K\"{a}hler potential are 
\begin{equation}
S_\mathrm{kin} =
\mathrm{N}\int \dd^4 x\; \dd^2\theta \dd^2 \bar{\theta} 
\sum_{i=1}^{3} 
\mathrm{tr} \left[ 
\Phi_i^\dagger \Phi_i
\right]
=
\mathrm{N}\int \dd^4 x\; 
\sum_{i=1}^{3} 
\mathrm{tr} \left[ 
\phi_i^\dagger \square \phi_i -
\I \bar{\psi}_i\bar{\sigma}^\mu \partial_\mu \psi_i +
F_i^\dagger F_i
\right] ~,
\end{equation}
while the interaction terms are
\begin{equation}
\begin{split}
S_\mathrm{int} 
&=
\mathrm{N} \cdot \I \xi 
\int \dd^4 x\; \dd^2\theta \;
\mathrm{tr} \left[
\Phi_1 \Phi_2 \Phi_3
\right]_{\bar{\theta} = 0}
+
\mathrm{N} \cdot \I \xi
\int \dd^4 x\; \dd^2\bar{\theta} \;
\mathrm{tr} \left[
\Phi^\dagger_1 \Phi^\dagger_2 \Phi^\dagger_3
\right]_{\theta = 0}\\
&=
\mathrm{N} \cdot \I \xi 
\int \dd^4 x \;
\mathrm{tr} \left[
\phi_1 \phi_2 F_3 + 
\phi_1 F_2 \phi_3 +
F_1 \phi_2 \phi_3 - 
\phi_1 \psi_2 \psi_3 -
\phi_2 \psi_3 \psi_1 -
\phi_3 \psi_1 \psi_2
\right]\\
& \phantom{=} +
\mathrm{N} \cdot \I \xi
\int \dd^4 x \;
\mathrm{tr} \left[
\phi^\dagger_1 \phi^\dagger_2 F^\dagger_3 + 
\phi^\dagger_1 F^\dagger_2 \phi^\dagger_3 +
F^\dagger_1 \phi^\dagger_2 \phi^\dagger_3 - 
\phi^\dagger_1 \bar{\psi}_2 \bar{\psi}_3 -
\phi^\dagger_2 \bar{\psi}_3 \bar{\psi}_1 -
\phi^\dagger_3 \bar{\psi}_1 \bar{\psi}_2
\right] ~.
\end{split}
\end{equation}
The auxiliary fields $F_i$ and $F^\dagger_i$ are not dynamical and we may eliminate them with the help of their equations of motion:
\begin{equation}
\begin{aligned}[c]
F_1^A = - \I \xi \, \phi^*_{2,B} \phi^*_{3,C} \cdot \mathrm{tr}\left[ T^A T^B T^C \right] ~,\\
F_2^A = - \I \xi \, \phi^*_{3,B} \phi^*_{1,C} \cdot \mathrm{tr}\left[ T^A T^B T^C \right] ~,\\
F_3^A = - \I \xi \, \phi^*_{1,B} \phi^*_{2,C }\cdot \mathrm{tr}\left[ T^A T^B T^C \right] ~,
\end{aligned}
\qquad
\begin{aligned}[c]
F^{*,A}_1 = - \I \xi \, \phi_{2,B} \phi_{3,C} \cdot \mathrm{tr}\left[ T^A T^B T^C \right] ~,\\
F^{*,A}_2 = - \I \xi \, \phi_{3,B} \phi_{1,C} \cdot \mathrm{tr}\left[ T^A T^B T^C \right] ~,\\
F^{*,A}_3 = - \I \xi \, \phi_{1,B} \phi_{2,C} \cdot \mathrm{tr}\left[ T^A T^B T^C \right] ~.
\end{aligned}
\end{equation}
We observe that the r.\,h.\,s.\ do not have the form of commutators anymore, in comparison to the equations of motions from the undeformed action of $\mathcal{N} = 4$ in \eqref{eq:Auxiliary_SYM_F}.
As shown in \cite{Fokken:2013aea}, it implies that integrating out the auxiliary fields may produce double-trace terms by the completeness relation of the generators of the adjoint representation of the gauge group $\SU{\mathrm{N}}$, as displayed in \eqref{eq:Adjoint_generators}.
Thus, we obtain the on-shell component action
\begin{equation}
\begin{split}
S
=
\mathrm{N}\int \dd^4 x\;  
\mathrm{tr} 
\left\lbrace
\sum_{i=1}^{3}
\left[ 
\phi_i^\dagger \square \phi_i -
\I \bar{\psi}_i\bar{\sigma}^\mu \partial_\mu \psi_i
\right]
+
\xi^2
\left[
\phi_1 \phi_2 \phi^\dagger_1 \phi^\dagger_2 +
\phi_3 \phi_1 \phi^\dagger_3 \phi^\dagger_1 +
\phi_2 \phi_3 \phi^\dagger_2 \phi^\dagger_3
\right]
\right.\\
\left.
- 
\I \xi
\left[
\phi_1 \psi_2 \psi_3 +
\phi_2 \psi_3 \psi_1 +
\phi_3 \psi_1 \psi_2 
\right]
- 
\I \xi
\left[
\phi^\dagger_1 \bar{\psi}_2 \bar{\psi}_3 +
\phi^\dagger_2 \bar{\psi}_3 \bar{\psi}_1 +
\phi^\dagger_3 \bar{\psi}_1 \bar{\psi}_2
\right]
+
\mathcal{L}_\mathrm{dt}
\right\rbrace ~, \label{eq:DS_components}
\end{split}
\end{equation}
and $\mathcal{L}_\mathrm{dt}$ describes the double-trace interaction terms, which survive in the double-scaling limit.
They have the form
\begin{equation}
\mathcal{L}_\mathrm{dt}
=
-\frac{\xi^2}{\mathrm{N}}
\left\lbrace
\mathrm{tr}
	\left[
	\phi_1 \phi_2
	\right]
\mathrm{tr}
	\left[
	\phi_1^\dagger \phi_2^\dagger
	\right]
+
\mathrm{tr}
	\left[
	\phi_1 \phi_3
	\right]
\mathrm{tr}
	\left[
	\phi_1^\dagger \phi_3^\dagger
	\right]
+
\mathrm{tr}
	\left[
	\phi_2 \phi_3
	\right]
\mathrm{tr}
	\left[
	\phi_2^\dagger \phi_3^\dagger
	\right]
\right\rbrace ~.
\label{eq:DoubleTraceTerms}
\end{equation}
It is satisfying to see that they do not need to be added due to radiative corrections to the single-trace component action, as seen on the example of the bi-scalar fishnet theory in \eqref{eq:Fishnet_DoubleTraceTerms}; instead they naturally and directly emerge from the $\mathcal{N} = 1$ superspace formulation.
However, we also notice that not all double-trace terms appear from the superspace approach, but only the first one in \eqref{eq:Fishnet_DoubleTraceTerms_mixedflavor}.
The double-trace terms are however not needed in our below analysis, since they do not appear in the thermodynamic limit of vacuum diagrams and the superconformal symmetry makes them redundant in the computation of exact scaling dimensions, similar to section \ref{subsec:ExactAnomalousDimensionBiScalarFN}.
Furthermore, we stress that the single-trace action \eqref{eq:DS_components} is equivalent to the $\chi$-CFT \cite{Gurdogan:2015csr}, when we identify all couplings at the same value $\xi$, after changing the convention $\xi \rightarrow -\xi$.
In \cite{Kazakov:2018gcy}, it was shown that the couplings do not run at the supersymmetric point of the $\chi$-CFT and thus the super brick wall theory \eqref{eq:DS_components} with double-trace terms \eqref{eq:DoubleTraceTerms} is free of divergences.

\subsection[Introducing a spectral parameter into the $\mathcal{N} = 1$ action]{Introducing a spectral parameter into the $\mathbf{\mathcal{N} = 1}$ action}
The double-scaled $\chi$-CFTs are expected to inherit integrability from their $\mathcal{N} = 4$ SYM ``parent theory'' \cite{Gurdogan:2015csr}, which should include the special case \eqref{eq:DSbetaDeformation}.
However, putting quantum integrability to good use always requires the introduction of a suitable spectral parameter.
In general, it is not easy to find it in an integrable, planar, conformal QFT.
This is certainly the case for $\mathcal{N} = 4$ SYM, where it somewhat mysteriously first appears when converting local composite operators into quantum spin chains.
Impressively, in the much simpler setting of the fishnet model, it has recently been shown in \cite{Kazakov:2018qbr} that the correct spectral parameter may be directly introduced by deforming the model's action, at the cost of giving up the locality of the resulting ``QFT''. 
Unfortunately, this construction has not yet been achieved for general $\chi$-CFTs.
We will now show that in the special case of the $\mathcal{N} = 1$ $\chi$-CFT, a suitable deformation may nevertheless be found by deforming its $\mathcal{N} = 1$ action.
%Integrability requires to widen the scope to the theory space containing the object of study, in our case \eqref{eq:DSbetaDeformation}. 
%This integrable fiber is parameterized by the spectral parameter and for a fixed value we find our theory under investigation.
%However, it is also interesting to consider theories at other points of the spectral parameter.
%For us, the spectral parameter will enter in the exponents of the super propagators.
%Once we set the exponent to $1$ we recover the double-scaled $\beta$-deformation \eqref{eq:DSbetaDeformation}.
%In order to make the exponent physical \cite{Alfimov:2023vev}, we have to modify the kinetic terms of the component fields, or in the language of the chiral superfield, we have to modify the canonical K\"{a}hler potential.
%Its diagonal part gets enhanced by an extra d'Alembertian raised to the spectral parameter, for each of the three flavors,
%The deformed action is
We propose
\begin{equation}
\begin{split}
S_{\boldsymbol{\omega}} ~=~
& S_{\mathrm{kin},\boldsymbol{\omega}} + S_{\mathrm{int},\boldsymbol{\omega}}\\
~=~
& \mathrm{N}\int \dd^4 x\; \dd^2\theta \dd^2 \bar{\theta}
\left\lbrace 
\sum_{i=1}^{3} 
\mathrm{tr} \left[ 
\Phi_i^\dagger \square^{\omega_i} \Phi_i
\right]
+																																
\I \xi \cdot \bar{\theta}^2 \;
\mathrm{tr}\left[ \Phi_1 \Phi_2 \Phi_3 \right]
+
\I \xi																															\cdot \theta^2 \;
\mathrm{tr}\left[ \Phi_1^\dagger \Phi_2^\dagger \Phi_3^\dagger \right]
\right\rbrace ~,
\end{split}
\label{eq:DSbetaDeformation_gen}
\end{equation}
where $\boldsymbol{\omega}$ is a shorthand notation for the deformation parameters $\omega_1$, $\omega_2$ and $\omega_3$.
It will be related to the model's spectral parameters, see \eqref{eq:gen_superPropagator_graphical} below.
The mass dimension of the chiral superfield gets deformed to $\left[\Phi_i\right] = [\Phi^\dagger_i ] = \frac{(D-2\mathcal{N})-2\omega_i}{2}\vert_{D=4,\mathcal{N}=1} = 1 - \omega_i$. For the interaction to be marginal, we require $3 - (\omega_1 +\omega_2 + \omega_3) = D - \mathcal{N}$, which is for $D=4$, $\mathcal{N}=1$ equivalent to the relation $\omega_1 +\omega_2 + \omega_3 = 0$. 
One recovers the original theory \eqref{eq:DSbetaDeformation} by setting all $\omega_i = 0$.
The kinetic terms of the component fields are
\begin{equation}
\begin{split}
S_{\mathrm{kin},\boldsymbol{\omega}} 
&=
\mathrm{N}\int \dd^4 x\; \dd^2\theta \dd^2 \bar{\theta} 
\sum_{i=1}^{3} 
\mathrm{tr} \left[ 
\Phi_i^\dagger \square^{\omega_i} \Phi_i
\right]\\
&=
\mathrm{N}\int \dd^4 x\; 
\sum_{i=1}^{3} 
\mathrm{tr} \left[ 
\phi_i^\dagger \square^{1+\omega_i} \phi_i -
\I \bar{\psi}_i \bar{\sigma}^\mu \square^{\omega_i} \partial_\mu \psi_i +
F_i^\dagger \square^{\omega_i} F_i
\right] ~,
\end{split}
\label{eq:DSbetaDeformation_gen_kin}
\end{equation}
and we use their non-local kinetic operators to construct the superpropagator according to the derivation in section \ref{subsec:TheGeneralizedSuperpropagator}.

\subsection{Exact all-loop anomalous dimension}
\label{sec:ExactAnomalousDimensions_SBW}
The exact computation of the anomalous dimension via the diagonalization of graph-building operators using the super chain relations is also possible in one instance for the super brick wall theory.
The program already presented in sections \ref{subsec:ExactAnomalousDimensionBiScalarFN} and \ref{sec:ExactAnomalousDimensionsFrom4PtCorrelationFunctions}, where (super) conformal symmetry played a decisive role for the construction of the eigenfunctions, the (super) conformal triangles.
For the superconformal algebra $\mathfrak{sl}(4 | 1)$ in four dimensions with $\mathcal{N} = 1$ supersymmetry, some of the superconformal triangles are constructed in \cite{Fitzpatrick:2014oza,Khandker:2014mpa,Li:2016chh}.
However, the triangles with the R-charge we need in the example below are not know to us.
Thus, we are in the same situation as in the two-magnon case of the superfishnet theory, which is studied in section \ref{subsec:TwoMagnonCase}.
There, the assumption of the existence of a complete basis of superconformal eigenfunctions of the graph-building operator admittedly seems ad-hoc.
Therefore, we provide here more evidence that the conjectured completeness relation and the resulting anomalous dimensions are correct, by reproducing a result from the literature \cite{Kazakov:2018gcy} on the dynamical fishnet theory, or $\chi$-CFT, from the assumptions on R-charged superconformal triangles applied to the super brick wall theory \eqref{eq:DSbetaDeformation}.
Hence, we switch from the three-dimensional superconformal setup of the superfishnet theory to the four-dimensional superconformal brick wall theory \cite{Kade:2024ucz}.
This theory is the double-scaled $\beta$-deformation of $\mathcal{N} = 4$ SYM, which is the supersymmetric case of the so-called dynamical fishnet theory \cite{Kazakov:2018gcy}.

We use the conventions of \cite{Kade:2024ucz} for four-dimensional $\mathcal{N} = 1$ superspace and consider the supergraph expansion of the correlator
\begin{equation}
\begin{split}
&\left\langle 
\mathrm{tr}
\left[
\Phi_1 (z_1)
\Phi_1 (z_2)
\right]
\mathrm{tr}
\left[
\Phi_1^\dagger (z_3)
\Phi_1^\dagger (z_4)
\right]
\right\rangle \\
= &
\adjincludegraphics[valign=c,scale=1]{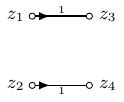}
+
\xi^4
\adjincludegraphics[valign=c,scale=1]{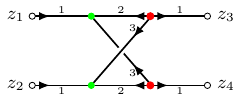}
+
\xi^8
\adjincludegraphics[valign=c,scale=1]{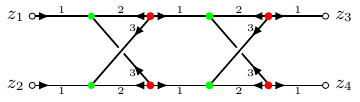} \\
&+
\xi^{12}
\adjincludegraphics[valign=c,scale=1]{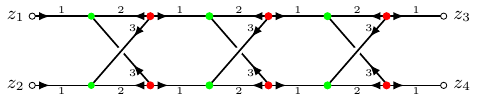} 
+
\cdots
+
\left(
z_3 
\leftrightarrow
z_4
\right) 
\end{split}
\label{eq:0MagnonPertExpansion_SBW}
\end{equation}
in the large-$\mathrm{N}$ limit.
The super brick wall theory's chiral, non-unitary, cubic vertices allow only ladder-like, cylindrical supergraphs.
Note that in \eqref{eq:0MagnonPertExpansion_SBW}, the upper-left and lower-right vertex of each rung is drawn with the wrong flavor orientation.
This is for illustrational reasons only; to be more precise, the connecting superpropagator should close after winding around the compact direction of the cylinder and respect the super Feynman rules \eqref{eq:betaDef_FeynmanRules_vertices}.

The graph-building operators in the diagrams \eqref{eq:0MagnonPertExpansion_SBW} are 
\begin{equation}
\begin{aligned}[c]
\mathbb{H}
=
\adjincludegraphics[valign=c,scale=1]{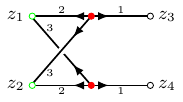} ~,
\end{aligned}
\qquad\qquad
\begin{aligned}[c]
\mathbb{P}
=
\adjincludegraphics[valign=c,scale=1]{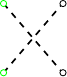} ~,
\end{aligned}
\label{eq:0MagnonGraphBuilder_SBW}
\end{equation}
and we will show that $\mathbb{H}$ and $\mathbb{H} \circ \mathbb{P}$ have the same eigenvalue on the eigenfunction of our interest.
With the graph-building operators, we can formally write the correlation function \eqref{eq:0MagnonPertExpansion_SBW} as
\begin{equation}
\left\langle 
\mathrm{tr}
\left[
\Phi_1 (z_1)
\Phi_1 (z_2)
\right]
\mathrm{tr}
\left[
\Phi_1^\dagger (z_3)
\Phi_1^\dagger (z_4)
\right]
\right\rangle 
= 
\adjincludegraphics[valign=c,scale=1]{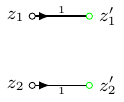}
\circ 
\left[
\frac{\left( 1 + \mathbb{P} \right)}{1 - \xi^4 \mathbb{H}}
\right] ~.
\label{eq:0MagnonPertExpansion_SBW_1}
\end{equation}
The eigenvalues of $\mathbb{H}$ corresponding to the eigenfunction $\Psi_u = \tfrac{\bar{\theta}_{12}^2}{\left[ x_{12}^2 \right]^u}$ can be calculated similarly to \eqref{eq:2Magnon_ev_1}, with the difference that we are in four-dimensional $\mathcal{N} = 1$ superspace and that we have to convolute the four-dimensional kite diagram $\mathsf{kite}^{(4)} ( x^2, u ) = \frac{I^{(4)}(u)}{\left[ x^2 \right]^{u}}$ \cite{Gromov:2018hut} with the two more superpropagators.
We observe that the permutation graph-builder $\mathbb{P}$ has a trivial action on the eigenfunction $\Psi_u$ since the eigenfunction is symmetric under the exchange of the external points.
Furthermore, the eigenfunction carries R-charge such that we expect the parameter $u$ to be related to the scaling dimension of the eigenfunction, which is as in the two-magnon case given by $\Delta_{\Psi_{u}} = \Delta - \Delta_1 - \Delta_2 + R_1 + R_2$.
Now the R-charge of $\mathcal{O}_1 = \mathcal{O}_2 = \Phi_1^\dagger$ is $R_1 = R_2 = - \frac{1}{2}$, which implies $u= 2 - \frac{\Delta}{2}$.
For the graph-builder $\mathbb{H}$ we find the following eigenvalue equation,
\begin{equation}
\begin{split}
\Psi_u \circ \mathbb{H} ~
&=
\adjincludegraphics[valign=c,scale=1]{figures/superFN/2ptfctn/zeromagnon_SBW/diagonalization_left/Builder_H_SBW_ev.pdf}
=
- 4 u \left( 1 - u \right)
I^{(4)} (u) \cdot
\adjincludegraphics[valign=c,scale=1]{figures/superFN/2ptfctn/zeromagnon_SBW/diagonalization_left/Builder_H_SBW_ev_1.pdf}\\
&=
16\, u \left( 1 - u \right)\; 
I^{(4)} (u)\;
r (2 - u , u + 1 , 1)\,
r (2 - u , u , 1) \cdot
\Psi_u ~.
\end{split}
\label{eq:0Magnon_SBW_ev_1}
\end{equation}
In the last step, we used the four-dimensional $\mathcal{N} = 1$ versions of the super chain relations \eqref{eq:ChainRelChiral_Aux} and \eqref{eq:ChainRelChiral}.
The eigenvalue corresponding to $\Psi_u$ is therefore 
\begin{equation}
\mathrm{E}^\mathrm{SBW}_0 (u)
=
16\, u \left( 1 - u \right) \cdot
I^{(4)} (u) \cdot
r (2 - u , u + 1 , 1)\,
r (2 - u , u , 1) ~,
\end{equation}
where the factor from the kite diagram reads \cite{Gromov:2018hut}
\begin{equation}
I^{(4)} (u) 
= 
\frac{1}{2 u-2}
\left[
\psi ^{(1)}\left(\tfrac{u-1}{2}\right)
- \psi ^{(1)}\left(\tfrac{1-u}{2}\right)
+ \psi ^{(1)}\left(\tfrac{2-u}{2}\right)
- \psi ^{(1)}\left(\tfrac{u}{2}\right)
\right] ~,
\end{equation}
with the second derivative of the logarithm of the gamma function $\psi ^{(1)} (z) = \frac{\dd^2}{\dd z^2} \mathrm{log}\, \Gamma (z)$.
We observe again that the action of $\mathbb{P}$ on $\Psi_u$ is trivial.
Using the functional relation of $\psi ^{(1)} (z)$ and rescaling the coupling of the super brick wall theory $\xi \rightarrow \frac{\xi}{2\pi}$, the pole condition of \eqref{eq:0MagnonPertExpansion_SBW_1}, $1 = \xi^4 \mathrm{E}^\mathrm{SBW}_0 (2 - \tfrac{\Delta}{2})$, is equivalent to the result (4.28) in \cite{Kazakov:2018gcy} of the dynamical fishnet theory \eqref{eq:Chi_SYM_action} with $\xi_1 = \xi_2 = \xi_3 = \xi$, specified to the supersymmetric, spinless case $\omega , S = 0$ and $\Delta = 2 + 2\I \nu$ therein.

\section{Vacuum graphs in the thermodynamic limit}
\label{sec:VacuumGraphsThermoLimit_susy}

\begin{figure}[h]
\includegraphics[scale=1]{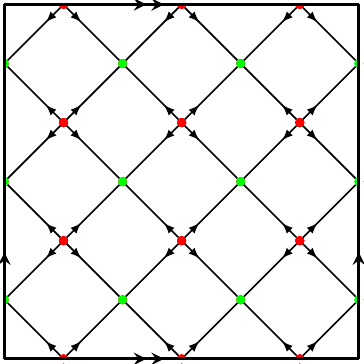}
\hspace{1cm}
\includegraphics[scale=1]{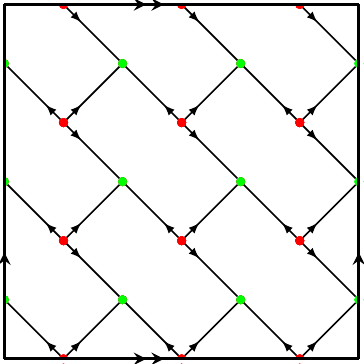}  \centering
\caption{The toroidal super vacuum graphs exemplify a contribution to the free energy of the superfishnet theory \eqref{eq:Action_superfishnet_Phi} (left) and to the super brick wall theory \eqref{eq:DSbetaDeformation} (right).
One notices that the regular fishnet- (left) and brick wall pattern (right). 
In the planar limit $\mathrm{N}\rightarrow\infty$, the leading order results in such toroidal diagrams, with arbitrary height and length.
The doubly periodic boundary conditions of the torus are implemented into the figure:  
top and bottom lines are identified, and so are the left and right boundaries.}
\label{fig:betaDef_vacuumgraph}
\end{figure}

We turn to the study of super vacuum diagrams of the superfishnet \eqref{eq:Action_superfishnet_Phi} and super brick wall theory \eqref{eq:DSbetaDeformation} and their non-local deformations \eqref{eq:Action_genSuperfishnet_Phi} and \eqref{eq:DSbetaDeformation_gen}, respectively.
Examples of a vacuum supergraphs in the planar limit are shown in fig.\ \ref{fig:betaDef_vacuumgraph}. 
The regular pattern is due to the highly constraining Feynman rules, notably, the interaction vertices \eqref{eq:FeynmanRules_vertices} and \eqref{eq:betaDef_FeynmanRules_vertices}. 
The vertices are chiral and demand a very particular order of the flavor of the superfields.
Diagrammatically, the non-local deformations \eqref{eq:Action_genSuperfishnet_Phi} and \eqref{eq:DSbetaDeformation_gen} reproduce the same graphs as in fig.\ \ref{fig:betaDef_vacuumgraph} because the superpotential is the same as in the undeformed theories.
However, the propagator weights differ by their exponent and the propagators have to be multiplied by the prefactor $c_0(1 + \omega_i )$ as indicated in \eqref{eq:gen_superPropagator_graphical}.
Our goal is to repeat the analysis of section \ref{sec:IntegrableVacuumDiagramsAndTheCriticalCoupling} for super vacuum graphs and obtain their thermodynamic limit and the critical couplings of the supersymmetric theories.

In the non-supersymmetric case, we argued in section \ref{subsec:IntegrableVacuumDiagramsFromTheGeneralizedPartitionFunction} that vacuum diagrams are infrared divergent because they scale proportional to the volume of spacetime. 
Similarly, superspace vacuum graphs are proportional to the a priori ill-defined $\int \dd^D x\cdot\int \dd^2 \theta\, \dd^2 \bar{\theta} = \infty \cdot 0$.
(The zero stemming from the fermionic integration is just a manifestation of the well-known statement that the vacuum energy of supersymmetric field theories is zero.)
In our supersymmetric models at hand, we proceed in the same way as in the non-supersymmetric case and leave one of the superspace points in the vacuum supergraphs un-integrated in order to obtain a well-defined density. 

Starting from the super weight \eqref{eq:gen_superWeight_graphical}, a superpropagator without the factor $c_0(1 + \omega_i )$, we can identify\footnote{ 
Note that we could have chosen another row matrix kernel with the chiralities interchanged, i.\,e.\ with anti-chiral (green) external vertices and chiral (red) internal ones.
}
a generalized row-matrix
\begin{equation}
T_N(U)
=
\adjincludegraphics[valign=c,scale=0.7]{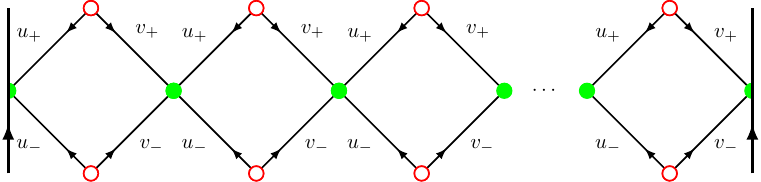}
\label{eq:gen_RowMatrix_omega}
\end{equation}
It is building up the vacuum diagrams in fig.\ \ref{fig:betaDef_vacuumgraph} after e.\,g.\ fixing the parameter matrix, c.\,f.\ \eqref{eq:Parametermatrix_U} to $U^\mathrm{SFN}=\left(\begin{smallmatrix}u_+ & v_+ \\ u_- & v_- \end{smallmatrix}\right)=\left(\begin{smallmatrix} \sfrac{1}{2} - \omega_1 & \sfrac{1}{2} - \omega_4 \\ \sfrac{1}{2} - \omega_2 & \sfrac{1}{2} - \omega_3 \end{smallmatrix}\right)$ for the non-locally deformed superfishnet theory and $U^\mathrm{SBW} = \left(\begin{smallmatrix}u_+ & v_+ \\ u_- & v_- \end{smallmatrix}\right)=\left(\begin{smallmatrix} 0 & 1 - \omega_1 \\ 1 - \omega_3 & 1 - \omega_2 \end{smallmatrix}\right)$ for the deformed super brick wall theory.
In the limit $\omega_i \rightarrow 0 $, we find the vacuum graphs of the undeformed theories.
In contrast to the non-supersymmetric case in section \ref{subsec:IntegrableVacuumDiagramsFromTheGeneralizedPartitionFunction}, we only consider homogeneous row-matrices here.
Due to the lack of a superspace STR, see the discussion in section \ref{subsec:SuperconfStarIntegral}, the row-matrices cannot be shown to commute, and this is why we do not call them super transfer matrices, but row-matrices.

Formally, we may write a generalized $M\x N$ toroidal vacuum supergraph as 
\begin{equation}
Z_{MN}(\mathbf{u}) = \mathrm{tr}\left[ T_N(\mathbf{u})^M \right],
\label{eq:gen_vacuumdiagram}
\end{equation}
which graphically represents $M$ generalized row matrices of length $N$ stacked on top of each other and identified periodically by the trace. 
However, the normalization factors $c_0(1 + \omega_i )$ are not included in \eqref{eq:gen_vacuumdiagram} and we will have to include them to describe physical superspace vacuum diagrams.
To reinstate them, we consider a normalized row matrix\footnote{Remember that $\sum_i \omega_i = 0$, where we have four $\omega_i$'s in the deformed superfishnet theory and three in the deformed super brick wall theory.
}, indicated by the hat,
\begin{equation}
\begin{array}{ccc}
\hat{T}_N^\mathrm{SFN}
=
\left[
\prod_{i=1}^4
c_0(1 + \omega_i )
\right]^N
T_N(U^\mathrm{SFN}) &
\mathrm{and} &
\hat{T}_N^\mathrm{SBW}
=
\left[
\prod_{i=1}^3
c_0(1 + \omega_i )
\right]^N
T_N(U^\mathrm{SBW}) ~.
\end{array}
\end{equation}
Thus, we obtain the physical vacuum supergraphs of the supersymmetric theories by staking up the row-matrices \eqref{eq:gen_RowMatrix_omega} and taking the trace.
We find $\hat{Z}_{MN}^\mathrm{SFN} = \mathrm{tr} (\hat{T}_N^\mathrm{SFN})^M $ and $\hat{Z}_{MN}^\mathrm{SBW} = \mathrm{tr} (\hat{T}_N^\mathrm{SBW})^M $.

We aim to calculate the critical coupling $\xi_\mathrm{cr}$. It is defined as the radius of convergence of the expansion of the free energy of the superfishnet and super brick wall theory and their series are
\begin{equation}
\hat{Z}^\mathrm{SFN} 
= 
\sum_{M,N=1}^\infty 
\hat{Z}_{MN}^\mathrm{SFN}
(\I \xi)^{2MN} 
~~~~ \mathrm{and} ~~~~
\hat{Z}^\mathrm{SBW} 
= 
\sum_{M,N=1}^\infty 
\hat{Z}_{MN}^\mathrm{SBW}
(- \xi)^{2MN} ~.
\end{equation}
The different signs in front of the coupling have their origin in the Feynman rules \eqref{eq:FeynmanRules_vertices} and \eqref{eq:betaDef_FeynmanRules_vertices}.
The critical couplings are the radii of convergence
\begin{subequations}
\begin{align}
\xi_\mathrm{cr}^\mathrm{SFN} &
=
\left[
\lim_{M,N\rightarrow \infty}
\left|
- \hat{Z}_{MN}^\mathrm{SFN}
\right|^{\frac{1}{MN}}
\right]^{-\frac{1}{2}}
=
\left[
\prod_{i=1}^4
c_0(1 + \omega_i )
\lim_{M,N\rightarrow \infty}
\left|
Z_{MN} (U^\mathrm{SFN})
\right|^{\frac{1}{MN}}
\right]^{-\frac{1}{2}} ~, \\
\xi_\mathrm{cr}^\mathrm{SBW} &
=
\left[
\lim_{M,N\rightarrow \infty}
\left|
\hat{Z}_{MN}^\mathrm{SBW}
\right|^{\frac{1}{MN}}
\right]^{-\frac{1}{2}}
=
\left[
\prod_{i=1}^3
c_0(1 + \omega_i )
\lim_{M,N\rightarrow \infty}
\left|
Z_{MN}  (U^\mathrm{SBW})
\right|^{\frac{1}{MN}}
\right]^{-\frac{1}{2}} ~.
\end{align}\label{eq:CriticalCoupling_from_K}%
\end{subequations}
The only yet unknown quantity is the thermodynamic limit of the bare vacuum graphs, which are defined in \eqref{eq:gen_vacuumdiagram}.
Hence, we wish to determine the thermodynamic limit 
\begin{equation}
K(U) ~:=~
\lim_{M,N\rightarrow \infty} \vert Z_{MN}(U)\vert^{\frac{1}{MN}}
\label{eq:gen_freeEnergy}
\end{equation}
and then evaluate them at $U^\mathrm{SFN}$ and $U^\mathrm{SBW}$.
We will calculate the limit \eqref{eq:gen_freeEnergy} using the method of inversion relations, in relation to the non-supersymmetric case in section \eqref{subsec:ThermodynamicLimitOfVacuumDiagramsAndCriticalCoupling}.

\subsection{Inversion relations}
\label{subsec:InversionRelations_susy}
The super x-unity relation \eqref{eq:SuperXUnity} allows us to find four different forms of the inverse of the row matrix $T_N(U)$ by the same moves as explained in \eqref{eq:TransferMatrixAnnihilation}. 
We find
\begin{equation}
T_N(U)
\circ
T_N(U_{\mathrm{inv}})
=
F_N \cdot \mathbb{1}_N
\label{eq:RowMatrixInversion}
\end{equation}
to hold for 
\begin{subequations}
\begin{align}
{\scriptstyle
U_\mathrm{inv}
=
\left(\begin{smallmatrix} -u_- & D - 1 -v_- \\ D - 1 - u_+ & -v_+ \end{smallmatrix}\right)
}
~~ & {\scriptstyle \mathrm{and}~~
F_N ~= ~ \left[ 16\pi^{2D}\; a_0(u_+)\, a_0(D - 1 - u_+)\, a_0(v_-)\, a_0(D - 1 - v_-) \right]^N
} ~, \\
{\scriptstyle
U_\mathrm{inv}
=
\left(\begin{smallmatrix} -u_- & D - 1 - v_- \\ -u_+ & D - 1 - v_+ \end{smallmatrix}\right)
}
~~ & {\scriptstyle \mathrm{and}~~
F_N ~ = ~ \left[ 16\pi^{2D}\; a_0(v_+)\, a_0(D - 1 - v_+)\, a_0(v_-)\, a_0(D - 1 - v_-) \right]^N 
} ~, \\
{\scriptstyle
U_\mathrm{inv}
=
\left(\begin{smallmatrix} D - 1 - u_- & -v_- \\ D - 1 - u_+ & -v_+ \end{smallmatrix}\right)
}
~~ & {\scriptstyle \mathrm{and}~~
F_N ~ = ~ \left[ 16\pi^{2D}\; a_0(u_+)\, a_0(D - 1 - u_+)\, a_0(u_-)\, a_0(D - 1 - u_-) \right]^N 
} ~,\\
{\scriptstyle
U_\mathrm{inv}
=
\left(\begin{smallmatrix} D - 1 - u_- & -v_- \\ -u_+ & D - 1 - v_+ \end{smallmatrix}\right)
}
~~ & {\scriptstyle \mathrm{and}~~
F_N ~=~ \left[ 16\pi^{2D}\; a_0(u_-)\, a_0(D - 1 - u_-)\, a_0(v_+)\, a_0(D - 1 - v_+) \right]^N 
} ~.\label{eq:RowMatrixInversion_4}
\end{align}\label{eq:RowMatrixInversion}%
\end{subequations}
We recall that $D$ is the dimension of the bosonic part of superspace, i.\,e.\ we have $D=3$ for the superfishnet theory and $D=4$ for the super brick wall theory.
For example, \eqref{eq:RowMatrixInversion_4} for $D=4$ can be obtained from the super x-unity relation \eqref{eq:SuperXUnity} by the steps
\begin{subequations}
{\allowdisplaybreaks
\begin{align}
& T_{N} 
\left(\begin{smallmatrix}u_+ & v_+ \\ u_- & v_- \end{smallmatrix}\right)
\circ
T_{N}
\left(\begin{smallmatrix} 3 - u_- & -v_- \\ -u_+ & 3 - v_+ \end{smallmatrix}\right)
=
\adjincludegraphics[valign=c,scale=0.5]{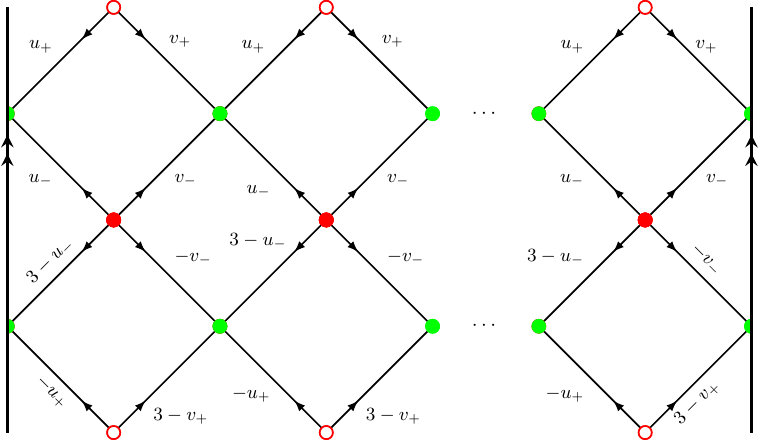}
\label{eq:TransferMatrixAnnihilation1_super}\\
%%%%%%%%%%
~&\stackrel{\eqref{eq:SuperXUnity_red}}{=}~
\adjincludegraphics[valign=c,scale=0.5]{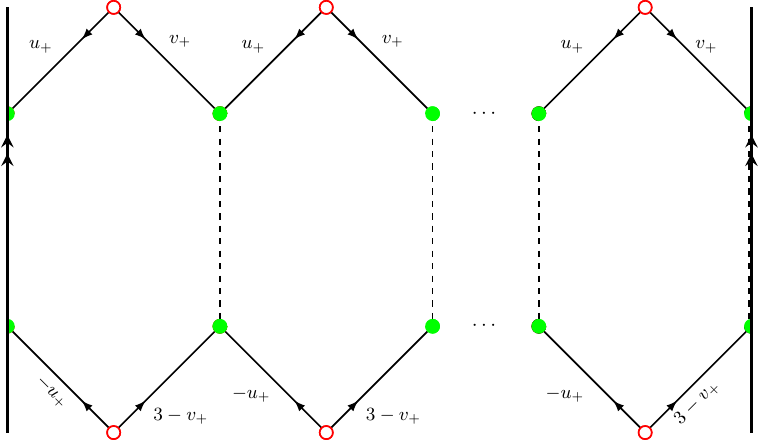} 
\cdot 
\left[
- 4 \pi^4 a_0 (u_-) a_0 (3 - u_-)
\right]^N 
\\
%%%%%%%%%%
~&\stackrel{\phantom{\eqref{eq:SuperXUnity_green}}}{=}~
\adjincludegraphics[valign=c,scale=0.5]{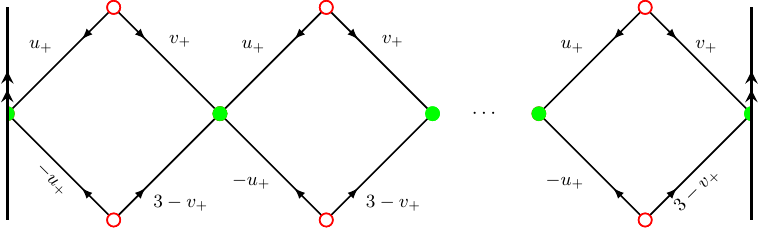}
\cdot
\left[
- 4 \pi^4 a_0 (u_-) a_0 (3 - u_-)
\right]^N 
\\
%%%%%%%%%%
~&\stackrel{\eqref{eq:SuperXUnity_green}}{=}~
\adjincludegraphics[valign=c,scale=0.5]{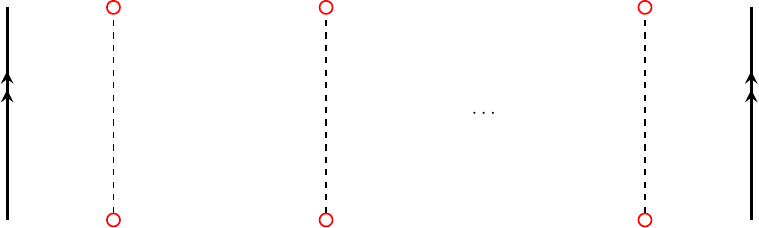} 
\cdot
\left[
16 \pi^8\; a_0 (u_-) a_0 (3 - u_-)\;
 a (v_+) a (3 - v_+)
\right]^N
\\
%%%%%%%%%%
~&\stackrel{\phantom{\eqref{eq:SuperXUnity_green}}}{=}
\left[
16 \pi^8\; a_0 (u_-) a_0 (3 - u_-)\;
 a (v_+) a (3 - v_+)
\right]^N \cdot \mathbb{1}_N
\label{eq:TransferMatrixAnnihilationEnd_super} ~ ,
\end{align}\label{eq:TransferMatrixAnnihilation_super}%
}%
\end{subequations}
and the other equations of \eqref{eq:RowMatrixInversion} can be obtained by using variations of \eqref{eq:SuperXUnity} with left- and right external points interchanged.

\begin{figure}[h]
\includegraphics[scale=0.3]{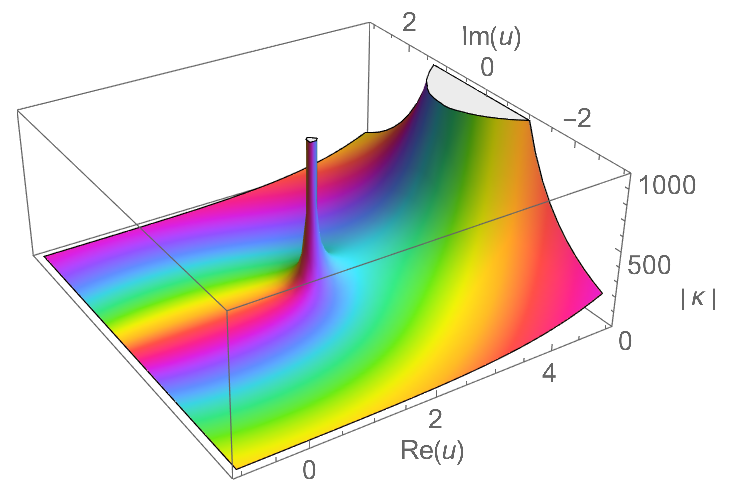} 
\centering
\caption{The plot shows the magnitude of the function $\kappa (u)$ from \eqref{eq:kappa_4DN1_collaps} for four-dimensional $\mathcal{N}=1$ superspace and complex $u$. It contributes to the critical coupling of the super brick wall theory. We observe that there is a characteristic pole at $u=2$. This is an expected property for a free energy in the thermodynamic limit, which has a pole at the value of the crossing parameter. For four dimensions, this is $\frac{D}{2} = 2$. Furthermore, the coloring indicates the phase of the complex number $\kappa (u)$. In the physical strip $[ 0 ,2)$, the graph is red, which corresponds to a zero-phase and real values of $\kappa (u)$. For the application to the super brick wall theory, we related the critical coupling to the value at $u=1$.
}
\label{fig:plot_of_kappa_D4N1}
\end{figure}

Projecting \eqref{eq:gen_vacuumdiagram} on the eigenvector corresponding to the maximal eigenvalue dominating the thermodynamic limit \cite{Kade:2023xet},
\begin{equation}
K (\mathbf{u})
=
\lim_{M,N\rightarrow \infty} 
\vert \mathrm{tr}\left[ T_N(\mathbf{u})^M \right] \vert^{\frac{1}{MN}}
=
\lim_{N\rightarrow \infty} 
\vert \Lambda_{\mathrm{max},N}(\mathbf{u}) \vert^{\frac{1}{N}} ~,
\end{equation}
we can turn \eqref{eq:RowMatrixInversion} into four functional relations for $K(U)$, which are of the form $K(U) K(U_\mathrm{inv}) = F_N^{\sfrac{1}{N}}$. 
Based on the observation that $K(U)$ corresponds to a rhombus of four superpropagators, according to \eqref{eq:gen_freeEnergy} and \eqref{eq:gen_vacuumdiagram}, we make the ansatz $K(U) = \kappa (u_+)\kappa (u_-)\kappa (v_+)\kappa (v_-)$.
This is in analogy to \eqref{eq:K_ThermodynamicLimit_Ansatz}.
We find that $\kappa(u)$ has to satisfy
\begin{equation}
\kappa(u)\kappa(-u) = 1
~~~
\mathrm{and}
~~~
\kappa(u)\kappa(D - 1 - u) = 4\pi^D\; a_0(u)\, a_0(D - 1 - u) ~.
\label{eq:InversionRelations_kappa}
\end{equation}
We can construct a solution with the help of appendix \ref{sec:SolvingInversionRelations}, where we plug the two functional relations \eqref{eq:InversionRelations_kappa} iteratively into each other and require the solution to have no poles in the physical interval $\left[ 0, \frac{D}{2}\right)$, which indicates the maximal eigenvalue \cite{StroganovInvRelLatticeModels}.
We find
\begin{subequations}
\begin{align}
\kappa (u) 
\stackrel{D=3}{=}
2^{\frac{3 u}{2}+1} \pi ^{\frac{3 u}{2}-\frac{1}{2}} 
\frac{\Gamma \left(\frac{3}{2}-u\right) \Gamma \left(\frac{u}{2}+\frac{1}{4}\right)}{\Gamma \left(\frac{1}{4}\right)} &
\prod _{k=1}^{\infty } 
\frac
{\Gamma (2 k - u + \frac{3}{2} ) \Gamma ( 2 k + u ) \Gamma (2 k - \frac{3}{2} )}
{\Gamma (2 k + u - \frac{3}{2} ) \Gamma ( 2 k - u ) \Gamma (2 k + \frac{3}{2} )} ~, \label{eq:kappa_3DN2} \\
\kappa (u) 
\stackrel{D=4}{=}
12^{\frac{u}{3}} \pi ^{\frac{4 u}{3}} 
\frac
{\Gamma \left(\frac{u+1}{3}\right) \Gamma (2 -u) }
{\Gamma \left(\frac{1}{3}\right)} &
\prod _{k=1}^{\infty } 
\frac
{\Gamma (3 k - u + 2 ) \Gamma ( 3 k + u ) \Gamma (3 k - 2 )}
{\Gamma (3 k + u - 2 ) \Gamma ( 3 k - u ) \Gamma (3 k + 2 )} ~. \label{eq:kappa_4DN1}
\end{align}
\label{eq:kappa_susy}%
\end{subequations}
The expressions resemble a lot the findings of the non-supersymmetric case \eqref{eq:Kappagenfinalresult}, despite having the one here is derived from a more exotic superspace computation.
The only imprint of supersymmetry is the factor in front of the counting parameter in the gamma functions: it is $D-1$, which is the scaling dimension of the superspace measure $\dd^D x \; \dd^2 \theta \, \dd^2 \bar{\theta}$.
A special value is $\kappa (0) = 1$ for any $D$, which reflects the trivial statement that a superpropagator with exponent zero should not contribute to the critical coupling.
For the case of the four-dimensional $\mathcal{N} = 1$ super brick wall result \eqref{eq:kappa_4DN1}, we make an interesting observation.
With the help of the functional relation of the gamma function, we observe that the infinite product collapses to the expression
\begin{equation}
\kappa (u) 
\stackrel{D=4}{=}
2^{\frac{2 u}{3}} 3^{\frac{4 u}{3}-2} \pi ^{\frac{4 u}{3}}
\frac{\Gamma (2-u) \Gamma \left(\frac{u}{3}\right) \Gamma \left(\frac{u+1}{3}\right)}{\Gamma (u) \Gamma \left(1-\frac{u}{3}\right) \Gamma \left(\frac{4}{3}-\frac{u}{3}\right)}
\label{eq:kappa_4DN1_collaps}
\end{equation}
with the special value $\kappa (1) = \left(\frac{2\pi^2}{3} \right)^{2/3} \Gamma (\frac{1}{3})$.
This is a rare case, where we are able to study a function $\kappa (u)$ analytically as a function of $u$, since the general representation as an infinite product is unpractical in this regard.
A plot of the function \eqref{eq:kappa_4DN1_collaps} is presented in fig.\ \ref{fig:plot_of_kappa_D4N1}.
It has a characteristic pole at the value of the crossing parameter $u = 2$, which is the value for which the bosonic part of the superpropagator becomes close to a delta function, up to the damping factor $a(\varepsilon)$, c.\,f.\ \eqref{eq:Delta_bosonic}.

Finally, we find the critical coupling for the non-locally deformed theories to be
\begin{equation}
\xi_\mathrm{cr}^\mathrm{SFN}
=
\left[
\frac{1}{4} \prod_{i = 1}^4 a_0(1 - \omega_i) \kappa (\tfrac{1}{2} - \omega_i )
\right]^{-1/2}
~~~~ \mathrm{and} ~~~~
\xi_\mathrm{cr}^\mathrm{SBW}
= 
\left[
\prod_{i=1}^3
a_0( 1 + \omega_i )\,
\kappa\left( 1 - \omega_i \right)
\right]^{-1/2} ~,
\label{eq:CritCoup_nonlocal}
\end{equation}
where we plugged the function $\kappa (u)$ into $K(U)$ and eventually into \eqref{eq:CriticalCoupling_from_K}.
We used the definition of the factor $c_0(v)$ from \eqref{eq:Factor_cEll}.
For generic deformation parameters $\omega_i$, it is difficult to evaluate the infinite product \eqref{eq:kappa_susy}.
However, by the constraint $\sum_i \omega_i = 0$, we are able to construct some cases where we turn some elements in $U^\mathrm{SFN}$ or $U^\mathrm{SBW}$ to zero, so that the contribution of some propagators vanishes and the graphical link disappears.

Of course, of most importance are the undeformed superfishnet and super brick wall theory, which we access by choosing all deformation parameters to be zero, $\omega_i = 0$.
In this case, the critical couplings 
\begin{subequations}
\begin{align}
\xi_\mathrm{cr}^\mathrm{SFN}
&=
\frac{2}{\Gamma (\frac{1}{2})^2} \cdot
\kappa (\tfrac{1}{2})^{-2}
=
\frac{\left(\frac{2}{\pi}\right)^{3/2}}{\Gamma \left(\frac{1}{4}\right)^2} 
=
\frac{1}{ \pi^3 \sqrt{2}\, | \eta ( \I )| ^{2} } ~, \\
\xi_\mathrm{cr}^\mathrm{SBW}
&=
\kappa\left( 1 \right)^{-3/2}
=
\frac{3}{2 \pi ^2 \;\Gamma (\frac{1}{3})^{3/2}}
=
\frac{3^{9/8}}{4\pi^3\, | \eta ( \e^{\frac{\I \pi }{3}} )|}
\end{align}
\label{eq:Result_CriticalCoupling_SUSY}%
\end{subequations}
follow from \eqref{eq:CritCoup_nonlocal}.
The expression in terms of the Dedekind eta values is obtained by the identities of the gamma function values under \eqref{eq:ResultsCriticalCoupling_NonSusy}.
In comparison to \eqref{eq:ResultsCriticalCoupling_NonSusy}, we notice that the exponents of the Dedekind eta values are significantly lower.

%%%----------------------------------------------------------------------------------------
%%%	THESIS CONTENT - PART 7 - Conclusions
%%%----------------------------------------------------------------------------------------
\chapter{Conclusions and outlook}
\label{chpt:Conclusions}

To conclude, we give a summary over the results of this thesis and comment on future research directions.
By using the correspondence to integrable lattice models, we were able to derive several exact results in integrable quantum field theories.
On the one hand, we applied the method of inversion relations to obtain the critical coupling of the bi-scalar fishnet theory, the fishnet-deformation of ABJM, a six-dimensional cubic scalar theory, the brick wall theory, the fermionic fishnet theory, and of two supersymmetric theories, the superfishnet and super brick wall model.
We found the numerical curiosity that all the values of the critical couplings are related to values of the Dedekind eta function.
Furthermore, employing the chain relations, we were able to diagonalize the supergraph-building operators for three classes of four-point correlation functions, which were the zero- and two-magnon cases in the superfishnet and zero-magnon cases of the super brick wall theory.
This gave us access to the exact anomalous dimension of various single-trace operators, and in the zero-magnon superfishnet case, we could even find the all-loop scaling dimension of operators containing derivatives.
Moreover, this allowed us to determine the OPE coefficient associated to the four-point function exactly, and we found the latter to all orders in the coupling.
Additionally, we made conceptual progress by extending the notion of integrable Feynman graphs to boundary integrability.
With the two boundary-star-triangle relations, we found two criteria for integrable boundary conditions of Feynman graphs.
The two solutions we presented allow us to model a trivial boundary as well as an external spacetime point.
Finally, we report novel box-type boundary conditions for the six-vertex model.
In the case of a square box, we demonstrated that its bulk- and boundary-integrability allow us to deduce a recursion relation for the partition function.
We found a function satisfying it and conjecture that this is describing the partition function for box-boundary conditions at any lattice size.

\subsubsection{Future research directions}
Based on our findings, there are many future research directions.
First, the conjecture of the partition function for the box-boundary conditions of the six-vertex model has to be proven or discarded.
Thanks to Monte-Carlo simulations, one can check the numerical predictions of the conjectured expression and perform valuable preliminary cross-checks \cite{Sylju_sen_2004}.
A proof could follow the lines of the domain-wall boundary conditions \cite{izergin1987partition,AGIzergin_1992} and examine the degree of the partition function in the spectral parameter, since its dependence is polynomial.
The degree of the polynomial determines the number of coefficients which have to be fixed by the recursion relation.
The box boundary can be used to model numerous different boundary conditions by individually choosing the K-matrix of each side.
Here, we only looked at the arrow-reflecting box, but one can also solve the bYBE for arrow-sourcing or -sinking walls and combine the boundary conditions of all four sides.
Furthermore, one can extend the conjecture to a rectangular box or study the thermodynamic limit of the partition function, similar to the domain-wall boundary conditions and its connection to matrix models \cite{korepin2000thermodynamic,zinn2000six}.

Second, the framework of boundary-integrable Feynman graphs should be applied to a physical, integrable QFT.
Possible applications might be the study of two- and four-point functions in the checkerboard theory \cite{Alfimov:2023vev}, which could be achieved by the canonical solution presented in this thesis.
The existence of other solutions of the bYBE is also an interesting question.
Imaginable are Wilson lines or other defects in the context of fishnet theories \cite{Gromov:2021ahm}, which couple to the bulk Feynman diagrams, when computing quantities like the quark-antiquark potential.
Furthermore, one could combine the box-boundary conditions with boundary-integrable Feynman diagrams.
Using the canonical solution on all four walls, one can hope to find an explanation by lattice models of why Basso-Dixon integrals can be expressed in determinant form \cite{Basso:2017jwq,Derkachov:2018rot}.

Third, the study of the superfishnet- and super brick wall theory has just begun and many follow-up investigations are possible.
Most importantly is the quest to find a supersymmetric star-triangle relation in order to construct an R-matrix and explicitly construct commuting transfer matrices, similar to the fishnet case \cite{Chicherin:2012yn,Gromov:2017cja}
One can approach the problem systematically by deriving an R-matrix from non-compact representations of the theories' superconformal algebra $\mathfrak{osp} (2 \vert 4)$ or $\mathfrak{sl} (4 \vert 1)$, respectively.
For the one-dimensional superconformal algebra, this was done in \cite{Derkachov:2001sx,Derkachov:2005hx,Belitsky:2006cp}.
Instrumentally, a supersymmetric analog of the principle series representation of the superconformal algebra should be constructed to find an Lax operator, which can be used to solve the RLL relations and obtain the R-matrix.
Using non-compact oscillator representations for the superalgebras provides another angle of attack to this problem \cite{Frassek:2023tka}. 
After having established the R-matrix, many investigations of the bi-scalar theory could be uplifted to the superfishnet theory.
Interesting would be the TBA for two-point functions \cite{Basso:2018agi,Basso:2019xay}, supersymmetric Basso-Dixon diagrams \cite{Basso:2017jwq,Derkachov:2020zvv,Derkachov:2021rrf,Basso:2021omx,Loebbert:2024fsj}, the construction of a holographic superfishchain \cite{Gromov:2019aku,Gromov:2019bsj,Gromov:2019jfh,Gromov:2021ahm,iakhibbaev2023generalising}, the Yangian bootstrap of supergraphs \cite{Chicherin:2017frs,Corcoran:2021gda} and the interplay of super Feynman graphs with supersymmetric curves \cite{Duhr:2022pch,Duhr:2023eld,Duhr:2024hjf}.

Fourth, the idea of the superspace formalism for integrable Feynman diagrams could be extended to the full $\mathcal{N}=4$ SYM.
The theory can be formulated on a different kind of superspace, the harmonic superspace \cite{Galperin:1985bj,Galperin:2001seg}.
The Feynman rules have been established in \cite{Chicherin:2016fac} and one can imagine that the generalization of the propagator exponents yields useful integral relations such as the chain relation or even the STR.
This could be a promising attempt, despite the appearance of gauge supermultiplets.

Fifth, the appearance of Dedekind eta values in the critical couplings' results might hint to the existence of a string theory computation via the AdS/CFT correspondence.
The toroidal partition functions of two-dimensional worldsheet CFTs typically give rise to Dedekind eta functions \cite{DiFrancesco:1997nk}, which depend on the modulus of the worldsheet torus and could reproduce our findings for the critical coupling.
At first glance, this intuitively makes sense: the toroidal vacuum (super) graphs become dense in the thermodynamical limit and form a continuous surface.
This surface should be interpreted as the string's worldsheet.
Moreover, the Dedekind eta functions obtained in this dissertation are evaluated at two very special points of the complex structure modulus of the torus, the orbifold singularities in the fundamental domain $\tau = \I$ and $\tau = \e^{\I \pi /3}$.
In our integrability-based analysis, these values are determined by the physical values of the spectral parameters of the vacuum graphs.
However, in a two-dimensional CFT computation, we expect an integral over the whole fundamental domain of moduli space.
This would suggest the existence of a mechanism that stabilizes the modulus of the complex structure at the particular values.

Sixth, the whole program of identifying a statistical lattice model within double-scaled $\gamma$-deformations of holographic theories could be applied to the $\mathrm{AdS}_3 \mathrm{/CFT}_2$-correspondence.
Its integrability on both sides of the AdS/CFT-correspondence is reviewed in \cite{Sfondrini:2014via}.
One can imagine to deform the Higgs-Branch of the two-dimensional CFT \cite{OhlssonSax:2014jtq} or to perform the deformation on the dual symmetric-orbifold CFT directly.
This would yield a unique window into the triality of integrable models, conformal field theories, and holographic dual string theories, since for particular examples of $\mathrm{AdS}_3$-backgrounds the holographic duality is made manifest on the level of the partition function \cite{Eberhardt:2017pty,Eberhardt:2019niq}. 

To end this thesis, we would like to speculate about the extension of the ideas which were reviewed and newly formulated in this dissertation into the final goal, an explicit triality between gauge theory, string theory, and statistical models.
On the level of strongly deformed versions of $\mathcal{N}=4$ SYM, the lattice model perspective can be made explicit.
One of this dissertation's novel contributions is to initiate the study of the duality by supersymmetric Feynman graphs.
However, besides the fishchain theory, the relations to a holographic string theory are still unclear, especially from the lattice model side.
A major challenge is to undo the double-scaling limit, which seems to be the fundamental for the emergence of the lattice structure in Feynman graphs.
Therefore, we expect that overcoming this problem is the long-term priority.
It could lead to a remarkable triality between supersymmetric gauge theories, superstring theories and statistical models with spins valued in a non-compact representation of a superconformal algebra on a regular lattice.
Possibly, the latter can be thought of as some highly generalized version of ice.

\appendix
\chapter{Proof of integral relations}
\label{app:ProofOfUniquenessRelations}
In this appendix, we give a detailed derivation of the fundamental integral relations.
They are the scalar STR, the fermionic STR, the super chain relations and an auxiliary relation for the proof of the super x-unity relation.

\section{Proof of the scalar uniqueness relation}
\label{appsec:ProofOfTheScalarUniquenessRelation}
Fist, we proof the bosonic uniqueness relation \eqref{eq:CFTUniqueness_Scalar}, which is interpreted as a STR in \eqref{eq:CFTUniqueness_Scalar_Diagram}.
We begin on the l.\,h.\,s.\ and the star-integral 
\begin{equation}
\mathsf{star}
:=
\int \dd^D x_0 ~ 
\frac{1}{\left[ x_{10}^2 \right]^{u_1}}
\frac{1}{\left[ x_{20}^2 \right]^{u_2}}
\frac{1}{\left[ x_{30}^2 \right]^{u_3}} ~.
\end{equation}
For the moment, we leave $u_1$, $u_2$ and $u_3$ unconstrained and we will find below that they have to add up to the dimension $D$ in order to be able to find the equality to a triangle graph.
We introduce Schwinger parameters by the relation 
\begin{equation}
\frac{1}{\left[ x^2 \right]^{u}}
=
\frac{1}{\Gamma (u)} 
\int_0^\infty \dd s ~
s^{u - 1} \e^{- s\, x^2},
\label{eq:SchwingerTrick}
\end{equation}
which follows from the integral representation of the gamma function.
We find for the star-integral the expression
\begin{equation}
\mathsf{star}
=
\frac{1}{\Gamma (u_1) \Gamma (u_2) \Gamma (u_3)}
\int_0^\infty
\dd s_1 \dd s_2 \dd s_3 ~
s_1^{u_1 - 1} s_2^{u_2 - 1} s_3^{u_3 - 1}
\int \dd^D x_0 ~
\e^{ - s_1\, x_{10}^2 - s_2\, x_{20}^2 - s_3\, x_{30}^2 } ~.
\end{equation}
It allows us to perform the Gaussian $x_0$-integration after expanding the exponent.
Concretely, we evaluate the integral as
\begin{equation}
\int \dd^D x_0 ~
\e^{ - s_1\, x_{10}^2 - s_2\, x_{20}^2 - s_3\, x_{30}^2 }
=
\e^{- \frac{1}{S} \left( s_1 s_2 x_{12}^2 + s_1 s_3 x_{13}^2 + s_2 s_3 x_{23}^2 \right)}
\left( \tfrac{\pi}{S} \right)^{\frac{D}{2}}
\end{equation}
where we used the abbreviation $S := s_1 + s_2 + s_3$.
In order to perform the integration over the Schwinger parameters, we substitute them by $t_1$, $t_2$, and $t_3$ that are related to $s_1$, $s_2$, and $s_3$ by
\begin{equation}
\begin{array}{lll}
s_i = \frac{T}{t_i} ~, & S = s_1 + s_2 + s_3 = \frac{T^2}{t_1 t_2 t_3} ~, & \mathrm{det} \left(\frac{\partial s_i }{\partial t_j} \right) = \frac{T^3}{t_1^2 t_2^2 t_3^2} ~, \\
t_j = \frac{s_1 s_2 s_3}{S \cdot s_j} ~, & T = t_1 t_2 + t_2 t_3 + t_1 t_3 = s_1 s_2 s_3 ~. &
\end{array}
\label{eq:STR_Substitution}
\end{equation}
After the substitution the star-integral reads
\begin{equation}
\begin{split}
\mathsf{star}
~=~
&\frac{\pi^{\frac{D}{2}}}{\Gamma (u_1) \Gamma (u_2) \Gamma (u_3)}\\
&\cdot
\int_0^\infty
\dd t_1 \dd t_2 \dd t_3 ~
T^{u_1 + u_2 + u_3 - D}\;
t_1^{\frac{D}{2} - u_1 - 1}
t_2^{\frac{D}{2} - u_2 - 1}
t_3^{\frac{D}{2} - u_3 - 1}
\e^{- t_3 x_{12}^2 - t_2 x_{13}^2 - t_1 x_{23}^2} ~.
\end{split}
\end{equation}
Under the assumption $u_1 + u_2 + u_3 = D$, the expression factorizes in three integrals over the individual parameters $t_j$, such that we reverse the Schwinger trick \eqref{eq:SchwingerTrick} to find the uniqueness relation \eqref{eq:CFTUniqueness_Scalar}
\begin{equation}
\begin{split}
\int \dd^D x_0 ~ 
\frac{1}{\left[ x_{10}^2 \right]^{u_1}}
\frac{1}{\left[ x_{20}^2 \right]^{u_2}}
\frac{1}{\left[ x_{30}^2 \right]^{u_3}}
 ~ = ~ &
\pi^{\frac{D}{2}}
\frac
{\Gamma (\frac{D}{2} - u_1) \Gamma (\frac{D}{2} - u_2) \Gamma (\frac{D}{2} - u_3)}
{\Gamma (u_1) \Gamma (u_2) \Gamma (u_3)} \\
&\cdot
\frac{1}{\left[ x_{23}^2 \right]^{\frac{D}{2} - u_1}}
\frac{1}{\left[ x_{13}^2 \right]^{\frac{D}{2} - u_2}}
\frac{1}{\left[ x_{12}^2 \right]^{\frac{D}{2} - u_3}} ~.
\end{split}
\end{equation}

\section{Proof of the fermionic uniqueness relation}
\label{appsec:ProofOfTheFermionicUniquenessRelation}
The fermionic STR \eqref{eq:STR_Fermion_medial} is proven in a similar manner as the bosonic STR in \eqref{appsec:ProofOfTheScalarUniquenessRelation}.
However, we have to rewrite the fermionic propagators as bosonic ones with a derivative acting on them.
The fermionic star is the l.\,h.\,s.\ of \eqref{eq:STR_Fermion_medial} and reads
\begin{equation}
\mathsf{fstar}
:=
\int \dd^D x_0 ~ 
\frac{1}{\left[ x_{10}^2 \right]^{u_1}}
\frac{x_{10, \alpha \dot{\alpha}}}{\left|x_{10}\right|}
\frac{1}{\left[ x_{20}^2 \right]^{u_2}}
\frac{\bar{x}_{20}^{\dot{\alpha} \beta}}{\left|x_{20}\right|}
\frac{1}{\left[ x_{30}^2 \right]^{u_3}} ~,
\end{equation}
with $u_1 + u_2 + u_3 = D$.
We can turn the additional fermion propagators into derivatives by the useful equality
\begin{equation}
\frac{1}{\left[ x_{10}^2 \right]^{u}}
\frac{x_{10, \alpha \dot{\alpha}}}{\left|x_{10}\right|}
=
\frac{1}{1 - 2u}
\partial_{1, \alpha \dot{\alpha}}
\frac{1}{\left[ x_{10}^2 \right]^{u - \frac{1}{2}}} ~,
\end{equation}
which leaves us with 
\begin{equation}
\begin{split}
\mathsf{fstar}
=
\frac{1}{(1 - 2u_1) (1 - 2u_2)}
\partial_{1, \alpha \dot{\alpha}}
\partial_2^{\dot{\alpha} \beta}
\int \dd^D x_0 ~ 
\frac{1}{\left[ x_{10}^2 \right]^{u_1 - \frac{1}{2}}}
\frac{1}{\left[ x_{20}^2 \right]^{u_2 - \frac{1}{2}}}
\frac{1}{\left[ x_{30}^2 \right]^{u_3}} ~.
\end{split}
\end{equation}
Note that the scalar integral crucially cannot be turned into a triangle by the scalar STR \eqref{eq:CFTUniqueness_Scalar}, since the exponents only add up to $D - 1$.
Still, one can introduce Schwinger parameters by \eqref{eq:SchwingerTrick}, which leaves us with a Gaussian integral and performing it, gives
\begin{equation}
\mathsf{fstar}
=
\frac{1}{4}
\int_0^\infty \dd s_1\, \dd s_2\, \dd s_3 ~
\frac{
s_1^{(u_1 - \frac{1}{2}) - 1}
s_2^{(u_2 - \frac{1}{2}) - 1}
s_3^{u_3 - 1}
}{ \Gamma (u_1 + \frac{1}{2}) \Gamma (u_2 + \frac{1}{2}) \Gamma (u_3) }
\left(
\frac{\pi}{S}
\right)^{\frac{D}{2}}
\partial_{1, \alpha \dot{\alpha}}
\partial_2^{\dot{\alpha} \beta}\,
\mathsf{exp} ~.
\label{eq:STRFermionic_proof_1}
\end{equation}
Here we use again the abbreviation $S = s_1 + s_2 + s_3$ and we denote
\begin{equation}
\mathsf{exp}
:=
\e^{-\frac{1}{S} 
\left(
s_1 s_2 \cdot x_{12}^2 +
s_1 s_3 \cdot x_{13}^2 +
s_2 s_3 \cdot x_{23}^2 
\right)} ~.
\end{equation}
By usage of $x_{\alpha \dot{\alpha}} \bar{x}^{\dot{\alpha} \beta} = x^2 \delta_{\alpha}^{\beta}$ and $\partial_{\alpha \dot{\alpha}} \bar{x}^{\dot{\alpha} \beta} = D \cdot \delta_{\alpha}^{\beta}$, one can show that acting with the derivatives on $\mathsf{exp}$ gives
\begin{equation}
\partial_{1, \alpha \dot{\alpha}}
\partial_2^{\dot{\alpha} \beta}\,
\mathsf{exp}
=
4
\frac{s_1 s_2}{S}
\left[
\delta_{\alpha}^{\beta}
\left( 
\frac{D}{2}
+
\sum_{i=1}^3
s_i 
\frac{\partial}{\partial s_i}
\right)
+
s_3 
\cdot
x_{13,\alpha \dot{\alpha}}
\bar{x}_{23}^{\dot{\alpha} \beta}
\right]
\mathsf{exp} ~.
\label{eq:STRFermionic_proof_2}
\end{equation}
The r.\,h.\,s.\ can be plugged into \eqref{eq:STRFermionic_proof_1} and the terms with the $s_i$-derivatives can be reformulated by integration by parts,
\begin{subequations}
\begin{align}
\int_0^\infty 
\frac{s_1^{u_1 - \frac{1}{2}}}{S^{\frac{D}{2} + 1}}
\cdot
s_1
\frac{\partial}{\partial s_1}
\mathsf{exp}
& =
- \int_0^\infty 
\frac{s_1^{u_1 - \frac{1}{2}}}{S^{\frac{D}{2} + 1}}
\left[
u_1 + 
\frac{1}{2} - 
\frac{s_1}{S} \left( \frac{D}{2} + 1 \right)
\right]
\mathsf{exp} ~, \\
\int_0^\infty 
\frac{s_2^{u_2 - \frac{1}{2}}}{S^{\frac{D}{2} + 1}}
\cdot
s_2
\frac{\partial}{\partial s_2}
\mathsf{exp}
& =
- \int_0^\infty 
\frac{s_2^{u_2 - \frac{1}{2}}}{S^{\frac{D}{2} + 1}}
\left[
u_2 + 
\frac{1}{2} - 
\frac{s_2}{S} \left( \frac{D}{2} + 1 \right)
\right]
\mathsf{exp} ~, \\
\int_0^\infty 
\frac{s_3^{u_3 - 1}}{S^{\frac{D}{2} + 1}}
\cdot
s_3
\frac{\partial}{\partial s_3}
\mathsf{exp}
& =
- \int_0^\infty 
\frac{s_3^{u_3 - 1}}{S^{\frac{D}{2} + 1}}
\left[
u_3 - 
\frac{s_3}{S} \left( \frac{D}{2} + 1 \right)
\right]
\mathsf{exp} ~.
\end{align}
\end{subequations}
We find that the expression in the round brackets in \eqref{eq:STRFermionic_proof_2} does not contribute to \eqref{eq:STRFermionic_proof_1}, where we recall that $u_1 + u_2 + u_3 = D$.
Thus, we have
\begin{equation}
\mathsf{fstar}
=
\pi^{\frac{D}{2}}\,
x_{13,\alpha \dot{\alpha}}
\bar{x}_{23}^{\dot{\alpha} \beta}
\int_0^\infty \dd s_1\, \dd s_2\, \dd s_3 ~
\frac{
s_1^{u_1 - \frac{1}{2}}
s_2^{u_2 - \frac{1}{2}}
s_3^{u_3}
}{ \Gamma (u_1 + \frac{1}{2}) \Gamma (u_2 + \frac{1}{2}) \Gamma (u_3) } \,
S^{- \frac{D}{2} - 1}\,
\mathsf{exp} ~,
\end{equation}
and next we can make the same substitution as in the scalar case \eqref{eq:STR_Substitution}.
Therefore, we obtain
\begin{equation}
\begin{split}
\mathsf{fstar}
& =
\pi^{\frac{D}{2}}\,
x_{13,\alpha \dot{\alpha}}
\bar{x}_{23}^{\dot{\alpha} \beta}
\int_0^\infty \dd t_1\, \dd t_2\, \dd t_3 ~
\frac{
t_1^{\frac{D}{2} - u_1 - \frac{1}{2}}
t_2^{\frac{D}{2} - u_2 - \frac{1}{2}}
t_3^{\frac{D}{2} - u_3 - 1}
}{ \Gamma (u_1 + \frac{1}{2}) \Gamma (u_2 + \frac{1}{2}) \Gamma (u_3) } \,
\e^{- t_3 x_{12}^2 - t_2 x_{13}^2 - t_1 x_{23}^2} \\
& =
r_{\frac{1}{2}} (u_3 , u_1 , u_2)
\cdot
\frac{1}{\left[ x_{12}^2 \right]^{\frac{D}{2} - u_3}}
\cdot
\frac{1}{\left[ x_{13}^2 \right]^{\frac{D}{2} - u_2}}
\frac{x_{13, \alpha \dot{\alpha}}}{\left|x_{13}\right|}
\cdot
\frac{1}{\left[ x_{23}^2 \right]^{\frac{D}{2} - u_1}}
\frac{\bar{x}_{23}^{\dot{\alpha} \beta}}{\left|x_{23}\right|} ~,
\end{split}
\end{equation}
after reversing the Schwinger trick \eqref{eq:SchwingerTrick}.
This is the r.\,h.\,s.\ of \eqref{eq:STR_Fermion_medial} and completes the proof of the fermionic spin-$\frac{1}{2}$ STR.

\section{Details of the super-integral calculations}

\subsection{Super chain relation}
\label{subsec:SuperChainRule}
We will present the derivation of the super chain relations \eqref{eq:ChainRelAntiChiral} in three-dimensional $\mathcal{N} = 2$ superspace as well as four-dimensional $\mathcal{N} = 1$ superspace.
The computation to proof \eqref{eq:ChainRelChiral} works analogously.
We start with the integral 
\begin{equation}
\left.
\int \dd^D x_0\, \dd^2\bar{\theta}_0
\frac{1}{\left[x_{1\bar{0}}^2 \right]^{u_1}}
\frac{1}{\left[x_{2\bar{0}}^2 \right]^{u_2}}
\right|_{\substack{\theta_0 = 0 \\ \bar{\theta}_{1,2}=0}}
\end{equation}
and represent the fermionic dependency via shift operators\footnote{
A very useful relation for shift operators is
\begin{equation*}
\e^{\I \theta \sigma^\mu \bar{\theta}\, \partial_\mu}
\left(
f (x) \cdot g(x)
\right)
=
\e^{\I \theta \sigma^\mu \bar{\theta}\, \partial_\mu}
f (x) \;
\e^{\I \theta \sigma^\mu \bar{\theta}\, \partial_\mu}
g (x) ~.
\end{equation*}
}
, c.\ f.\ \eqref{eq:gen_superPropagator}. 
Afterwards, we can execute the integral over the bosonic subspace by \eqref{eq:ChainRuleScalar} to find
\begin{equation}
\begin{split}
\int \dd^2\bar{\theta}_0\;
\e^{- 2\I \theta_1 \sigma^\mu \bar{\theta}_0 \partial_{1,\mu}}
\e^{- 2\I \theta_2 \sigma^\nu \bar{\theta}_0 \partial_{2,\nu}}\;
\int \dd^D x_0\, 
\frac{1}{\left[x_{10}^2 \right]^{u_1}}
\frac{1}{\left[x_{20}^2 \right]^{u_2}}\\
=
\int \dd^2\bar{\theta}_0\;
\e^{- 2\I \theta_{12} \sigma^\mu \bar{\theta}_0 \partial_{1,\mu}}
\frac{r (D - u_1 - u_2  ,u_1, u_2)}{\left[ x_{12}^2 \right]^{u_1 + u_2 - \frac{D}{2}}} ~.
\end{split}
\end{equation}
We used the fact that $\partial_{2,\mu}$ can be replaced by $-\partial_{1,\mu}$, when acting on a function of $x_{12}$.
The Gra\ss mann integral can now be executed, which amounts to picking the second order in the expansion of the exponential. 
Consecutively using \eqref{eq:ThetaSquare_ThetaCube} for $D=4$ $\mathcal{N}=1$ superspace and \eqref{eq:ThetaSquare_ThetaCube_D3N2} for $D=3$ $\mathcal{N}=2$ superspace, leaves us with 
\begin{equation}
\theta_{12}^2 
\square_1 
\frac{r (D - u_1 - u_2  ,u_1, u_2)}{\left[ x_{12}^2 \right]^{u_1 + u_2 - \frac{D}{2}}}
=
\left\lbrace 
\begin{matrix}
4 \, r_0 ( 2 - u_1 - u_2 , u_1 , u_2 ) 
\frac{\theta_{12}^2 }{\left[ x_{12}^2 \right]^{u_1 + u_2 - \frac{1}{2}}}\\
- 4 \, r_0 ( 3 - u_1 - u_2 , u_1 , u_2 )
\frac{\theta_{12}^2 }{\left[ x_{12}^2 \right]^{u_1 + u_2 - 1}}
\end{matrix}
\right. 
 ~,
\end{equation}
where we used the functional relation of the gamma function after performing the derivative by the equation $\square_{1} [ x_{12}^2 ]^{-u} = - 4u (\frac{D}{2} - u -1) [ x_{12}^2 ]^{-u-1}$.
As in \eqref{eq:ChainRelAntiChiral}, the upper case is related to the three-dimensional $\mathcal{N} = 2$ superspace and the lower case corresponds to the four-dimensional $\mathcal{N} = 1$ superspace.

\subsection{An auxiliary relation for the construction of super x-unity}
\label{subsec:auxRelForXUnity}
We present the derivation of \eqref{eq:3ptFctnOnePointIntegrated} by direct calculation.
Again, the upper case refers to three-dimensional $\mathcal{N} = 2$ superspace and the lower case to the four-dimensional $\mathcal{N} = 1$ superspace.
The diagram reads
\begin{equation}
\left.
\int \dd^D x_0\; \dd^2\theta_0\,\dd^2\bar{\theta}_0 \;\delta^{(2)}(\theta_0)
\frac{1}{\left[ x_{2\bar{0}} \right]^{u_2}}
\frac{1}{\left[ x_{3\bar{0}} \right]^{D - 1 - u_1 - u_2}} ~
\int \dd^Dx_1\; \dd^2\theta_1\,\dd^2\bar{\theta}_1 \;\delta^{(2)}(\bar{\theta}_1)
\frac{1}{\left[ x_{1\bar{0}} \right]^{u_1}} \right|_{\bar{\theta}_{2,3} = 0} ~.
\label{eq:AuxRelDiagram}
\end{equation}
The fermionic part of the superspace integral over $z_1$ can be performed, and we find
\begin{equation}
\int \dd^D x_1\; \dd^2\theta_1\,\dd^2\bar{\theta}_1 \;\delta^{(2)}(\bar{\theta}_1)
\frac{1}{\left[ x_{1\bar{0}} \right]^{u_1}}
=
\left\lbrace 
\begin{matrix}
4 \\
- 4 
\end{matrix}
\right\rbrace
\cdot
u_1 (\tfrac{D}{2} - u_1 - 1)\;
\bar{\theta}^2_0\;
\int \dd^D x_1\;
\frac{1}{\left[ x_{10} \right]^{u_1 + 1}} ~.
\label{eq:AuxRel_1}
\end{equation}
Here we used again the relation $\square_{1} [ x_{12}^2 ]^{-u} = - 4u (\frac{D}{2} - u -1) [ x_{12}^2 ]^{-u-1}$.
Plugging \eqref{eq:AuxRel_1} into \eqref{eq:AuxRelDiagram}, we can perform the $z_0$ superspace integral.
The fermionic integrations evaluate the fermionic delta functions $\theta_0^2$ and $\bar{\theta}_0^2$, and the bosonic integral is performed by the bosonic star-triangle relation in $D$ dimensions \eqref{eq:CFTUniqueness_Scalar}.
We obtain for \eqref{eq:AuxRelDiagram}
\begin{equation}
\left\lbrace 
\begin{matrix}
4 \\
- 4 
\end{matrix}
\right\rbrace
\cdot
u_1 (\tfrac{D}{2} - u_1 - 1) 
r_0(u_2, D - 1 - u_1 - u_2, u_1 +1)
\int \dd^D x_1\; 
\frac{1}{\left[ x_{13} \right]^{\frac{D}{2} - u_2}}
\frac{1}{\left[ x_{23} \right]^{\frac{D}{2} - u_1 - 1}}
\frac{1}{\left[ x_{12} \right]^{u_1 + u_2 + 1 - \frac{D}{2}}} ~.
\end{equation}
Next, we use the bosonic chain relation \eqref{eq:ChainRuleScalar} to evaluate the integration over $x_1$ and we find the dependence on the external points vanishing and \eqref{eq:AuxRelDiagram} turning into a factor
\begin{equation}
\left\lbrace 
\begin{matrix}
4 \\
- 4 
\end{matrix}
\right\rbrace
\cdot
u_1 
(\tfrac{D}{2} - u_1 - 1) 
r_0(u_2, D - 1 - u_1 - u_2, u_1 +1)
r_0(D - u_1 - 1, u_1 + u_2 + 1 - \tfrac{D}{2}, \tfrac{D}{2} - u_2)
 ~.
\end{equation}
Finally, the result of \eqref{eq:3ptFctnOnePointIntegrated} can be obtained by using the explicit form of the $r_0$-factor \eqref{eq:Factor_rEll} and the functional relation of the gamma function.

\chapter{Perturbative calculations in bi-scalar fishnet theory}
\label{app:PerturbativeCalculationsInFishnetTheory}

\section{Renormalization of the double-trace couplings}
\label{sec:RenormalizationOfDoubleTraceCouplings}
We give a detailed derivation of the beta function of the double-trace couplings of bi-scalar fishnet theory as stated in section \ref{subsec:DoubleTraceTermsRenormalizationAndConformalFixedPoint}.
Our calculation is up to three loops\footnote{We thank G.\ Korchemsky for correspondence on this topic.}, however the fixed-point was obtained up to seven loops in \cite{Korchemsky:2018hnb} using Mellin-Barnes techniques.
In this section we use momentum-space Feynman diagrams and regularized the divergences by dimensional regularization.
Therefore, our spacetime is assumed to have $D = 4 - 2\varepsilon$ dimensions and once all the divergences are renormalized, we can take the limit $\varepsilon \rightarrow 0$.
Furthermore, we study the bi-scalar fishnet theory \eqref{eq:Fishnet_biscalar_action} on a spacetime with Euclidean signature.
In this section, we make use of momentum-space Feynman rules.

\subsection{Regularized amplitudes}
\label{subsec:RegularizedAmplitudes}
Formally, we introduce an counter term $\delta^{(L)}$ for the $\alpha_1^2$ coupling by 
\begin{equation}
\alpha_1^2 \rightarrow \alpha_{1,\mathrm{R}}^2 := \mu^\varepsilon \alpha_1^2 ~+~ \mu^\varepsilon \alpha_1^2 \cdot \delta^{(L)} := \mu^\varepsilon \alpha_1^2 Z^{(L)},
\label{eq:RenormalizedCouplingAlpha}
\end{equation}
and $Z^{(L)}=1+\delta^{(L)}$. 
In addition, we introduced a scale $\mu$, which vanished in the limit $D \rightarrow 4$.
The value of $\delta^{(L)}$, and therefore the value of $Z^{(L)}$, have to be determined loop-order by loop-order and the superscript $L$ indicates the loop level under examination. 
The condition for the tuning of $\delta^{(L)}$ is the finiteness of the four-point amplitude
\begin{equation}
A(p_1, p_2, p_3, p_4)
=
\left\langle 
\mathrm{tr}
\left( 
\phi_1 (p_1) \phi_1 (p_2) 
\right)
\mathrm{tr}
\left( 
\phi_1^\dagger (p_3) \phi_1^\dagger (p_4) 
\right)
\right\rangle
\label{eq:Fishnet_4ptForRen}
\end{equation}
up to $L$-loops, when one includes diagrams that contain the counter term vertex. 
The diagrams containing it are of lower loop order than $L$, but since diagrams without counter term vertices are divergent as $\frac{1}{\varepsilon^L}$ in dimensional regularization, $\delta^{(L)}$ admits an expansion
\begin{equation}
\delta^{(L)} ~=~ \dfrac{(z_{1,1}+z_{1,2}+z_{1,3}+...)}{\varepsilon} ~+~ \dfrac{(z_{2,2}+z_{2,3}+z_{2,4}+...)}{\varepsilon^2} ~+~ ... ~.
\end{equation}
The coefficients $z_{n,k}$ are determined by requiring the cancellation of the $\frac{1}{\varepsilon^n}$-divergences of the $k$-loop diagrams. 
Furthermore, the coefficients depend on the renormalized couplings, since they correspond to diagrams containing vertices proportional to the renormalized couplings, i.\,e.\ $z_{n,k}= z_{n,k}(\xi_\mathrm{R}^2,\alpha_{1,\mathrm{R}}^2)$.
At this point, we recall that the coupling $\xi^2$ is not renormalized, since diagrams like the one shown in \eqref{eq:Fishnet_missingXiRenormalization} are not possible.
However, we have to include the scale $\mu$ to keep the coupling dimensionless and we write $\xi_\mathrm{R}^2 = \mu^\varepsilon \xi^2$.

In order to keep track of the overall order of the couplings, it is handy to introduce a bookkeeping parameter $w$ and rescale the bare couplings as
\begin{equation}
\begin{array}{cc}
\xi^2 \rightarrow w\cdot \xi^2 ~, &
\alpha_{1}^2 \rightarrow w\cdot \alpha_{1}^2.
\end{array}
\label{eq:FishnetRenRescaling}
\end{equation}
Then, for the fishnet theory, the $L$-loop diagrams scale as $w^{L+1}$, which implies for for the coefficients 
\begin{equation}
z_{n,k} \rightarrow z_{n,k}w^{k}.
\end{equation}
We recall that $k$ refers to the number of loops at which the factor $z_{n,k}$ has to compensate the divergences.
However, special to fishnet theory with its quartic interactions, $k$ is related to the coupling-order, which is the number of vertices in a graph.
The reason can be found in the topology of the Feynman graphs (the skeleton ones, not the fatgraph).
We consider coordinate-space Feynman diagrams for this reasoning.
For a graph with $L$ loops it Eulers formula $L=E-V+1$ for connected graphs holds\footnote{This form can be derived from the general one, $\chi = b_0 - b_1 + b_2$, \eqref{eq:EulerCharacteristic}. Skeleton graphs are drawn on the plane, therefore $\chi = 2$. The number of vertices is the zeroth Betti number $b_0 = V$, the number of propagators the first, $b_1 = E$, and the number of faces $b_2$ is the number of loops plus the external face surrounding the graph. Therefore, we have $b_2 = L + 1$.}, where the number of vertices is the sum of four external ones (which spawn external edges) and internal ones $V= 4 + V_{\mathrm{int}}$. 
The number of propagators is half the number of open half-edges, which are the four external ones and, specific for the fishnet theory with quartic coupling, four times the number of internal vertices.
Thus, the number of propagators is $E=\frac{1}{2}(4+4\cdot V_{\mathrm{int}})$. 
Putting everything together, one finds $L= V_{\mathrm{int}}-1$, i.e. all diagrams with the same number of vertices belong to the same loop order. 
Hence, the exponent of the bookkeeping parameter $w$ keeps track of the number of vertices and loops.
The rescaled renormalization factor for L-loop renormalization can be written formally as the double sum
\begin{equation}
Z^{(L)} ~=~ 1+\delta^{(L)} = 1 + \sum_{k=1}^{L}  \sum_{n=1}^k \dfrac{1}{\varepsilon^n} z_{nk} w^{k} 
~~~ \mathrm{with}  ~~~
\delta_k ~=~ \sum_{n=1}^{k} \dfrac{1}{\varepsilon^n}  z_{nk} ~.
\label{eq:ZLExpansion2}
\end{equation}
Therefore, the rescaling \eqref{eq:FishnetRenRescaling} labels the different coupling/loop orders in the expansion of the renormalized coupling \eqref{eq:RenormalizedCouplingAlpha},
\begin{equation}
\alpha_{1,\mathrm{R}}^2 = \mu^\varepsilon \alpha_1^2 w + \mu^\varepsilon \alpha_1^2 \delta_1 w^2 + \mu^\varepsilon \alpha_1^2 \delta_2 w^3 + ... ~.
\label{eq:AlphaRinTermsOfDelta}
\end{equation}
Via the coefficients $\delta_k$, the diagram, which consist of only one $\alpha_{1,\mathrm{R}}^2$ vertex compensates the divergences of multiple higher-loop diagrams that are all of order $k+1$ in $w$. 
The ``bare'' higher-loop diagrams need to have $k+1$ vertices to match the order in $w$, since they only involve the first term on the r.\,h.\,s.\ of \eqref{eq:AlphaRinTermsOfDelta}, and by $L= V_{\mathrm{int}}-1$, those have $L=k$ loops. 
Of course, all coefficients $\delta_k$ participate at every higher Loop order $L\geq k$ as well, but at loop order $L=k$ they appear fist and exclusively at this order in the single vertex diagram.
Thus, they are determined the easiest at this loop level.

To determine the $\delta_k$'s, we follow the paradigm that after the transition of the bare coupling to the renormalized one, one requires to have a finite four point correlation function \eqref{eq:Fishnet_4ptForRen}. 
Prior to renormalizing the coupling (i.e. performing \eqref{eq:RenormalizedCouplingAlpha}), the $L$-loop four point functions receives contributions from diagrams with loops $\leq L$
\begin{equation}
A_L ~=~ \sum_{\ell = 0}^L A^{(\ell)} ~,
\end{equation}
where $A^{(\ell)}$ consists of all ``bare'' $\ell$-loop diagrams. 
Since, the number of loops is linked to the number of vertices and therefore to the power of $w$ (after introducing the bookkeeping parameter but not the renormalized coupling) by $L=V_{\mathrm{int}-1}$, the divergent $\ell$-loop contributions are
\begin{equation}
A^{(\ell)} ~=~ w^{\ell +1}\sum_{n=0}^\ell \frac{1}{\varepsilon^n} a^{(\ell)}_{n}(\alpha_1^2,\xi^2) ~.
\end{equation}
Replacing $\alpha_1^2$ by the renormalized coupling $\alpha_{1,\mathrm{R}}^2$ yields the finite amplitude
\begin{equation}
A^{(\ell)}_\mathrm{R} ~=~ w^{\ell +1}\sum_{n=0}^\ell \frac{1}{\varepsilon^n} \sum_{k=0}^\infty a^{(\ell)}_{\mathrm{R}, nk}(\alpha_1^2,\xi^2; \delta_k) w^k 
~~~ \mathrm{with} ~~~
A_{\mathrm{R},L} ~=~  \sum_{\ell=0}^L A^{(\ell)}_\mathrm{R} ~,
\end{equation}
where the coefficients $a^{(\ell)}_{\mathrm{R}, nk}$ compensate the $\frac{1}{\varepsilon^n}$ divergences.
Thus, renormalization up to $L$-loops means that all terms of $A_{\mathrm{R},L}$ of order smaller of equal to $w^{L+1}$ have to be finite. 
This statement can be made precise by introducing the operator $\kappa$, which just keeps the divergent part of any $\varepsilon$-expansion\footnote{This operator could also be defined differently to apply another scheme, e\,g.\ for the $\bar{\mathrm{MS}}$-scheme the $\kappa$-operator should be replaced by $\kappa^{\bar{\mathrm{MS}}}\left[\bullet\right] =\kappa\left[\bullet\right] + \gamma_\mathrm{E}$}, e\,g.\ $\kappa \left[\frac{2}{\varepsilon^2} +  \frac{7}{\varepsilon} + 4 + \varepsilon^5 \right] =\frac{2}{\varepsilon^2} +  \frac{7}{\varepsilon}$.
The central constraint, imposed by finiteness, that allows to determine the orders of the counter term $\delta_k$ order by order is
\begin{equation}
\kappa \left[ A_{\mathrm{R},L} \right] ~\stackrel{!}{=}~ 0 + \mathcal{O}\left( w^{L+2} \right) \, .
\label{eq:RenFinitnessEq}
\end{equation}
Essentially, this has to hold for every order in $w$ separately, thus this equation furnishes $L$ equations for all the $L$ unknowns $\delta_1,..., \delta_L$.
In the following, we renormalize $\alpha_1^2$ from \eqref{eq:Fishnet_4ptForRen} up to three loops, which yields $\delta_1$, $\delta_2$ and $\delta_3$.

\subsubsection*{Tree Level}
The renormalization procedure explained above will now be applied. The $0$-loop contribution, i.\,e.\ just the double-trace vertex, is
\begin{equation}
A^{(0)} ~ = ~ 4\cdot (4\pi)^2 \, \alpha_1^2   \longrightarrow
A^{(0)}_\mathrm{R} ~=~ 4\cdot (4\pi)^2 \, \mu^\varepsilon \left( \alpha_1^2 w + \alpha_1^2 \delta_1 w^2 + \alpha_1^2 \delta_2 w^3 +\mathcal{O}(w^4) \right) \, .
\label{eq:A0R}
\end{equation}
Formally one can now ``renormalize'' up to $0$-loop order by imposing finiteness by \eqref{eq:RenFinitnessEq}
\begin{equation}
\kappa \left[ A_{\mathrm{R},0}\right] = \kappa \left[ A^{(0)}_\mathrm{R} \right] = \kappa \left[ 4\cdot (4\pi)^2 \, \mu^\varepsilon  \alpha_1^2 w \right] +\mathcal{O}(w^2) = 0 + \mathcal{O}(w^2) \, .
\end{equation}
However, since there are not any divergences the condition \eqref{eq:RenFinitnessEq} is trivially satisfied and has not to be imposed. 
Note that the $\delta_k$'s are not appearing here because they are of a higher order in $w$.
Despite of containing divergent parts, they are not seen by the $\kappa$-operator.

\subsubsection{\boldmath One-loop level}
The first non-trivial order is at $1$-loop. 
We encounter the integral \cite{Korchemsky:2018hnb}
\begin{equation}
\begin{split}
\pi (s) ~ := ~ & \int \dfrac{\text{d}^{4-2\varepsilon}\ell}{\I (2\pi)^{4-2\varepsilon}} \dfrac{1}{\ell^2 (k-\ell)^2}
~ = ~ \dfrac{(-s/\mu^2)^{-\varepsilon}}{(4\pi)^{2-\varepsilon}}\dfrac{\Gamma (\varepsilon)\; \Gamma (1-\varepsilon)^2}{\Gamma (2-2\varepsilon)}\\
~ = ~ & \dfrac{1}{16\pi^2 \varepsilon} + \dfrac{2-\gamma_\mathrm{E} + \Log{\frac{4\pi \mu^2}{-s}}}{16\pi^2} + \mathcal{O}(\varepsilon) \, ,
\end{split}\label{eq:BubbleInt}
\end{equation} 
with $k^2 = s$ and we use the abbreviations $s_{ij}=(p_i + p_j)^2$ and $\gamma_\mathrm{E}$ for the Euler-Mascheroni constant.
The one-loop contribution is 
\begin{equation}
\begin{split}
A^{(1)} ~=~ &
\left[ \begin{gathered}\includegraphics[scale=0.8]{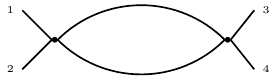}\end{gathered} + (p_1 \leftrightarrow p_2)\right] +
\begin{gathered}\includegraphics[scale=0.8]{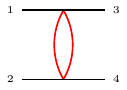} 
\end{gathered} +
\begin{gathered}\includegraphics[scale=0.8]{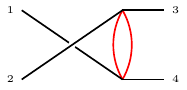} 
 \end{gathered}\\
~=~ & 8\cdot \left[ (4\pi)^2 \, \alpha_1^2\right]^2 \cdot \pi (s_{12}) 
~ + ~ \left[ (4\pi)^2 \, \xi^2\right]^2 \cdot \pi (s_{13})
~ + ~ \left[ (4\pi)^2 \, \xi^2\right]^2 \cdot \pi (s_{23})\\
~=~ & \dfrac{32\pi^2\, (\xi^4+4\, \alpha_1^4)}{\varepsilon} 
~+~ 32\pi^2\cdot \left( \xi^4+4\, \alpha_1^4\right) \left[ 2-\gamma_\mathrm{E} + \Log{4\pi} \right] \\
& ~-~ 16\pi^2\cdot \left[ 8\, \alpha_1^4 \, \Log{\frac{-s_{12}}{\mu^2}} + \xi^4 \, \Log{\frac{-s_{13}}{\mu^2}} +\xi^4 \, \Log{\frac{-s_{23}}{\mu^2}}\right] 
~+~ \mathcal{O}(\varepsilon) \, .
\label{eq:A1}
\end{split}
\end{equation}
We equipped the diagrams with the appropriate symmetry factors.
Inserting  the renormalized coupling \eqref{eq:AlphaRinTermsOfDelta} (although only the lowest order in $w$ is of importance here) for $\alpha_{1}^2$ and $\xi^2 \rightarrow \xi^2_\mathrm{R}$ yields
\begin{equation}
\begin{split}
A^{(1)}_{\mathrm{R}}
~=~ & \mu^{2\varepsilon}\dfrac{32\pi^2\, (\xi^4+4\, \alpha_1^4)}{\varepsilon}w^2 
~+~ 32\pi^2\cdot\mu^{2\varepsilon} \left( \xi^4+4\, \alpha_1^4\right) \left[ 2-\gamma_\mathrm{E} + \Log{4\pi} \right] w^2 \\
& ~-~ 16\pi^2\cdot \mu^{2\varepsilon} \left[ 8\, \alpha_1^4 \, \Log{\frac{-s_{12}}{\mu^2}} + \xi^4 \, \Log{\frac{-s_{13}}{\mu^2}} +\xi^4 \, \Log{\frac{-s_{23}}{\mu^2}}\right] w^2  \\
& ~+~ \mathcal{O}(\varepsilon) ~+~ \mathcal{O}(w^3) \, .
\end{split}
\end{equation}
The $1$-loop amplitude then is obtained by adding the relevant orders from the tree level \eqref{eq:A0R} to the one-loop contribution.
We find
\begin{align}
A_{\mathrm{R},1}&  = 4\cdot (4\pi)^2 \, \mu^\varepsilon \left( \alpha_1^2 w + \alpha_1^2 \delta_1 w^2\right) +\mu^{2\varepsilon}\dfrac{32\pi^2\, (\xi^4+4\, \alpha_1^4)}{\varepsilon}w^2 +\mathcal{O}(\varepsilon^0)+\mathcal{O}(w^3)
+O\left(\varepsilon \right) \, .
\label{eq:AR1}
\end{align}
Imposing finiteness by \eqref{eq:RenFinitnessEq} gives the defining equation for $\delta_1$,
\begin{equation}
\kappa\left[ A_{\mathrm{R},1} \right] = 4\cdot (4\pi)^2 \, \alpha_1^2 \kappa\left[\delta_1\right] w^2 + \dfrac{32\pi^2\, (\xi^4+4\, \alpha_1^4)}{\varepsilon}w^2 +\mathcal{O}(w^3) \stackrel{!}{=} 0 + \mathcal{O}(w^3) \, .
\end{equation}
Since the $\delta_k$'s are purely divergent, we have
\begin{equation}
\delta_1 = \kappa\left[\delta_1\right] = -\frac{\xi^4+4\, \alpha_1^4}{2\alpha_{1}^2}\frac{1}{\varepsilon} ~.
\label{eq:Fishnet_delta1_bare}
\end{equation}
Note that here it makes not difference if the renormalized or bare couplings appear because we can neglect higher orders in $w$ and $\varepsilon$, since $\xi_\mathrm{R}^2 = \xi^2+\mathcal{O}(\varepsilon)$ and $\alpha_{1,\mathrm{R}}^2=\alpha_1^2 + \mathcal{O}(w)+\mathcal{O}(\varepsilon)$. 
This allows to reexpress \eqref{eq:Fishnet_delta1_bare} as 
\begin{equation}
\delta_1 = -\frac{\xi_\mathrm{R}^4+4\, \alpha_{1,\mathrm{R}}^4}{2\alpha_{1,\mathrm{R}}^2}\frac{1}{\varepsilon} ~.
\end{equation}
With that and \eqref{eq:ZLExpansion2}, we find the first contribution to the renormalization factor
\begin{equation}
z_{11}=-\frac{\xi_\mathrm{R}^4+4\, \alpha_{1,\mathrm{R}}^4}{2\alpha_{1,\mathrm{R}}^2} ~.
\end{equation}

\subsubsection{\boldmath Two-loop level}
At two loops, we need another Feynman integral \cite{Korchemsky:2018hnb}, 
\begin{equation}
\begin{split}
V (s) ~ := ~ & \int \dfrac{\text{d}^{4-2\varepsilon}\ell}{\I (2\pi)^{4-2\varepsilon}} \dfrac{\pi (\ell^2)}{(\ell + p_1)^2 (\ell - p_2)^2}
~ = ~ \dfrac{(-s/\mu^2)^{-2\varepsilon}}{(4\pi)^{4-2\varepsilon}}\dfrac{\Gamma (\varepsilon)\;\Gamma (2\varepsilon)\; \Gamma (1-\varepsilon)^2\; \Gamma (1-2\varepsilon)^2}{\Gamma (2-2\varepsilon)\;\Gamma (2-3\varepsilon)}\\
~ = ~ & \dfrac{1}{512\pi^4 \varepsilon^2} + \dfrac{5-2\gamma_\mathrm{E} + 2\,\Log{\frac{4\pi \mu^2}{-s}}}{512\pi^4\varepsilon} + \mathcal{O}(\varepsilon^0) \, .
\end{split}\label{eq:HVInt}
\end{equation}
The bare $2$-loop contribution is made up by the diagrams 
\begin{equation}
\begin{split}
A^{(2)} ~=~ & \left[ \begin{gathered}\includegraphics[scale=0.8]{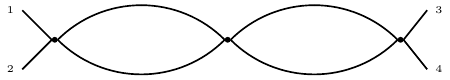} 
\end{gathered} + \begin{gathered}\includegraphics[scale=0.8]{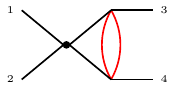} 
\end{gathered} + \begin{gathered}\includegraphics[scale=0.8]{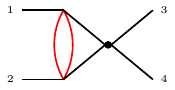} 
 \end{gathered} \right] \\
& ~+~ (p_1 \leftrightarrow p_2)\\
~=~ & 16\cdot \left[ (4\pi)^2 \, \alpha_1^2\right]^3 \cdot \pi (s_{12})^2 
~+~ 2\cdot 4\; \left[ (4\pi)^2 \, \xi^2\right]^2 (4\pi)^2 \, \alpha_1^2 \cdot V(s_{12})\\
~=~ & \dfrac{64\pi^2\cdot \alpha_1^2 \left( \xi^4+4\, \alpha_1^4\right)}{\varepsilon^2}\\
& ~+~ \dfrac{64\pi^2\cdot \alpha_1^2 \left( \xi^4+4\, \alpha_1^4\right)}{\varepsilon}\left[ 5-2\gamma_\mathrm{E} -2\Log{\frac{-s_{12}}{4\pi\mu^2}} \right]
~-~ \dfrac{64\pi^2\cdot \alpha_1^2 4\, \alpha_1^6}{\varepsilon} ~+~ \mathcal{O}(\varepsilon^0) \, .
\end{split}
\label{eq:A2}
\end{equation}
After introducing the renormalized couplings, one finds
\begin{equation}
\begin{split}
A^{(2)}_\mathrm{R} ~=~ & \frac{64 \pi ^2 \alpha _1^2 w^3 \left(4 \alpha _1^4+\xi ^4\right)}{\varepsilon ^2} \\
& -\frac{64 \pi ^2 \alpha _1^2 w^3 \left[8 \alpha _1^4 \left(\log \left(\frac{s}{\mu ^2}\right)+\gamma_\mathrm{E} -2\right)+\xi ^4 \left(2 \log \left(\frac{s}{\mu ^2}\right)+2 \gamma_\mathrm{E} -5\right)\right]}{\varepsilon } +\mathcal{O}\left(\varepsilon ^0\right) + \mathcal{O}\left(w^4\right) .
\end{split}
\end{equation}
Notice that $\delta_1$ and $\delta_2$ are in terms $\mathcal{O}(w^4)$ in the equation above but they enter $A_{\mathrm{R},2}$ by $A^{(1)}_\mathrm{R}$ and $A^{(0)}_\mathrm{R}$ at the order $\mathcal{O}(w^3)$, respectively. 
The sum yields the divergent part of the $2$-loop amplitude
\begin{equation}
\begin{split}
\kappa\left[A_{\mathrm{R},2}\right] ~=~ & 
w^3 \left\lbrace\frac{64 \pi ^2 \alpha _1^2 \left(4 \alpha _1^4+\xi ^4\right)}{\varepsilon ^2}\right.\\
&\left. -\frac{64 \pi ^2 \alpha _1^2 \left[-4 \alpha _1^2 \delta _1+8 \alpha _1^4 \left(\log \left(\frac{s}{\mu ^2}\right)+\gamma_\mathrm{E} -2\right)+\xi ^4 \left(2 \log \left(\frac{s}{\mu ^2}\right)+2 \gamma_\mathrm{E} -5\right)\right]}{\varepsilon }\right. \\
& \left. -256 \pi ^2 \alpha _1^4 \delta _1 \left(\log \left(\frac{s}{\mu ^2}\right)+\gamma_\mathrm{E} -2\right)+64 \pi ^2 \alpha _1^2\delta _2\right\rbrace\\
& + w^2 \left\lbrace 64 \pi ^2 \alpha _1^2 \delta _1+\frac{32 \pi ^2 \left(4 \alpha _1^4+\xi ^4\right)}{\epsilon }\right\rbrace .
\end{split}
\end{equation}
With this expression one can solve the system of equations imposed by the the vanishing of the orders $w^3$ and $w^4$ separately by \eqref{eq:RenFinitnessEq}. 
With $\delta_1=\kappa \left[\delta_1\right]$ and $\delta_2=\kappa \left[\delta_2\right]$, the solutions are
\begin{align}
\delta_1 = -\frac{\xi^4+4\, \alpha_1^4}{2\alpha_{1}^2}\frac{1}{\varepsilon} 
~~~~\mathrm{and} ~~~~
\delta_2 = -\frac{\xi ^4}{\varepsilon} +\frac{\xi ^4 + 4 \alpha _1^4}{\varepsilon^2} ~.
\end{align}
Finally, one can replace the bare couplings by the renormalized ones.
As before, the corrections are of higher order in $w$ than the one corresponding to the $\delta_k$'s or $z_{nk}$'s and we find
\begin{subequations}
\begin{align}
\delta_1 = -\frac{\xi_\mathrm{R}^4+4\, \alpha_{1,\mathrm{R}}^4}{2\alpha_{1,\mathrm{R}}^2}\frac{1}{\varepsilon} 
~~~ & \Leftrightarrow ~~~ z_{11}= -\frac{\xi_\mathrm{R}^4+4\, \alpha_{1,\mathrm{R}}^4}{2\alpha_{1,\mathrm{R}}^2} ~,\\
\delta_2 = -\frac{\xi_\mathrm{R} ^4}{\varepsilon} +\frac{\xi_\mathrm{R}^4 + 4 \alpha _{1,\mathrm{R}}^4}{\varepsilon^2}
~~~ & \Leftrightarrow ~~~ z_{12}= -\xi_\mathrm{R}^4 , ~~z_{22}=\xi_\mathrm{R} ^4 + 4 \alpha_{1,\mathrm{R}}^4 ~.
\end{align}
\end{subequations}

\subsubsection{\boldmath Three-loop level}
At three loops, the expression for the amplitude-contribution becomes rather lengthy.
We present the Feynman graphs used and the result for the renormalization factor $Z^{(3)}$.
We find the renormalization coefficients of the lowest order to be
\begin{equation}
\begin{split}
Z
=&
1
-
\frac{4 \alpha_{1,\mathrm{R}}^4+\xi_\mathrm{R}^4}{2 \alpha_{1,\mathrm{R}}^2 \varepsilon }
-
\frac{\xi_\mathrm{R}^4}{\varepsilon }
+
\frac{\xi_\mathrm{R}^8-4 \alpha_{1,\mathrm{R}}^4 \xi_\mathrm{R}^4}{6 \alpha_{1,\mathrm{R}}^2 \varepsilon }
+
\frac{4 \alpha_{1,\mathrm{R}}^4+\xi_\mathrm{R}^4}{\varepsilon ^2}
+
\frac{20 \alpha_{1,\mathrm{R}}^4 \xi_\mathrm{R}^4+\xi_\mathrm{R}^8}{6 \alpha_{1,\mathrm{R}}^2 \varepsilon ^2}\\
&-
\frac{16 \alpha_{1,\mathrm{R}}^4 \xi_\mathrm{R}^4+48 \alpha_{1,\mathrm{R}}^8+\xi_\mathrm{R}^8}{6 \alpha_{1,\mathrm{R}}^2 \varepsilon ^3} ~,
\label{eq:Alpha1ZFactor}
\end{split}
\end{equation}
where the last term is a three-loop contribution.
There are four different classes of Feynman diagrams with three loops, which contribute the the correlation function \eqref{eq:Fishnet_4ptForRen}.
We use the abbreviation $S := \log \frac{s}{4 \pi  \mu ^2}$.
\begin{itemize}
\item
The chain of three bubble integrals is the single one \eqref{eq:BubbleInt}, raised to the third power,
\begin{equation}
\begin{split}
G_{1} ~:=~ & \begin{gathered}\includegraphics[scale=0.8]{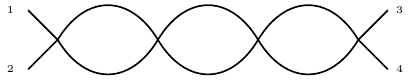} \end{gathered} \\
~=~ &  16 \cdot \left[ (4\pi)^2 \alpha_1^2 \right]^4 \cdot \pi (s_{12})^3 	
~=~ \alpha _1^8 \cdot 8^{2 \varepsilon +3} \pi ^{3 \varepsilon +2}\frac{ \Gamma (1-\varepsilon )^6 \Gamma (\varepsilon )^3}{\Gamma (2-2 \varepsilon )^3}\left(\frac{s}{\mu ^2}\right)^{-3 \varepsilon }\\	
~=~ & \frac{512 \pi ^2 \alpha _1^8}{\varepsilon ^3}
 -\frac{1536 \pi ^2 \alpha _1^8 \left[S +\gamma_\mathrm{E} -2\right]}{\varepsilon ^2} \\
 & -\frac{128 \pi ^2 \alpha _1^8 \left[-18 S^2 -36 (\gamma_\mathrm{E} -2) S -18 \gamma_\mathrm{E} ^2+\pi ^2+72 \gamma_\mathrm{E} -96\right]}{\varepsilon }\\
 & + \mathcal{O}\left( \varepsilon^0\right) .
\end{split}
\label{eq:ThreeBubbles}
\end{equation}

\item
The combination of a bubble integral \eqref{eq:BubbleInt} and the two-loop integral \eqref{eq:HVInt} is 
\begin{equation}
\begin{split}
	G_{2} ~:=~ & \begin{gathered}\includegraphics[scale=0.8]{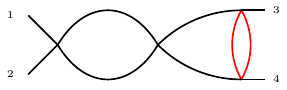}\end{gathered} ~=~ D_{\mathrm{HVV}} \\
					~=~ &  4 \cdot \left[ (4\pi)^2 \alpha_1^2 \right]^2 \left[ (4\pi)^2 \xi^2 \right]^2 \cdot \pi (s_{12}) V(s_{12}) \\	
					~=~ & \alpha _1^4 \;\xi ^4 \cdot 2^{6 \varepsilon +7} \pi ^{3 \varepsilon +2}\frac{ \Gamma (1-2 \varepsilon )^2 \Gamma (1-\varepsilon )^4 \Gamma (\varepsilon )^2 \Gamma (2 \varepsilon ) }{\Gamma (2-3 \varepsilon ) \Gamma (2-2 \varepsilon )^2}\left(\frac{s}{\mu ^2}\right)^{-3 \varepsilon }\\	
~=~ & \frac{64 \pi ^2 \alpha _1^4 \xi ^4}{\varepsilon ^3}
-\frac{64 \pi ^2 \alpha _1^4 \xi ^4 \left(3 S+3 \gamma_\mathrm{E} -7\right)}{\varepsilon ^2}\\
& +\frac{16\pi ^2\xi ^4\left[ 54 S^2 +36 (3 \gamma_\mathrm{E} -7) S  +54 \gamma_\mathrm{E} ^2+\pi ^2 -252 \gamma_\mathrm{E} + 396\right]}{3 \varepsilon } \\
& +\mathcal{O}\left( \varepsilon^0\right) .
\end{split}
\end{equation}

\item 
The diagram with a $\alpha_1^2$ vertex left and right is a special case of a so-called kite diagram.
It is determined in \cite{Grozin:2012} via a hypergeometric function.
\begin{equation}
\begin{split}
	G_{3} ~:=~ & \begin{gathered}\includegraphics[scale=0.8]{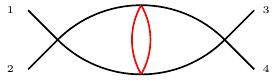}\end{gathered} \\
					~=~ & 8\cdot \left(4\pi\alpha_1\right)^4 \left(4\pi\xi\right)^4 \left(-\mu^{2}\right)^{2\varepsilon}\\
					 & \cdot \int \dfrac{\dd^{4-2\varepsilon}k_1}{\I\left( 2\pi\right)^{4-2\varepsilon}}
					\dfrac{\dd^{4-2\varepsilon}k_2}{\I\left( 2\pi\right)^{4-2\varepsilon}} 
					\dfrac{1}{k_1^2\, \left(k_1-p\right)^2}\,\pi \left( \left(k_1-k_2\right)^2\right)\dfrac{1}{k_2^2\, \left(k_2-p\right)^2} \\
					~=~ & 8\cdot \alpha _1^4 \xi ^4 \cdot 2^{6 \varepsilon +5} \pi ^{3 \varepsilon +2} \cdot \varepsilon\;  \frac{\Gamma (1-2 \varepsilon ) \Gamma (1-\varepsilon )^4 \Gamma (\varepsilon ) \Gamma (3 \varepsilon -1)}{(3 \varepsilon -2) \Gamma (3-4 \varepsilon ) \Gamma (2-2 \varepsilon ) \Gamma (\varepsilon +1)} \left(\frac{\mu ^2}{s}\right)^{3 \varepsilon } \\
					&\cdot \left[  _3F_2\left(\begin{smallmatrix}
					1, ~ 2-3 \varepsilon , ~ 2-2 \varepsilon \\ ~ 3-4 \varepsilon ,  ~ 3-3 \varepsilon ~
					\end{smallmatrix} \vert\, 1 \right) +(3 \varepsilon -2) \frac{\Gamma (\varepsilon ) \Gamma (2 \varepsilon -1) \Gamma (3-4 \varepsilon )}{\Gamma (1-\varepsilon )}~\cos (2 \pi  \varepsilon ) \right] \\		
~=~ &  \frac{128 \pi ^2 \alpha _1^4 \xi ^4}{3 \varepsilon ^3}-\frac{128\pi ^2 \alpha _1^4 \xi ^4 \left[3 S +3 \gamma_\mathrm{E} -7\right]}{3 \varepsilon ^2}\\
& -\frac{32 \pi ^2 \alpha _1^4 \xi ^4 \left[-18 S^2 -12 (3 \gamma_\mathrm{E} -7) S -18 \gamma_\mathrm{E} ^2+\pi ^2+84 \gamma_\mathrm{E} -124\right]}{3 \varepsilon } \\
& + \mathcal{O}\left( \varepsilon^0 \right) .
\end{split}
\label{eq:TheEyeDiag}
\end{equation}

\item 
The last diagram is a variation of a box-integral, where two propagators have non-unit exponents.
It is presented in \cite{tarasov2017massless}.
\begin{equation}
\begin{split}
	G_{4} ~:=~ & \begin{gathered}\includegraphics[scale=0.8]{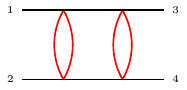}\end{gathered} ~+~ (p_1\leftrightarrow p_2)\\
	~=~ &  \left( 4\pi\xi \right)^8 \left( \mu^2 \right)^{-2\varepsilon}
	\int\dfrac{\dd^{4-2\varepsilon}k}{\I\left( 2\pi\right)^{4-2\varepsilon}} \pi \left( \left(k-p_1\right)^2\right) \dfrac{1}{k^2\, \left(k-p\right)^2} \pi \left( \left(k-p_3\right)^2\right)\\
					~=~ & (4 \pi )^8\xi ^8 \cdot \I^{2 \varepsilon } (4 \pi )^{3\varepsilon -6} \left(\frac{s}{\mu ^2}\right)^{-2 \varepsilon -\delta} \left(\frac{t}{\mu ^2}\right)^{-3 \varepsilon -\delta} \left(\frac{ \Gamma (1-\varepsilon )^2 \Gamma (\varepsilon )}{\Gamma (2-2 \varepsilon )}\right)^2\\
					&\left[ \left(\frac{s}{\mu ^2}\right)^{2 \varepsilon +\delta} \text{c}_1(1,\varepsilon +\delta,1,\varepsilon\vert 4-2 \varepsilon ) ~ \, _3F_2\left( \begin{smallmatrix}1,~1,~3 \varepsilon +\delta \\ 2 \varepsilon +1,~2 \varepsilon +1 +\delta\end{smallmatrix}\vert -\frac{s}{t}\right)\right.\\
					&  + \left(\frac{s}{\mu ^2}\right)^{\delta} \left(\frac{t}{\mu ^2}\right)^{2\varepsilon} \cdot\text{c}_2(1,\varepsilon ,1,\varepsilon +\delta \vert 4-2 \varepsilon ) ~\, _2F_1\left(\begin{smallmatrix}  1-2 \varepsilon ,~\varepsilon +\delta \\1+\delta \end{smallmatrix}\vert -\frac{s}{t}\right) \\
					& \left. + \left(\frac{t}{\mu ^2}\right)^{2\varepsilon+\delta} \cdot\text{c}_2(1,\varepsilon +\delta,1,\varepsilon +\delta \vert 4-2 \varepsilon ) ~\, _2F_1\left(\begin{smallmatrix}  1-2 \varepsilon -\delta ,~\varepsilon  \\1-\delta \end{smallmatrix}\vert -\frac{s}{t}\right)\right] \\
					& + (t \rightarrow u)\\
					 ~=~ & \frac{32 \pi ^2 \xi ^8}{3 \varepsilon ^3} -\frac{32 \pi ^2 \xi ^8 \left(3 S +3 \gamma_\mathrm{E} -8\right)}{3 \varepsilon ^2} \\
					&+\frac{ 8 \pi ^2 \xi ^8 \left[18 S^2 +12 (3 \gamma_\mathrm{E} -8) S +18 \gamma_\mathrm{E} ^2+3 \pi ^2-96 \gamma_\mathrm{E} +176\right]}{3 \varepsilon } \\
		&+ \mathcal{O}\left( \varepsilon^0\right) + \mathcal{O}\left( \delta\right) .
\end{split}
\end{equation}	
With $\nu_{ijkl} = \nu_i + \nu_j +\nu_k + \nu_l$, $\nu_{ijk} = \nu_i + \nu_j +\nu_k$ and $\nu_{ij} = \nu_i + \nu_j$, we use the abbreviations 
\begin{subequations}
\begin{align}
\text{c}_1(\nu_1,\nu_2,\nu_3,\nu_4 \vert d ) ~=~ & \frac{\I^d \Gamma \left(\frac{d}{2}-\nu _{123}\right) \Gamma \left(\frac{d}{2}-\nu _{134}\right) \Gamma \left(-\frac{d}{2}+\nu _{1234}\right)}{\Gamma \left(\nu _2\right) \Gamma \left(\nu _4\right) \Gamma \left(d-\nu _{1234}\right)}\\
\text{c}_2(\nu_1,\nu_2,\nu_3,\nu_4 \vert d ) ~=~ & \frac{\I^d \Gamma \left(\nu _2-\nu _4\right) \Gamma \left(\frac{d}{2}-\nu _{12}\right) \Gamma \left(\frac{d}{2}-\nu _{23}\right) \Gamma \left(-\frac{d}{2}+\nu _{123}\right)}{\Gamma \left(\nu _1\right) \Gamma \left(\nu _2\right) \Gamma \left(\nu _3\right) \Gamma \left(d-\nu _{1234}\right)},
\end{align}
\end{subequations}	
where $c_2$ is divergent for $\nu_2 = \nu_4$, however this is compensated in $G_4$ by the introduction of $\delta$ by modifying $\nu_2 \rightarrow \nu_2 +\delta$, then expanding $\delta$ around zero shows that the divergence gets canceled by the hypergeometric function and the final result is finite (modulo the divergences in $\varepsilon \rightarrow 0$).
\end{itemize}

\subsection{Beta function of the double-trace coupling}
The beta function describes how the coupling $\alpha_{1,\mathrm{R}}^2$ runs with the scale $\mu$.
However, we want the bi-scalar fishnet theory to be a CFT.
And indeed, we have to make sure that the coupling $\alpha_{1,\mathrm{R}}^2$ is scale-independent.
This implies that we have to tune the double-trace coupling to a zero of the beta function.
Here we determine the beta function and its zeros.
By definition, we have
\begin{equation}
\beta_{\alpha_1^2} ~:=~ \mu \dfrac{\mathrm{d}}{\mathrm{d}\mu} \alpha_{1,\mathrm{R}}^2,
\end{equation} 
where, in dimensional regularization, $\alpha_{1,\mathrm{R}}^2=\mu^\varepsilon \alpha_1^2 Z^{(L)}$ has a non-trivial $\mu$-dependence also through $Z^{(L)}$. 
Performing the derivative, we obtain the following expressions for the beta function
\begin{subequations}
\begin{align}
\beta_{\alpha_1^2} ~=~ & \varepsilon\cdot \mu^\varepsilon Z^{(L)} \alpha_1^2 ~+~ \alpha_1^2 \mu^\varepsilon \cdot \mu \dfrac{\mathrm{d}Z^{(L)}}{\mathrm{d}\mu}\label{eq:BetaFctBare}\\
~=~ & \varepsilon \cdot \alpha_{1,\mathrm{R}}^2 ~+~ \alpha_{1,\mathrm{R}}^2 \cdot \frac{\mu}{Z^{(L)}}\dfrac{\mathrm{d}Z^{(L)}}{\mathrm{d}\mu}, \label{eq:BetaFctRen}\\
~=~ & \varepsilon \cdot \alpha_{1,\mathrm{R}}^2 ~+~ \beta_{\alpha_1^2}^{D=4} ~,
\end{align}
\end{subequations}
expressed in the bare- and renormalized coupling, respectively. 
In the last line, the second term was denoted as $\beta_{\alpha_1^2}^{D=4}$, since it is indeed the four dimensional beta function, i.\,e.\ $\beta_{\alpha_1^2} \stackrel{\varepsilon\rightarrow 0}{\longrightarrow} \beta_{\alpha_1^2}^{D=4}$. 
Thus, for its explicit calculation, knowledge about $Z^{(L)}$ is decisive.

The beta function of $\xi^2$ formally plays a role as well, despite of being zero when $\varepsilon\rightarrow 0$. 
It is 
\begin{equation}
\beta_{\xi^2} ~=~ \mu \dfrac{\mathrm{d}}{\mathrm{d}\mu} \xi_{\mathrm{R}}^2 = \varepsilon \cdot \xi_{\mathrm{R}}^2 ~,
\end{equation}
and $Z_{\xi^2}=1$ because there are no diagrams, which could renormalize this interaction, see \eqref{eq:Fishnet_missingXiRenormalization}. 
Consequently, we have $\beta_{\xi^2}^{D=4}=0$. 
However, $\beta_{\xi^2}$ is required for the calculation for the beta function of $\alpha_1^2$, since $Z^{(L)}$ depends on $\xi^2_{\mathrm{R}}$.
Using the chain rule for derivatives in \eqref{eq:BetaFctRen}, we obtain
\begin{equation}
\beta_{\alpha_1^2} ~=~ \varepsilon \cdot \alpha_{1,\mathrm{R}}^2 ~+~ \alpha_{1,\mathrm{R}}^2 \cdot \frac{1}{Z^{(L)}}\left[ \dfrac{\partial Z^{(L)}}{\partial\xi_\mathrm{R}^2} \beta_{\xi^2} + \dfrac{\partial Z^{(L)}}{\partial\alpha_{1,\mathrm{R}}^2} \beta_{\alpha_{1}^2} \right] .
\label{eq:BetaAlpha1InTermsOfZ}
\end{equation} 
This equation has to be solved for $\beta_{\alpha_1^2}$. 
To do so, we first use the fact that beta functions are finite for $\varepsilon \rightarrow 0$ and we make the ansatz $\beta_{\alpha_1^2}= \varepsilon \beta_{\alpha_1^2}^{(1)} + \beta_{\alpha_1^2}^{(0)}$. 
Next, we multiply both sides of \eqref{eq:BetaAlpha1InTermsOfZ} by $Z^{(L)}$ and use the expansion \eqref{eq:ZLExpansion2} at $L\rightarrow \infty$. 
Finally, comparing terms proportional to the same powers of $\varepsilon$ gives
\begin{subequations}
\begin{align}
\mathrm{Order}~\varepsilon: ~~~ & \beta_{\alpha_1^2}^{(1)} ~=~ \alpha_{1,\mathrm{R}}^2 \label{eq:BetaAlpha1Order1}\\
\mathrm{Order}~\varepsilon^0: ~~~ & \beta_{\alpha_1^2}^{(0)} ~=~ 
\sum_{k=1}^\infty \alpha_{1,\mathrm{R}}^2 \left[ \xi_\mathrm{R}^2 \dfrac{\partial z_{1k}}{\partial \xi_\mathrm{R}^2} + \alpha_{1,\mathrm{R}}^2 \dfrac{\partial z_{1k}}{\partial \alpha_{1,\mathrm{R}}^2}\right]  w^k  ~=~ \beta_{\alpha_1^2}^{D=4} ~. \label{eq:BetaAlpha1Order0}
\end{align}
\end{subequations}
Since the beta function is finite, the other orders $\frac{1}{\varepsilon^m}$ yield further equations involving the coefficients $z_{nk}$, which can be used as consistency checks for the calculation of the diagrams.
Note that the beta function depends through \eqref{eq:BetaAlpha1Order0} only on the coefficients $z_{1k}$, that are those corresponding to the $\frac{1}{\varepsilon}$-poles of the $k$-loop diagrams. 
In the end, the full beta function is
\begin{equation}
\beta_{\alpha_1^2} ~=~ \varepsilon \cdot \alpha_{1,\mathrm{R}}^2 ~+~ \sum_{k=1}^\infty \alpha_{1,\mathrm{R}}^2 \left[ \xi_\mathrm{R}^2 \dfrac{\partial z_{1k}}{\partial \xi_\mathrm{R}^2} + \alpha_{1,\mathrm{R}}^2 \dfrac{\partial z_{1k}}{\partial \alpha_{1,\mathrm{R}}^2}\right]  w^k ~.
\label{eq:BetaAlpha1Expansion}
\end{equation}
Only the second term, which is $\beta_{\alpha_1^2}^{D=4}$, survives in the limit $\varepsilon \rightarrow 0$. 
The introduction of the bookkeeping parameter can always be undone by setting $w\rightarrow 1$.

For the three-loop result of the renormalization factor \eqref{eq:Alpha1ZFactor}, we find the double-trace beta function of bi-scalar fishnet theory,
\begin{equation}
\beta_{\alpha_1^2}^{3-\mathrm{loop}} 
~=~ \varepsilon \cdot \alpha_{1,\mathrm{R}}^2 
~-~ \dfrac{1}{2}\left( \xi_\mathrm{R}^4 + 4 \alpha_{1,\mathrm{R}}^4 \right)
~-~  2\alpha_{1,\mathrm{R}}^2 \xi_\mathrm{R}^4
~+~ \dfrac{1}{2}\left( -4\alpha_{1,\mathrm{R}}^4 \xi_\mathrm{R}^4 + \xi_\mathrm{R}^8 \right) ~.
\end{equation}
The four-dimensional beta function has to be zero for conformal symmetry on the quantum level.
This is fulfilled, if the fixed-point of the coupling $\alpha_{1,\mathrm{R}*}$ satisfies
\begin{equation}
\beta_{\alpha_1^2}^{3-\mathrm{loop},D=4} 
~=~ 
~-~ \dfrac{1}{2}\left( \xi_\mathrm{R}^4 + 4 \alpha_{1,\mathrm{R}*}^4 \right)
~-~  2\alpha_{1,\mathrm{R}*}^2 \xi_\mathrm{R}^4
~+~ \dfrac{1}{2}\left( -4\alpha_{1,\mathrm{R}*}^4 \xi_\mathrm{R}^4 + \xi_\mathrm{R}^8 \right) =0 ~.
\end{equation}
This equation has two solutions for $\alpha_{1,\mathrm{R}*}^2$, namely
\begin{equation}
\alpha_{1,\mathrm{R}*}^2 ~=~ \dfrac{-\xi_\mathrm{R}^4 \pm \sqrt{-\xi_\mathrm{R}^4+\xi_\mathrm{R}^8+\xi_\mathrm{R}^{12}}}{2(1+\xi_\mathrm{R}^4)} 
~=~ \pm \dfrac{\I}{2}\xi_\mathrm{R}^2 -\dfrac{\xi_\mathrm{R}^4}{2} \mp \dfrac{3\I}{4}\xi_\mathrm{R}^6 + \mathcal{O}(\xi_\mathrm{R}^8) ~,
\label{eq:Alpha1FixedPointExpansion3L}
\end{equation}
which are the first three orders of the expansion of the fixed points, when expanding for small $\xi_\mathrm{R}^2$. 
Note that at one of the fixed points \eqref{eq:Alpha1FixedPointExpansion3L}, the coupling renormalization $Z^{(3)}$ is not $1$ but 
\begin{equation}
Z^{(3)} ~=~ 1 + \left[ \xi_\mathrm{R}^4 - \dfrac{4\I}{3} \xi_\mathrm{R}^6 +\mathcal{O}(\xi_\mathrm{R}^{10}) \right] \dfrac{1}{\varepsilon} + \mathcal{O}(\tfrac{1}{\varepsilon^2})
\end{equation}
for small $\xi_\mathrm{R}^2$.
Finally, we remark that from now on, section \ref{subsec:DoubleTraceTermsRenormalizationAndConformalFixedPoint} and below, we will drop the subscript and call the renormalized couplings again $\alpha_1^2$ and $\xi^2$.

\section[The anomalous dimension of $\mathrm{tr} ( \phi_1^2 )$]{\boldmath The anomalous dimension of $\mathrm{tr} ( \phi_1^2 )$}
\label{sec:PerturbativeCalculationAnomalousDimension}
At the fixed-points of the double-trace couplings, in particular $\alpha_1^2 = \alpha_{1,*}^2$, see \eqref{eq:Alpha1FixedPointExpansion3L}, the bi-scalar fishnet theory \eqref{eq:Fishnet_biscalar_action} is a CFT.
Conformal symmetry implies that the properly renormalized two-point functions of an operator $\mathcal{O}$ take the following form:
\begin{equation}
\begin{split}
\corr{\mathcal{O}(x)\bar{\mathcal{O}}(y)}_\mathrm{R} 
&= 
\mu^{\gamma (\xi^2)}\sqrt{Z_\mathcal{O}}^2 \corr{\mathcal{O}(x)\bar{\mathcal{O}}(y)} 
= 
\dfrac{v(\xi^2)\mu^{\gamma (\xi^2)}}{\left|x-y \right|^{2\Delta_0 + \gamma (\xi^2)} }\\
& = 
\dfrac{v_0}{\left|x-y\right|^{2\Delta_0}}
\left[ 
1-\frac{\xi ^2 \left( v_0 \gamma_1 L - v_1 \right)}{v_0}+\frac{\xi ^4 \left( v_0 \gamma_1^2 L^2 -2 \gamma_1 v_1 L -  v_0 \gamma_2 L + v_2 \right)}{2 v_0} + \mathcal{O}\left(\xi^6\right)
\right] . 
\end{split}\label{eq:2ptFormalExpansion}
\end{equation}
The classical dimension $\Delta_0$, the normalization of the two-point function $v(\xi^2)=\sum_{k=0}^\infty \frac{v_k}{k!} \xi^{2k}$ and the anomalous dimension $\gamma(\xi^2)=\sum_{k=1}^\infty \frac{\gamma_k}{k!} \xi^{2k}$ are all related to the operator $\mathcal{O}$ under investigation. 
The latter two are functions of the coupling $\xi^2$ and thus their expansion coefficients $v_k$ and $\gamma_k$ can be determined by perturbative calculations up to a certain loop order. 
Here, the abbreviation $L:= \log{\left[\mu^2(x-y)^2\right]}$ is used and the energy scale $\mu$ adjusts to the mass dimensions in the expansion in $\xi^2$.
The wave-function renormalization factor $Z_\mathcal{O}$ is specific for the operator and has to be determined at the case at hand \cite{Pomoni:2009joh}.

Our operator of interest is $\mathcal{O}=\tr{\phi_1\phi_1}$, which has a classical dimension $\Delta_0=2$ and the goal of this section is to compute its anomalous dimension and two-point normalization, $\gamma = \gamma (\xi^2)$ and $v = v(\xi^2)$, respectively, via perturbation theory. 
In order to determine their expansion coefficients, one has to match the formal expansion \eqref{eq:2ptFormalExpansion} to the expansion in Feynman diagrams at the double-trace couplings' fixed point. 
This will be done in the following up to 3-loop diagrams and with the only appearing double-trace coupling $\alpha_1^2$ untuned.
Just at the end, when both expressions are getting compared, the perturbation expansion will be evaluated at its fixed-point $\alpha_{1,*}^2$.

The expressions for the diagrams can be obtained by using directly the coordinate-space Feynman rules, or by Fourier transforming the respective momentum-space Feynman diagrams via the useful relation (see \eqref{eq:GenPropagatorGWeight})
\begin{equation}
\int \frac{\dd^D p}{(2\pi)^D}\e^{\I\; p\cdot x} \left(p^2\right)^\alpha ~=~ 4^\alpha \pi^{-D/2}\frac{\Gamma (\frac{D}{2}+\alpha)}{\Gamma (-\alpha)} \frac{1}{\left(x^2\right)^{\frac{D}{2}+\alpha}} ~.
\end{equation}
We will follow the latter strategy such that we can use the Feynman diagrams appearing in the amplitude \eqref{eq:Fishnet_4ptForRen} in section \ref{subsec:RegularizedAmplitudes} for $p_2 = p_4 = 0$ and $p_1 = p_3 = p$ (this amputates two external propagators) and we Fourier transform them.
For now, we set $y=0$, knowing that we can reinstate it at a non-zero value by shifting $x$ in the Fourier transformed diagrams.
We find the following expressions for the diagrams in dimensional regularization ($D=4-2\varepsilon$):
\begin{itemize}
\item Tree level: the lowest-order contribution is the Fourier transform of the first diagram in \eqref{eq:A1} and we use the cartoon $\mathrm{O}$ as an abbreviation for the two-point coordinate-space diagram. 
The tree level contribution is 
\begin{equation}
A_{\mathrm{tree}} = \left(\frac{\e^{-\gamma_\mathrm{E}}}{\pi}\right)^{2\varepsilon} \mathrm{O} 
= \frac{1}{(2\pi)^{4-2\varepsilon}}\frac{\Gamma (1-\varepsilon)^2}{\left(x^2\right)^{2-2\varepsilon}} ~.
\label{eq:AnomalDimDiagTree}
\end{equation}
Since it is finite for $\varepsilon \rightarrow 0$ and is contained in higher-order diagrams, it will be factored out in all of following diagrams to simplify the calculation. 
Note that a one-loop diagram in momentum-space is the finite tree level contribution in coordinate-space, when Fourier transforming it.
The tree level diagram \eqref{eq:AnomalDimDiagTree} produces the first factor before the expansion in \eqref{eq:2ptFormalExpansion}. 
Here and in the following, factors like $\left(\e^{-\gamma_\mathrm{E}}/\pi\right)^{\left(l+2\right)\varepsilon}$ ensure the absence of terms proportional to the Euler-Mascheroni constant $\gamma_\mathrm{E}$, which are getting captured by a redefinition of $\mu$ in this way.

\item One-loop level: the only diagram contributing is $\mathrm{O}\makebox[4pt]{O}$ , which is the cartoon for the amputated version of the first diagram in \eqref{eq:A2}:
\begin{equation}
\left(\frac{\e^{-\gamma_\mathrm{E}}}{\pi}\right)^{3\varepsilon} \mathrm{O}\makebox[4pt]{O} 
= 2\cdot (4\pi)^{2}\alpha_1^2\cdot\frac{A_{\mathrm{tree}}}{8 \pi ^2 }\left[\frac{1}{\varepsilon }+L+1+\mathcal{O}\left(\varepsilon\right)\right] \, .
\label{eq:AnomalDimDiagOO}
\end{equation}

\item There are two diagrams contributing at two-loops:
\begin{itemize}
\item Adding another bubble to the one-loop diagram gives the two loop diagram $\mathrm{O}\makebox[4pt]{O}\makebox[8pt]{O}$, related to \eqref{eq:ThreeBubbles} after the Fourier transform, which evaluates to 
\begin{equation}
\left(\frac{\e^{-\gamma_\mathrm{E}}}{\pi}\right)^{4\varepsilon} \mathrm{O}\makebox[4pt]{O}\makebox[8pt]{O} =
4\cdot (4\pi)^4 \alpha_1^4 \cdot\frac{A_{\mathrm{tree}}}{256 \pi ^4}
\left[\frac{3}{\varepsilon ^2}+\frac{6L+6}{\varepsilon}+6 L^2+12 L+3\zeta_2+ \mathcal{O} \left(\varepsilon\right)\right] \, .
\label{eq:AnomalDimDiagOOO}
\end{equation}

\item The first diagram containing a single-trace interaction (two vertices to be precise) is $\raisebox{1pt}{<} \makebox[3pt]{\textcolor{red}{O}} \makebox[7pt]{\raisebox{1pt}{>}}$, related to \eqref{eq:TheEyeDiag}, with the regularized expression
\begin{equation}
\left(\frac{\e^{-\gamma_\mathrm{E}}}{\pi}\right)^{4\varepsilon} \raisebox{1pt}{<} \makebox[3pt]{\textcolor{red}{O}} \makebox[7pt]{\raisebox{1pt}{>}} 
= 
1\cdot (4\pi)^4 \xi^4 \cdot \frac{A_{\mathrm{tree}}}{256 \pi ^4}
\left[\frac{1}{ \varepsilon ^2}+\frac{2 L+3}{\varepsilon } + 2 L^2+ 6 L+\zeta_2 + 3 + \mathcal{O} \left(\varepsilon\right)\right] \, .
\label{eq:AnomalDimDiag<O>}
\end{equation}
\end{itemize}

\item The three-loop level consist of two diagrams as well:
\begin{itemize}
\item The (bubble)$^4$ diagram is illustrated by the cartoon $\mathrm{O}\makebox[4pt]{O}\makebox[8pt]{O}\makebox[4pt]{O} $ and has the $\varepsilon$-expansion
\begin{equation}
\begin{split}
\left(\frac{\e^{-\gamma_\mathrm{E}}}{\pi}\right)^{5\varepsilon} \mathrm{O}\makebox[4pt]{O}\makebox[8pt]{O}\makebox[4pt]{O} 
=
8\cdot(4 \pi )^6\alpha _1^6\cdot \frac{A_{\mathrm{tree}}}{1024 \pi ^6}
\left[\frac{1}{ \varepsilon ^3}+\frac{3 (L+1)}{ \varepsilon ^2}+\frac{\frac{9}{2} L^2 + 9 L + \frac{3}{2}\zeta_2}{ \varepsilon }\right.\\
\left.+\left(\frac{9}{2}L^3+\frac{27}{2}L^2+\frac{3}{4}L + 53 \zeta (3)+\frac{3}{4}-40\right)+ \mathcal{O} \left(\varepsilon \right)\right] \, .
\end{split}\label{eq:AnomalDimDiagOOOO}
\end{equation}

\item The composition of the diagram $\raisebox{1pt}{<} \makebox[3pt]{\textcolor{red}{O}} \makebox[7pt]{\raisebox{1pt}{>}}$ with a bubble gives the second three-loop diagram with an expansion
\begin{equation}
\begin{split}
\left(\frac{\e^{-\gamma_\mathrm{E}}}{\pi}\right)^{5\varepsilon} \raisebox{1pt}{<} \makebox[3pt]{\textcolor{red}{O}} \makebox[7pt]{\raisebox{1pt}{>}}\makebox[4pt]{O}
= 
4\cdot(4 \pi )^6\alpha _1^2 \xi^4\cdot \frac{A_{\mathrm{tree}}}{1024 \pi ^6}
\left[
\frac{1}{3\varepsilon ^3}+\frac{3 L+4}{3\varepsilon ^2}+\frac{\frac{9}{2} L^2+12 L+\frac{3}{2}\zeta_2+4}{3\varepsilon }\right. \\
\left. +\frac{18 L^3+72 L^2+3 \left(16+\pi ^2\right) L+4 \left(67 \zeta (3)+\pi ^2-44\right)}{12}+ \mathcal{O} \left(\varepsilon\right)
\right] \, .
\end{split}\label{eq:AnomalDimDiag<O>O}
\end{equation}
\end{itemize}
\end{itemize}
To obtain the perturbative three-loop expression for the two-point function \eqref{eq:2ptFormalExpansion} (terms of quartic order in the couplings and higher are out of reach at this order), one has to sum up the diagrams (\ref{eq:AnomalDimDiagTree}-\ref{eq:AnomalDimDiag<O>O}). 
This results in a divergent expression due to the $1/\varepsilon^3$, $1/\varepsilon^2$ and $1/\varepsilon$ poles in the individual diagrams. 
Those divergences are subject to two $Z$-factors for renormalization, the coupling renormalization of \eqref{eq:Alpha1ZFactor} and the wave-function renormalization of the operator $\tr{\phi_1\phi_1}$. 
The latter one has to be additionally introduced, the coupling renormalization alone is not capable of capturing all divergences in the correlation function.
We determine it now by first making the ansatz
\begin{equation}
Z_\mathcal{O}= 1  
+\frac{\tau _{13} + \tau _{12} + \tau _{11} }{\varepsilon }
+\frac{\tau _{23} + \tau _{22}}{\varepsilon ^2}
+\frac{\tau _{33} }{\varepsilon ^3} 
\label{eq:OperatorRenZ}
\end{equation}
with coefficients $\tau_{nk}$ of order $k$ in the couplings, which have to be determined by requiring a finite, renormalized two-point function. 
We illustrate this at the perturbative expression at two-loops (the three loop expression is very long),
\begin{equation}
\begin{split}
\corr{\mathcal{O}(x)\bar{\mathcal{O}}(y)}_\mathrm{R} =& ~ A_\mathrm{tree} \cdot \\
\left\lbrace 
1 \right.
+w &\left(\frac{4 \alpha _1^2+\tau _{11}}{\varepsilon }+4 (L+1) \alpha _1^2\right)\\
+w^2 &\left[  
\frac{-\xi ^4+4 \alpha _1^4+4 \alpha _1^2 \tau _{11}+\tau _{22}}{\varepsilon ^2}
 +\frac{\xi ^4+16 (L+1) \alpha _1^4+4 (L+1) \alpha _1^2 \tau _{11}+\tau _{12}}{\varepsilon }\right.\\
&\left. +\left(\left(L^2+4 L+3\right) \xi ^4+\frac{4}{3} \left(15 L^2+30 L+\pi ^2\right) \alpha _1^4+\frac{1}{3} \left(6 L^2+12 L+\pi ^2\right) \alpha _1^2 \tau _{11}\right)\right.\\
&\left.\left.
+ \mathcal{O} \left(\varepsilon\right)
\right]+ \mathcal{O} \left(w^3\right) \right\rbrace \, .
\end{split}\label{eq:AUpTo2Loops}
\end{equation}
Here, the counting parameter for the coupling order $w$ is reinstalled. 
The coupling is renormalized here, i.\,e.\ it enters the calculation multiplied by the first two coupling-orders of $Z$ from \eqref{eq:Alpha1ZFactor}. 
The coefficients of the wave-function renormalization $Z_\mathcal{O}$, the $\tau$'s, can be determined by requiring the poles to vanish in \eqref{eq:AUpTo2Loops}. 
When considering the analog three-loop expression, one can even determine the $\tau_{n3}$'s.
We find
\begin{equation}
\begin{array}{lll}
\tau_{11}= -4 \alpha _1^2 ~, &
\tau _{12}= -\xi ^4 ~, &
\tau _{13}= -\frac{4}{3} \alpha _1^2 \xi ^4 ~,\\
 & \tau _{22}= 12 \alpha _1^4+\xi ^4 ~, &
\tau _{23}= \frac{20}{3} \alpha _1^2 \xi ^4 ~,\\
 & &\tau _{33}= -\frac{16}{3} \alpha _1^2 \left(6 \alpha _1^4+\xi ^4\right) ~.
\end{array}
\end{equation}
This yields the three-loop wave-function renormalization of the operator $Z_\mathcal{O}$ by \eqref{eq:OperatorRenZ}.

As a remark, the necessity of the double-trace couplings becomes evident when requiring finiteness of the renormalized two-point function. 
Without double-trace couplings, one would obtain \eqref{eq:AUpTo2Loops} with $\alpha_1^2\rightarrow 0$. 
This is 
\begin{equation}
\begin{split}
\corr{\mathcal{O}(x)\bar{\mathcal{O}}(y)}_{\mathrm{R},\alpha_1^2\rightarrow 0} =
A_\mathrm{tree}&\left\lbrace 
1+w \left( \frac{\tau _{11}}{\varepsilon }+ \mathcal{O} \left(\varepsilon \right)\right)\right.
+w^2 \left[ \frac{\xi ^4+\tau _{22}}{\varepsilon ^2}+\frac{(2 L+3) \xi ^4+\tau _{12}}{\varepsilon } \right. \\
& ~~~~~~~~~ + \left. \frac{1}{6} \left(12 L^2+36 L+\pi ^2+18\right) \xi ^4+ \mathcal{O} \left(\varepsilon \right)\right]\\
& ~~~~~~~~ + \left. \mathcal{O} \left(w^3\right)\right\rbrace \, ,
\end{split}
\end{equation}
and requiring the poles to vanish would imply $\tau _{12}=-(2 L+3) \xi ^4$. 
But this would mean that the corresponding counter-term is non-local due to the dependence on $L=\log{\left[\mu^2(x-y)^2\right]}$.

Let us get back to the conformal case where we tune the double-trace coupling to the fix-point \eqref{eq:Alpha1FixedPointExpansion3L}. 
Then the perturbative two-loop function \eqref{eq:AUpTo2Loops} simplifies to
\begin{equation}
\corr{\mathcal{O}(x)\bar{\mathcal{O}}(y)}_\mathrm{R,fixed-point} =
A_\mathrm{tree}\left[ 
1+2 i (L+1) \xi ^2 + \left(-2 L^2 -4 L+1\right)\xi ^4+ \mathcal{O} \left(\xi ^6\right)+ \mathcal{O} \left(\varepsilon\right)
\right] \, ,
\label{eq:AUpTo2LoopFixPoint}
\end{equation}
after setting $w\rightarrow 1$ (there is only the $\xi^2$-coupling left, so the counting parameter $w$ is superfluous). Comparing \eqref{eq:AUpTo2LoopFixPoint} and \eqref{eq:2ptFormalExpansion} yields the coefficients $v_0$, $v_1$, $v_3$ and $\gamma_1$ and $\gamma_2$. 
However, the calculation can be extend to three loops (it is not shown for brevity).
It yields even more coefficients and the anomalous dimension and the normalization of the two-point function at three loops are
\begin{align}
\gamma (\xi^2) & = -2\I \xi^2 + \frac{1}{2}\cdot 0 \xi^4 + \frac{1}{6}\cdot 6\I \xi^6 + O(\xi^8) 
= -2\I \xi^2 + \I \xi^6 + \mathcal{O} (\xi^8)\\
\begin{split}
v (\xi^2) & = 1 + 2\I \xi^2 + \frac{1}{2}\cdot 2 \xi^4 + \frac{1}{6}\cdot 2\I\left( -75 + 40 \pi^2 - 308\zeta_3\right)\xi^6+ \mathcal{O} (\xi^8)\\
& = 1 + 2\I\xi^2 + \xi^4 +\I \left( -25 + 80\zeta_2 - \frac{308}{3}\zeta_3 \right)\xi^6 + \mathcal{O} (\xi^8).
\end{split}\label{eq:Fishnet_ResultAnomalousDimPerturbative}
\end{align}
The anomalous dimension coincides with the lowest orders of the expansion of $-\sqrt{2-2\sqrt{1+4\xi^4}}$,  which is the exact result obtained by the eigenvalue of the graph-building operator in \eqref{eq:Fishnet_ScalingDimensions_S0}.
The normalization starts involving values of the zeta function $\zeta_i$ at the three-loop order.

\chapter{Solving inversion relations}
\label{sec:SolvingInversionRelations}

In this appendix we present a procedure for determining a solution of the inversion relations \cite{StroganovInvRelLatticeModels}, which is applied to \eqref{eq:InversionRelationsKappa} and furnishes the solution \eqref{eq:Kappagenfinalresult}. 
Furthermore, we give details on the special functions necessary for the computation of the free energy of the eight-vertex model, which is presented in section \ref{sec:8Vmodel}.

\section{General case}
We call the function we are looking for $S(\alpha)$, since the method of solving inversion relations was pioneered for S-matrices in Sine-Gordon QFTs \cite{Zamolodchikov:1978xm}.
The procedure was applied in \cite{Bazhanov:2016ajm} to many other integrable edge-interaction models.

Let us consider the relations in a fractional form\footnote{The fractional form is equivalent to the multiplicative one in \eqref{eq:InversionRelationsKappa}. 
It can be shown by first multiplying \eqref{eq:InversionGeneralized2} at $\alpha$ with itself at $\alpha \rightarrow -\alpha$. Then one uses the unitarity relation \eqref{eq:InversionGeneralized1} for the denominator, which reduces it to one. 
Finally, we re-parameterize $\beta := \eta - \alpha$ and find \eqref{eq:InversionRelationsKappa} after choosing $f_0 (\beta) = \frac{\pi^\beta}{\Gamma (\beta + \ell)}$ with $\eta = \frac{D}{2}$.\label{footnote:InversionRelationsFractionalVsMultiplicative}} with the function $S(\alpha)$, which we would like to determine, with a crossing parameter $\eta$ and a seed function $f_0(\alpha)$,
\begin{subequations}
\begin{align}
S(\alpha)S(-\alpha) &= 1  &&\mathrm{(Unitarity)}\label{eq:InversionGeneralized1}\\
\frac{S(\eta - \alpha)}{S(\alpha)} &= \frac{f_0(\eta - \alpha)}{f_0(\alpha)} &&\mathrm{(Crossing)} ~.\label{eq:InversionGeneralized2}
\end{align}
\label{eq:InversionGeneralized}%
\end{subequations}
We will construct the solution of \eqref{eq:InversionGeneralized} in an iterative fashion \cite{Shankar:1977cm,Bombardelli:2016scq,Bazhanov:2016ajm} by first making an ansatz for one of the two equations in \eqref{eq:InversionGeneralized} and restricting it via the other, but again only up to some factor. This factor will be partially fixed by plugging the ansatz back in the first equation. Repeating this procedure infinitely many times yields the solution. 
Concretely, it looks like this:
\begin{enumerate}
\item 
Start with an ansatz for the crossing equation \eqref{eq:InversionGeneralized2}. 
There are two obvious choices, $S(\alpha) = f_0 (\alpha)$ or $S(\alpha) = \frac{1}{f_0 (\eta - \alpha)}$. 
we choose the latter, which we call the \textit{crossing-choice}. 
It consists of constructing the ansatz for the crossing equation always in such a way that the sought functions (here $S(\alpha)$, later $f_{2n}(\alpha)$) make up for the remaining part of the crossing equation with a minus sign in front of $\alpha$ in the argument. 
After this choice, one multiplies by a function $f_1$ to make up for the remaining freedom in \eqref{eq:InversionGeneralized2}. 
It has to satisfy a crossing equation on its own and the first step in the iteration looks like
\begin{equation}
S(\alpha) =  \frac{1}{f_0 (\eta - \alpha)} f_1(\alpha) ~~\mathrm{with}~~ \frac{f_1(\eta - \alpha)}{f_1(\alpha)} =1 ~~\mathrm{(Crossing }~ f_1 \mathrm{)} ~.
\label{eq:SolvingInversionStep1}
\end{equation}
Generally a step of the iteration consists of taking the ansatz from the last step, require it to satisfy a crossing/unitarity relation by making a choice which factor to compensate, multiplying by a new function, and observing that it has to satisfy a crossing/unitarity relation. 
This relation will then be used in the second-to-next step.

\item
Now we use the unitarity relation \eqref{eq:InversionGeneralized1} by plugging in $S(\alpha)$ as obtained in \eqref{eq:SolvingInversionStep1}. 
We obtain
\begin{equation}
\frac{f_1(\alpha) f_1(-\alpha)}{f_0(\eta-\alpha) f_0(\eta+\alpha)} =1 ~.
\end{equation}
For $f_1$ to be a solution, we make an ansatz by the so-called \textit{unitarity-choice}. 
It refers to the rule that ansätze for solutions of unitarity equations, $f_{2n-1}$, always make up for that part of the equation, which carries a $+\alpha$ in the argument. 
This yields
\begin{equation}
f_1(\alpha)= f_0(\eta + \alpha) f_2(\alpha) ~~\mathrm{with}~~ f_2(\alpha)f_2(-\alpha)=1 
~~\mathrm{(Unitarity }~ f_2 \mathrm{)} ~.
\label{eq:SolvingInversionStep2}
\end{equation}
Again, one has the freedom to multiply a function $f_2(\alpha)$, which however has to satisfy a unitarity relation.

\item
Next we take the crossing equation for $f_1$ in \eqref{eq:SolvingInversionStep1} and plug in the new expression for $f_1$ from \eqref{eq:SolvingInversionStep2}. 
This yields
\begin{equation}
\frac{f_0(2\eta - \alpha) f_2(\eta - \alpha)}{f_0(\eta + \alpha) f_2(\alpha)} = 1 ~.
\end{equation}
Making the crossing-choice means constructing $f_2$ such that it captures the part with the $-\alpha$ in the argument, i.\,e.\ $f_0(2\eta - \alpha)$. 
Yet, one still has the freedom to multiply $f_0(2\eta - \alpha)$ with another function $f_3$, satisfying a crossing equation on its own.
Thus we set
\begin{equation}
f_2(\alpha) = f_0(2\eta - \alpha)f_3(\alpha) ~~\mathrm{with}~~ \frac{f_3(\eta - \alpha)}{f_3(\alpha)} =1 ~~\mathrm{(Crossing }~ f_3 \mathrm{)} ~.
\label{eq:SolvingInversionStep3}
\end{equation}

\item
This step is again a unitarity-step, similar to the second one. 
Now take the unitarity relation for $f_2$ in \eqref{eq:SolvingInversionStep2} and plug in the ansatz for $f_2$ from \eqref{eq:SolvingInversionStep3}. 
One finds
\begin{equation}
f_0(2\eta - \alpha)f_0(2\eta + \alpha)f_3(\alpha)f_3(-\alpha)=1 ~.
\end{equation}
Making the unitarity-choice refers to compensating $f_0(2\eta + \alpha)$ by $f_3(\alpha)$. 
Still, there is the freedom for another factor $f_4$, which again satisfies a unitarity relation, and we obtain
\begin{equation}
f_3(\alpha)=\frac{1}{f_0(2\eta + \alpha)}f_4(\alpha) ~~\mathrm{with}~~ f_4(\alpha)f_4(-\alpha)=1 
~~\mathrm{(Unitarity }~ f_4 \mathrm{)} ~.
\label{eq:SolvingInversionStep4}
\end{equation} 

\item
The way the steps continue hopefully became apparent by now, but we show yet another step, namely crossing step.
Take the crossing equation for $f_3$ in \eqref{eq:SolvingInversionStep3} and plug it in the ansatz from the previous step, which gives
\begin{equation}
\frac{f_0(2\eta + \alpha)f_4(\eta - \alpha)}{f_0(3\eta - \alpha)f_4(\alpha)} = 1 ~.
\end{equation}
The crossing-choice dictates that $f_4(\alpha)$ compensates $f_0(3\eta - \alpha)$ and another factor $f_5$ can be multiplied if it satisfies a crossing equation itself, therefore it yields
\begin{equation}
f_4(\alpha) = \frac{1}{f_0 (3\eta - \alpha)}f_5 (\alpha) ~~\mathrm{with}~~ \frac{f_5(\eta - \alpha)}{f_5(\alpha)} =1 ~~\mathrm{(Crossing }~ f_5 \mathrm{)} ~.
\end{equation}
\end{enumerate}
After doing some more steps, one quickly realizes that the pattern continues and the $n$-th step is very similar to the $(n-4)$-th step, only with two more $\eta$'s in the argument of $f_0$. 
Explicitly, those five steps show that the solution, up to now, is 
\begin{equation}
S(\alpha) = \frac{1}{f_0(\eta - \alpha)} f_0(\eta + \alpha) f_0(2\eta - \alpha) \frac{1}{f_0(2\eta + \alpha)} \frac{1}{f_0(3\eta - \alpha)} \cdot f_5(\alpha) ~.
\end{equation}
After infinitely many steps, we can observe that the procedure yields an infinite product, namely
\begin{equation}
S(\alpha)  = \prod_{k=1}^\infty  \frac{f_0\left((2k-1)\cdot\eta + \alpha\right) f_0(2k\cdot\eta - \alpha)}{f_0\left((2k-1)\cdot\eta - \alpha\right) f_0(2k\cdot\eta + \alpha)} ~.
\label{eq:AAlphaVersion1}
\end{equation}
In a further step, the factors may be regrouped to give another representation,
\begin{equation}
S(\alpha)  =\frac{1}{f_0(\eta - \alpha)} \prod_{k=1}^\infty  \frac{f_0\left(2k\cdot\eta -\eta + \alpha\right) f_0(2k\cdot\eta - \alpha)}{f_0\left(2k\cdot\eta +\eta - \alpha\right) f_0(2k\cdot\eta + \alpha)} ~.
\end{equation}
Finally, using the telescope product $\prod_{k=1}^\infty\frac{f_0((2k+1)\eta)}{f_0((2k-1)\eta)} = \frac{1}{f_0(\eta)}$, we obtain
\begin{equation}
S(\alpha)  =\frac{f_0(\eta)}{f_0(\eta - \alpha)} \prod_{k=1}^\infty  \frac{f_0\left(2\eta k -\eta + \alpha\right) f_0(2\eta k - \alpha)f_0(2\eta k + \eta)}{f_0\left(2\eta k +\eta - \alpha\right) f_0(2\eta k + \alpha)f_0(2\eta k - \eta)} ~.
\label{eq:BazhanovGeneralizedf0}
\end{equation}
We comment on the role of the choices (crossing- and unitarity-choice). 
In this derivation, we picked the $+\alpha$ arguments in the unitarity relations and $-\alpha$ arguments in the crossing ones. 
Exchanging this assignments would have given a similar expression \eqref{eq:AAlphaVersion1} but with the arguments of $f_0$ multiplied with a minus sign.
The latter choices would yield poles in the physical strip $\left[ 0 , \eta \right)$, which corresponds to a solution not referring to the maximal eigenvalue.  
Picking for both, unitarity and crossing, the $+\alpha$ arguments or both times the $-\alpha$ would not expand the product and only yield the relations \eqref{eq:InversionGeneralized}, and hence just a trivial statement. 
The result \eqref{eq:BazhanovGeneralizedf0} is in accordance with \cite{Bazhanov:2016ajm}.

In conclusion, we will apply the general solution \eqref{eq:BazhanovGeneralizedf0} to our set of inversion relations \eqref{eq:InversionRelationsKappa}. 
By the transition to the fractional form \eqref{eq:InversionGeneralized}, as described in footnote \ref{footnote:InversionRelationsFractionalVsMultiplicative}, we identify the seed function as
\begin{equation}
f_0(u) 
~=~
\frac{\pi^u}{\Gamma (u + \ell)}
\end{equation}
by comparing \eqref{eq:InversionRelationsKappa} with the general relations \eqref{eq:InversionGeneralized}. Then by \eqref{eq:BazhanovGeneralizedf0} we find the solution
\begin{equation}
\kappa_\ell (u) = 
\pi^{u} \frac{\Gamma \left( \frac{D}{2} - u + \ell \right) }{\Gamma \left( \frac{D}{2} + \ell \right)}
\prod_{k=1}^{\infty} \frac{\Gamma \left( Dk + \frac{D}{2} - u + \ell \right) \Gamma \left( Dk + u + \ell \right)\Gamma \left( Dk - \frac{D}{2} + \ell \right) }{\Gamma \left( Dk - \frac{D}{2} + u + \ell \right) \Gamma \left( Dk - u + \ell \right)\Gamma \left( Dk + \frac{D}{2} + \ell \right)}
\label{eq:kappaiFreeEnergy}
\end{equation}
where we used the QFT crossing parameter $\eta =\frac{D}{2}$.
This is the solution presented in \eqref{eq:Kappagenfinalresult}.

\section{Theta functions, elliptic gamma functions and the eight-vertex model}
\label{sec:ThetaFctEllipticGammaFctAndFreeEnergy8V}
We use the notation of \cite{whittaker1996course} for the theta functions.
They read
\begin{subequations}
\begin{align}
\vartheta_1 (z \vert q)
&=
2 \sum_{n=0}^\infty
(-1)^n
q^{( n + \frac{1}{2} )^2}
\mathrm{sin}
[ (2n + 1) z ]
=
2\, G\, q^{\frac{1}{4}}\, \mathrm{sin} (z)
\prod_{n=1}^\infty
\left(
1 - q^{2n} \e^{2\I z}
\right)
\left(
1 - q^{2n} \e^{- 2\I z}
\right) ~, \\
\vartheta_2 (z \vert q)
&=
2 \sum_{n=0}^\infty
q^{( n + \frac{1}{2} )^2}
\mathrm{cos}
[ (2n + 1) z ]
=
2\, G\, q^{\frac{1}{4}}\, \mathrm{cos} (z)
\prod_{n=1}^\infty
\left(
1 + q^{2n} \e^{2\I z}
\right)
\left(
1 + q^{2n} \e^{- 2\I z}
\right) ~, \\
\vartheta_3 (z \vert q)
&=
1
+
2 \sum_{n=1}^\infty
q^{n^2}
\mathrm{cos}
(2nz)
=
G
\prod_{n=1}^\infty
\left(
1 + q^{2n - 1} \e^{2\I z}
\right)
\left(
1 + q^{2n - 1} \e^{- 2\I z}
\right) ~, \\
\vartheta_4 (z \vert q)
&=
1
+
2 \sum_{n=1}^\infty
(-1)^n
q^{n^2}
\mathrm{cos}
(2nz)
=
G
\prod_{n=1}^\infty
\left(
1 - q^{2n - 1} \e^{2\I z}
\right)
\left(
1 - q^{2n - 1} \e^{- 2\I z}
\right)
\end{align}
\label{eq:Def_thetafunctions}%
\end{subequations}
with $G : = \prod_{n=1}^\infty (1 - q^{2n})$ and the elliptic nome $q$.
They have a representation as an infinite product.
When one faces multiple infinite products, the elliptic gamma functions are elegant functions to describe their properties.
Using the convention by Rains \cite{rains2007limits}, we have the elliptic gamma function of $r$-th order
\begin{equation}
\Gamma^{(r)}
(z \vert q_1, ... , q_r)
=
\prod_{n_1, ... , n_r = 0}^\infty
\frac{
1 
-
q_1^{n_1 + 1} \cdots q_r^{n_r + 1} z^{-1} 
}{
\left(
1
-
q_1^{n_1} \cdots q_r^{n_r} z
\right)^{(-1)^r}
}
=
\mathrm{exp}
\left[
\sum_{k=1}^\infty
\frac{
(-1)^r z^k -  q_1^{k} \cdots q_r^{k} z^{-k}
}{
k
\cdot
\prod_{i=1}^r
(1 - p_i^k)
}
\right]
\label{eq:Def_ellipticGamma}
\end{equation}
and the equality between the first and second representation can be shown by taking the logarithm of the former, and expanding the logarithm of the nominator and denominator.
Then the $r$ sums can be performed with the help of the geometric series.
We are particularly interested in the first- and second-order elliptic gamma functions and they satisfy the helpful relations
\begin{subequations}
\begin{align}
&\Gamma^{(1)} (q \cdot z \vert q)
=
\Gamma^{(1)} ( z^{-1} \vert q) ~, &&
\Gamma^{(1)} (z \vert q)
=
\Gamma^{(1)} (z \vert q^2)
\Gamma^{(1)} (q \cdot z \vert q^2) ~, \\
&\Gamma^{(2)} (p\, q \cdot z \vert q , p)
=
\frac{1}{
\Gamma^{(2)} ( z^{-1} \vert q , p)
} ~, &&
\Gamma^{(2)} (p \cdot z \vert q , p)
=
\Gamma^{(1)} (z \vert q )
\Gamma^{(2)} (z \vert q , p) ~.
\end{align}
\label{eq:EllipitcGamma_handyrelations}%
\end{subequations}
Next, we derive the eight-vertex model's free energy from the generic solution of inversion relations \eqref{eq:BazhanovGeneralizedf0}.
The inversion relations for the eight-vertex model \eqref{eq:InvRel_8V} may be brought into the fractional form \eqref{eq:InversionGeneralized}, if we redefine $\kappa ( u ) = \mathrm{w} (u) \tilde{\kappa} ( u )$.
We find the seed function to be $f_0 ( u ) = \mathrm{w} (u)^{-1}$ and, with the help of the general solution \eqref{eq:BazhanovGeneralizedf0}, the free energy of the eight-vertex model is found as
\begin{equation}
\kappa (u)
=
\mathrm{w} (u)
\frac{\mathrm{w}(\eta - u)}{\mathrm{w}(\eta )} 
\prod_{k=1}^\infty  
\frac{
\mathrm{w} (2\eta k + \eta - u ) \mathrm{w} (2\eta k + u ) \mathrm{w} ( 2\eta k - \eta )
}{
\mathrm{w} ( 2\eta k - \eta + u ) \mathrm{w} ( 2\eta k - u ) \mathrm{w} (2\eta k + \eta)
} ~.
\label{eq:8V_freeEnergy_w}
\end{equation}
The function $\mathrm{w} (u)$ is given in \eqref{eq:8V_unitarity_seed} as $\mathrm{w} (u) = - \I \rho \cdot\Theta (0) \Theta (\I (\eta + u) ) H (\I (\eta + u) )$.
It is a lengthy calculation to plug this into \eqref{eq:8V_freeEnergy_w} and express it in terms of $\Gamma^{(1)}$ and $\Gamma^{(2)}$.
It is convenient to consider $\mathrm{log} \, \kappa (u)$ and to use the product representation of the functions $H (u)$ and $\Theta (u)$ by the theta functions \eqref{eq:Def_thetafunctions}.
One of the two infinite products from the definition of the theta functions and the solution \eqref{eq:8V_freeEnergy_w}, which turn into sums after taking the logarithm, can be eliminated by the geometric series.
The remaining one is used to identify the elliptic gamma functions in the second representation of \eqref{eq:Def_ellipticGamma}.
Eventually, using \eqref{eq:EllipitcGamma_handyrelations} and choosing $\rho = \Gamma^{(1)} (q \vert q^2 )^2$ gives the form presented in \eqref{eq:8V_freeEnergy_kappa}.

\chapter{Superspace notations}
\label{app:SuperspaceNotations}

\section[Four-dimensional $\mathcal{N} = 1$ superspace]{Four-dimensional $\mathbf{\mathcal{N} = 1}$ superspace}
\label{appsec:TheFourDimensionalN1Superspace}

\subsection{Spinor algebra}
The notation follows \cite{Wess:1992cp} and for the sake of completeness we list important relations here as well.
The Pauli matrices are defined as
\begin{equation}
\begin{aligned}[c]
\sigma^0 &= \begin{pmatrix} -1 & 0 \\ 0 & -1\end{pmatrix} ~,\\
\sigma^1 &= \begin{pmatrix} 0 & 1 \\ 1 & 0 \end{pmatrix} ~,
\end{aligned}
\qquad\qquad
\begin{aligned}[c]
\sigma^2 &= \begin{pmatrix} 0 & - \I \\  \I & 0\end{pmatrix} ~,\\
\sigma^3 &= \begin{pmatrix} 1 & 0 \\ 0 & -1 \end{pmatrix} ~.
\end{aligned}
\end{equation}
Small Greek indices of anti-commuting spinors are raised and lowered depending on their chirality,
\begin{equation}
\begin{aligned}[c]
\psi^\alpha = \varepsilon^{\alpha \beta} \psi_{\beta} ~,\\
\psi_\alpha = \varepsilon_{\alpha \beta} \psi^{\beta} ~,
\end{aligned}
\qquad\qquad
\begin{aligned}[c]
\bar{\psi}^{\dot{\alpha}} = \varepsilon^{\dot{\alpha} \dot{\beta}} \bar{\psi}_{\dot{\beta}} ~,\\
\bar{\psi}_{\dot{\alpha}} = \varepsilon_{\dot{\alpha} \dot{\beta}} \bar{\psi}^{\dot{\beta}} ~,
\end{aligned}
\end{equation}
with the epsilon tensor squaring to identity as $\varepsilon^{\alpha \beta} \varepsilon_{\beta \gamma} = \delta^\alpha_\gamma$. 
Its non-zero components are $\varepsilon_{21} = \varepsilon^{12} = 1$ and $\varepsilon_{12} = \varepsilon^{21} = -1$.

Spinor bilinears are 
\begin{equation}
\begin{aligned}[c]
\psi \chi &= \left( \psi \chi \right) = \psi^\alpha \chi_\alpha = - \psi_\alpha \chi^\alpha ~,\\
\bar{\psi} \bar{\chi} &= \left( \bar{\psi} \bar{\chi} \right) = \bar{\psi}_{\dot{\alpha}} \bar{\chi}^{\dot{\alpha}} = - \bar{\psi}^{\dot{\alpha}} \bar{\chi}_{\dot{\alpha}} = \left( \psi \chi \right)^\dagger ~,
\end{aligned}
\qquad\qquad
\begin{aligned}[c]
\psi \sigma^\mu \bar{\chi} = \psi^\alpha \sigma^\mu_{\alpha \dot{\alpha}} \bar{\chi}^{\dot{\alpha}} ~,\\
\left( \psi \sigma^\mu \bar{\chi} \right)^\dagger = \chi \sigma^\mu \bar{\psi} ~,
\end{aligned}
\label{eq:4DN1_SpinorBilinears}
\end{equation}
and we use the bracket notation $\left( \psi \chi \right) = \left( \chi \psi \right)$ if the index contraction is unclear.
Denoting $\bar{\sigma}^{\mu,\dot{\alpha}\alpha} = \varepsilon^{\alpha \beta} \sigma^{\mu}_{\beta\dot{\beta}} \varepsilon^{\dot{\alpha}\dot{\beta}}$, one finds the relations
\begin{equation}
\begin{aligned}[c]
\sigma^{\mu}_{\alpha\dot{\alpha}} \bar{\sigma}_\mu^{\dot{\beta}\beta} = - 2\, \delta_\alpha^\beta \delta_{\dot{\alpha}}^{\dot{\beta}} ~,\\
\mathrm{tr} \left[ \sigma^\mu \bar{\sigma}^\nu \right] = - 2\, \eta^{\mu\nu} ~,
\end{aligned}
\qquad\qquad
\begin{aligned}[c]
\left[ \sigma^\mu \bar{\sigma}^\nu + \sigma^\nu \bar{\sigma}^\mu \right]_\alpha^{~\beta} = - 2\, \eta^{\mu\nu} \delta_\alpha^\beta ~, \\
\left[ \bar{\sigma}^\mu \sigma^\nu + \bar{\sigma}^\nu \sigma^\mu \right]^{\dot{\alpha}}_{~\dot{\beta}} = - 2\, \eta^{\mu\nu} \delta^{\dot{\alpha}}_{\dot{\beta}} ~ ,
\end{aligned}
\label{eq:SigmaMatricesRelations}
\end{equation}
with the Lorentzian metric $\eta^{\mu \nu} = \mathrm{diag} ( -1, 1, 1, 1 )$.
 
For coinciding spinors and in particular for spinorial Gra\ss mann numbers we further denote $\theta^2 := \left( \theta \theta \right)$ and $\bar{\theta}^2 := \left( \bar{\theta} \bar{\theta} \right)$ and we have the helpful relations
\begin{equation}
\begin{aligned}[c]
\theta^\alpha \theta^\beta &= - \frac{1}{2} \varepsilon^{\alpha \beta} \theta^2 ~,\\
\bar{\theta}^{\dot{\alpha}} \bar{\theta}^{\dot{\beta}} &= \phantom{-} \frac{1}{2} \varepsilon^{\dot{\alpha} \dot{\beta}} \bar{\theta}^2 ~.
\end{aligned}
\qquad\qquad
\begin{aligned}[c]
\theta \sigma^\mu \bar{\theta}\; \theta \sigma^\nu \bar{\theta} = - \frac{1}{2} \theta^2 \bar{\theta}^2 \eta^{\mu\nu} ~,
\end{aligned}
\label{eq:ThetaSquare_ThetaCube}
\end{equation}
The covariant super derivatives and supersymmetry generators are given by
\begin{subequations}
\begin{align}
D_\alpha 
=
\partial_\alpha + \I \sigma^\mu_{\alpha \dot{\alpha}} \bar{\theta}^{\dot{\alpha}} \partial_\mu ~,
\hspace{1cm}
\bar{D}_{\dot{\alpha}} 
=
- \bar{\partial}_{\dot{\alpha}} - \I \theta^\alpha \sigma^\mu_{\alpha \dot{\alpha}} \partial_\mu  ~, \label{eq:SuperCovariantDerivatives}%
\\
Q_\alpha 
=
\partial_\alpha - \I \sigma^\mu_{\alpha \dot{\alpha}} \bar{\theta}^{\dot{\alpha}} \partial_\mu ~,
\hspace{1cm}
\bar{Q}_{\dot{\alpha}} 
=
- \bar{\partial}_{\dot{\alpha}} + \I \theta^\alpha \sigma^\mu_{\alpha \dot{\alpha}} \partial_\mu  ~,
\label{eq:SuperCharges_4DN1}%
\end{align}
\end{subequations}
with $\partial_\alpha = \frac{\partial}{\partial \theta^\alpha}$, $\bar{\partial}_{\dot{\alpha}} = \frac{\partial}{\partial \bar{\theta}^{\dot{\alpha}}}$ and $\partial_\mu = \frac{\partial}{\partial x^\mu}$.

\subsection{Berezin integral}
\label{sec:BerezinIntegral_D4N1}
Integration over the fermionic part of superspace picks out the quadratic component in the Gra\ss mann spinors, such that
\begin{equation}
\int \dd^2 \theta \; \theta^2  =  1
~~ \mathrm{and} ~~
\int \dd^2 \bar{\theta} \; \bar{\theta}^2  =  1 ~.
\end{equation}
Furthermore, squares of the Gra\ss mann spinors, appearing in a Berezinian alongside other functions of the integration Gra\ss mann variable, act as delta distributions on fermionic superspace, which we denote by $\delta^{\left(2\right)} \left( \theta_{12} \right) = \theta_{12}^2$ and $\delta^{\left(2\right)} \left( \bar{\theta}_{12} \right) = \bar{\theta}_{12}^2$ with $\theta_{12}= \theta_1 - \theta_2$ and $\bar{\theta}_{12}= \bar{\theta}_1 - \bar{\theta}_2$.
Concerning the mass dimension, Gra\ss mann bilinears have the same values assigned as bosonic coordinates, i.\ e.\ $\left[ \theta^\alpha \right] = \left[ \bar{\theta}^{\dot{\alpha}} \right] = \frac{\left[ x^\mu \right]}{2} = -\frac{1}{2}$.
This implies for the measure of the Berezin integral the mass dimensions $\left[ \dd^2 \theta \right] = \left[ \dd^2 \bar{\theta} \right] = 1$.

Note that integration and derivation for Gra\ss mann numbers is equivalent, or in formulas
\begin{subequations}
\begin{align}
\int \dd^2 \theta \; f(\theta)  = \left[ - \frac{1}{4} D^2 f(\theta) \right]_{\theta = 0} ~, &
\hspace{1cm}
\int \dd^2 \bar{\theta} \; f(\bar{\theta})  = \left[ - \frac{1}{4} \bar{D}^2 f(\bar{\theta}) \right]_{\bar{\theta} = 0}  ~,\\
\int \dd^2 \theta\, \dd^2 \bar{\theta} \; f(\theta,\bar{\theta}) 
& = \left[ \frac{1}{16} D^2 \bar{D}^2 f(\theta,\bar{\theta}) \right]_{\theta, \bar{\theta} = 0 } ~.
\end{align}
\end{subequations}

\section[Three-dimensional $\mathcal{N} = 2$ superspace]{Three-dimensional $\mathbf{\mathcal{N} = 2}$ superspace}
\label{appsec:TheThreeDimensionalN2Superspace}
We follow the notation of \cite{Benna:2008zy} for the spinor conventions.
The antisymmetric $\varepsilon_{\alpha \beta}$ has the components $\varepsilon^{12} = - \varepsilon_{12} = 1$, such that $\varepsilon^{\alpha \beta} \varepsilon_{\beta \gamma} = \delta^\alpha_\gamma$.
The three-dimensional gamma matrices are $\gamma^{\mu\; \beta}_{~\alpha} = ( \I \sigma^2 , \sigma^1 , \sigma^3 )$ and they satisfy the Clifford algebra $\gamma^{\mu\; \beta}_{~\alpha} \gamma^{\nu\; \gamma}_{~\beta} = \eta^{\mu\nu} \delta^\gamma_\alpha + \varepsilon^{\mu\nu\rho} \gamma^{~~ \gamma}_{\rho\,\alpha}$ with the Lorentzian spacetime metric $\eta^{\mu\nu} = \mathrm{diag}(-1, 1, 1)$.
Lowering an index gives $\gamma_{\alpha\beta}^\mu = \varepsilon_{\beta \delta} \gamma^{\mu\; \delta}_{~\alpha} = ( - \mathbb{1} , - \sigma^3 , \sigma^1 )$, which is symmetric in the spinor indices $\gamma_{\alpha\beta}^\mu = \gamma_{\beta\alpha}^\mu$.
Furthermore, the trace $\gamma^{\mu\; \alpha}_{~\alpha} = 0$ vanishes.

Spinor indices of anti-commuting fermions and Gra\ss mann numbers are raised and lowered with the antisymmetric epsilon tensor according to 
\begin{equation}
\begin{aligned}[c]
\psi^\alpha = \varepsilon^{\alpha \beta} \psi_{\beta} ~,\\
\psi_\alpha = \varepsilon_{\alpha \beta} \psi^{\beta} ~,
\end{aligned}
\qquad\qquad
\begin{aligned}[c]
\bar{\psi}^{\alpha} = \varepsilon^{\alpha \beta} \bar{\psi}_{\beta} ~,\\
\bar{\psi}_{\alpha} = \varepsilon_{\alpha \beta} \bar{\psi}^{\beta} ~,
\end{aligned}
\end{equation}
and the spinor bilinears are 
\begin{equation}
\begin{aligned}[c]
\psi \chi & = \psi^\alpha \chi_\alpha = - \psi_\alpha \chi^\alpha ~,\\
\bar{\psi} \bar{\chi} & = \bar{\psi}^{\alpha} \bar{\chi}_{\alpha} = - \bar{\psi}_{\alpha} \bar{\chi}^{\alpha} ~,
\end{aligned}
\qquad\qquad
\begin{aligned}[c]
\psi \gamma^\mu \bar{\chi} = \psi^\alpha \gamma^\mu_{\alpha \beta} \bar{\chi}^{\beta} ~.
\end{aligned}
\end{equation}
We stress the difference of the conventions here to the four-dimensional $\mathcal{N} = 1$ case \eqref{eq:4DN1_SpinorBilinears}.
Consequently, the squares of spinorial Gra\ss mann numbers, for example, are denoted by $\theta^2 := \theta^\alpha \theta_\alpha$ and $\bar{\theta}^2 := \bar{\theta}^\alpha \bar{\theta}_\alpha$ and we have the frequently used relations
\begin{equation}
\begin{aligned}[c]
\theta^\alpha \theta^\beta &= - \frac{1}{2} \varepsilon^{\alpha \beta} \theta^2 ~,\\
\bar{\theta}^{\alpha} \bar{\theta}^{\beta} &= - \frac{1}{2} \varepsilon^{\alpha \beta} \bar{\theta}^2 ~.
\end{aligned}
\qquad\qquad
\begin{aligned}[c]
\theta \gamma^\mu \bar{\theta}\; \theta \gamma^\nu \bar{\theta} = \frac{1}{2} \theta^2 \bar{\theta}^2 \eta^{\mu\nu} ~,
\end{aligned}
\label{eq:ThetaSquare_ThetaCube_D3N2}
\end{equation}
which differ again in some signs to the $D=4$ $\mathcal{N}=1$ conventions \eqref{eq:ThetaSquare_ThetaCube}.
Other relations, like the more general Fierz identities, can be found in the appendix of \cite{Benna:2008zy}.
The covariant super derivatives and supersymmetry generators read
\begin{subequations}
\begin{align}
D_\alpha 
=
\partial_\alpha + \I\, \gamma^\mu_{\alpha \beta} \bar{\theta}^{\beta} \partial_\mu ~,
\hspace{1cm}
\bar{D}_{\alpha} 
=
- \bar{\partial}_{\alpha} - \I\, \theta^\beta \gamma^\mu_{\beta \alpha} \partial_\mu  ~, \label{eq:SuperCovariantDerivatives}%
\\
Q_\alpha 
=
\partial_\alpha - \I\, \gamma^\mu_{\alpha \beta} \bar{\theta}^{\beta} \partial_\mu ~,
\hspace{1cm}
\bar{Q}_{\alpha} 
=
- \bar{\partial}_{\alpha} + \I\, \theta^\beta \gamma^\mu_{\beta \alpha} \partial_\mu  ~.
\end{align}
\end{subequations}
Since the three-dimensional gamma matrices are real, the supercharges $Q_\alpha$ and $\bar{Q}_{\alpha}$ are counted individually and the group element corresponding to the supercharges generate the $\mathcal{N} = 2$ fermionic subspace.
It is described by the coordinates $\theta^\alpha$ and $\bar{\theta}^\alpha$.
In contrast, the $\mathcal{N}=1$ four-dimensional supercharges \eqref{eq:SuperCharges_4DN1} combine to a single complex supercharge and the coordinates are $\theta^\alpha$ and $\bar{\theta}^{\dot{\alpha}}$ as well.
This explains why we only have to adjust the bosonic coordinate within $z= (x, \theta , \bar{\theta})$ and drop the dot in the spinor indices, when performing calculations in three-dimensional $\mathcal{N} = 2$ superspace instead of four-dimensional $\mathcal{N}=1$ superspace \cite{Park:1997bq,Park:1999cw}.

The three-dimensional $\mathcal{N} = 2$ chiral superfields have the expansions
\begin{subequations}
\begin{align}
\Phi (z)
&=
\phi (x_+) + \sqrt{2}\; \theta \psi (x_+) + \theta^2 F(x_+) ~, \\
\Phi^\dagger (z)
&=
\phi^\dagger (x_-) - \sqrt{2}\; \bar{\theta} \bar{\psi} (x_-) - \bar{\theta}^2 F^\dagger(x_-) ~,
\end{align}\label{eq:Superfield_Components_3DN2}%
\end{subequations}
with $x^\mu_\pm = x^\mu \pm \I \theta \gamma^\mu \bar{\theta}$.
Berezin integration of the Gra\ss mann spinors works analogously to the four-dimensional $\mathcal{N} = 1$ case, which is presented in section \ref{sec:BerezinIntegral_D4N1}.

%----------------------------------------------------------------------------------------
%	BIBLIOGRAPHY
%----------------------------------------------------------------------------------------

%\bibliographystyle{unsrt}
%\bibliographystyle{./JHEP}
\bibliography{bib/Bibliography}

%----------------------------------------------------------------------------------------

\end{document}